\documentclass[10pt,aps,pra,onecolumn,amssymb,amsfonts,floatfix,superscriptaddress,longbibliography,notitlepage,nofootinbib]{revtex4-1}
\usepackage[utf8]{inputenc}
\pdfoutput=1
\usepackage{color}
\usepackage{float}
\usepackage{xcolor}
\usepackage{mathtools,amsmath,amsthm,amsfonts,amssymb}
\usepackage{commath}
\usepackage{physics}
\usepackage{bm}
\usepackage[pdftex]{graphicx}
\usepackage[colorlinks=false]{hyperref}
\usepackage{enumitem}

\begin{document}

\title{Classical and Quantum Properties of the Spin-Boson Dicke Model: Chaos, Localization, and Scarring}

\author{David Villase{\~n}or}
\email{d.v.pcf.cu@gmail.com}
\affiliation{Instituto de Ciencias Nucleares, Universidad Nacional Aut\'onoma de M\'exico, Apdo. Postal 70-543, C.P. 04510 Mexico City, Mexico}
\affiliation{Instituto de Investigaciones en Matem\'aticas Aplicadas y en Sistemas, Universidad Nacional Aut\'onoma de M\'exico, C.P. 04510 Mexico City, Mexico}
\affiliation{CAMTP - Center for Applied Mathematics and Theoretical Physics, University of Maribor, Mladinska 3, SI-2000 Maribor, Slovenia, European Union}
\author{Sa{\'u}l Pilatowsky-Cameo}
\affiliation{Center for Theoretical Physics, Massachusetts Institute of Technology, Cambridge, Massachusetts 02139, USA}
\author{Jorge Ch{\'a}vez-Carlos}
\affiliation{Department of Physics, University of Connecticut, Storrs, Connecticut 06269, USA}
\affiliation{Departamento de F\'isica, Cinvestav, Apdo. Postal 14-740, C.P. 07000 Mexico City, Mexico}
\author{Miguel A. Bastarrachea-Magnani}
\affiliation{Departamento de F\'isica, Universidad Aut\'onoma Metropolitana-Iztapalapa, Av. Ferrocarril San Rafael Atlixco 186, C.P. 09310 Mexico City, Mexico}
\author{Sergio Lerma-Hern{\'a}ndez}
\affiliation{Facultad de F\'isica, Universidad Veracruzana, Campus Arco Sur, Paseo 112, C.P. 91097 Xalapa, Mexico}
\author{Lea F. Santos}
\email{lea.santos@uconn.edu}
\affiliation{Department of Physics, University of Connecticut, Storrs, Connecticut 06269, USA}
\author{Jorge G. Hirsch}
\email{hirsch@nucleares.unam.mx}
\affiliation{Instituto de Ciencias Nucleares, Universidad Nacional Aut\'onoma de M\'exico, Apdo. Postal 70-543, C.P. 04510 Mexico City, Mexico}

\begin{abstract}
This review article describes major developments associated with the Dicke model, from its introduction in the 1950s to explain the transition from a normal to a superradiant phase to its modern applications in quantum many-body physics. Over the decades, this interacting spin-boson model has played a central role in the study of collective light-matter interactions, chaos, and quantum phase transitions. We focus on properties and phenomena that are best understood when seen from both the classical and quantum perspectives, with particular emphasis on the emergence of chaos, localization, and scarring. While our primary emphasis is on the isolated model, we also discuss recent advances in the open Dicke model, where environmental couplings are needed for describing realistic experimental platforms and exploring new regimes of quantum dynamics.
\end{abstract}

\maketitle

\tableofcontents

\section{Introduction}
\label{sec:Introduction}
Light-matter interaction is central to the development of quantum technologies, since various quantum devices rely on optical means for the preparation, detection, and control of matter excitations. Several approaches have been proposed to elucidate the microscopic description of light coupled to many-body quantum systems within the framework of cavity quantum electrodynamics (QED)~\cite{Scully2008Book}. The path to a simplified description of light-matter interactions at the quantum level was put forth in the 1930s by I. I. Rabi, who proposed a fundamental model comprising a single atom under the two-level approximation interacting with a single-mode electromagnetic field~\cite{Rabi1936,Rabi1937}. In the 1950s, R. H. Dicke proposed an extension of the Rabi model to explain collective effects in light-matter interactions, leading to the concept of superradiance, i.e., the enhancement of photon emission due to cooperative effects~\cite{Gross1982}. The Dicke model consists of $\mathcal{N}$ two-level particles collectively interacting with a single-mode electromagnetic field~\cite{Dicke1954}. In the 1960s,  E. T. Jaynes and F. W. Cummings considered a full quantum version of the Rabi model and introduced a simplification via the rotating-wave approximation~\cite{Jaynes1963}. The collective  equivalent to the Jaynes-Cummings model is an integrable version of the Dicke model known as the Tavis-Cummings model~\cite{Tavis1968}. These four models, Rabi, Dicke, Jaynes-Cummings, and Tavis-Cummings models, have become archetypal models in quantum optics.

This review article concerns physical phenomena associated with the isolated Dicke model, including phase transitions~\cite{Wang1973,Hepp1973AP,Hioe1973,Hepp1973PRA,Emary2003PRE,Emary2003PRL,Bastarrachea2014PRAa}, chaos~\cite{DeAguiar1992,Chavez2016,Bastarrachea2014PRAb,Bastarrachea2014JP,Bastarrachea2015,Bastarrachea2016PRE,Bastarrachea2017PS}, scarring~\cite{DeAguiar1991,Furuya1992,Bakemeier2013,Villasenor2020,Pilatowsky2021NC,Pilatowsky2021NJP}, localization~\cite{Wang2020,Villasenor2021,Pilatowsky2022Q}, nonequilibrium dynamics~\cite{Bakemeier2013,Lerma2018AIPCP,Lerma2018JPA,Chavez2019,Lerma2019,Pilatowsky2020}, entanglement~\cite{Furuya1998,Song2008,Song2012,Kloc2017,Lerose2020}, thermalization~\cite{Villasenor2023}, multifractality~\cite{Bastarrachea2024}, and more~\cite{Pilatowsky2022PRE,Wang2024}. We review and discuss the connections among these properties, often resorting to the quantum-classical correspondence for deeper insights. 

Even though we concentrate on the isolated Dicke model, we also mention and provide references for driven and dissipative extensions of the model, as well as the various experimental realizations, covering the earliest implementations in the context of superradiance~\cite{Skribanowitz1973}, quantum phase transition~\cite{Baumann2010}, and  contemporary setups that utilize ultracold atoms~\cite{Schneble2003,Baumann2010,Nagy2010,Keeling2010,Baumann2011,Ritsch2013,Klinder2015,Kollar2017}, trapped ions~\cite{Harkonen2009,Cohn2018,Safavi2018}, superconducting circuits~\cite{Blais2004,Casanova2010,Mezzacapo2014,Jaako2016,DeBernardis2018}, and others~\cite{Dimer2007,Baden2014,Zhang2018}.
We also refer the reader to other existing review articles that are complementary to ours, such as those by B.~M. Garraway~\cite{Garraway2011}, by P. Kirton et al.,~\cite{Kirton2019}, by A. Le Boit\'e~\cite{LeBoite2020}, and by J. Larson and T. Mavrogordatos~\cite{Larson2021}. 

A main feature of the Dicke model is its normal-to-superradiant ground-state quantum phase transition, that takes place in the thermodynamic limit and has been extensively investigated~\cite{Emary2003PRE,Emary2003PRL,Castanos2011b}. At a finite temperature, the model also exhibits a thermal phase transition, which received a lot of attention in the 1970s~\cite{Wang1973,Hepp1973AP,Hioe1973,Hepp1973PRA,Carmichael1973,Comer1974}. More recently, the analysis of quantum phase transitions was extended to higher energy levels in what became known as excited-state quantum phase transitions~\cite{PerezFernandez2011,Brandes2013,Bastarrachea2014PRAb,Bastarrachea2014JP,Kloc2017,Kloc2017JPA,Cejnar2021}, which is reviewed here.

The Dicke model has a well-defined classical limit, which was much explored in studies of the chaotic regime of the model~\cite{DeAguiar1992}. The analysis eventually covered both classical and quantum chaos~\cite{Emary2003PRE,Emary2003PRL,Chavez2016,Bastarrachea2014JP,Bastarrachea2014PRAb}, and the quantum-classical correspondence was soon extended to explain other properties of the system, such as localization, quantum scarring, and equilibration dynamics.

Quantum localization is associated with the absence of diffusion~\cite{Chirikov1981}, a well-known example being the Anderson localization~\cite{Anderson1958}, where interference effects reduce diffusion in disordered systems. Localized phases have potential applications in quantum technologies, because they prevent quantum information scrambling and can assist with information storage. In the case of the Dicke model, different localization measures have been introduced to quantify the spread of coherent states and eigenstates over both the phase space and the Hilbert space~\cite{Wang2020,Villasenor2021,Pilatowsky2022Q}.

Quantum scarring arises from the influence of measure-zero classical structures originating in the corresponding phase space~\cite{Heller1984,Heller1986PROC,Heller1991COLL}. In the classical domain, trajectories of chaotic systems that typically fill the available phase space may coexist with unstable periodic orbits of measure zero. These classical unstable periodic orbits can manifest in the structure of quantum states as concentrated regions of high probability, termed quantum scars~\cite{Heller1984}, which confine the dynamics of chaotic systems, such as the Dicke model~\cite{Pilatowsky2021NC}. Early studies about scarring in the Dicke model were made in the 1990s~\cite {DeAguiar1991,Furuya1992}, where families of unstable periodic orbits were identified. New families were later found in Refs.~\cite{Villasenor2020,Pilatowsky2021NC,Pilatowsky2021NJP}, and the connections between localization and scarring in the Dicke model were explored in Refs.~\cite{Pilatowsky2021NC,Pilatowsky2022Q}. 

Another direction of the analysis of isolated chaotic systems refers to their equilibration process and conditions for thermalization~\cite{Borgonovi2016,Dalessio2016}. Equilibration refers to the process where an observable stabilizes around an asymptotic value after a transient time, with fluctuations decreasing as the system size increases. Thermalization occurs when the asymptotic value of the observable approaches the predictions from statistical mechanics.  In the context of the Dicke model, thermalization was studied in Ref.~\cite{Villasenor2023}.

Quantum dynamics in the isolated Dicke model was investigated at short and long times in an attempt to connect it with the presence of chaos and anticipate the onset of thermalization. The out-of-time-ordered correlator~\cite{Maldacena2016PRD,Hashimoto2017}, for example, which quantifies the degree of noncommutativity in time between two Hermitian operators, was shown to grow exponentially fast in the chaotic Dicke model with a rate that coincides with the classical Lyapunov exponent~\cite{Chavez2019}, although this happens also in the regular regime at a critical point~\cite{Pilatowsky2020}. The fast initial evolution of the entanglement entropy in the Dicke model was analyzed in Refs.~\cite{Furuya1998,Song2008,Song2012,Lerose2020}, while the long-time dynamics and the onset of dynamical manifestations of   spectral correlations via the emergence of the correlation hole were investigated in Refs.~\cite{Lerma2019,Villasenor2020} using the survival probability (probability for finding the initial state later in time)~\cite{Lerma2018AIPCP,Lerma2018JPA}.

This review article is organized as follows.
In Sec.~\ref{sec:DickeModel}, we describe the Dicke model, including some extensions and available experimental realizations. We introduce its integrable limits, elucidate the quasi-integrability of its low-lying spectrum under an adiabatic approximation, and discuss numerical solutions of the quantum Hamiltonian. 
Next, in Sec.~\ref{sec:ClassicalDickeModel}, we introduce the classical limit of the Dicke model and explain how to identify, via the analysis of its energy surface and the semiclassical density of states, both ground-state and excited-state quantum phase transitions. We also describe phase-space representations of quantum states via the Wigner and Husimi quasiprobability distributions. 
In Sec.~\ref{sec:ChaosTheory}, we introduce the notion of classical chaos and show results for the Dicke model using Poincar\'e sections and Lyapunov exponents. We also discuss quantum chaos using the spectral properties of the model and establish the quantum-classical correspondence. 
In Sec.~\ref{sec:ChaosDynamics}, we focus on the dynamical signatures of quantum chaos. In particular, we present results for the evolution of out-of-time-ordered correlators and entanglement entropies, as well as for the survival probability using as initial states random and coherent states. 
Section~\ref{sec:QuantumLocalization} covers quantum localization, reviewing localization measures used for the Dicke model in the Hilbert space (R\'enyi entropies) and in the phase space (R\'enyi occupations).
Section~\ref{sec:QuantumScarring} is devoted to quantum scarring. We present the Dicke model's fundamental families of periodic orbits emanating from stationary points, show measures to quantify the degree of scarring, and explain how the R\'enyi occupations can be used as tools to detect unstable periodic orbits. We also discuss the connections between scarring, localization, and quantum ergodicity.
Section~\ref{sec:Conclusions} provides conclusions and outlines future perspectives for the study of the Dicke model.

\section{Dicke model}
\label{sec:DickeModel}
How light quanta interact with matter is a fundamental question in quantum optics. In 1936 and 1937, I.~I. Rabi proposed a simplified model to address the problem of coherent light-matter interaction by considering a single quantum emitter under the two-level approximation and a single mode of a classical radiation field. This model became known as the Rabi model~\cite{Rabi1936,Rabi1937} and introduced the idea of Rabi oscillations. In 1963, E.~T. Jaynes and F.~W. Cummings considered the interaction between a two-level atom and a quantized radiation field and were able to find an exact solution by discarding the fast oscillating (counter-rotating) terms of the interaction Hamiltonian, a procedure known as rotating-wave approximation (RWA)~\cite{Allen1975Book,Scully2008Book}. Since then, the Jaynes-Cummings model has become a standard tool in quantum optics~\cite{Jaynes1963,Scully2008Book}.

Moving beyond a single emitter, a breakthrough in understanding light-matter interaction came in 1954 with a work by R.~H. Dicke. To explain the emission of coherent radiation, he proposed to treat a gas of two-level molecules as a single quantum system with the molecules interacting with a common radiation field, introducing the concept of superradiance~\cite{Dicke1954}. The rate of the radiation emitted by  a group of $\mathcal{N}$ emitters collectively interacting with light is proportional to $\mathcal{N}^2$ rather than $\mathcal{N}$, thus, the enhanced emission is deemed superradiant~\cite{Gross1982}. The model became known as the Dicke model. In 1968, M. Tavis and F.~W. Cummings provided an integrable version of the Dicke model, the Tavis-Cummings model, where the counter-rotating term is absent~\cite{Tavis1968}. 

In 1973, Y.~K. Wang and F.~T. Hioe~\cite{Wang1973}, K. Hepp and E.~H. Lieb~\cite{Hepp1973AP,Hepp1973PRA}, H.~J. Carmichael, C.~W. Gardiner and D.~F. Walls~\cite{Carmichael1973}, followed by G.~C. Duncan~\cite{Comer1974}, extended a previous analysis performed in 1969~\cite{Mallory1969}. They showed that, in the thermodynamic limit,  the Dicke model has a phase transition between a normal and a superradiant phase. This transition persists at  zero temperature  when the coupling between the atoms and the field reaches a critical value~\cite{Hillery1985}.  Since then, various other models have emerged from the Dicke model tailored to the needs of specific systems and approximations~\cite{Larson2021,MarquezPeraca2020}. 

The chaotic dynamics in the classical version of the  Dicke model was first explored in the 1970s~\cite{Belobrov1976} and 1980s~\cite{Fox1987}, and then studied in the 1990s by M.~A.~M. de Aguiar, K. Furuya, C.~H. Lewenkopf, M.~C. Nemes, and G.~Q. Pellegrino~\cite{DeAguiar1992,Furuya1998}. In 2003, C. Emary and T. Brandes revealed the presence of quantum chaos in the spectral energy fluctuations~\cite{Emary2003PRE,Emary2003PRL}.

The classical Hamiltonian for a set of $\mathcal{N}$ particles with identical mass and charge interacting with an electromagnetic field within a cavity can be described using the minimal-coupling Hamiltonian~\cite{Scully2008Book},
\begin{equation}
    H_{\text{mc}} = \sum_{i=1}^{\mathcal{N}} \left\{
    \frac{1}{2m}\left[\mathbf{p}_{i}-e\mathbf{A}(\mathbf{q}_{i},t)\right]^{2}+e\mathcal{U}(\mathbf{q}_{i},t)+\mathcal{V}(|\mathbf{q}_{i}|) \right\} +
    \frac{1}{2}\int dV\left[\varepsilon_{0}|\mathbf{E}(\mathbf{q},t)|^{2}+\frac{1}{\mu_{0}}|\mathbf{B}(\mathbf{q},t)|^{2}\right] ,
    \label{Eq:firstH}
\end{equation}
where $e$ and $m$ are the charge and mass of each particle, $\mathbf{q}_{i}$ and $\mathbf{p}_{i}$ are the canonical position and momentum for each particle, $\mathbf{A}(\mathbf{q}_{i},t)$ and $\mathcal{U}(\mathbf{q}_{i},t)$ are the vector and scalar potentials of the external field, $\mathcal{V}(|\mathbf{q}_{i}|)$ is an electrostatic potential, $\mathbf{E}$ and $\mathbf{B}$ represent the electric and magnetic fields, $\varepsilon_{0}$ and $\mu_{0}$ are the permittivity and permeability of free space, and $V$ is the cavity's volume. The first term in Eq.~\eqref{Eq:firstH} includes the energy associated with atom excitations and the light-matter interaction. The second term   accounts for the radiation field. Upon quantization, the formulation of the Dicke model is based on a series of conventional approximations, as follows~\cite{Klimov2009Book}:

\begin{itemize}
    \item All emitters are single electron atoms.
    \item Only one-photon transitions are considered. 
    \item All atoms are sufficiently far apart from each other, so the interactions among them can be neglected. 
    \item In each atom, only two atomic levels are considered, which interact with a nearly resonant mode of the electromagnetic field inside the cavity. This is called the two-level approximation. Notice that this assumption  breaks the electromagnetic gauge symmetry, which leads to different versions of the Hamiltonian depending on the gauge choice~\cite{Nahmad-Achar2013,MarquezPeraca2020,Garziano2020,Stokes2019}.
    \item The volume containing the atoms is smaller than the wavelength of the single mode of electromagnetic radiation, that is, all atoms interact with the same electromagnetic field, and the electromagnetic potentials are position-independent. This is called the long-wave approximation~\cite{Scully2008Book}. As a result, the light-matter coupling strength is the same for all atoms, so that they respond collectively and coherently.
\end{itemize}

Under the aforementioned approximations, the Hamiltonian for the Dicke model takes the simple form
\begin{equation}
    \hat{H}_{\text{D}} = \hat{H}_{\text{F}}+\hat{H}_{\text{A}}+\hat{H}_{\text{I}},
    \label{Eq:secondH}
\end{equation}
where 
\begin{equation}
    \hat{H}_{\text{F}} = \hbar\omega\hat{a}^{\dag}\hat{a}
\end{equation}
represents the energy of the single mode of the confined radiation resonant with the two-level transition, $\omega$ is the frequency of the field, $\hat{a}^{\dag}$ and $\hat{a}$ are its  creation and annihilation operators that follow the bosonic commutation relation $[\hat{a},\hat{a}^{\dagger}]=\hat{\mathbb{I}}$. The second term in Eq.~\eqref{Eq:secondH}, 
\begin{equation}
  \hat{H}_{\text{A}} = \hbar\omega_{0}\hat{J}_{z}, 
\end{equation}
contains the energy of the collective two-level atomic system, $\omega_{0}$ is the frequency for the two-level atomic transition and the operators $\hat{J}_{\mu}=\frac{1}{2}\sum_{n=1}^{\mathcal{N}}\hat{\sigma}_{\mu}^{(n)}$ ($\mu=x,y,z$) identify collective pseudospin operators for a set of $\mathcal{N}$ two-level systems where $\hat{\sigma}_{x,y,z}$ are the Pauli matrices. $\hat{J}_{z}$ represents the relative population between single atoms in the ground or excited state of the two-level transition, and $\hat{J}_{\pm}=\sum_{n=1}^{\mathcal{N}}\hat{\sigma}_{\pm}^{(n)}$ are the raising and lowering collective pseudospin operators defined as $\hat{J}_{\pm}=\hat{J}_{x}\pm i\hat{J}_{y}$. The pseudospin operators follow the standard commutation relations for angular momentum $[\hat{J}_{\mu},\hat{J}_{\nu}]=i\hbar\epsilon_{\mu\nu\kappa}\hat{J}_{\kappa}$. The last term in Eq.~\eqref{Eq:secondH},
\begin{equation}
    \label{eq:hint}
    \hat{H}_{\text{I}} = \frac{2 \hbar \gamma}{\sqrt{\mathcal{N}}}\left(\hat{a}^{\dag}+\hat{a}\right)\hat{J}_{x} = \frac{\hbar\gamma}{\sqrt{\mathcal{N}}}\left(\hat{a}^{\dag}+\hat{a}\right)\left(\hat{J}_{+}+\hat{J}_{-}\right),
\end{equation}
includes the light-matter interaction strength with a coupling parameter $\gamma$ that corresponds to half the Rabi splitting. Setting $\hbar=1$, one gets the standard form of the Dicke Hamiltonian that is considered throughout this review,
\begin{equation}
    \label{eqn:dicke_hamiltonian}
    \hat{H}_{\text{D}} = \omega\hat{a}^{\dag}\hat{a}+\omega_{0}\hat{J}_{z}+\frac{\gamma}{\sqrt{\mathcal{N}}}\left(\hat{a}^{\dag}+\hat{a}\right)\left(\hat{J}_{+}+\hat{J}_{-}\right) .
\end{equation}
This Hamiltonian commutes with  $\hat{\mathbf{J}}^{2}=\hat{J}_{x}^{2}+\hat{J}_{y}^{2}+\hat{J}_{z}^{2}$ and with the parity operator (see Sec.~\ref{Sec:IntegrabilityDicke}). The classical limit of the Dicke model reveals chaotic properties (see Sec.~\ref{subs:classicalChaos}), which get  reflected in the spectrum of the quantum Hamiltonian (see Sec.~\ref{subs:quantchaos}).

The collective system of two-level atoms interacting with radiation can also be seen as a simplified description of a collective system of qubits (pseudospin) coupled to a quantum resonator (boson). This has allowed the Dicke Hamiltonian to go beyond the field of atomic physics and quantum optics, becoming a paradigmatic description of the spin-boson interaction, a simple picture to describe collective interactions, and a pivotal step to understand the underlying physics in QED and relevant systems in the emerging fields of quantum technologies, including quantum computing, quantum information processing, and quantum communication~\cite{Garraway2011,Kirton2019,LeBoite2020,Larson2021}.

\subsection{Regimes and limits of the Dicke model}
\label{subsec:RegimesDicke}

Depending on the strength of the interaction and the number of two-level systems, the Dicke model can lead to important limiting scenarios. The {\it strong coupling} regime, for example, is achieved when the light-matter interaction strength $\gamma$ is larger than the damping strength $\kappa$ of the system ($\gamma\gg\kappa$)~\cite{Khitrova2006}. This regime allows for the observation and control of coherent dynamics, which is important for the emergence of new experimental architectures in quantum information technologies and related fields.  Increasing  the light-matter interaction strength $\gamma$ to exceed the energy scales of the atoms and the field leads to the {\it ultra-strong coupling} regime ($\gamma\approx 0.1\omega$). It has attracted attention in the past decades, allowing increased control of quantum systems and enabling applications such as lasers, quantum sensing, and quantum information processing~\cite{Kockum2019,FornDiaz2019,MarquezPeraca2020}. Other regimes also exist for even stronger light-matter interactions, such as the deep-strong coupling ($\gamma \geq \omega$)~\cite{Casanova2010,Bayer2017,Yoshihara2017} and the extreme or very-strong coupling regime ($\gamma \geq 10.0 \omega$)~\cite{Khurgin2001, Brodbeck2017}.

In the strong coupling regime, one can take the RWA and eliminate the so-called non-resonant or counter-rotating terms, $\hat{a}^{\dag}\hat{J}_{+}$ and $\hat{a}\hat{J}_{-}$, in the Dicke Hamiltonian [Eq.~\eqref{eqn:dicke_hamiltonian}]. This amounts to disregarding contributions oscillating very fast in time, such that their temporal average goes to zero. In this limit, the Dicke model becomes the Tavis-Cummings model~\cite{Tavis1968},
\begin{equation}
\label{eqn:tavis_cummings_hamiltonian}
    \hat{H}_{\text{TC}} = \omega\hat{a}^{\dag}\hat{a}+\omega_{0}\hat{J}_{z}+\frac{\gamma}{\sqrt{\mathcal{N}}}\left(\hat{a}^{\dag}\hat{J}_{-}+\hat{a}\hat{J}_{+}\right),
\end{equation}
which is an integrable variant of the Dicke model. The Tavis-Cummings model is a valuable tool for benchmarking results of the Dicke model. 

Two other important limits of the Dicke model are obtained when one considers a single two-level system ($\mathcal{N}=1$). The Dicke model becomes the quantum Rabi model~\cite{Rabi1936,Rabi1937,Braak_2016},
\begin{equation}
    \hat{H}_{\text{R}} = \omega\hat{a}^{\dag}\hat{a}+\frac{\omega_{0}}{2}\hat{\sigma}_{z}+\gamma\left(\hat{a}^{\dag}+\hat{a}\right)\left(\hat{\sigma}_{+}+\hat{\sigma}_{-}\right),
\end{equation}
and the Tavis-Cummings model transforms into the Jaynes-Cummings  model~\cite{Jaynes1963,DeBernardis2024},
\begin{equation}
    \hat{H}_{\text{JC}} = \omega\hat{a}^{\dag}\hat{a}+\frac{\omega_{0}}{2}\hat{\sigma}_{z}+\gamma\left(\hat{a}^{\dag}\hat{\sigma}_{-}+\hat{a}\hat{\sigma}_{+}\right).
\end{equation}
The properties of both Rabi and Jaynes-Cummings models have been extensively studied (see the review article in Ref.~\cite{Larson2021}).

There are different perturbative approximations to the Jaynes-Cummings and the Rabi model. An example is the  Schrieffer-Wolf transformation to the Jaynes-Cummings model in the dispersive regime, $\gamma \ll \omega_{0}-\omega$, that leads to the AC Stark Hamiltonian~\cite{Blais2004},
\begin{equation}
    \hat{H}_{\text{ACS}} = \left[1+\frac{\gamma^{2}}{\omega(\omega_{0}-\omega)}\hat{\sigma}_{z}\right]\omega\hat{a}^{\dagger}\hat{a}+\left[1+\frac{\gamma^{2}}{\omega_{0}(\omega_{0}-\omega)}\right]\frac{\omega_{0}}{2}\hat{\sigma}_{z}. 
\end{equation}

Additionally, the ultra-strong coupling regime can be divided into a perturbative region, $0.3 \omega\leq \gamma \lesssim \omega $, and a non-perturbative region, $\omega \lesssim \gamma$ ~\cite{Rossatto2017}. In the perturbative regime, one can get a Bloch-Siegert-like Hamiltonian~\cite{Klimov2009Book}, 
\begin{equation}
    \hat{H}_{\text{BS}} =
    \left(1+\frac{\omega_{\text{BS}}}{\omega}\hat{\sigma}_{z}\right)\omega\hat{a}^{\dagger}\hat{a}+
    \left(1+\frac{\omega_{\text{BS}}}{\omega_{0}}\right)\frac{\omega_{0}}{2}\hat{\sigma}_{z}+
    \left[f\left(\hat{a}^{\dagger}\hat{a}\right)\hat{\sigma}_{-}\hat{a}^{\dagger}+\text{h.c.}\right],
\end{equation}
from second-order perturbation theory, where $\omega_{\text{BS}}=\gamma^{2}/(\omega+\omega_{0})$ is the Bloch-Siegert shift, and $f(x)=-\gamma(1-\omega_{\text{BS}}x)$.  

A generalization of the Dicke model is the Hopfield Hamiltonian~\cite{Hopfield1958,Garziano2020},
\begin{equation}
    \hat{H}_{\text{H}} = \sum_{\mathbf{k}}\left[\omega_{\mathbf{k}}\hat{a}^{\dagger}_{\mathbf{k}}\hat{a}_{\mathbf{k}}+\omega_{0\mathbf{k}}\hat{b}^{\dagger}_{\mathbf{k}}\hat{b}_{\mathbf{k}}+\gamma\left(\hat{b}^{\dagger}_{\mathbf{k}}\hat{a}_{\mathbf{k}}+\hat{a}^{\dagger}_{\mathbf{k}}\hat{b}_{\mathbf{k}}\right)\right] ,
\end{equation}
that considers multimode cavity photons $\hat{a}_{\mathbf{k}}$ and matter excitations $\hat{b}_{\mathbf{k}}$ with momentum $\mathbf{k}$, and dispersion relations $\omega_{\mathbf{k}}$ and $\omega_{0\mathbf{k}}$, respectively. The Hopfield model is a customary description of polaritons, i.e., the light-matter superimposed quasiparticles emerging in the strong coupling regime, that have gained inertia in the last decades in the field of microcavity semiconductors~\cite{Carusotto2013,Deng2010}. The model has also been employed to describe the ultra-strong coupling regime~\cite{Ciuti2005}. Within the picture of the Hopfield Hamiltonian, the low-lying energy states of the Dicke model can be interpreted as polaritons~\cite{Baksic2014}.

\subsubsection{Extensions and variants of the Dicke model}

Many extensions and generalizations of the Dicke model have been studied. Most of them emerge by relaxing some of the conventional approximations discussed in the beginning of Sec.~\ref{sec:DickeModel}. In the following, we include a representative sample of these cases. 

A general representation of the Dicke Hamiltonian that allows for different strengths in the resonant and non-resonant terms is
\begin{equation}
    \label{eq:HD_generalized}
    \hat{H}_{\text{GD}} = \omega\hat{a}^{\dag}\hat{a}+\omega_{0}\hat{J}_{z}+\frac{1}{\sqrt{\mathcal{N}}}\left[\gamma_{-}\left(\hat{a}^{\dag}\hat{J}_{-}+\hat{a}\hat{J}_{+}\right)+\gamma_{+}\left(\hat{a}^{\dag}\hat{J}_{+}+\hat{a}\hat{J}_{-}\right)\right],
\end{equation}
which has been called either generalized, unbalanced, or anisotropic Dicke model~\cite{DeAguiar1992,Buijsman2017}. The Dicke model corresponds to $\gamma_{-}=\gamma_{+}=\gamma$ and the Tavis-Cummings model to $\gamma_{-}=\gamma$ and $\gamma_{+}=0$ [Eqs.~\eqref{eqn:dicke_hamiltonian} and~\eqref{eqn:tavis_cummings_hamiltonian}, respectively]. The generalized Dicke model exhibits a rich phase space and the presence of novel critical phenomena~\cite{Bastarrachea2016JSTAT,Kloc2017,Das2022,Das2023a}.

Including collective interactions between the two-level systems leads to the extended Dicke model~\cite{Chen2006,Abdel-Rady2017,Salah2018,Sinha2020},
\begin{equation}
    \hat{H}_{\text{ED}} = \hat{H}_{\text{D}} + \frac{1}{\mathcal{N}}\left(\eta_{z}\hat{J}_z^2 + \eta_{x}\hat{J}_x^2\right) ,
    \label{Eq:Dicke-Lipkin}
\end{equation}
where $\eta_{x,z}$ are the strengths of the collective interactions. This Hamiltonian is a combination of the Dicke and the Lipkin-Meshkov-Glick Hamiltonians~\cite{Lipkin1965,Meshkov1965}. The Lipkin-Meshkov-Glick term $\left(\eta_{z}\hat{J}_z^2 + \eta_{x}\hat{J}_x^2\right)$ can also be interpreted as a spin-1/2 model with all-to-all couplings~\cite{Santos2016}.
Generally speaking, the presence of matter interactions creates a first-order phase transition~\cite{Lee2004,Chen2008b,Chen2010,Rodriguez2011,Rodriguez2018}, shifts of the critical coupling and of the scaling behavior of the geometric phase in the standard normal-to-superradiant phase transition~\cite{Chen2006,Li2013,Jaako2016}, and gives rise to a richer phase diagram~\cite{Robles2015,Rodriguez2018,Herrera2022}. Recently, it has been pointed out that a bound luminosity state exists in the semiclassical dynamics of the model in Eq.~\eqref{Eq:Dicke-Lipkin}~\cite{Seidov2023}. A richer extension is the Dicke-Ising model \cite{Schellenberger2024}.

Modifications of the bosonic field can also be implemented. An example is the two-photon Dicke model, 
\begin{equation}
    \label{eqn:Dicke_2photon}
    \hat{H}_{\text{TPD}} = \omega\hat{a}^{\dag}\hat{a}+\omega_{0}\hat{J}_{z}+\frac{\gamma}{\sqrt{\mathcal{N}}}\left(\hat{a}^{\dag 2}+\hat{a}^2\right)\left(\hat{J}_{+}+\hat{J}_{-}\right) ,
\end{equation}
which exhibits the collapse of the energy spectrum at a critical value of the coupling parameter $\gamma$~\cite{Felicetti2015,Duan2016}. This collapse is similar to the one obtained when transitioning from a harmonic to an inverted oscillator, with the flat potential at the critical point~\cite{Armenta2020}. At the critical coupling parameter, the eigenenergy spectrum consists of both discrete energy levels and a continuous energy spectrum~\cite{Duan2016,Lo2020,Lo2021}. The two-photon Dicke model predicts superradiance phenomena~\cite{Garbe2017,Chen2018} and the squeezing of photonic quadratures when approaching the boundary between the superradiant phase and the unbounded region~\cite{Banerjee2022}. However, recent studies have demonstrated that the superradiant phase is a fictitious concept in this system, attributed to finite size effects~\cite{Ramirez2026Arxiv}. Moreover, classical and quantum signatures allow one to distinguish the chaotic and regular behavior in this model~\cite{S.Wang2019,Ramirez2025}.

The deformed Dicke model~\cite{Corps2022} includes a direct coupling to an external bosonic reservoir. There is also an extended Dicke model that includes terms containing nonlinear operators that correspond to the real and imaginary parts of the square of the field amplitude inside a nonlinear material~\cite{Hillery1987}.

Adding quadratic terms to the field of the Tavis-Cummings model gives~\cite{Rodriguez2010} 
\begin{equation}
    \hat{H}_{\text{SQ}} = \hat{H}_{\text{TC}} + G \left(\hat{a}^{\dagger 2} + \hat{a}^2\right),
\end{equation}
where $G$ is the strength of the two photon term. This modification induces a symmetry breaking, that leads to a quantum phase transition without the ultra-strong coupling requirement, and allows for an optical switching from the normal phase to the superradiant phase by increasing the pump field intensity~\cite{Zhu2020}. It has been predicted that 
a suitable choice of the parameters of the non-linear optical medium could make possible the use of a low intensity laser to access the superradiant region experimentally~\cite{Guerra2020}. 

Adding quadratic self-interactions to both the field and the two-level ensemble, a new Hamiltonian can be created, such as the  generalized Tavis-Cummings model~\cite{Rodriguez-Lara_2013},
\begin{equation}
    \hat{H}_{\text{GTC}} = \hat{H}_{\text{TC}} + g \left(\hat{a}^{\dagger 2} \hat{a}^2 + \hat{J}_z^2\right).
\end{equation}

It is also possible to consider multimode or multilevel Dicke and Tavis-Cummings models, where superradiant and critical phenomena transitions have been studied~\cite{Hayn2012,Baksic2013,Castanos2014,Cordero2017}. Related to this subject, Ref.~\cite{Valencia2023} discusses a general framework to describe a cavity QED simulator with $N$-level bosonic atoms. Another direction for the extensions of the Dicke model is to take the spatial dependence of the phase of the electromagnetic field  into account, which has been used to show new superradiant phases~\cite{Li2006}. Recent works~\cite{Das2024} have extended the study of the thermal and quantum phase transitions for the disordered Dicke model by considering coupling strengths generated from random distributions.

\subsubsection{Experimental realizations of the Dicke model}

A first major interest in the experimental realization of the Dicke model was for the observation of the superradiant phenomena. Given the algebraic simplicity of the model, realizations of Dicke-like Hamiltonians have been sought in several platforms, from ultracold atoms in optical lattices~\cite{Nagy2010,Liu2011,Yuan2017} to superconducting qubits~\cite{Jaako2016,Yang2017,DeBernardis2018,Pilar2020}. However, there is a plethora of systems where superradiant-related effects have been proposed to exist, including nuclei~\cite{Auerbach2011}, systems involving graphene~\cite{Cong2016}, bidimensional materials~\cite{Hagenmuller2012,Chirolli2012}, and quantum dots~\cite{Scheibner2007}. 

The challenges to reach large light-matter interactions in atomic physics due to the small dipole moment of the atomic transitions~\cite{Raimond2001} hindered experimental realizations of the Dicke model for decades. Even though superradiance was verified in pumped atomic gases twenty years after Dicke's proposal~\cite{Skribanowitz1973}, it was not until 2010, when a non-equilibrium Dicke model was implemented in the translational degrees of freedom of a Bose-Einstein condensate within an optical cavity~\cite{Baumann2010}, that the superradiant quantum phase transition was observed experimentally in terms of the self-organization of the condensate~\cite{Nagy2010}. Since then, many tunable quantum many-body systems have emerged as promising setups for realizing the Dicke model and its variants. Several platforms have reached ultra-strong light-matter coupling, including superconducting circuits, microcavity semiconductors (inorganic and organic), and optomechanical systems. For a full description of the state-of-the-art, see Refs.~\cite{FornDiaz2019,Kockum2019,MarquezPeraca2020}.

Among the contemporary experimental proposals for the realization of the Dicke model, there are setups with ultracold atoms~\cite{Schneble2003,Baumann2010,Nagy2010,Keeling2010,Baumann2011,Ritsch2013,Klinder2015,Kollar2017}, trapped ions~\cite{Harkonen2009,Cohn2018,Safavi2018}, cavity-assisted Raman transitions~\cite{Dimer2007,Baden2014,Zhang2018,Kroeze2018}, spin-orbit coupled Bose-Einstein condensates~\cite{Hamner2014},  and superconducting circuits~\cite{Blais2004,Casanova2010,Mezzacapo2014,Jaako2016,DeBernardis2018}. Focusing on the ultra-strong-coupling regime, Ref.~\cite{Dareau2018} shows a mechanical analog of the Dicke model by coupling the spin of individual neutral atoms to their quantized motion in an optical trapping potential. Spectroscopic evidence for a Dicke superradiant phase transition was achieved with ErFeO$_3$ in Ref.~\cite{Kim2024}. A recent proposal for the detection of ground-state virtual photons employs unconventional ``light fluxonium''-like superconducting quantum circuit implemented by superinductors to allow an efficient conversion of virtual photons into real ones. This would enable their detection with resources available to present-day quantum technologies~\cite{Giannelli2024}.

In the classical limit, the standard and deformed Dicke models have been implemented in the laboratory using resonant circuits~\cite{Quiroz2020,Quiroz2023}, where  signatures of the ground-state phase transition and of the presence of chaos were observed.

An attractive application of the Dicke Hamiltonian is the realization of Dicke quantum batteries, which are believed to exhibit quantum advantage in their charging power~\cite{Ferraro2018,Andolina2019,Andolina2019b}. In these systems, coherence promotes coherent work, and together with entanglement, inhibits incoherent work~\cite{Shi2022}. Their self-discharge time has been extended using molecular triplets~\cite{Tibben2024}.

\subsection{Integrable limits of the Dicke model}
\label{Sec:IntegrabilityDicke}

The Dicke model has more degrees of freedom than conserved quantities, so it does not have an exact analytical solution. The classical Hamiltonian is non-integrable and the quantum Hamiltonian exhibits quantum chaos (see Secs.~\ref{subs:classicalChaos} and~\ref{subs:quantchaos}, respectively). 

The quantum Hamiltonian of the Dicke model commutes with the Casimir operator of the pseudospin algebra,  $\hat{\mathbf{J}}^{2}=\hat{J}_{x}^{2}+\hat{J}_{y}^{2}+\hat{J}_{z}^{2}$, so the pseudospin length $j$ is a conserved quantity and the Hilbert space of the system is divided into subspaces of fixed $j$, where the symmetric subspace has the maximum value $j=\mathcal{N}/2$. This subspace contains the collective excitations of the $\mathcal{N}$ two-level systems, 
with all atoms sharing the same phase, and includes the ground state of the whole system. Thus, if one restricts the study to this subspace, as often done, the Hamiltonian describes a system with two (classical) degrees of freedom, one for the boson (field) and one for the pseudospin (collective two-level system), which are the building blocks of the Hilbert space of this system. The quantum Hamiltonian additionally commutes with the parity operator,
\begin{equation}
    \label{eqn:parity_operator}
    \hat{\Pi}=e^{i\pi\hat{\Lambda}}=e^{i\pi\left(\hat{a}^{\dagger}\hat{a}+\hat{J}_{z}+j\,\hat{\mathbb{I}}\right)},
\end{equation}
where $\hat{\Lambda}=\hat{a}^{\dagger}\hat{a}+\hat{J}_{z}+j\,\hat{\mathbb{I}}$ is the operator representing the total number of excitations. The parity symmetry is not enough to achieve integrability in the classical limit, because the symmetry is not continuous. 

One way to reach integrability in the classical sense is to apply the RWA to the Dicke Hamiltonian, which leads to the integrable Tavis-Cummings Hamiltonian~\cite{Tavis1968}, as previously defined. The Tavis-Cummings Hamiltonian commutes with $\hat{\Lambda}$, so the total number of excitations is conserved. This additional symmetry ensures the integrability of the model. Integrability can also be achieved by setting to zero one of the three relevant parameters of the Dicke Hamiltonian in Eq.~\eqref{eqn:dicke_hamiltonian}, that is, $\gamma=0$ or $\omega=0$ or $\omega_0=0$. The three quantum Hamiltonians associated with these cases are discussed below.

\subsubsection{Non-interacting Hamiltonian}
\label{sec:zg}

Integrability is trivially achieved in the non-interacting limit, where $\gamma \to 0$. Because the two degrees of freedom are uncoupled, the classical  Hamiltonian is integrable. In this case, the quantum Hamiltonian in Eq.~\eqref{eqn:dicke_hamiltonian} becomes 
\begin{equation}
    \label{eqn:fock_hamiltonian}
    \hat{H}_{\text{D}}(\omega,\omega_{0},0) = \omega\hat{a}^{\dag}\hat{a}+\omega_{0}\hat{J}_{z},
\end{equation}
which commutes with both operators $\hat{n}=\hat{a}^{\dag}\hat{a}$ and $\hat{J}_{z}$. The eigenbasis of $\hat{H}_{\text{D}}(\omega,\omega_{0},0)$ is the tensor product between the Fock states $|n\rangle$, associated with the bosonic field, and the angular momentum states $|j,m_{z}\rangle$ of the collective two-level system. We call this eigenbasis the {\it Fock basis},
\begin{equation}
    \label{eqn:fock_basis}
    |n\rangle\otimes|j,m_{z}\rangle \equiv |n;j,m_{z}\rangle,
\end{equation}
which satisfies the textbook relations~\cite{Zhang1990}
\begin{gather}
    \hat{a}^{\dag}\hat{a}~|n;j,m_{z}\rangle = n|n;j,m_{z}\rangle, \\
    \mathbf{\hat{J}}^{2}~|n;j,m_{z}\rangle = j(j+1)|n;j,m_{z}\rangle, \\
    \hat{J}_{z}~|n;j,m_{z}\rangle = m_{z}|n;j,m_{z}\rangle.
\end{gather}

\subsubsection{Zero energy splitting}
\label{sec:zsf}

An integrable limit is also achieved when one takes $\omega_{0}\to 0$. In this case, the quantum Hamiltonian is 
\begin{equation}
    \label{eqn:ceb_hamiltonian}
    \hat{H}_{\text{D}}(\omega,0,\gamma) = \omega\left[\hat{a}^{\dag}\hat{a}+G\left(\hat{a}+\hat{a}^{\dagger}\right)\hat{J}_{x}\right] ,
\end{equation}
where $G=2\gamma/\left(\omega\sqrt{\mathcal{N}}\right)$. To find the exact eigenbasis in this limit, one needs to take a displacement over the creation (annihilation) operator $\hat{a}^{\dag}$ ($\hat{a}$) using~\cite{Chen2008,Bastarrachea2011,Bastarrachea2014PS}
\begin{equation}
    \label{eqn:operator_displacement}
    \hat{A} = \hat{a}+G\hat{J}_{x},
\end{equation}
which leads to the transformed Hamiltonian
\begin{equation}
\label{eqn:coherent_hamiltonian}
    \hat{H}_{\text{D}}(\omega,0,\gamma) = \omega\left(\hat{A}^{\dag}\hat{A}-G^{2}\hat{J}_{x}^{2}\right).
\end{equation}
Because $[\hat{A},\hat{J}_{x}]=0$, the Hamiltonian commutes with both operators $\hat{N}=\hat{A}^{\dag}\hat{A}$ and $\hat{J}_{x}$. The basis that exactly solves the Hamiltonian is the tensor product between the eigenstates of the operator $\hat{N}=\hat{A}^{\dag}\hat{A}$, associated with the modified field that follows the rules of the Heisenberg-Weyl algebra, and the rotated Dicke states $|j,m_{x}\rangle$, related to the collective two-level system. The new vacuum state $|N=0;j,m_{x}\rangle$ of the modified bosonic operator $\hat{N}$ is defined as $\hat{A}|N=0;j,m_{x}\rangle=0$. Hence, the new vacuum state is an eigenstate of the annihilation operator $\hat{a}$,
\begin{equation}\label{eq:vac}
    \hat{a}|N=0;j,m_{x}\rangle = -Gm_{x}|N=0;j,m_{x}\rangle=\alpha_{m_{x}}|\alpha_{m_{x}}\rangle\otimes|j,m_{x}\rangle,
\end{equation}
with eigenvalue $\alpha_{m_{x}}=-Gm_{x}$. This is a coherent state when spanned in the Fock basis, $|N=0;j,m_x\rangle=\hat{D}(\alpha_{m_x})|n=0\rangle\otimes |j,m_x\rangle$, where $\hat{D}(\alpha)=\exp\left(\alpha\hat{a}^{\dagger}-\alpha^{*}\hat{a}\right)$ is the displacement operator. The whole eigenbasis of $\hat{H}_{D}(\omega,0,\gamma)$ can be obtained by repeatedly applying $\hat{A}^{\dagger}$,
\begin{equation}
\label{eq:Nvac}
|N;j,m_{x}\rangle = \frac{1}{\sqrt{{N}!}}\left( \hat{A}^{\dagger} \right)^{N} \hat{D}(\alpha_{m_x})|N=0\rangle\otimes |j,m_x\rangle = \frac{1}{\sqrt{N!}} \left( \hat{a}^{\dagger} - \alpha_{m_{x}} \right)^{N} \hat{D}(\alpha_{m_{x}}) |n=0\rangle\otimes|j,m_x\rangle,
\end{equation}
where Eq.~\eqref{eqn:operator_displacement} and the eigenvalue equation $\hat{J}_{x}|j,m_{x}\rangle=m_{x}|j,m_{x}\rangle$ were used. From the commutation relation $\hat{D}(\alpha_{m_{x}})\hat{a}^{\dagger} = \left(\hat{a}^{\dagger}-\alpha_{m_x}\right)\hat{D}(\alpha_{m_x})$, it follows that
\begin{equation}
    \label{eqn:efficient_basis}
   |N;j,m_x\rangle  = |N,\alpha_{m_x}\rangle\otimes |j,m_x\rangle,
\end{equation}
where the state 
\begin{equation}
\label{eqn:displacedFockStates}
   |N,\alpha_{m_{x}}\rangle \equiv \hat{D}(\alpha_{m_{x}})|N\rangle
\end{equation} 
is the $N$-th eigenstate 
$|N\rangle =\frac{1}{\sqrt{N!}}\left(\hat{a}^{\dagger}\right)^N|n=0\rangle$, of the bosonic operator $\hat{n}=\hat{a}^\dagger\hat{a}$,
displaced by the complex number $\alpha_{m_{x}}$. These states are called displaced Fock states~\cite{deOliveira1990}. The states $|N;j,m_{x}\rangle$ satisfy the relations
\begin{gather}
    \hat{A}^{\dag}\hat{A}|N;j,m_{x}\rangle = N|N;j,m_{x}\rangle, \label{eq:AdAeig} \\
    \mathbf{\hat{J}}^{2}|N;j,m_{x}\rangle = j(j+1)|N;j,m_{x}\rangle, \\
    \hspace{1cm}\hat{J}_{x}|N;j,m_{x}\rangle = m_{x}|N;j,m_{x}\rangle.
\end{gather}
It was demonstrated in Ref.~\cite{Chen2008} that the eigenbasis of $\hat{H}_{\text{D}}(\omega,0,\gamma)$ in Eq.~\eqref{eqn:efficient_basis} is an efficient basis to diagonalize the full Dicke Hamiltonian $\hat{H}_{\text{D}}(\omega,\omega_0,\gamma)$. This is why we call it the {\it efficient basis}, and employ it in many of our computations. 

The eigenbasis in this limit is often evoked when $\gamma/\omega_{0}\to \infty$. It is considered an approximate exact solution in the superradiant phase for large light-matter interactions and for other perturbative approaches~\cite{Emary2003PRE}. Notice that this basis is the result of the application of a combined unitary transformation of the form $\hat{U}=\exp[-G\hat{J}_{x}\left(\hat{a}^{\dagger}-\hat{a}\right)]$ to the vacuum state, which is equivalent to simultaneously displacing the annihilation (creation) operator and rotating the pseudospin. This kind of transformation has been called the ``polaron frame''~\cite{Alcalde2012,Jaako2016,DeBernardis2018,Pilar2020}, and connects with other methods~\cite{LeBoite2020}, including the generalized RWA~\cite{Irish2007}. This transformation has also been used for solving the two-photon Rabi model~\cite{Duan2015,Duan2016}.

\subsubsection{Zero bosonic frequency}
\label{sec:zbf}

Integrability is achieved when one takes $\omega\to 0$, when the Hamiltonian in Eq.~\eqref{eqn:dicke_hamiltonian} becomes
\begin{equation}
\hat{H}_{\text{D}}(0,\omega_{0},\gamma) = \omega_{0}\left(\hat{J}_{z}+G_{0}\hat{q}
\hat{J}_{x}\right) ,
\end{equation}
where $\hat{q}=\left(\hat{a}^{\dag}+\hat{a}\right)/\sqrt{2}$ and $G_{0}=2\sqrt{2}\gamma/\left(\omega_{0}\sqrt{\mathcal{N}}\right)$. Given that there is no explicit dependence on $\hat{J}_{y}$, we can make a rotation in terms of $\hat{q}$ to diagonalize the system, such that
\begin{gather}
\hat{J}_{x} = \cos\hat{\phi}\hat{J}'_{x} + \sin\hat{\phi}\hat{J}'_{z}, 
\label{eqn:rotation1}
\\
\hat{J}_{z} = -\sin\hat{\phi}\hat{J}'_{x} + \cos\hat{\phi}\hat{J}'_{z},
\label{eqn:rotation}
\end{gather}
where $\left(\cos\hat{\phi},\sin\hat{\phi}\right) = \left(\hat{\mathbb{I}},G_{0}\hat{q}\right)/\sqrt{1+G_{0}^{2}\hat{q}^{2}}$, satisfying  $\cos^{2}\hat{\phi} + \sin^{2}\hat{\phi} = \hat{\mathbb{I}}$. This can be done, because $\hat{q}$ commutes with the pseudospin operators $\hat{\mathbf{J}}$. The Hamiltonian transforms into
\begin{equation}
\hat{H}_{\text{D}}(0,\omega_{0},\gamma)=\omega_{0}\sqrt{\hat{\mathbb{I}}+G_{0}^{2}\hat{q}^{2}}\hat{J}'_{z}. 
\end{equation}
In this case, the eigenstates are the tensor product between Dicke states $|j,m_{z}'\rangle$ in the rotated frame and real space eigenstates $|q\rangle$, that is,
\begin{equation}
|q\rangle\otimes|j,m'_{z}\rangle=|q;j,m'_{z}\rangle,
\label{eqn:exactsolution2}
\end{equation}
satisfying the relations
\begin{gather}
\hat{J}'_{z}|q;j,m'_{z}\rangle=m'_{z}|q;j,m'_{z}\rangle,\\
\hat{\mathbf{J}}^2|q;j,m'_{z}\rangle=j(j+1)|q;j,m'_{z}\rangle,\\
\hat{q}|q;j,m'_{z}\rangle=q|q;j,m'_{z}\rangle.
\end{gather}
The spectrum is now continuous. When $q=0$, the effect of $\hat{J}_{x}$ vanishes, and the ground-state is just that of the atoms in the $z$-direction. On the contrary, for $q\to\pm\infty$, the operator $\hat{J}_{x}$ dominates.

\subsection{Quasi-integrability of the Dicke model, the adiabatic approximation}

The Born-Oppenheimer approximation (BOA) is widely  employed in atomic and molecular physics to separate the fast dynamics of the electrons from  that of the much heavier atomic nuclei. Roughly, this approximation consists of treating the slow variables as constant (adiabatically changing) parameters for the Hamiltonian of the fast variables. After solving this  effective Hamiltonian for the fast variables, one unfreezes  the slow variables and considers a second Hamiltonian for them, where the   fast variables are replaced   by their respective expectation values. A similar separation of slow bosons (pseudospin) and  fast pseudospin (bosons) variables can be performed in the Dicke model, depending on their parameters $\omega$, $\omega_0$ and $\gamma$. The latter two integrable limits of the Dicke model,  $\omega_0 \rightarrow 0$ and $\omega \rightarrow 0$, are extreme cases where the BOA becomes exact and  can be used to obtain, by using the BOA,  very good approximations to the exact solution of the complete Dicke Hamiltonian at low energies, either by considering  slow  bosons  with respect to the pseudospin or vice versa.

\subsubsection{Fast pseudospin and slow bosons}

Let us first consider the situation where the pseudospin dynamics is much faster than that of the bosonic variables. To treat this case, we use the exact solution for the zero bosonic frequency ($\omega\rightarrow 0$) in Eq.~\eqref{eqn:exactsolution2} to express a general state as
\begin{equation}
    |\Psi\rangle=\sum_{m'_z}\psi_{m_z'}(q)|q;j,m'_z\rangle.
\end{equation}
In terms of the components $\psi_{m_z'}(q)$, the eigenvalue equation for the whole Dicke Hamiltonian can be written as~\cite{Relano2016EPL} 
\begin{equation}
    \label{eqn:BOAfastpseudospin}
    \left(\hat{H}_{\text{eff}}-E\right)\psi_{m'_z}(q)
     + \omega\sum_{m''_z}\left[\langle q;j,m'_z| \hat{p}|q;j,m''_z\rangle \hat{p}\psi_{m''_z}(q)+\frac{\psi_{m''_z}(q)}{2}\langle q;j,m'_z| \hat{p}^2|q;j,m''_z\rangle\right]=0 ,
\end{equation}
where
\begin{equation}
    \label{eqn:BOAfastpseudospinHamiltonian}
    \hat{H}_{\text{eff}} = \frac{\omega}{2}\left(\hat{p}^2+q^2\right)+\omega_{0}\sqrt{1+G_0^2 q^2}m'_z ,
\end{equation}
and $\hat{p}=-id/dq$. When $\omega=0$, we retrieve the exact solution in Eq.~\eqref{eqn:exactsolution2}. For finite $\omega$, we can still obtain a good approximation by considering the effective Hamiltonian for the bosonic variables in Eq.~\eqref{eqn:BOAfastpseudospinHamiltonian} only, and neglecting the second term in Eq.~\eqref{eqn:BOAfastpseudospin}, which indeed holds if $\langle q;j,m'_z| \hat{p}|q;j,m''_z\rangle\approx 0$. This is the BOA for the Dicke model, which corresponds to the case where the dynamics of the bosonic variables is very slow compared to the dynamics of the pseudospin. A detailed analysis of the  non-adiabatic coefficients, $\langle q;j,m'_z|\hat{p}|q;j,m''_z\rangle$, shows that the BOA is a good approximation not only for $\omega\ll \omega_0$, but whenever $\omega^2 \omega_0\ll 8 \gamma^3$ is fulfilled~\cite{Relano2016EPL}. Therefore, even in the resonant case, $\omega=\omega_0$, the fast-pseudospin BOA provides a good approximation if the coupling is large enough, $\gamma\gg \omega$. 

\begin{figure}[t!]
    \centering
    \begin{tabular}{cc}
       \includegraphics[height=0.25\textwidth]{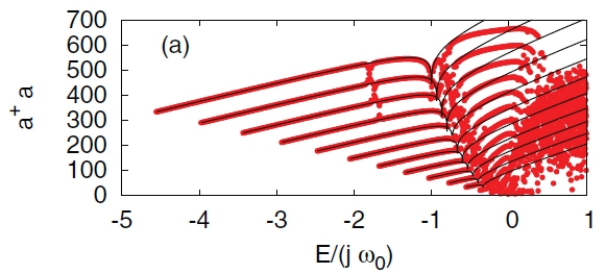}  & 
       \includegraphics[height=0.25\textwidth]{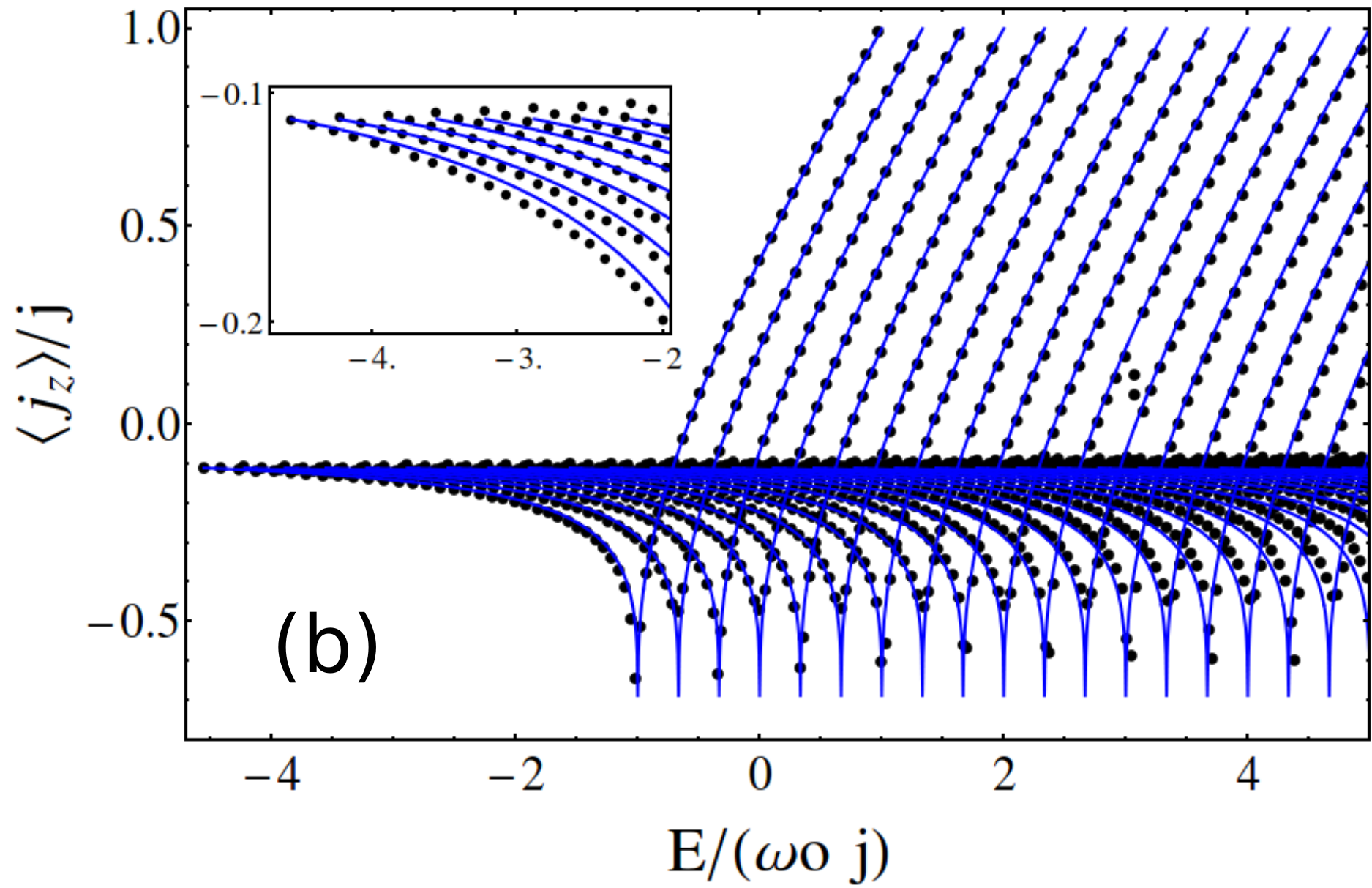}
    \end{tabular}
    \caption{(a) Expectation values of the number of photons $\hat{a}^{\dag}\hat{a}$ for eigenstates of the Dicke Hamiltonian versus their respective eigenvalues. The eigenvalues are scaled to the value $j \omega_{0}$. The red dots are numerical exact results and the black solid lines are obtained with the BOA for fast pseudospin. Hamiltonian parameters: $\omega=\omega_{0}/5$, $\gamma=3\gamma_{\text{c}}$, and $j=15$. The critical coupling parameter is $\gamma_{\text{c}}=\sqrt{\omega\omega_{0}}/2$. Figure taken from Ref.~\cite{Relano2016EPL}.
    (b) Expectation values of the $z$-component of pseudospin $\hat{J}_{z}$. The expectation values are scaled to the system size $j$ and the eigenvalues to the value $\omega_{0} j$. The black dots are numerical exact results and the blue solid lines are obtained from the BOA for fast bosons. The inset magnifies the low-energy region. Hamiltonian parameters: $\omega=20\omega_{0}$, $\gamma=3\gamma_{\text{c}}$, and $j=60$. The critical coupling parameter is $\gamma_{\text{c}}=\sqrt{\omega\omega_{0}}/2$. Figure taken from Ref.~\cite{Bastarrachea2017JPA}.
    }
    \label{fig:BOA}
\end{figure}

It is worth emphasizing that the fast pseudospin BOA consists of two successive diagonalizations. The first one is obtained by considering in the complete Dicke Hamiltonian the bosonic variables, $\hat{a}=(\hat{q}+i\hat{p})/2$ and $\hat{a}^\dagger$, as fixed (adiabatically changing) parameters, leading to a Hamiltonian that is solved by the rotation in Eqs.~\eqref{eqn:rotation1} and~\eqref{eqn:rotation}. After this first diagonalization, one derives the effective Hamiltonian for the bosonic variables given by Eq.~\eqref{eqn:BOAfastpseudospinHamiltonian}. This second Hamiltonian is equivalent to a non-relativistic particle moving in an effective potential (the adiabatic potential),  
\begin{equation}
    V_{m'_z}^{\text{ad}}(q)=\frac{\omega}{2}q^2+\omega_{0}\sqrt{1+G_0^2 q^2}m'_z , 
\label{eqn:adiabaticpotential}
\end{equation}
which depends on  the quantum numbers $m'_z$  from  the first diagonalization. One gets an approximation of the exact solution by diagonalizing this second Hamiltonian. Figure~\ref{fig:BOA}(a) shows the degree of accuracy of the adiabatic approximation compared to exact results obtained numerically. Generally, the BOA describes very well the exact results in an energy interval that extends from the ground state up to a very high excitation energy.

The fast-pseudospin BOA applied to the Dicke and related models has been discussed in several works~\cite{Liberti2006-2,Larson2012}. For example, in Ref.~\cite{Sainz_2008}, an expansion of the adiabatic potential in Eq.~\eqref{eqn:adiabaticpotential} was obtained as a particular case of a generic Hamiltonian describing an interaction between two quantum systems  in the case where the characteristic frequency of one  system  is 
lower than the corresponding frequency of the second one. The BOA for fast pseudospin has been used to gain insight into different physical aspects of the Dicke model, such as the scaling of its ground-state entanglement~\cite{Liberti2006}, the tunneling-driven splitting between its ground and first excited state~\cite{ChenChen2007}, and its dynamical phase diagram~\cite{Lewis-Swan2021}.

\subsubsection{Fast bosons and slow pseudospin}

When the pseudospin dynamics is much slower than the bosonic dynamics, we can also use the BOA. Within this approximation, the eigenvalue equation for the whole Dicke Hamiltonian is once again solved in two consecutive diagonalizations. In this case, the first diagonalization consists in considering the pseudospin variables as fixed (adiabatically changing) parameters and solving for the bosonic variables using the displacement in  Eq.~\eqref{eqn:operator_displacement}. This leads to the following  Hamiltonian~\cite{Bastarrachea2017JPA}
\begin{equation}
    \hat{H}_{\text{B}}= \omega \left(\hat{A}^\dagger \hat{A}-G^2 \hat{J}_x^2\right)+\omega_0 \hat{J}_z, 
\end{equation}
whose eigenvalues $\hat{A}^\dagger \hat{A}|N\rangle=N|N\rangle$ [see Eq.~\eqref{eq:AdAeig}] define an effective Hamiltonian for the slow pseudospin variables,
\begin{equation}
    \hat{H}_{\text{P}}=\omega N +\omega_0 \hat{J}_z-\omega G^2 \hat{J}_x^2 ,
\end{equation}
which is a Lipkin-Meshkov-Glick Hamiltonian~\cite{Lipkin1965} with an added constant $\omega N$.
The diagonalization of this effective Hamiltonian is the second step of the BOA and allows one to obtain an excellent approximation to the exact results [see Fig.~\ref{fig:BOA}(b)]. This approximation has been applied to the Rabi model in Refs.~\cite{Irish2005,Irish2007} and the Dicke model, including leakage of bosons, in Ref.~\cite{Keeling2010}.

\subsubsection{Regions of validity of the adiabatic approximation}

\begin{figure}[t!]
    \centering
    \includegraphics[width=0.6\textwidth]{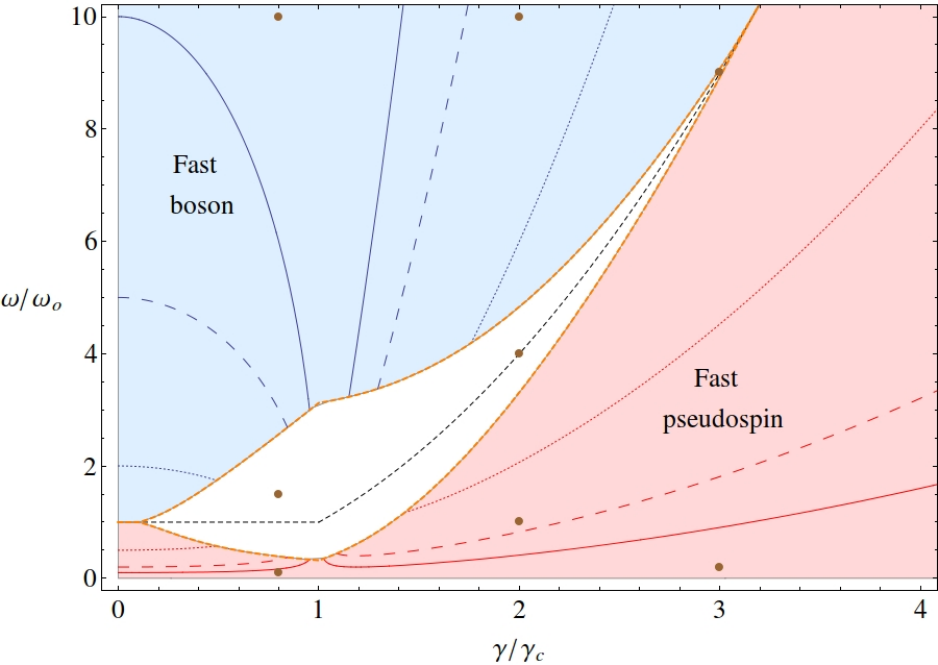}
    \caption{Parameter space of the Dicke model where the fast-pseudospin (red region) and fast-bosons (blue region) BOAs are expected to work. The dotted, dashed, and solid lines represent the ratios $\omega_{\text{P}}^{\text{(BOA)}}/\omega_{\text{B}}^{\text{(BOA)}}=2,5$, and $10$ in the red region and $\omega_{\text{B}}^{\text{(BOA)}}/\omega_{\text{P}}^{\text{(BOA)}}=2,5$ and $10$ in the blue region, respectively. The two BOAs are expected to work better as these ratios increase. The critical coupling parameter is $\gamma_{\text{c}}=\sqrt{\omega\omega_{0}}/2$. Figure taken from Ref.~\cite{Bastarrachea2017JPA}.
    }
    \label{fig:BOAvalidity}
\end{figure}

To determine the regions in the parameter space of the Dicke Hamiltonian, where the BOAs are valid, at least for the low energy region, the large $j$ semi-classical limit is considered. In this semi-classical limit, the frequency of the bosonic, $\omega_{\text{B}}^{\text{BOA}}$, and of the pseudospin, $\omega_{\text{P}}^{\text{BOA}}$, variables can be analytically calculated, respectively, within the fast-pseudospin BOA and the fast-bosons BOA. These frequencies must satisfy the following relations
\begin{align}
    \frac{\omega_{\text{P}}^{\text{BOA}}}{\omega_{\text{B}}^{\text{BOA}}} \gg 1  \ \ \ \  & \text{for the validity of the fast-pseudospin BOA,} \label{eqn:BOAcond1a} \\
    \frac{\omega_{\text{P}}^{\text{BOA}}}{\omega_{\text{B}}^{\text{BOA}}} \ll 1  \ \ \ \ &\text{for the validity of the fast-bosons BOA}.   \label{eqn:BOAcond1b}
\end{align} 
On the other hand, the two  fundamental frequencies, $\omega_\pm$ ( $\omega_+\geq\omega_-$), of the Dicke Hamiltonian for small excitations  around its  ground-state can also be calculated analytically  in the  limit $j\rightarrow\infty$~\cite{Brandes2013,Bastarrachea2014PRAb,Pilatowsky2021NJP}. If the BOAs are a good description of the Dicke Hamiltonian,  in addition to the conditions in Eqs.~\eqref{eqn:BOAcond1a} and~\eqref{eqn:BOAcond1b}, the following requirements must hold
\begin{align}
     \omega_{\text{P}}^{\text{BOA}} \approx \omega_+ \ \ \ \ \text{and} \ \ \  \omega_{\text{B}}^{\text{BOA}} \approx \omega_-  \ \ \ \ &\text{for the validity of the fast pseudospin BOA,} \label{eqn:BOAcond2a} \\
    \omega_{\text{P}}^{\text{BOA}} \approx \omega_- \ \ \ \ \text{and} \ \ \  \omega_{\text{B}}^{\text{BOA}} \approx \omega_+ \ \ \ \ &\text{for the validity of the fast bosons BOA}. \label{eqn:BOAcond2b}
\end{align}
Strictly speaking, these equalities are satisfied only in the integrable limits. However, they are approximately fulfilled in ample regions of the parameter space. The diagram in Fig.~\ref{fig:BOAvalidity} is obtained by considering a tolerance of $5 \%$ for the equalities given by Eqs.~\eqref{eqn:BOAcond2a} and~\eqref{eqn:BOAcond2b}~\cite{Bastarrachea2017JPA}. The red region indicates the parameters for which the conditions in Eqs.~\eqref{eqn:BOAcond1a} and~\eqref{eqn:BOAcond2a} (with a $5 \%$ tolerance) are met, while the blue region signals the parameter region where Eqs.~\eqref{eqn:BOAcond1b} and~\eqref{eqn:BOAcond2b} (also with a $5 \%$ tolerance) are simultaneously satisfied. A rough but simple estimate for the validity of either BOA is as follows. For $\gamma<\gamma_{\text{c}}$, where $\gamma_{\text{c}}=\sqrt{\omega\omega_0}/2$ is the critical coupling parameter of the Dicke model for the normal-to-superradiant quantum phase transition (see Sec.~\ref{subsec:QPT}), the condition for the validity of the fast-pseudospin BOA is $\omega/\omega_0\ll 1$
and for the fast-bosons BOA, it is $\omega/\omega_0\gg1$ , whereas for $\gamma>\gamma_{\text{c}}$ the conditions change to $\omega/\omega_0\ll \gamma^{2}/\gamma_{\text{c}}^{2}$ for the fast-pseudospin BOA and $\omega/\omega_0\gg \gamma^{2}/\gamma_{\text{c}}^{2}$ for the fast-bosons BOA.

\subsection{Numerical solutions to the Dicke Hamiltonian}

To obtain the Hamiltonian spectrum of the complete Dicke model, one has to resort to numerical solutions. To solve the time-independent Schr\"{o}dinger equation $\hat{H}_{\text{D}}|E_{k}\rangle = E_{k}|E_{k}\rangle$, where $E_{k}$ and $|E_{k}\rangle$ are its eigenvalues and eigenvectors, respectively, we need to choose a basis to write the matrix representation of the Hamiltonian. A natural choice is the Fock basis in Eq.~\eqref{eqn:fock_basis},  which is the set of exact eigenstates of the Hamiltonian in Eq.~\eqref{eqn:fock_hamiltonian}.
In this basis, the elements of the Dicke Hamiltonian matrix are~\cite{Bastarrachea2011}
\begin{equation}
    \langle n';j,m'_{z}|\hat{H}_{\text{D}}|n;j,m_{z}\rangle = \left(\omega n+\omega_{0}m_{z}\right)\delta_{n',n}\delta_{m'_{z},m_{z}} + h(n',n;j,m'_{z},m_{z}),
\end{equation}
where
\begin{equation}
    h(n',n;j,m'_{z},m_{z}) = \frac{\gamma}{\sqrt{\mathcal{N}}}\left(\sqrt{n+1}\delta_{n',n+1}+\sqrt{n}\delta_{n',n-1}\right)  \left[c_{+}(j,m_{z})\delta_{m'_{z},m_{z}+1} + c_{-}(j,m_{z})\delta_{m'_{z},m_{z}-1}\right],
\end{equation}
and $c_{\pm}(j,m_{z}) = \sqrt{j(j+1)-m_{z}(m_{z}\pm1)}$.

Since the bosonic subspace is unbounded, the resulting Hamiltonian matrix has an infinite dimension. One needs to truncate the Hilbert space for a given number of bosons $n_{\max}$, where $n_{\max}$  is an eigenvalue of the  number operator $\hat{n}=\hat{a}^{\dag}\hat{a}$. This results in a matrix with the finite dimension $\mathcal{D}_{\text{F}}=(2j+1)(n_{\max}+1)$. With the truncation, we  face the {\it convergence} problem. Truncating the Hilbert space makes the eigenstates of the Hamiltonian incomplete, so one has to ensure that all or most of the wave function is contained in the truncated Hilbert space after diagonalization. This comes with its challenges, as the convergence depends not only on the energy regime, but also on the number $\mathcal{N}$ of two-level systems and the light-matter interaction strength $\gamma$. As either $\mathcal{N}$ or $\gamma$ increases,  $n_{\max}$ has to be increased accordingly, given that more bosons are required to describe a larger and stronger correlated system exactly.

In the Fock basis, a lower bound for the value of $n_{\max}$ that ensures the convergence of the full collective ground state can be found using coherent states and adding 3 times the quadratic deviation. In the superradiant phase, it takes the form~\cite{Bastarrachea2014PS}
\begin{equation}
    n_{\max} \geq \mathcal{N}\gamma^{2} \left( 1-\frac{\gamma_{\text{c}}^{4}}{\gamma^{4}}\right)\left(1 + 3 \left[\mathcal{N}\gamma^{2}\left(1-\frac{\gamma_{\text{c}}^{4}}{\gamma^{4}}\right)\right]^{-\frac{1}{2}}\right).
\end{equation} 
Just for the convergence of the ground state, the minimum value of $n_{\max}$  scales linearly in $\mathcal{N}$ and quadratically in $\gamma$.  Consequently, a detailed description of higher energy states demands significant computational resources and processing time, which renders the diagonalization problem in the Fock basis impracticable. Hence, for studies where a large number of eigenstates is necessary, such as the analysis of the dynamical features of the system, one should look for alternative methods. An option is to use the efficient basis, as described next.

\subsubsection{Efficient basis}

To handle high energies, the efficient basis in Eq.~\eqref{eqn:efficient_basis}, which corresponds to the exact eigenstates of the Dicke Hamiltonian when $\omega_0=0$, is a better alternative than the Fock basis. The efficient basis was first employed in Ref.~\cite{Chen2008} to obtain an exact numerical solution of the ground state, and it was later shown to be useful also for excited states~\cite{Bastarrachea2014PS,Bastarrachea2014PRAa}.

In the efficient basis, the matrix elements of the Dicke Hamiltonian are given by
\begin{equation}
    \langle N';j,m_{x}'|\hat{H}_{\text{D}}|N;j,m_{x}\rangle = \omega\left(N-G^{2}m_{x}^{2}\right)\delta_{N',N}\delta_{m_{x}',m_{x}} + H(N',N;j,m_{x}',m_{x}), 
\end{equation}
where
\begin{gather}
    H(N',N;j,m_{x}',m_{x}) = -\frac{\omega_{0}}{2}\left[C_{+}(N',N;j,m_{x}',m_{x})\delta_{m_{x}',m_{x}+1} + C_{-}(N',N;j,m_{x}',m_{x})\delta_{m_{x}',m_{x}-1}\right]
\end{gather}
and $C_{\pm}(N',N;j,m_{x}',m_{x}) = 
 \sqrt{j(j+1)-m_{x}(m_{x}\pm1)} \langle N',\alpha_{m_{x}'}|N,\alpha_{m_{x}\pm1}\rangle$ are functions containing the overlaps between displaced Fock states [Eq.~\eqref{eqn:displacedFockStates}], given by~\cite{deOliveira1990}
\begin{align}
    \langle N',\alpha_{m_{x}'}|N,\alpha_{m_{x}}\rangle & = \langle N' |\hat{D}(G{m_x'}-Gm_x)|N\rangle  \\
    & = \sqrt{\frac{N!}{N'!}}
    [G(m_x'-m_x)]^{N'-N}e^{-G^{2}(m_x'-m_x)^{2}/2}L_N^{(N'-N)}\left
    [G^{2}(m_x'-m_x)^{2}\right] , \nonumber
\end{align}
where $L_N^{(N'-N)}$ is the associated Laguerre polynomial.
These overlaps can also be written as a series expansion~\cite{Bastarrachea2011,Bastarrachea2014PRAb},
\begin{equation}
    \langle N',\alpha_{m_{x}'}|N,\alpha_{m_{x}}\rangle  = \left\{\begin{array}{ll} (-1)^{N}D_{N',N} & \text{if } m_{x}>m_{x}' \\ (-1)^{N'}D_{N',N} & \text{if } m_{x}<m_{x}' \\ \delta_{N',N} & \text{if } m_{x}=m_{x}' \end{array}\right. ,
\end{equation}
where
\begin{equation}
    D_{N',N} = e^{-G^{2}/2}\sum_{k=0}^{\min(N',N)}\frac{\sqrt{N'!N!}(-1)^{-k}G^{N'+N-2k}}{(N'-k)!(N-k)!k!}.
\end{equation}

Similarly to the Fock basis, a truncation value $N_{\max}$ is required for the efficient basis, where $N_{\max}$  is an eigenvalue of the modified bosonic number operator $\hat{N}=\hat{A}^{\dag}\hat{A}$. The resulting dimension of the truncated Hilbert space is $\mathcal{D}_{\text{E}}=(2j+1)(N_{\max}+1)$.  It was found~\cite{Chen2008,Bastarrachea2011,Bastarrachea2012,Bastarrachea2014PS,Hirsch2014} that in the efficient basis, the value of $N_{\max}$ necessary to obtain a converged spectrum for thousands of converged excited states ($\mathcal{N}>60$ and arbitrary values of $\gamma$) is drastically smaller than in the Fock basis. 

The convergence criterion amounts to keeping the tails of the eigenstates very small for a given numerical tolerance~\cite{Bastarrachea2014PS,Hirsch2014}. The efficient basis is $\gamma$-dependent, because of its functional dependence through the displacement in Eq.~\eqref{eqn:operator_displacement}, so it automatically adapts to describe the superradiant phase and the normal phase (recovering the non-interacting limit, when $\gamma\to 0$). The efficient basis is better adapted than the Fock basis to describe the polaritonic nature of the exact solutions, i.e., the light-matter superposition, as it describes strong correlated states.

\begin{figure}[t!]
    \centering
    \includegraphics[width=0.9\textwidth]{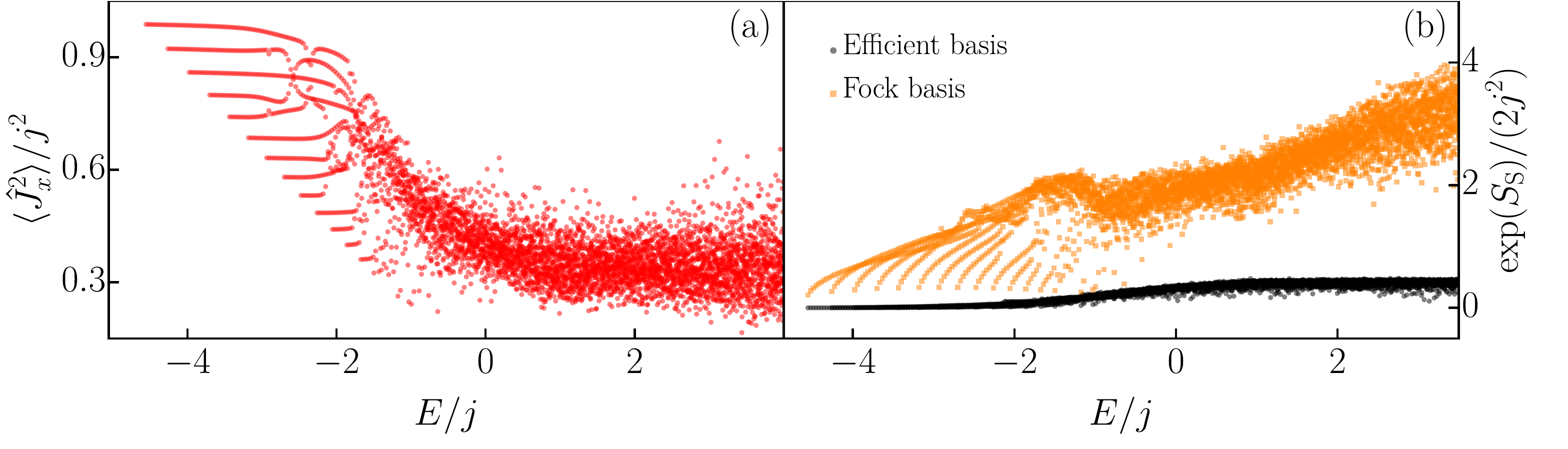}
    \caption{(a) Expectation values of the operator $\hat{J}_x^2$ for eigenstates of the Dicke Hamiltonian with positive parity. The expectation values are scaled to the value $j^{2}$ and the eigenvalues to the system size $j$.
    (b) Exponential of the Shannon entropy [Eq.~\eqref{eqn:shannon_entropy}] for eigenstates of the Dicke Hamiltonian with positive parity. The exponential of the entropy is scaled to the value $2j^{2}$ and the eigenvalues to the system size $j$. The Shannon entropy was computed using the parity projected efficient basis [Eq.~\eqref{eqn:ecbp_basis}] (black dots) for the positive parity sector and the Fock basis [Eq.~\eqref{eqn:fock_basis}] (orange squares) for the same parity sector [see Eq.~\eqref{eqn:Fock_parity}], respectively. The truncation value used with the efficient basis is $N_{\max}=300$ and with the Fock basis, it is $n_{\max}=3N_{\max}=900$.
    Hamiltonian parameters: $\omega = \omega_0 = 1$, $\gamma = 3\gamma_{\text{c}}$, and $j = 30$. The critical coupling parameter is $\gamma_{\text{c}}=\sqrt{\omega\omega_{0}}/2$.
    }
    \label{fig:Peres_lattice_Jx2}
\end{figure}

To visually demonstrate the advantage of using the efficient basis for the Dicke Hamiltonian, we show in Fig.~\ref{fig:Peres_lattice_Jx2}(a) the expectation value of the operator $\hat{J}_{x}^{2}$, which is calculated for $N_{\text{c}}=6206$ converged eigenstates using the truncation $N_{\max}=300$. We choose $\hat{J}_{x}^{2}$, because it is proportional to the interacting term in the Dicke Hamiltonian and, as discussed before, as the coupling strength increases and we move to higher energies, the number of photons needed for the convergence of the eigenstates increases. Therefore, the operator $\hat{J}_{x}^{2}$ is  a good choice for exhibiting the advantages of the efficient basis. To reproduce Fig.~\ref{fig:Peres_lattice_Jx2}(a) using the Fock basis, we would need a truncation value $n_{\max}$ about three times larger than $N_{\max}$~\cite{Bastarrachea2014JP,Villasenor2023}.

To better explain this statement, we show in Fig.~\ref{fig:Peres_lattice_Jx2}(b) the Shannon entropy,
\begin{equation}
    \label{eqn:shannon_entropy}
    S_{\text{S}}=-\sum_{x,y}p_{x,y}\ln(p_{x,y})  ,  
\end{equation}
for the Dicke Hamiltonian eigenstates $|E_{k}\rangle=\sum_{x,y}c_{x,y}|x;j,y\rangle$ using the efficient and the Fock basis. The Shannon entropy measures the level of delocalization of the eigenstates in a chosen basis. In Eq.~\eqref{eqn:shannon_entropy}, $p_{x,y}=|c_{x,y}|^{2}=|\langle x;j,y|E_{k}\rangle|^{2}$,  where  $(x,y)=(n,m_{z})$ identifies the Fock basis and $(x,y)=(N,m_{x})$ corresponds  to the efficient basis. We see that the eigenstates are composed of a much lower number of states of the efficient basis than of the Fock basis in all energy ranges. In Fig.~\ref{fig:Peres_lattice_Jx2}(b), the truncation used with the Fock basis was 3 times larger than the truncation of the efficient basis, $n_{\max}=3N_{\max}=900$.

For other methods to address analytically and numerically the Dicke Hamiltonian and its variants, see Refs.~\cite{Kirton2019,LeBoite2020}.

\subsubsection{Parity projected efficient basis}

The Dicke Hamiltonian commutes with the parity operator $\hat{\Pi}$ in Eq.~\eqref{eqn:parity_operator}, which separates the Hilbert space into two subspaces. The Fock basis is an eigenbasis of the parity operator and satisfies the eigenvalue equation
\begin{equation}
    \label{eqn:Fock_parity}
    \hat{\Pi}|n;j,m_{z}\rangle = (-1)^{n+m_{z}+j}|n;j,m_{z}\rangle = p|n;j,m_{z}\rangle,
\end{equation}
where each eigenvalue $p=\pm1$ identifies the positive and negative parity, respectively. However, the efficient basis states in Eq.~\eqref{eqn:efficient_basis} are not eigenstates of the parity operator. To account for this discrete symmetry, it is necessary to build up an efficient basis with parity projected states~\cite{Bastarrachea2014PRAb}. It can be shown that because of Eq.~\eqref{eq:Nvac} and the properties of Glauber coherent states, we have the following relations
\begin{gather}
    \hat{\Pi}|j,m_{x}\rangle = |j,-m_{x}\rangle, \\
    \hat{\Pi}(\hat{a}^{\dagger})^{k}|\alpha\rangle=(-1)^{k}(\hat{a}^{\dagger})^{k}|-\alpha\rangle .
\end{gather}
Hence, as $\alpha_{m_{x}}=-Gm_{x}$ the global result of applying the parity operator to the efficient basis states is
\begin{equation}
    \hat{\Pi}|N;j,m_{x}\rangle = (-1)^{N}|N;j,-m_{x}\rangle.
\end{equation}
By representing the parity operator in the subspace $\{|N;j,\pm m_{x}\rangle\}$
\begin{equation}
    \hat{\Pi} = (-1)^{N}\left(\begin{array}{cc}0 & 1\\1 & 0\end{array}\right),
\end{equation}
we obtain its eigenvectors 
\begin{equation}
    \label{eqn:ecbp_basis}
    |N;j,m_{x};p\rangle = \frac{|N;j,m_{x}\rangle+p(-1)^{N}|N;j,-m_{x}\rangle}{\sqrt{2(1+\delta_{m_{x},0})}}
\end{equation}
and eigenvalues $p=\pm1$, which satisfy the eigenvalue equation $\hat{\Pi}|N;j,m_{x};p\rangle=p|N;j,m_{x};p\rangle$. The basis in Eq.~\eqref{eqn:ecbp_basis} is then called {\it parity projected efficient basis}.

\subsection{Open Dicke model}

Significant efforts have been dedicated to studying quantum systems that include driving or coupling to an environment. External drives and dissipation yield a rich set of dynamical responses~\cite{Nagy2010,Dasgupta2015,Gelhausen2018,Sinha2021} that include the presence of steady states~\cite{Nagy2011}, formation of time crystals~\cite{Gong2018,Zhu_2019,Mattes2023}, and thermalization~\cite{Ray2016}. Under quasiperiodic  drives, the anisotropic Dicke model features a prethermal plateau that increases  with the driving frequency before heating to an infinite-temperature state~\cite{Das2023b}. In the limit of few emitters, energies have been shown to collapse as the driving field acting on a damped Tavis-Cummings model is increased, which marks the onset of a dissipative quantum phase transition~\cite{Karmstrand2024}. Other studies have explored dissipative quantum phase transitions in both the Dicke model~\cite{Larson2017,kirton2017,Kirton2019,Boneberg2022,Prasad2022,Tong2025} and the Rabi model~\cite{Hwang2018,Lyu2024}. The importance of dissipation has prompted further research into dissipative systems, particularly in the areas of classical and quantum chaos~\cite{Villasenor2024,Villasenor2024PRL,Ferrari2025,Mondal2026}, thermalization~\cite{Cipolloni2024,Singha2024,Almeida2026}, quantum evolution~\cite{Graefe2008,Graefe2010,Graefe2010JPA}, among others.

\subsubsection{Lindblad master equation}

A widely used and effective method for describing dissipative quantum systems is through Lindblad master equations in the Markov approximation~\cite{BreuerBook,CarmichaelBook1993,CarmichaelBook2002}. These equations incorporate dissipative effects by adding non-unitary terms or channels to the evolution of the system's state. This approach enables us to obtain spectra through diagonalization, perform quantum evolution, and link classical aspects with quantum aspects, thereby drawing a direct analogy with isolated systems~\cite{Haake1991Book}.

The Lindblad master equation~\cite{BreuerBook,CarmichaelBook1993,CarmichaelBook2002} for the open Dicke model (setting $\hbar=1$) is given by
\begin{equation}
    \label{eq:DickeLiouvillian}
    \frac{d\hat{\rho}}{dt} = \hat{\mathcal{L}}_{\text{D}}\hat{\rho} = -i[\hat{H}_{\text{D}},\hat{\rho}] + \kappa\left(2\hat{a}\hat{\rho}\hat{a}^{\dagger} - \{\hat{a}^{\dagger}\hat{a},\hat{\rho}\}\right),
\end{equation}
where $\hat{\rho}$ is the density operator of the system. The term $\hat{H}_{\text{D}}$ is the Dicke Hamiltonian presented in Eq.~\eqref{eqn:dicke_hamiltonian} and describes the unitary evolution. The bosonic operators $\hat{a}$ and $\hat{a}^{\dagger}$ represent the jump operators of the dissipative channel in Lindblad form that describes the nonunitary evolution. The parameter $\kappa$ is the cavity decay that regulates the dissipation strength, which is interpreted as the loss of photons within the cavity. The Dicke Liouvillian is a superoperator, $\hat{\mathcal{L}}_{\text{D}}\bullet=-i[\hat{H}_{\text{D}},\bullet] + \kappa\left(2\hat{a}\bullet\hat{a}^{\dagger} - \{\hat{a}^{\dagger}\hat{a},\bullet\}\right)$, which commutes with the parity superoperator $\hat{\mathcal{P}}\bullet=\hat{\Pi}\bullet\hat{\Pi} ^{\dagger}$, $[\hat{\mathcal{L}}_{\text{D}},\hat{\mathcal{P}}]=0$, and defines a weak symmetry~\cite{Buca2012,Albert2014,Lieu2020}. The eigenbasis of the parity superoperator identifies states of the system with well-defined parity, in analogy with the isolated system.

A quantum dissipative phase transition takes place at a critical coupling strength~\cite{Dimer2007,Kirton2019,Roses2020} in direct analogy with the isolated system (see Sec.~\ref{subsec:QPT}). For the dissipative system, the critical coupling is a function of the cavity decay 
\begin{equation}
    \gamma_{\text{c}}(\kappa)=\frac{\sqrt{\omega\omega_{0}}}{2}\sqrt{1+\frac{\kappa^{2}}{\omega^{2}}} ,
\end{equation}
defining two dissipative phases, a normal phase ($\gamma<\gamma_{\text{c}}(\kappa)$) and a superradiant phase ($\gamma>\gamma_{\text{c}}(\kappa)$). The case with null cavity decay $\kappa=0$ in the above equation recovers the critical coupling of the isolated system $\gamma_{\text{c}}=\sqrt{\omega\omega_{0}}/2$.

\subsubsection{Dissipative quantum chaos}

Significant contributions to understanding dissipative quantum chaos have been made through the study of the open Dicke model~\cite{Villasenor2024,Villasenor2024PRL}. The initial effort to characterize quantum chaos in dissipative quantum systems was inspired by the framework used for isolated quantum systems (see Sec.~\ref{subs:quantchaos}), where classical chaotic behavior is associated with spectral correlations of Ginibre ensembles~\cite{Ginibre1965}. This framework emerged as a conjecture known as the {\it conjecture of dissipative quantum chaos}~\cite{Grobe1988,Grobe1989}, later referred to as the Grobe-Haake-Sommers conjecture~\cite{Akemann2019}.

Although the original study of the conjecture focused on a dissipative kicked system, the link between chaotic motion (exemplified by the emergence of chaotic attractors) and spectral correlations (present in the eigenvalues of non-Hermitian spectra) was later taken for granted. Significant research has explored this conjecture to characterize quantum chaos in dissipative quantum many-body systems, which do not have a clear classical limit~\cite{Akemann2019,Jaiswal2019,Sa2020,Rubio2022,Garcia2023PRD,Akemann2025}. These studies aim to provide symmetry-based classifications for these systems~\cite{Hamazaki2020,Garcia2022,Kawabata2023,Sa2023}. 

Recent research on the open Dicke model has highlighted the breakdown of the conjecture of dissipative chaos~\cite{Villasenor2024PRL}. This finding was later confirmed in the original kicked system~\cite{Villasenor2025}, which also demonstrates a breakdown over a broad range of parameters. Since then, several significant descriptions have been proposed to characterize chaos in dissipative systems. Some approaches focus on the dynamical behavior by analyzing both transient and long-term motion~\cite{Ferrari2025,Mondal2026}, while others aim to systematically identify fingerprints associated with classical structures in the Liouvillian spectrum~\cite{Dutta2025,Cai2025,Li2025Arxiv}. The latest research indicates that the non-normality of quantum operators is a crucial aspect for developing an accurate test of dissipative quantum chaos~\cite{Naves2026}.

The search for a universal description of dissipative quantum chaos remains ongoing; however, key findings from the open Dicke model demonstrate the significance of this spin-boson system as a fundamental quantum system.

\subsubsection{Extensions, variants, and experimental realizations of the open Dicke model}

Further considerations can be included in the open Dicke model, such as two-photon interactions or driving, previously discussed in Sec.~\ref{subsec:RegimesDicke}. In the dissipative two-photon Dicke model, depending on the dissipation rate, a phase diagram can manifest stable, bistable, and unstable phases~\cite{Garbe2020}. Other studies have shown that the inclusion of two-photon loss restores stability in the two-photon Dicke model and superradiant states can emerge~\cite{Shah2025}. In the semiclassical limit, the anisotropic two-photon Dicke model with a dissipative bosonic field exhibits localized fixed points that reflect the spectral collapse of the isolated system~\cite{Li2022}.

An open two-component Dicke model presents a nonstationary phase, which is explained in terms of spontaneous breaking of the parity-time symmetry~\cite{Rodriguez2023}. The case of $\mathcal{N}$ identical three-level atoms interacting with a single-mode cavity field with dissipation can represent a shaken atom-cavity system~\cite{Skulte2021} and allow for the preparation of  coherent atomic states with high fidelity~\cite{Fan2023}. A large ensemble of driven  solid-state emitters strongly coupled to a nanophotonic resonator shows highly nonlinear optical emission~\cite{Lei2023}. An open generalized Dicke model with two dissimilar atoms in the regime of ultra-strongly-coupled cavity QED exhibits multiple resonances in the cavity spectra and anticrossing features~\cite{Akbari2023}. The presence of chaos and quantum scars has been reported in the periodically kicked Dicke model~\cite{Sinha2021}.

In an open system, a first-order dissipative phase transition could be observed in an optical cavity QED setting, where the stable co-existing phases are quantum states, exploiting the collective enhancement of the coupling between the atoms and the cavity field~\cite{Gabor2023}. Other implementations explored the dynamical critical properties of the dissipative phase transition by performing quenches across the phase boundary~\cite{Klinder2015PNAS}. It has been proposed that the open Dicke model should allow for the realization of a universal quantum heat machine that could function as an engine, refrigerator, heater, or accelerator, showing an improvement in its efficiency around the critical value of the phase transition parameter of the model~\cite{Xu2024}.

\section{Classical Dicke model and phase space}
\label{sec:ClassicalDickeModel}
The algebraic nature of the Dicke model allows one to define a semiclassical approximation based on a mean-field approximation that uses a variational wave function. The basic idea is to start with an arbitrary initial state and assume that the quantum dynamics will remain close to the classical trajectory as time passes~\cite{Bakemeier2013}. The usual procedure consists of taking the expectation value of the quantum Hamiltonian under coherent states, because these states minimize the Heisenberg uncertainty principle and can be considered as the most classically accessible quantum states~\cite{Perelomov1986Book}. Alternatively, this classical limit  can be derived by considering the path integral representation of the quantum propagator in terms of coherent states and performing a saddle-point approximation. The first-order term resulting from this approximation leads to a classical system with a Hamiltonian given by the expectation value of the quantum Hamiltonian in coherent states~\cite{Ribeiro2006}.

The Hilbert space of the Dicke model is a product space composed of a bosonic subspace $\mathcal{H}_{\text{B}}$ and an atomic subspace $\mathcal{H}_{\text{A}}$, that is $\mathcal{H}_{\text{D}}=\mathcal{H}_{\text{B}}\otimes\mathcal{H}_{\text{A}}$. In this way, the coherent states used to obtain the classical Dicke model are a tensor product between Glauber coherent states, $|\alpha\rangle$, associated with the bosonic subspace and Bloch coherent states, $|z\rangle$, associated with the atomic subspace. Properties of both coherent states are given in~\ref{app:CoherentStates}.

\subsection{Classical Dicke Hamiltonian}

The tensor product between  $|\alpha\rangle$ and $|z\rangle$ defines a Glauber-Bloch coherent state,
\begin{equation}
    \label{eqn:glauber_bloch_coherent_state}
    |\mathbf{x}\rangle = |\alpha\rangle\otimes|z\rangle = |q,p\rangle\otimes|Q,P\rangle,
\end{equation}
where the parameters $\alpha,z\in\mathbb{C}$ are defined as
\begin{gather}
    \label{eqn:bosonic_parameter}
    \alpha = \sqrt{\frac{j}{2}} \,(q+i p), \\
    \label{eqn:atomic_parameter}
    z = \frac{Q+i P}{\sqrt{4-Q^{2}-P^{2}}},
\end{gather}
which gives the representations
\begin{gather}
    |q,p\rangle = e^{-(j/4)\left(q^{2}+p^{2}\right)}e^{\left[\sqrt{j/2}\left(q+ip\right)\right]\hat{a}^{\dagger}}|0\rangle, \\
    |Q,P\rangle = \left(1-\frac{Q^2+P^2}{4}\right)^{j}e^{\left[\left(Q+iP\right)/\sqrt{4-Q^2-P^2}\right]\hat{J}_{+}}|j,-j\rangle,
\end{gather}
where $|0\rangle$ is the vacuum state of the field and $|j,-j\rangle$ represents all atoms in their ground state. The parameters $\alpha$ and $z$ are associated with canonical variables of phase space, $\mathbf{x}=(q,p;Q,P)$. The phase space of the Dicke model, $\mathcal{M}$, is four-dimensional. Both bosonic $(q,p)$ and atomic $(Q,P)$ variables satisfy the Poisson brackets $\{q,p\} = \{Q,P\} = 1$.

Taking the expectation value of the Dicke Hamiltonian [Eq.~\eqref{eqn:dicke_hamiltonian}] under Glauber-Bloch coherent states $|\mathbf{x}\rangle=|\alpha\rangle\otimes|z\rangle$, and scaling it by $j$, leads to
\begin{align}
    \label{Eq:hdx}
    h_{\text{D}}(\mathbf{x}) & = \frac{1}{j}\langle\mathbf{x}|\hat{H}_{\text{D}}|\mathbf{x}\rangle = \frac{1}{j}\left(\langle\mathbf{x}|\hat{H}_{\text{F}}|\mathbf{x}\rangle + \langle\mathbf{x}|\hat{H}_{\text{A}}|\mathbf{x}\rangle + \langle\mathbf{x}|\hat{H}_{\text{I}}|\mathbf{x}\rangle\right) = h_{\text{F}}(\mathbf{x}) + h_{\text{A}}(\mathbf{x}) + h_{\text{I}}(\mathbf{x}) \\
    & = \frac{1}{j}\left[\omega|\alpha|^{2}-j\omega_{0}\frac{1-|z|^{2}}{1+|z|^{2}}+\gamma\sqrt{2j}\frac{z+z^{\ast}}{1+|z|^{2}}\left(\alpha+\alpha^{\ast}\right)\right], \nonumber
\end{align}
where the field $h_{\text{F}}(\mathbf{x})$ and the atomic $h_{\text{A}}(\mathbf{x})$ Hamiltonians can be interpreted as the energy of two classical harmonic oscillators, and the atom-field interaction Hamiltonian $h_{\text{I}}(\mathbf{x})$ as the classical coupling energy between them,
\begin{gather}
    h_{\text{F}}(\mathbf{x}) = \frac{\omega}{2}\left(q^{2}+p^{2}\right), \\
    h_{\text{A}}(\mathbf{x}) = \frac{\omega_{0}}{2}\left(Q^{2}+P^{2}\right)-\omega_{0}, \\
    h_{\text{I}}(\mathbf{x}) = 2\gamma qQ\sqrt{1-\frac{Q^{2}+P^{2}}{4}}.
\end{gather}
Under the last semiclassical approximation, the corresponding classical Dicke Hamiltonian is then written as
\begin{equation}
    \label{eqn:classical_dicke_hamiltonian}
    h_{\text{D}}(\mathbf{x}) = \frac{\omega}{2}\left(q^{2}+p^{2}\right) + \frac{\omega_{0}}{2}\left(Q^{2}+P^{2}\right)-\omega_{0} + 2\gamma qQ\sqrt{1-\frac{Q^{2}+P^{2}}{4}},
\end{equation}
where the scaling parameter $j$ in Eq.~\eqref{Eq:hdx} is associated with the scaled classical energy (classical energy shell) given by 
\begin{equation}
    \label{eqn:classical_energy_shell}
    h_{\text{D}}(\mathbf{x}) = \epsilon = \frac{E}{j}.
\end{equation}
The scaling by $j$ ensures that the classical dynamics does not depend on the system size and it defines an effective Planck constant, $\hbar_{\text{eff}}=1/j$~\cite{Ribeiro2006}. 

Following the semiclassical approximation presented above, we can obtain a classical limit of the generalized Dicke Hamiltonian introduced in Eq.~\eqref{eq:HD_generalized}. We found the expression
\begin{equation}
    \label{eqn:generalized_classical_dicke_hamiltonian}
    h_{\text{GD}}(\mathbf{x}) = \frac{\omega}{2}\left(q^{2}+p^{2}\right) + \frac{\omega_{0}}{2}\left(Q^{2}+P^{2}\right)-\omega_{0} + [(\gamma_{-}+\gamma_{+})qQ + (\gamma_{-}-\gamma_{+})pP]\sqrt{1-\frac{Q^{2}+P^{2}}{4}},
\end{equation}
where the case $\gamma_{-}=\gamma_{+}=\gamma$ gives the usual classical Dicke Hamiltonian presented in Eq.~\eqref{eqn:classical_dicke_hamiltonian}. Moreover, the case $\gamma_{-}=\gamma$ and $\gamma_{+}=0$ identifies the integrable version of the classical Dicke model, the classical Tavis-Cummings Hamiltonian. Many classical aspects of the anisotropic Dicke Hamiltonian [Eq.~\eqref{eqn:generalized_classical_dicke_hamiltonian}] will be presented in Sec.~\ref{subs:classicalChaos}.

\subsection{Ground-state energy and quantum phase transition}
\label{subsec:QPT}

Around the  1970s, several authors pointed out the existence of a  phase transition  for the Dicke Hamiltonian~\cite{Mallory1969,Hepp1973AP,Hioe1973,Hepp1973PRA}.
This phase transition occurs in the thermodynamic limit ($\mathcal{N} \to \infty$ or, equivalently, $j \to \infty$) even   at zero temperature, leading to a quantum phase transition (QPT)  \cite{Hillery1985,Emary2003PRL,Emary2003PRE} when the coupling strength reaches the critical value $\gamma_{\text{c}}=\sqrt{\omega\omega_{0}}/2$. Below this value ($\gamma<\gamma_{\text{c}}$), the ground state of the system has no photons and no excited atoms on average, and it is called the {\it normal phase}. Above this value ($\gamma>\gamma_{\text{c}}$), the ground state has an average number of photons and excited atoms comparable to the total number of atoms within the system. This phase is called the {\it superradiant phase}. Superradiance means that the average photon emission is a non-zero collective emission~\cite{Gross1982}.

The origin of the QPT in the Dicke model can be understood using the classical domain, which coincides with the thermodynamic limit of the quantum model; this property distinguishes the QPT of the Dicke model from other QPTs arising from quantum fluctuations~\cite{Larson2017}. The lowest energy of the classical Dicke Hamiltonian is obtained by minimizing the classical energy surface $h_{\text{D}}(\mathbf{x})=\epsilon=E/j$, which gives the set of coordinates $\mathbf{x}_{\text{gs}}=(q_{\text{gs}},p_{\text{gs}};Q_{\text{gs}},P_{\text{gs}})$ that defines the surface with the minimum value $\epsilon_{\text{gs}}$. The minimization procedure consists in setting  the Hamilton's equations of motion [Eqs.~\eqref{eqn:hamilton_dicke_q}-\eqref{eqn:hamilton_dicke_P} with $\gamma_{-}=\gamma_{+}=\gamma$] equal to zero to determine the coordinates 
\begin{equation}
    \label{eqn:superradiant_phase_stationary_point}
    \mathbf{x}_{\text{gs}} = \left\{\begin{array}{ll} (0,0;0,0) & \text{if } \gamma<\gamma_{\text{c}} \\
    \left(-\dfrac{2\gamma}{\omega}\sqrt{1-\dfrac{\gamma_{\text{c}}^{4}}{\gamma^{4}}},0;\sqrt{2\left(1-\dfrac{\gamma_{\text{c}}^{2}}{\gamma^{2}}\right)},0\right) & \text{if } \gamma>\gamma_{\text{c}}
    \end{array}\right.  .
\end{equation}
These two solutions represent the two qualitatively different
phases. By substituting the coordinates $\mathbf{x}_{\text{gs}}$ in Eq.~\eqref{eqn:classical_dicke_hamiltonian}, the lowest energy can be found for both the normal and the superradiant phase,
\begin{equation}
    \label{eqn:classical_ground_state_energy}
    \epsilon_{\text{gs}} = h_{\text{D}}(\mathbf{x}_{\text{gs}}) = -\omega_{0}\left\{\begin{array}{ll} 1 & \text{if } \gamma<\gamma_{\text{c}} \\ \dfrac{1}{2}\left(\dfrac{\gamma^{2}}{\gamma_{\text{c}}^{2}}+\dfrac{\gamma_{\text{c}}^{2}}{\gamma^{2}}\right) & \text{if } \gamma>\gamma_{\text{c}}\end{array}\right..
\end{equation}

\begin{figure}[t!]
    \centering
    \begin{tabular}{c}
    (a) \hspace{0.35\textwidth} (b) \\
    \includegraphics[width=0.7\textwidth]{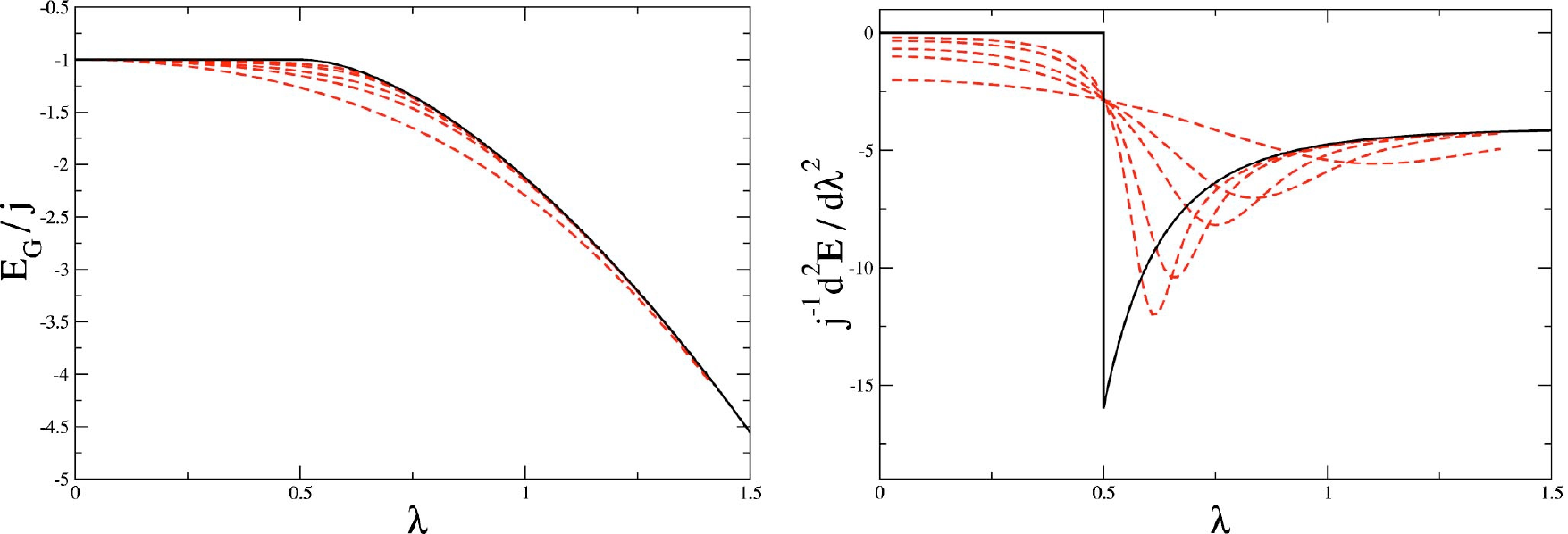} \\
    (c) \hspace{0.35\textwidth} (d) \\
    \includegraphics[width=0.7\textwidth]{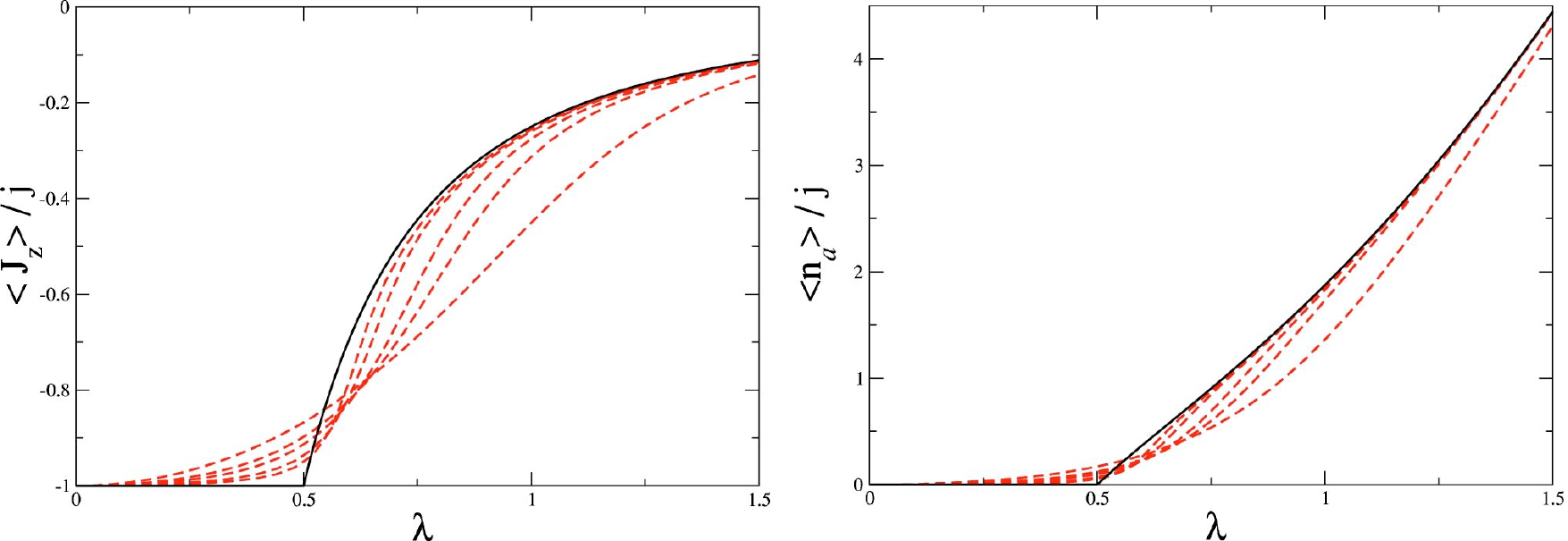}
    \end{tabular}
    \caption{(a) Ground-state energy of the Dicke Hamiltonian and (b) its second derivative as a function of the coupling parameter. Both quantities are scaled to the system size $j$.
    Expectation values of (c) the $z$-component of pseudospin $\hat{J}_{z}$ and (d) the number of photons $\hat{a}^{\dag}\hat{a}$ as a function of the coupling parameter. The expectation values are scaled to the system size $j$.
    The correspondence between the labels in panels (a)-(d) and the main text is: $\lambda=\gamma$. In each panel, from (a) to (d), the red dashed lines represent numerical implementations with different system sizes ($j=1/2,1,3/2,3$, and 5) and the black solid line represents the thermodynamic limit ($\mathcal{N}\to\infty$) of each quantity: (a) Eq.~\eqref{eqn:classical_ground_state_energy}, (b) second derivative of Eq.~\eqref{eqn:classical_ground_state_energy}, (c) Eq.~\eqref{eqn:classical_atoms}, and (d) Eq.~\eqref{eqn:classical_photons}.
    Hamiltonian parameters: $\omega=\omega_{0}=1$. Figures taken from Ref.~\cite{Emary2003PRE}.
    }
\label{fig:classical_ground_state_energy}
\end{figure}

Figure~\ref{fig:classical_ground_state_energy}(a) shows the behavior of the ground-state energy  as a function of the coupling parameter $\gamma$. The ground-state energy obtained numerically with the quantum Hamiltonian for different values of $j$ (red lines) is compared with the lowest classical energy in Eq.~\eqref{eqn:classical_ground_state_energy} (black line). In Fig.~\ref{fig:classical_ground_state_energy}(b), the discontinuity in the second derivative of the lowest classical energy at the critical coupling parameter $\gamma_{\text{c}}$ indicates that the QPT for the Dicke model is a second-order phase transition.

The expectation values of the $z$-component of pseudospin $\hat{J}_{z}$ and the number of photons $\hat{a}^{\dagger}\hat{a}$ in the thermodynamic limit ($\mathcal{N} \to \infty$) can be obtained using the mean-field approximation employed to define the classical Dicke Hamiltonian $h_{\text{D}}(\mathbf{x})$ with Glauber-Bloch coherent states, that is,
\begin{gather}
    j_{z}(\mathbf{x}) = \frac{1}{j}\langle \mathbf{x}|\hat{J}_{z}|\mathbf{x}\rangle = -\frac{1-|z|^{2}}{1+|z|^{2}} = \frac{Q^{2} + P^{2}}{2} - 1, \\
    n(\mathbf{x}) = \frac{1}{j}\langle \mathbf{x}|\hat{a}^{\dagger}\hat{a}|\mathbf{x}\rangle = \frac{|\alpha|^{2}}{j} = \frac{q^{2} + p^{2}}{2}.
\end{gather}
By substituting the coordinates $\mathbf{x}_{\text{gs}}$ from Eq.~\eqref{eqn:superradiant_phase_stationary_point} in the equations above,  we obtain the following results for the normal and superradiant phase, 
\begin{gather}
    \label{eqn:classical_atoms}
    j_{z} = j_{z}(\mathbf{x}_{\text{gs}}) = -\left\{\begin{array}{ll} 1 & \text{if } \gamma<\gamma_{\text{c}} \\ \dfrac{\gamma_{\text{c}}^{2}}{\gamma^{2}} & \text{if } \gamma>\gamma_{\text{c}}\end{array}\right., \\
    \label{eqn:classical_photons}
    n = n(\mathbf{x}_{\text{gs}}) = \dfrac{\omega_{0}}{\omega}\left\{\begin{array}{ll} 0 & \text{if } \gamma<\gamma_{\text{c}} \\ \dfrac{1}{2}\left(\dfrac{\gamma^{2}}{\gamma_{\text{c}}^{2}}-\dfrac{\gamma_{\text{c}}^{2}}{\gamma^{2}}\right) & \text{if } \gamma>\gamma_{\text{c}}\end{array}\right. .
\end{gather}
The behavior of $j_{z}$ and $n$ as a function of coupling parameter $\gamma$ is shown in Figs.~\ref{fig:classical_ground_state_energy}(c) and~\ref{fig:classical_ground_state_energy}(d), respectively. The figures show an abrupt change for both expectation values in the thermodynamic limit ($\mathcal{N}\to\infty$) at the critical value $\gamma_{\text{c}}$. For $\gamma < \gamma_{\text{c}}$, the expectation values are constant, while for $\gamma > \gamma_{\text{c}}$, they become proportional to the number of atoms within the system $\mathcal{N}$. This abrupt change for both quantities marks the QPT. The finite-size scaling of these and other quantities at the critical coupling parameter has been calculated in Ref.~\cite{Vidal2006EPL}.

\begin{figure}[t!]
    \centering
    \includegraphics[width=0.7\textwidth]{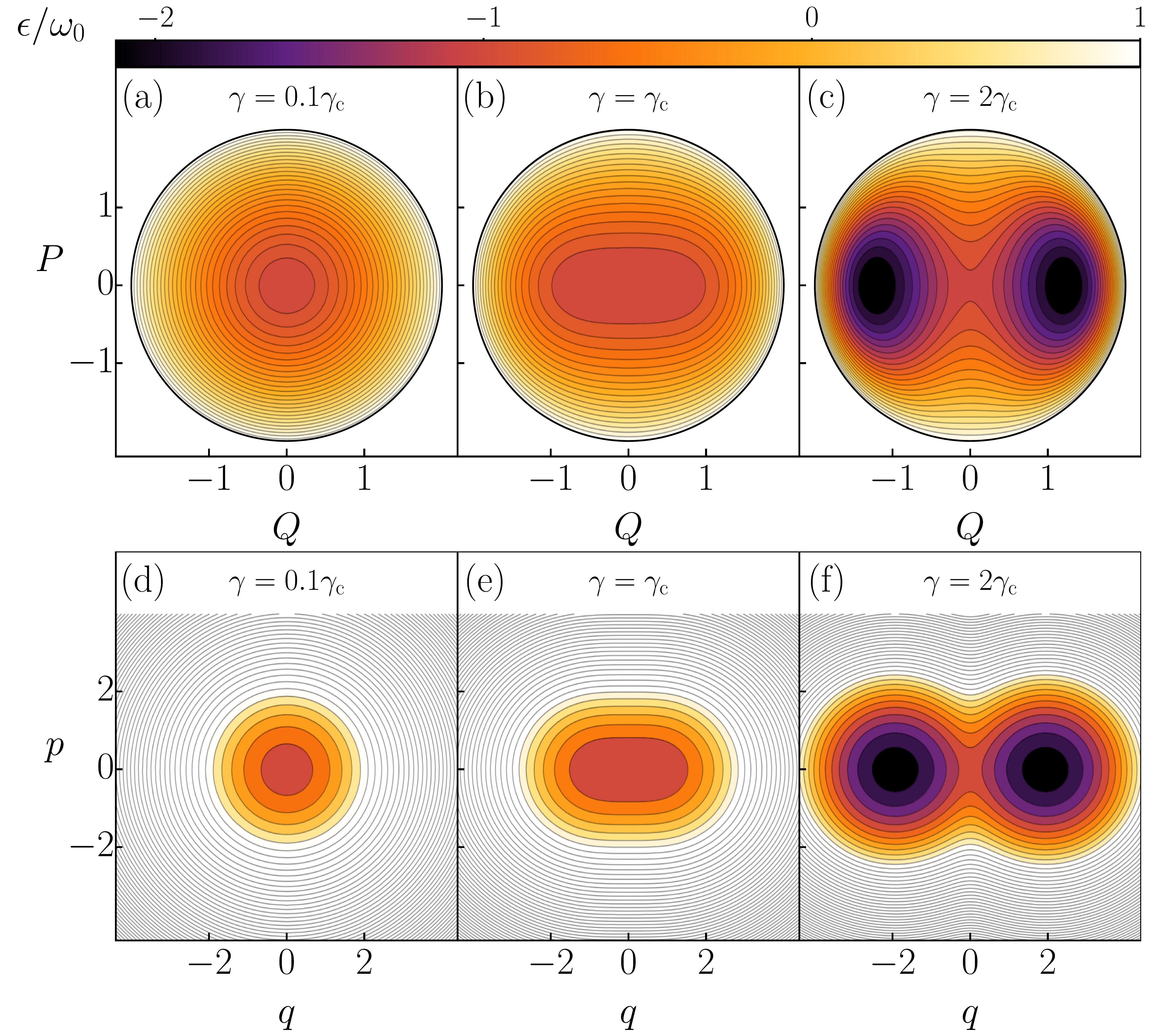}
    \caption{[(a)-(c)] Contour lines of the classical energy surface on the atomic plane $Q$-$P$ [Eq.~\eqref{eqn:classical_energy_QP}]. Each panel refers to a different coupling parameter: (a) $\gamma=0.1\gamma_{\text{c}}$, (b) $\gamma=\gamma_{\text{c}}$, and (c) $\gamma=2\gamma_{\text{c}}$. In each panel, from (a) to (c), the black solid border represents the available phase space at the classical energy shell $\epsilon=1$.
    [(d)-(f)] Contour lines of the classical energy surface on the bosonic plane $q$-$p$ [Eq.~\eqref{eqn:classical_energy_qp}]. Each panel refers to a different coupling parameter: (d) $\gamma=0.1\gamma_{\text{c}}$, (e) $\gamma=\gamma_{\text{c}}$, and (f) $\gamma=2\gamma_{\text{c}}$.
    The color scale at the top indicates low (darker tones) and high (lighter tones) energy values scaled to the atomic frequency $\omega_{0}$. The changes in the energy surfaces reflect the changes associated with the QPT. For the normal phase ($\gamma<\gamma_{\text{c}}$), the classical surface exhibits a single stable point, while the superradiant phase ($\gamma>\gamma_{\text{c}}$) is related with the appearance of two global minima. Hamiltonian parameters: $\omega=\omega_{0}=1$.
    }
    \label{fig:classical_energy_surface}
\end{figure}

To better visualize the properties of the QPT, we minimize $h_{\text{D}}(\textbf{x})$ over $(q,p)$ by setting $\partial h_{\text{D}}(\textbf{x})/\partial q=0$ and $\partial h_{\text{D}}(\textbf{x})/\partial p=0$ [see Eqs.~\eqref{eqn:hamilton_dicke_q} and~\eqref{eqn:hamilton_dicke_p} with $\gamma_{-}=\gamma_{+}=\gamma$]. In the same way, we minimize $h_{\text{D}}(\textbf{x})$ over $(Q,P)$ with $\partial h_{\text{D}}(\textbf{x})/\partial Q=0$ and $\partial h_{\text{D}}(\textbf{x})/\partial P=0$ [see Eqs.~\eqref{eqn:hamilton_dicke_Q} and~\eqref{eqn:hamilton_dicke_P} with $\gamma_{-}=\gamma_{+}=\gamma$]. By substituting the solution of each pair of equations in Eq.~\eqref{eqn:classical_dicke_hamiltonian}, we obtain a semiclassical expression for the minimum energy constrained to fixed values of the atomic $(Q,P)$ and the bosonic $(q,p)$ variables,
\begin{gather}
    \label{eqn:classical_energy_QP}
    \widetilde{h}_{\text{D}}(Q,P) = \omega_{0}\left[\frac{Q^{2}+P^{2}}{2}-1-\frac{\gamma^{2}Q^{2}}{2\gamma_{\text{c}}^{2}}\left(1-\frac{Q^{2}+P^{2}}{4}\right)\right], \\
    \label{eqn:classical_energy_qp}
    \widetilde{h}_{\text{D}}(q,p) = \frac{\omega}{2}\left(q^{2}+p^{2}\right)-\omega_{0}\sqrt{1+\frac{\omega\gamma^{2}q^{2}}{\omega_{0}\gamma_{\text{c}}^{2}}}.
\end{gather}
The above equations can be plotted as contour lines for different classical energy shells $\widetilde{h}_{\text{D}}(Q,P) = \epsilon$ and $\widetilde{h}_{\text{D}}(q,p) = \epsilon$. In Fig.~\ref{fig:classical_energy_surface}, three values of the coupling parameter $\gamma$ are considered. For $\gamma<\gamma_{\text{c}}$, where the ground state is in the normal phase, the classical projected energy surface exhibits  a stable point, while for $\gamma>\gamma_{\text{c}}$, where the ground state is in the superradiant phase, the energy surface shows two global minima. This change from one stable point to two stable points and one saddle point is called a Pitchfork bifurcation. The phase space in the atomic plane $Q$-$P$ is bounded and the Pitchfork bifurcation is inside it [Fig.~\ref{fig:classical_energy_surface}(c)]. On the contrary, the phase space in the bosonic plane $q$-$p$ is unbounded, but the Pitchfork bifurcation can also be observed with the appearance of two global minima in the superradiant phase [Fig.~\ref{fig:classical_energy_surface}(f)].

\subsubsection{No-go theorems}

There has been a long debate in the literature about whether the Dicke Hamiltonian provides a proper description of the collective excitations of two-level atoms interacting with one mode of the electromagnetic field inside a cavity. The main discussion has been about the need to include in the Hamiltonian the so-called diamagnetic term, a quadratic term in the electromagnetic potential, which appears when the minimal coupling between an electric charge and electromagnetic field is employed in the deduction of the cavity QED Hamiltonian. This term forbids the existence of a critical coupling, and therefore of a QPT, due to the oscillator strength sum rule~\cite{Rzazewski1975,Knight1978,Gawedzki1981,Keeling2007,DeLiberato2014,Vukics2014,Griesser2016,Kirton2019}.

In the last years, the ultra-strong-coupling regime has been reached in circuit QED experiments~\cite{Niemczyk2010,FornDiaz2010,Zhen2017,FornDiaz2017,Yoshihara2017,FornDiaz2019}, extending the debate over the proper Hamiltonian description of the system and the existence of the QPT. In circuit QED, instead of the electric dipole coupling, there is a capacitive coupling. The existence of Cooper pair boxes capacitively coupled to a resonator allows for the violation of the oscillator strength  sum rule, which explains the observation of the superradiant phase for these platforms~\cite{Nataf2010}.

\subsubsection{No singularities at the phase transition}

A proposal to realize the Dicke-model QPT in an optical
cavity QED system was put forward in Ref.~\cite{Dimer2007}. In Ref.~\cite{Nagy2010}, it was argued that the self-organization of atoms from a homogeneous distribution to a periodic pattern that happens for a critical driving strength in a laser-driven Bose-Einstein condensate in an optical cavity is equivalent to the QPT in the Dicke model. In both papers, following the treatment introduced in Ref.~\cite{Emary2003PRE}, the theoretical analysis was performed using a Holstein-Primakoff bosonic representation of the collective angular momentum operators truncated at first order to linearize the Hamiltonian. 

The technique introduced in Ref.~\cite{Emary2003PRE} provides a harmonic description of the Dicke Hamiltonian with an excitation energy vanishing at the QPT, in agreement with the numerical diagonalization of the full Dicke Hamiltonian. But this procedure has a  drawback: It predicts that, at the critical point, there are singularities in the expectation values of the number of photons, the number of atoms in excited states, and their fluctuations. However, it was shown in Refs.~\cite{Castanos2011c,Hirsch2013} that these observables exhibit a sudden increase from zero at the phase transition, as illustrated in Fig.~\ref{fig:classical_ground_state_energy}(d) for the number of photons, but they do not exhibit singularities. As the ground state expectation value of both the number of photons and of excited atoms scale with the number of atoms inside the cavity, it is natural that they would diverge in the thermodynamic limit, when this number goes to infinity. When these expectation values are scaled by the number of atoms in the system, as shown in Fig.~\ref{fig:classical_ground_state_energy},
the QPT is clearly seen in the derivatives of their expectation values, and no divergences are observed.

\subsection{Excited-state quantum phase transition}

In addition to the ground-state QPT, the Dicke model presents an excited-state quantum phase transition (ESQPT)~\cite{PerezFernandez2011,Brandes2013,Bastarrachea2014PRAb,Kloc2017,Kloc2017JPA}. ESQPTs are characterized by singularities in the discrete spectra of excited states~\cite{Cejnar2021}. They have been found in a variety of systems, including coupled atom–field systems, interacting bosonic systems describing collective motions of molecules, atomic nuclei and cold atoms, interacting Fermi and Bose–Fermi systems, and qubits with all-to-all interactions~\cite{Cejnar2021}. The  singularities in the energy spectrum associated with an  ESQPT  manifest in the dynamics of initial states  probing this critical energy.  For  the Dicke model,  these manifestations were  initially studied in  Ref.~\cite{PerezFernandez2011PRA}, and further  deepened and extended  to the anisotropic model in Refs.~\cite{Kloc2018, Kloc2021}.

In the case of the Dicke model, two ESQPTs are observed in Fig.~\ref{fig:semiclassical_density_states}. The ESQPT at the energy $\epsilon = 1$ occurs for any value of the coupling parameter. This transition is referred to as static and corresponds to the point where the whole phase space associated with the two-level atoms (the pseudospin sphere) becomes available to the system. The other ESQPT at the energy $\epsilon = -1$ only happens in the superradiant phase and is referred to as dynamic~\cite{Bastarrachea2014PRAa}. In the superradiant phase, the Dicke model develops a double well potential, as evidenced  by the appearance of the two minima in Figs.~\ref{fig:classical_energy_surface}(c) and~\ref{fig:classical_energy_surface}(f). The critical energy of the ESQPT in this case corresponds to the energy at the unstable local maximum of the double well. It is associated with  the separatrix in the phase space, which divides  the low-energy  region where classical trajectories are confined within the wells from the region in phase space where the trajectories move outside these wells. 

\begin{figure}[t!]
    \centering
    \includegraphics[width=0.7\textwidth]{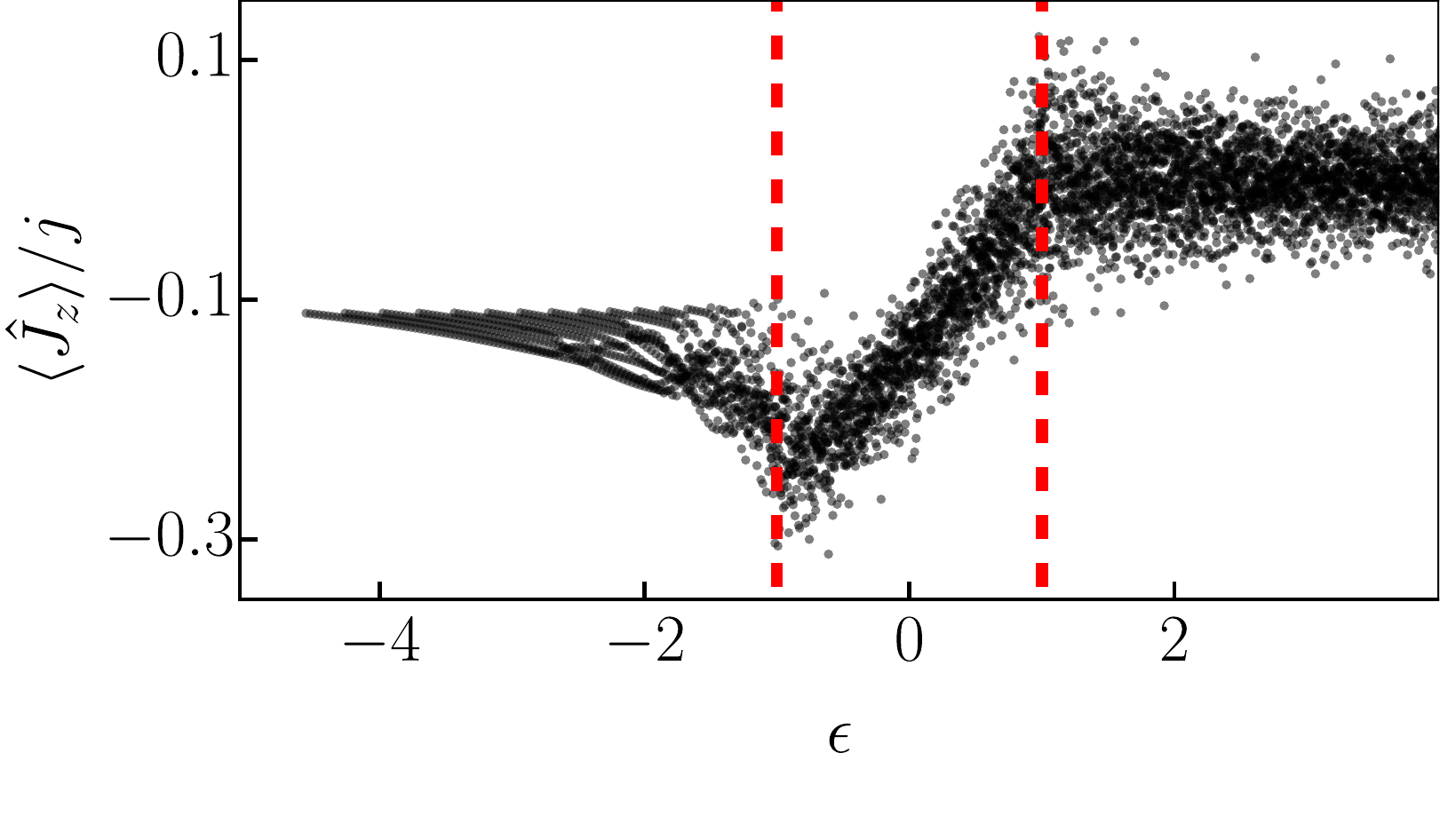}
    \caption{Expectation values of the $z$-component of pseudospin $\hat{J}_z$ for eigenstates of the Dicke Hamiltonian with positive parity. The expectation values are scaled to the system size $j$. The vertical red dashed lines represent the energies where the ESQPTs take place at $\epsilon = \mp 1$. Hamiltonian parameters: $\omega = \omega_0 = 1$, $\gamma = 3 \gamma_{\text{c}}$, and $j = 30$.
    }
    \label{fig:Peres_lattice_Jz}
\end{figure}

In Fig.~\ref{fig:Peres_lattice_Jz}, we show how the two ESQPTs of the Dicke model manifest in the expectation values of the $z$-component of pseudospin $\hat{J}_z$. Both ESQPTs are marked by dashed vertical lines. The first ESQPT is associated with a visible peak, where the lowest expectation value $\langle \hat{J}_z \rangle$ occurs. At the second ESQPT the full atomic phase space becomes available, and for energies $\epsilon \geq 1$, the expectation value $\langle \hat{J}_z \rangle$ fluctuates around zero, which is visualized as a horizontal region.

Studies of the Dicke model in resonance, where $\omega = \omega_0$, suggested that the emergence of quantum chaos was caused by the precursors of the ESQPT~\cite{Emary2003PRE,PerezFernandez2011,Bastarrachea2014PRAb}, but the detailed analysis of non-resonant cases and a close look at the vicinity of the QPTs in resonance showed that both phenomena, ESQPTs and chaos, respond to different mechanisms~\cite{Chavez2016}. When  $\omega > \omega_0$ there regular regions persist at energies higher than $\epsilon=-1$ in the superradiant phase ($\gamma > \gamma_{\text{c}}$).

\subsection{Semiclassical approximation to the density of states}

The singularities in the energy spectrum of the quantum  Dicke model can be unveiled by a  semiclassical approximation to its  density of states, which can be obtained using Weyl's law~\cite{Ozorio1988Book,Gutzwiller1990Book,Bastarrachea2014PRAa}. The density of states   of a quantum system with $d$ energy levels $E_{k}$ is defined as the ratio between the number of levels contained in an energy interval,
\begin{equation}
    \label{eqn:level_density}
    \rho(E) = \frac{\Delta d}{\Delta E} = \sum_{k}\delta(E-E_{k}),
\end{equation}
where $\delta$ is the Dirac delta function. 
Generally, the density of states can be divided into two terms, $\rho(E)=\nu(E)+\nu_{\text{f}}(E)$, where $\nu(E)$ represents a smooth function and $\nu_{\text{f}}(E)$ a fluctuating term~\cite{Wimberger2014Book}. The density of states in terms of the classical energy shell $\epsilon=E/j$ is defined as
\begin{equation}
    \label{eqn:level_density_epsilon}
    \rho(\epsilon) = \frac{\Delta d}{\Delta \epsilon} = j\frac{\Delta d}{\Delta E} = \sum_{k}\delta(\epsilon-\epsilon_{k}),
\end{equation}
with the same correspondence $\rho(\epsilon)=\nu(\epsilon)+\nu_{\text{f}}(\epsilon)$. We used, for simplicity, the same symbols for both densities, $\rho(E)$ and $\rho(\epsilon)$, which can be distinguished by their respective argument.

\begin{figure}[t!]
    \centering
    \includegraphics[width=0.9\textwidth]{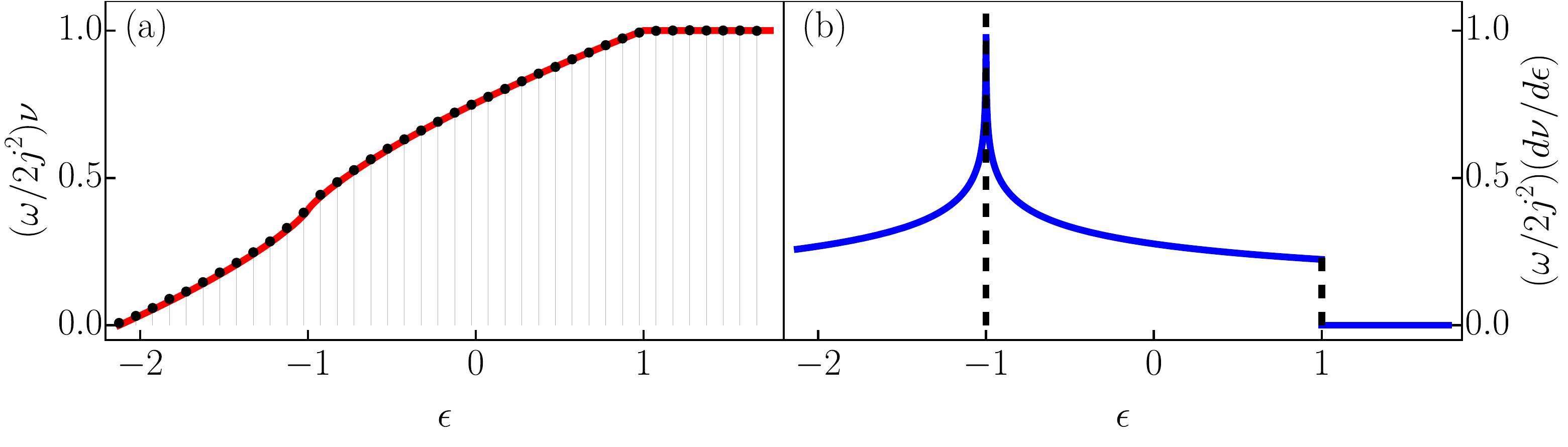}
    \caption{(a) Comparison between the semiclassical density of states [Eq.~\eqref{eqn:semiclassical_density_states}] (red solid line) and the histogram of the numerically computed energy eigenvalues of the Dicke Hamiltonian with positive parity (black dots). The density of states is scaled to the value $2j^{2}/\omega$. We obtained $N_{\text{c}}=24453$ converged energy eigenvalues using the parity projected efficient basis [Eq.~\eqref{eqn:ecbp_basis}]. (b) Derivative of the semiclassical density of states (blue solid line) scaled to the value $2j^{2}/\omega$. The vertical black dashed lines represent a logarithmic divergence at $\epsilon=-1$ and a discontinuity at $\epsilon=1$. Hamiltonian parameters: $\omega=\omega_{0}=1$, $\gamma=2\gamma_{\text{c}}$, and $j=100$.}
    \label{fig:semiclassical_density_states}
\end{figure}

The smooth term of the level density in Eq.~\eqref{eqn:level_density_epsilon} can be calculated  semiclassically by computing the available phase-space volume at the energy shell $\epsilon$ as
\begin{equation}
    \nu(\epsilon) = \frac{1}{(2\pi\hbar_{\text{eff}})^{2}}\int_{\mathcal{M}} d\mathbf{x}\,\delta[h_{\text{D}}(\mathbf{x})-\epsilon],
\end{equation}
where $\hbar_{\text{eff}}=1/j$, $h_{\text{D}}(\mathbf{x})$ is the classical Dicke Hamiltonian [Eq.~\eqref{eqn:classical_dicke_hamiltonian}], and the phase space $\mathcal{M}$ is four-dimensional with the coordinates $\mathbf{x}=(q,p;Q,P)$.
The integral can be solved using properties of the Dirac delta function~\cite{Bastarrachea2014PRAa}. The final expression for  $\nu(\epsilon)$ depends on the energy region as
\begin{equation}
    \label{eqn:semiclassical_density_states}
    \nu(\epsilon) =j\nu(E)= \dfrac{2j^{2}}{\omega}\left\{ \begin{array}{ll} \dfrac{1}{\pi} \int_{\,\xi_{\epsilon}^{-}}^{\xi_{\epsilon}^{+}} d\xi\,\cos^{-1}\left(\frac{\gamma_{\text{c}}}{\gamma}\sqrt{\frac{2(\xi-\epsilon/\omega_{0})}{1-\xi^{2}}}\right) & \text{if } \epsilon_{0}\leq\epsilon<-\omega_{0} 
    \\ \\
    \dfrac{1}{2}\left(1+\dfrac{\epsilon}{\omega_{0}}\right)+\dfrac{1}{\pi} \int_{\,\epsilon/\omega_{0}}^{\xi_{\epsilon}^{+}} d\xi\,\cos^{-1}\left(\frac{\gamma_{\text{c}}}{\gamma}\sqrt{\frac{2(\xi-\epsilon/\omega_{0})}{1-\xi^{2}}}\right) & \text{if } 
    -\omega_0\leq \epsilon\leq \omega_0
    \\ \\
    1 & \text{if } \epsilon>\omega_{0} \end{array} \right.,
\end{equation}
where
\begin{equation}
    \xi_{\epsilon}^{\pm} = -\frac{\gamma_{\text{c}}}{\gamma}\left(\frac{\gamma_{\text{c}}}{\gamma}\mp \sqrt{\frac{2(\epsilon-\epsilon_{0})}{\omega_{0}}}\right),
\end{equation}
and $\epsilon_{0}=\epsilon_{\text{gs}}$ is the ground-state energy in the thermodynamic limit in the superradiant phase ($\gamma>\gamma_{\text{c}}$), as presented in Eq.~\eqref{eqn:classical_ground_state_energy}. In Fig.~\ref{fig:semiclassical_density_states}(a), the semiclassical density of states [Eq.~\eqref{eqn:semiclassical_density_states}] is compared with the density of states computed numerically in the superradiant phase ($\gamma>\gamma_{\text{c}}$). The agreement is excellent. Moreover, the singularities in  the energy spectrum are signaled in Fig.~\ref{fig:semiclassical_density_states}(b) by a logarithmic divergence at $\epsilon=-1$ and a discontinuity at $\epsilon=1$ in the derivative of the semiclassical density of states.

\subsection{Quasiprobability distributions: Wigner and Husimi functions}

Quasiprobability distributions are representations of quantum states in phase space~\cite{Hillery1984}. They allow one to compare quantum and classical states on equal footing and provide both a powerful tool for classical simulations of quantum evolution and also a very conceptually appealing description of quantum dynamics.  For this reason, these distributions have been investigated since  the beginning  of quantum mechanics by E. Wigner in 1932~\cite{Wigner1932}, K. Husimi in 1940~\cite{Husimi1940}, R.~J. Glauber and E.~C.~G. Sudarshan~\cite{Glauber1963PRb,Sudarshan1963}
and  J.~R. Klauder~\cite{Klauder1963JMPa,Klauder1963JMPb}  in 1963.  

Quasiprobability distributions are real functions of the phase space variables, but as their name indicates, they are not proper probability distributions: they relax some of the properties that we usually associate with probability distributions~\cite{Lee1995}. For example, the Wigner function takes negative values, and the marginal distribution of the Husimi function over position or momentum is different from the distribution arising from the corresponding wave function. We will focus on these two quasiprobability distributions.

To understand the Wigner function, it is best to first understand the more general Wigner transform. The basic idea is that for any observable $\hat O$, one associates a real (but possibly negative) normalized function  of the phase space $O(\mathbf x)$, with the fundamental property that $\text{Tr}\left(\hat{O}_1 \hat{O}_2\right) =\int d \mathbf x \, O_1(\mathbf x) O_2(\mathbf x)$, with $\hbar_{\text{eff}}=1/j$. The Wigner function $\mathcal{W}_\rho$ is just the Wigner transform of a state $\hat \rho$, appropriately normalized. If we consider a pure state $\hat{\rho}=|\Psi\rangle\langle\Psi|$, the Wigner function allows one to compute expectation values by performing integrals in the phase space, $\langle\Psi|\hat{O}|\Psi\rangle=\int d \mathbf x \, \mathcal{W}_\Psi(\mathbf x) O(\mathbf x)$, just as one would do for classical distributions. We will not delve into the details on how to generally compute Wigner transforms (interested readers may see e.g., Ref.~\cite{Weinbub2018}). We give the explicit equations for the Wigner functions of coherent states in Sec.~\ref{subsec:wigandhuscoherent}, and describe how to utilize them for a semiclassical approximation to quantum dynamics called truncated Wigner approximation in Sec.~\ref{subsubsec:TWA}.

The Husimi quasiprobability distribution is defined as the expectation value of a quantum state under coherent states, introduced in Eq.~\eqref{eqn:glauber_bloch_coherent_state},
\begin{equation}
    \label{eqn:husimi_function}
    \mathcal{Q}_{\rho}(\alpha,z) = \langle \alpha|\otimes\langle z| \hat{\rho}|z\rangle\otimes|\alpha\rangle .
\end{equation}
The Husimi function satisfies the normalization condition
\begin{equation}
    \frac{2j + 1}{\pi^{2}} \iint \frac{d^{2}\alpha \, d^{2}z}{(1 + |z|^{2})^{2}} \, \mathcal{Q}_{\rho}(\alpha,z) = 1, 
\end{equation} 
and, unlike the Wigner function, never takes negative values.

\subsubsection{Wigner and Husimi functions for coherent states}
\label{subsec:wigandhuscoherent}

The Wigner and Husimi functions of Glauber-Bloch coherent states are given by normal distributions. The Wigner function of a Glauber coherent state $|\alpha_{0}\rangle$ with parameter $\alpha_{0}=\alpha(q_{0},p_{0})$ [see Eq.~\eqref{eqn:bosonic_parameter}] is given by
\begin{equation} 
    \label{eqn:glauber_wigner_function}
    \mathcal{W}_{\alpha_{0}}(\alpha) = \frac{j}{\pi}e^{-2|\alpha-\alpha_{0}|^{2}}.
\end{equation}
In the case of a Bloch coherent state $|z_{0}\rangle$ with parameter $z_{0}=z(\phi_{0},\theta_{0})$, where $(\phi,\theta)$ are the usual Bloch sphere coordinates related by $z=\tan(\theta/2)e^{-i\phi}$, the Wigner function takes on a more involved form~\cite{Yang2019}, 
\begin{equation} 
    \mathcal{W}_{z_{0}}(z) = \frac{(2j)!}{4\pi}\sum_{k=0}^{2j}\sqrt{\frac{2k+1}{(2j-k)!(2j+k+1)!}} P_{k}(\cos\Theta)  , 
\end{equation}
where $P_k(x)$ are Legendre polynomials and $\Theta$ is the angle between $z$ and $z_0$, when represented as points in the surface of the Bloch sphere, 
\begin{equation}
    \label{eqn:theta_distance}
    \cos\Theta = \cos\theta \, \cos\theta_{0} + \sin\theta \, \sin\theta_{0} \, \cos(\phi - \phi_{0}).
\end{equation}
Importantly, for large values of $j$, this  function is very well approximated by a normal distribution on the Bloch sphere, 
\begin{equation}
    \label{eqn:bloch_wigner_function}
    \mathcal{W}_{z_{0}}(z)\approx \frac{j}{\pi}e^{-j\Theta^{2}}.
\end{equation}
The Wigner function of Glauber-Bloch coherent states $|\mathbf{x}_{0}\rangle=|\alpha_{0}\rangle\otimes|z_{0}\rangle$ is simply the product of the Wigner functions in Eqs.~\eqref{eqn:glauber_wigner_function} and~\eqref{eqn:bloch_wigner_function},
\begin{equation} 
\label{eqn:glauber_bloch_wigner_function}
    \mathcal{W}_{\mathbf{x}_{0}}(\mathbf{x}) = \mathcal{W}_{\alpha_{0}}(\alpha) \, \mathcal{W}_{z_{0}}(z) 
     \approx \left(\frac{j}{\pi}\right)^{2}e^{-jD^{2}(\mathbf{x},\mathbf{x}_{0})},
\end{equation}
where
\begin{equation}
    \label{eqn:phase_space_separation}
    D(\mathbf{x},\mathbf{x}_{0}) = \sqrt{(q-q_{0})^{2} + (p-p_{0})^{2} + \Theta^{2}}
\end{equation}
is the phase-space distance between the points $\mathbf{x}_{0}$ and $\mathbf{x}$.

The Husimi functions of coherent states have a similar form. They can be easily obtained from the overlaps between coherent states, $|\langle\alpha|\alpha_0\rangle|^{2}$ and  $|\langle z|z_0\rangle|^{2} $  given  in  Eqs.~\eqref{eqn:glauber_coherent_state_square_norm} and~\eqref{eqn:bloch_coherent_state_square_norm}. One obtains  
\begin{gather}
    \label{eqn:glauber_husimi_function}
    \mathcal{Q}_{\alpha_{0}}(\alpha) = e^{-|\alpha-\alpha_{0}|^{2}}, \\
    \label{eqn:bloch_husimi_function}
    \mathcal{Q}_{z_{0}}(z) = \left[1-\frac{|z-z_{0}|^{2}}{(1+|z|^{2})(1+|z_{0}|^{2})}\right]^{2j} \approx e^{-(j/2)\Theta^{2}}.
\end{gather}

The Husimi function for the Glauber-Bloch coherent states $|\mathbf{x}_{0}\rangle=|\alpha_{0}\rangle\otimes|z_{0}\rangle$ is given by the product of Eqs.~\eqref{eqn:glauber_husimi_function} and~\eqref{eqn:bloch_husimi_function},
\begin{equation}
    \label{eqn:glauber_bloch_husimi_function}
    \mathcal{Q}_{\mathbf{x}_{0}}(\mathbf{x})  = \langle\mathbf{x}|\hat{\rho}_{\mathbf{x}_{0}}|\mathbf{x}\rangle =  \mathcal{Q}_{\alpha_{0}}(\alpha) \, \mathcal{Q}_{z_{0}}(z) \approx e^{-(j/2)D^{2}(\mathbf{x},\mathbf{x}_{0})} .
\end{equation}
Observe that both the Wigner and Husimi functions of coherent states are Gaussian distributions, but the variance of the Husimi function is twice that of the Wigner function.

\subsection{Quantum-classical correspondence using Husimi functions}
\label{subsubsec:husimi_functions}

In this section, we will describe how to utilize the Husimi function to visualize a quantum state in phase space, and compare with features of the classical Dicke model. Since the Husimi functions in the Dicke model are functions of four parameters, it is hard to visualize them directly. It is useful to project them or intersect them with two dimensional planes, as we describe in the following sections.

\subsubsection{Poincar\'e-Husimi functions}

The simplest way to visualize the Husimi function is to evaluate it along a two-dimensional surface embedded in the four-dimensional phase space. We consider the surface of fixed energy $\epsilon$ and $p=0$,
\begin{equation}
    h_{\text{D}}(q,p=0;Q,P) = \epsilon ,
    \label{Eq:hDp0}
\end{equation}
which is useful for states with small energy uncertainty (such as energy eigenstates or coherent states). In classical systems, the intersection of a trajectory in phase-space with some  surface is called a Poincar\'e section (see Sec.~\ref{subsubsec:poincare}), and, consequently, a Husimi function evaluated over the surface determined by Eq.~\eqref{Eq:hDp0} is called a Poincar\'e-Husimi function.

\begin{figure}[t!]
    \centering
    \includegraphics[width=0.9\textwidth]{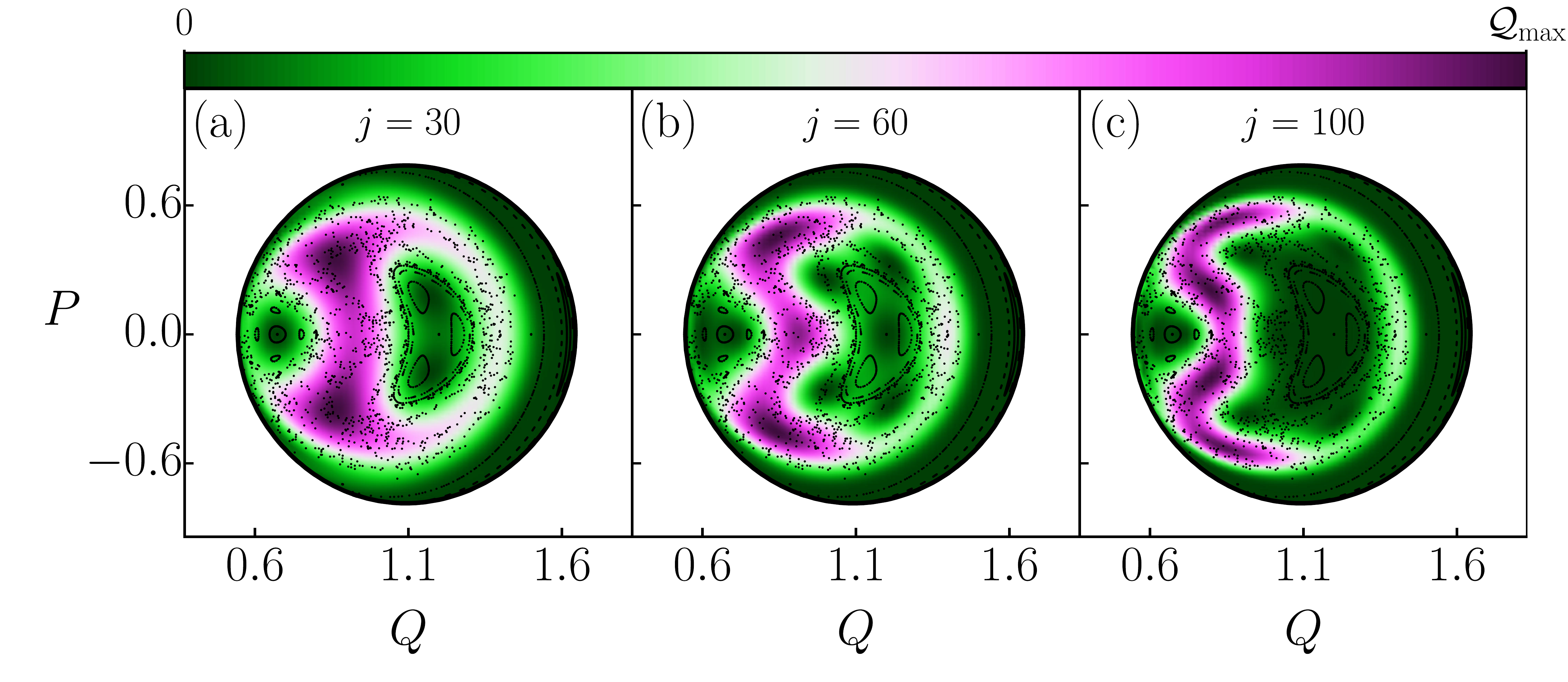}
    \caption{Poincar\'e-Husimi functions for an eigenstate of the Dicke Hamiltonian. The projections are shown on the atomic plane $Q$-$P$. Each panel refers to a different system size: (a) $j=30$, (b) $j=60$, and (c) $j=100$. The eigenstate chosen in each panel, from (a) to (c), has an eigenenergy close to the classical energy shell $\epsilon_{k} \sim \epsilon = -1.4$. The Husimi function is evaluated at the coordinates $\mathbf{x} = (q_{+}(\epsilon_{k}),0;Q,P)$ (see text for details). The color scale at the top indicates the concentration value of the Husimi function: dark green corresponds to zero, while dark purple represents higher values. In each panel, from (a) to (c), the black dots represent the Poincar\'e sections of classical trajectories and the black solid border represents the available phase space at the classical energy shell. The quantum-classical correspondence improves as the system size increases from panel (a) to panel (c). Hamiltonian parameters: $\omega=\omega_{0}=1$ and $\gamma=2\gamma_{\text{c}}$.
    }
    \label{fig:poincare_husimi}
\end{figure}

\begin{figure}[t!]
    \centering
    \includegraphics[width=0.6\textwidth]{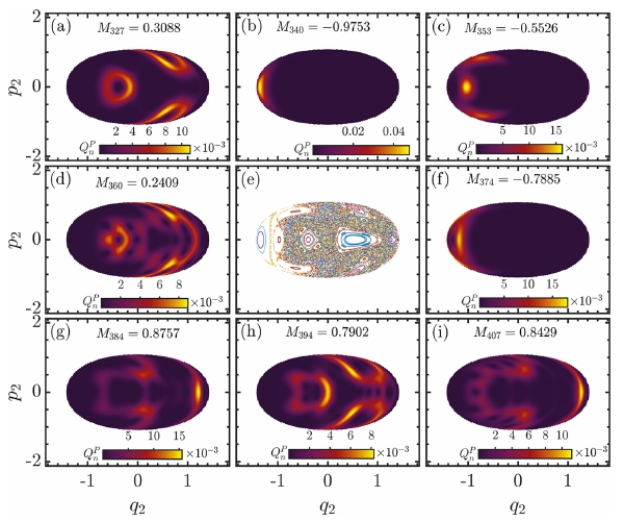}
    \caption{[(a)-(i)] Poincar\'e-Husimi functions for eigenstates of the Dicke Hamiltonian with positive parity. The projections are shown on the atomic plane $Q$-$P$. Each panel refers to a different eigenenergy $\epsilon_{k}$: (a) $\epsilon_{327}=-0.4391$, (b) $\epsilon_{340}=-0.4297$, (c) $\epsilon_{353}=-0.414$, (d) $\epsilon_{360}=-0.4067$, (f) $\epsilon_{374}=-0.3955$, (g) $\epsilon_{384}=-0.3856$, (h) $\epsilon_{394}=-0.3754$, and (i) $\epsilon_{407}=-0.3641$. The correspondence between the labels in panels (a)-(i) and the main text is: $q_{2}=Q$ and $p_{2}=P$. In each panel, from (a) to (i), the color scale indicates the concentration value of the Husimi function: dark purple corresponds to zero, while lighter tones represent higher values.
    (e) Poincar\'e section of classical trajectories computed at the classical energy shell $\epsilon=-0.4$.
    Hamiltonian parameters: $\omega=\omega_{0}=1$, $\gamma=0.47$, and $j=50$. Figure taken from Ref.~\cite{Wang2024}.
    }
    \label{fig:uniform_semiclassical_condensation}
\end{figure}

In Fig.~\ref{fig:poincare_husimi}, we show the Poincar\'e-Husimi function $\mathcal{Q}_{k}(\mathbf{x}) = |\langle E_{k}|\mathbf{x}\rangle|^{2}$ for an eigenstate $|E_{k}\rangle$ of the Dicke model with eigenenergy $E_{k}$, evaluated over the surface $\mathbf{x} = (q_{+}(\epsilon_{k}),p=0;Q,P)$, where $q_{+}$ represents the positive root of the second-degree Eq.~\eqref{Eq:hDp0}, with $\epsilon=\epsilon_{k}=E_{k}/j$. Each panel corresponds to a different system size $j$, as indicated. For comparison, Fig.~\ref{fig:poincare_husimi} also shows the Poincar\'e sections for classical trajectories (black dots) at the same energy $\epsilon_{k}$. The correspondence between the classical and quantum results improves as the system size $j$ increases from Fig.~\ref{fig:poincare_husimi}(a) to Fig.~\ref{fig:poincare_husimi}(c). This supports the principle of uniform semiclassical condensation for Wigner and Husimi functions~\cite{Robnik2020,Robnik2020SYN}, that states that in the semiclassical limit, eigenstates localized in mixed phase-space regions become exclusively localized around regular or chaotic regions. The Poincar\'e-Husimi function clusters over regions where the classical dynamics is chaotic (regular), which is signaled by a disordered (structured) Poincar\'e section. A recent study~\cite{Wang2024} has thoroughly verified this principle for the Dicke model. 
Figure~\ref{fig:uniform_semiclassical_condensation} shows some eigenstates of the Dicke model that are localized around regular or chaotic classical regions in phase space.

\subsubsection{Husimi functions of reduced density operators}

Another method to obtain a two-dimensional  visualization of the Husimi function is partially tracing out the atomic (A) or bosonic (B) degrees of freedom of the state before computing the Husimi function. Consider reduced density operators on the complementary spaces,
\begin{gather} 
    \hat{\rho}_{\text{B}} = \text{Tr}_{\text{A}}(\hat\rho) = \frac{2j + 1}{\pi} \iint \frac{d^{2}z}{(1 + |z|^{2})^{2}} \, \langle z |\hat{\rho}| z \rangle = \frac{2j + 1}{4\pi} \iint dQ \, dP \, \langle Q,P |\hat{\rho}| Q,P \rangle, 
    \label{Eq:rhoB}
    \\
    \hat{\rho}_{\text{A}} = \text{Tr}_{\text{B}}(\hat\rho) = \frac{1}{\pi} \iint d^{2}\alpha \, \langle \alpha |\hat{\rho}| \alpha \rangle = \frac{j}{2\pi} \iint dq \, dp \, \langle q,p |\hat{\rho}| q,p \rangle,
    \label{Eq:rhoA}
\end{gather}
where the changes of variables are performed with the Jacobians $\mathbf{J}(Q,P) = 4/(4 - Q^{2} - P^{2})^{2}$ in Eq.~\eqref{Eq:rhoB} and $\mathbf{J}(q,p) = j/2$ in Eq.~\eqref{Eq:rhoA}. By taking expectation values of the reduced density operators in the respective Glauber or Bloch coherent states, we  obtain the projected (reduced) Husimi functions over the bosonic $(q,p)$ or atomic $(Q,P)$ variables,
\begin{gather}
    \label{eqn:projected_exact_bosonic_husimi_function}
    \widetilde{\mathcal{Q}}_{\rho}(q,p) \equiv \langle \alpha |\hat{\rho}_{\text{B}}| \alpha \rangle = \langle q,p |\hat{\rho}_{\text{B}}| q,p \rangle = \frac{1}{C_{\text{A}}} \iint dQ \, dP \, \mathcal{Q}_{\rho}(\mathbf{x}), \\
    \label{eqn:projected_exact_atomic_husimi_function}
    \widetilde{\mathcal{Q}}_{\rho}(Q,P) \equiv \langle z |\hat{\rho}_{\text{A}}| z \rangle = \langle Q,P |\hat{\rho}_{\text{A}}| Q,P \rangle = \frac{1}{C_{\text{B}}} \iint dq \, dp \, \mathcal{Q}_{\rho}(\mathbf{x}) ,
\end{gather}
where  the normalization constants are $C_{\text{A}}=4\pi/(2j+1)$ and $C_{\text{B}}=2\pi/j$. Closed expressions for the projected Husimi function using the Fock basis were obtained in Refs.~\cite{DeAguiar1991,Furuya1992}.

\begin{figure}[t!]
    \centering
    \includegraphics[width=0.6\textwidth]{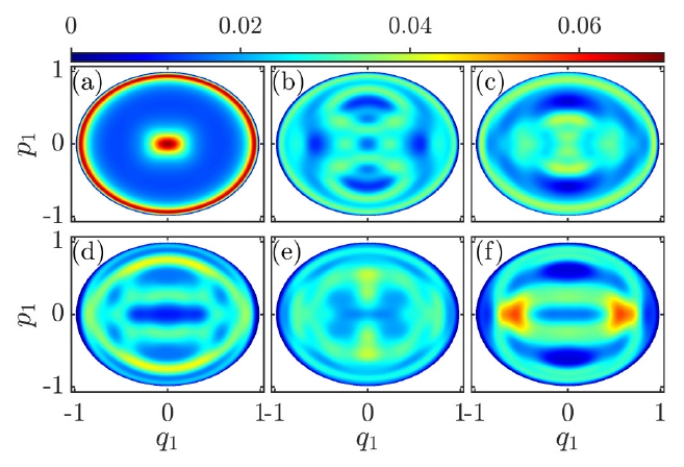}
    \caption{Projections of the Husimi function for eigenstates of the Dicke Hamiltonian with positive parity. The projections are shown on the atomic plane $Q$-$P$. The projections were computed using Eq.~\eqref{eqn:projected_exact_atomic_husimi_function}. Each panel refers to a different coupling parameter: (a) $\gamma=0.1$, (b) $\gamma=0.4$, (c) $\gamma=0.46$, (d) $\gamma=0.5$, (e) $\gamma=0.54$, and (f) $\gamma=0.7$. The correspondence between the labels in panels (a)-(f) and the main text is: $q_{1}=P/2$ and $p_{1}=Q/2$. The eigenstate chosen in each panel, from (a) to (f), has an eigenenergy $\epsilon_{k} \sim 0.8$. The color scale at the top indicates the concentration value of the Husimi function: dark blue corresponds to zero, while dark red represents higher values. Hamiltonian parameters: $\omega=\omega_{0}=1$ and $j=20$. Figure taken from Ref.~\cite{Wang2020}.
    }
    \label{fig:projected_husimi_function_scarred_eigenstates} 
\end{figure}

Figure~\ref{fig:projected_husimi_function_scarred_eigenstates} shows some examples of projected Husimi functions of selected eigenstates. Note that in all panels, there are visible structures in the eigenstates. For small coupling,  these structures are related to regular classical trajectories, while for large coupling, they  are related to quantum scars, as described in Sec.~\ref{sec:QuantumScarring}.

\subsubsection{Husimi functions projected over classical energy shells}

In the particular case of energy eigenstates, the Poincar\'e Husimi and the projections obtained from the reduced density matrix can be sharpened by only evaluating the projection along the classical energy shell $\mathcal{M}(\epsilon_k)=\{\mathbf{x}\, |\, h_{\text{D}}(\mathbf{x}) = \epsilon_{k} = E_k/j\}$. This generates a sharper image, and does not remove any information because the eigenstate is localized at the energy shell.

The energy shells are three-dimensional surfaces, so to obtain a two-dimensional representation, we consider the Husimi function evaluated at the classical energy shell $\epsilon=E/j$ and projected  into the atomic variables $(Q,P)$, so that~\cite{Pilatowsky2021NC}
\begin{equation}
    \label{eqn:projected_husimi_function}
    \widetilde{\mathcal{Q}}_{\epsilon,\rho}(Q,P) = \iint dq \, dp \, \delta[h_{\text{D}}(\mathbf{x})-\epsilon] \, \mathcal{Q}_{\rho}(\mathbf{x}),
\end{equation}
where $\delta$ is the Dirac delta function and $h_{\text{D}}(\mathbf{x})$ is the classical Dicke Hamiltonian in Eq.~\eqref{eqn:classical_dicke_hamiltonian}.

\begin{figure}[t!]
    \centering
    \includegraphics[width=0.6\textwidth]{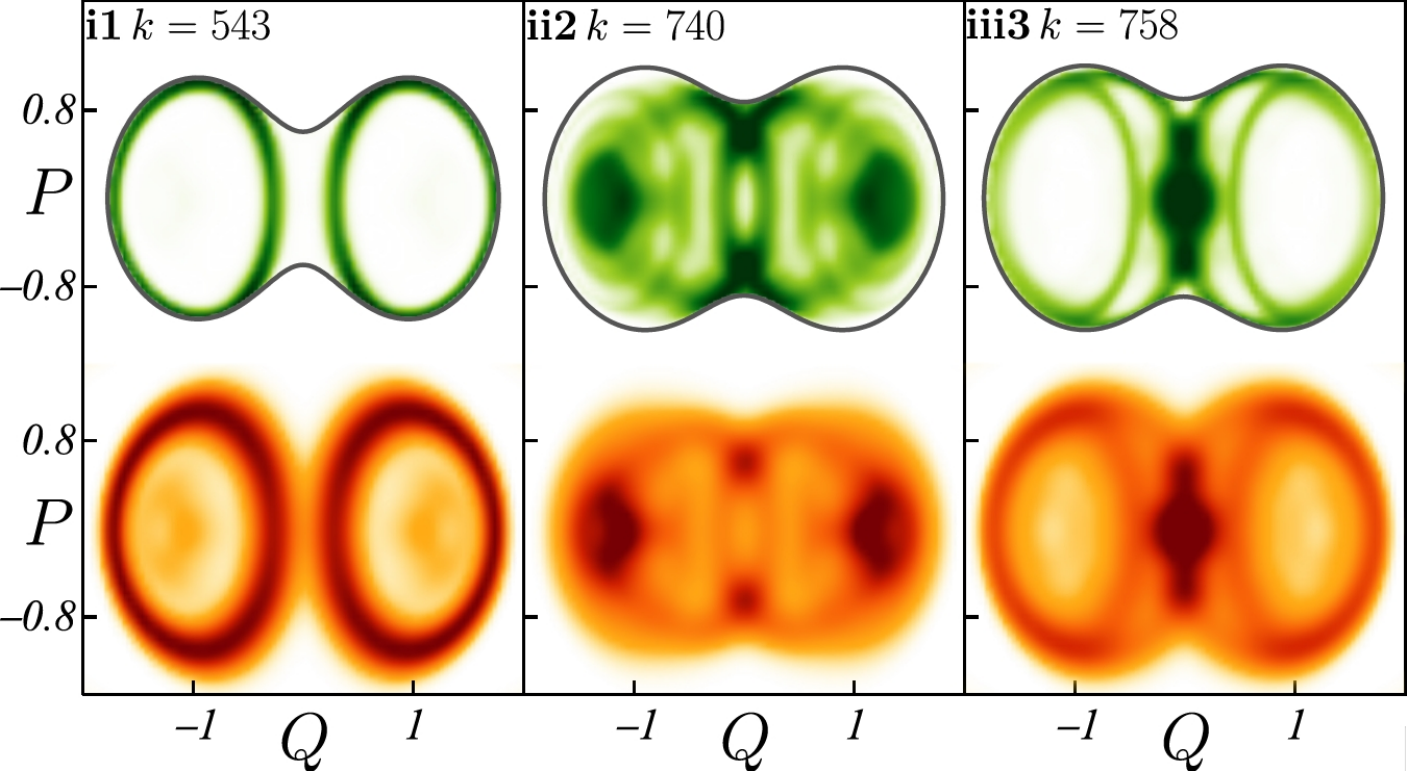}
    \caption{Projections of the Husimi function for eigenstates of the Dicke Hamiltonian with positive parity. The projections are shown on the atomic plane $Q$-$P$. The projections were computed with (top row) Eq.~\eqref{eqn:projected_husimi_function} and (bottom row) Eq.~\eqref{eqn:projected_exact_atomic_husimi_function}. Each panel, from (i1) to (iii3), refers to a different eigenenergy $\epsilon_{k}$ with spectrum label: (i1) $k=543$, (ii2) $k=740$, and (iii3) $k=758$. In each panel, from (i1) to (iii3) in the top row, the gray solid border represents the available phase space at each energy shell. Darker tones inside the projections represent higher concentration values of the Husimi function, while white corresponds to zero. Hamiltonian parameters: $\omega=\omega_{0}=1$, $\gamma=2\gamma_{\text{c}}$, and $j=30$. Figure taken from Ref.~\cite{Pilatowsky2021NJP}.
    }
    \label{fig:projected_husimi_function_scarred_eigenstates2}
\end{figure}

Figure~\ref{fig:projected_husimi_function_scarred_eigenstates2} shows a comparison between the projection method in Eq.~\eqref{eqn:projected_husimi_function} (top panels) and the Husimi function for the reduced density operator in Eq.~\eqref{eqn:projected_exact_atomic_husimi_function} (bottom panels). As it can be seen, more details are captured when only evaluating over the classical energy shell (green) as opposed to completely projecting all values with the reduced-density matrix method (orange). For the latter, the results are blurrier. In Sec.~\ref{sec:QuantumScarring}, we show how the energy-shell projection method can be used to identify quantum scars.

\subsubsection{Truncated Wigner approximation}
\label{subsubsec:TWA}

Even when using the efficient coherent basis, the system size is a limiting factor for the computation of the full-time dynamics of the Dicke model via exact diagonalization. However, if we are only interested in short-time dynamics, other methods are available. Because of the semiclassical nature of the large $j$ limit, the short-time quantum dynamics obeys classical equations of motion. The time up to which this is true is proportional to $j$. The idea of the truncated Wigner approximation (TWA) is to compute the short-time quantum dynamics using the classical equation of motion.

Imagine that we are interested in computing the expectation value of an observable $\hat{O}$ under a state $\hat\rho(t)$, as a function of time $t$. One can write this expectation value in terms of the Wigner transforms,
\begin{equation}
    \label{eq:expvalWignerFunction}
    \text{Tr}\left[\hat{\rho}(t) \hat{O}\right]=  \int_{\mathcal{M}} d\mathbf{x}\, \mathcal{W}_{\rho}(\mathbf{x},t) \, O(\mathbf{x}) ,
\end{equation}
where $\mathcal{W}_{\rho}(\mathbf{x},t) \equiv \mathcal{W}_{\rho(t)}(\mathbf{x})$ and $\hbar_{\text{eff}}=1/j$.
This looks like a classical expectation value in the phase space, but the quantum effects are still there, contained in small regions where the Wigner function is negative.
The time-evolution of the Wigner function is given by the Moyal bracket~\cite{Moyal1949},
\begin{equation}
 \frac{\partial \mathcal{W}_{\rho}(\mathbf{x},t)}{\partial t} = \left\{ \mathcal{W}_{\rho}(\mathbf{x},t) ,h_{\text{D}}(\mathbf{x})\right\}_{\text{M}},
\end{equation}
where $h_{\text{D}}(\mathbf{x})$ is the classical Dicke Hamiltonian given in Eq.~\eqref{eqn:classical_dicke_hamiltonian}. We will not go into the details of how the Moyal bracket is computed, but its key property is that it can be expanded in a series of $\hbar_{\text{eff}}=1/j$, and the first term corresponds to the usual Poisson bracket,
\begin{equation}
   \left\{ \mathcal{W}_{\rho}(\mathbf{x},t) ,h_{\text{D}}(\mathbf{x})\right\}_{\text{M}} = \left \{\mathcal{W}_{\rho}(\mathbf{x},t),h_{\text{D}}(\mathbf{x})\right \} + \mathcal{O}(\hbar_{\text{eff}}^2).
\end{equation}
The $\mathcal{O}(\hbar_{\text{eff}}^2)$ terms contain the quantum corrections to the classical evolution. In the TWA, we ignore those terms and are left with the classical Liouville equation. 

Under the Liouville evolution, the Wigner function is constant along the classical trajectories $\mathbf x(t)$ given by the classical Hamiltonian $h_{\text{D}}(\mathbf{x})$,
\begin{equation}
    \label{eqn:time_evolved_wigner_function}
    \mathcal{W}^{\text{TWA}}_{\rho}(\mathbf{x},t) = \mathcal{W}_{\rho}(\mathbf{x}(-t),t=0). 
\end{equation}
Thus, under the TWA, the Wigner function will not develop negative probability regions, and by selecting a coherent initial state, which has a completely positive Wigner function [see Eq.~\eqref{eqn:glauber_bloch_wigner_function}],  we may interpret the Wigner function as a classical probability distribution in phase space.  

To compute the expectation value of quantum observables under the TWA, one inserts $\mathcal{W}^{\text{TWA}}_{\rho}(\mathbf{x},t)$ into Eq.~\eqref{eq:expvalWignerFunction}. The Wigner transform $O$ of the observables $\hat{O}=\hat{q},\hat{p}, \hat{J}_{x,y,z}$ and their powers are shown in Table~\ref{tab:wignertrans}. Note that such mapping is linear, allowing one to easily calculate the Wigner transform of linear combinations of these observables. For details on how to compute the Wigner transform of products of $\hat{q}$ and $\hat{p}$, see Ref.~\cite{Polkovnikov2010}. For products of spin operators and higher powers, see Ref.~\cite{Vrilly1989}.

\begin{table}[]
\centering
\begin{tabular}{|c|c|}
\hline
Quantum observable & Wigner transform   \\ 
$\hat{O}$ & ${O}(q,p;Q,P)$  \\
\hline 
$\hat{q}^n$ & $q^n$   \\
$\hat{p}^n$ & $p^n$   \\
\hline 
$\hat{J}_{x,y,z}$ & $a_j\, c_{x,y,z} - 1$    \\

$\hat{J}_{x,y,z}^2$ & $\tfrac{b_j}{2}(c_{x,y,z} ^2 -\tfrac{1}{3}) + \frac{a_j^2}{3}$    \\
\hline
\end{tabular}
\caption{\label{tab:wignertrans}Wigner transform of some observables in the Dicke model, where 
$a_j=\sqrt{j(j+1)}$, $b_j=\sqrt{j(j+1)(2j-1)(2j+3)}$, $c_x=Q\sqrt{1-(Q^2 + P^2)/4}$,
$c_y=-P\sqrt{1-(Q^2 + P^2)/4}$, 
and $c_z=(Q^2 + P^2)/2 - 1$~\cite{Polkovnikov2010,Vrilly1989}.}
\end{table}

\section{Classical and quantum chaos}
\label{sec:ChaosTheory}
Depending on the interaction strength and the energy of the system, the classical Dicke model develops chaos in the sense of exhibiting a positive Lyapunov exponent and mixing. Signatures of this classical transition from a regular to a chaotic regime appear also in the quantum domain, where the eigenvalues become correlated and the eigenstates in most bases approach random vectors. These signatures characterize what became known as quantum chaos. Early studies on classical and quantum chaos in the Dicke model can be traced back to the 1970s \cite{Belobrov1976} and 1980s \cite{Graham1984,Fox1987}. Both classical (Sec.~\ref{subs:classicalChaos}) and quantum (Sec.~\ref{subs:quantchaos}) approaches are addressed in this section.

\subsection{Classical chaos and integrability}
\label{subs:classicalChaos}

Classical chaotic behavior arises in nonlinear dynamical systems, and one of its main characteristics is strong sensitivity to small changes in initial conditions. As a result, the precision required to describe the dynamics increases unboundedly over time. Despite the deterministic nature of these dynamical systems,  predictions become unfeasible at long time scales~\cite{Ott2002Book,Baker1996Book,Wimberger2014Book}. 

The phenomenon of chaos is associated with the loss of integrability. In regular systems with $N$ degrees of freedom, solutions can be derived from their $N$ independent constants of motion. Chaotic behavior exists in a realm between regular behavior (characterized by integrals of motion) and unpredictable stochastic behavior (characterized by complete randomness)~\cite{Goldstein2002Book}.

\subsubsection{Integrability of Hamiltonian systems}

For conservative Hamiltonian systems with $N$ degrees of freedom, there is a Hamiltonian function independent of time, $H = H(q_{i},p_{i})$, where $(q_{i},p_{i})$ are the canonical position-momentum variables that satisfy the Hamilton's equations of motion $\dot{q}_{i} = \partial H / \partial p_{i}$ and $\dot{p}_{i} = -\partial H / \partial q_{i}$ for $i=1,\ldots,N$~\cite{Reichl2004Book,Santhanam2022}. Integrability is defined by the existence of $N$ integrals of motion, whose Poisson brackets vanish. If these conditions are achieved by the Hamiltonian system, it is integrable, because it can be reduced to quadratures. This allows one to exactly transform the integrable system into action-angle variables with the actions being  invariants of motion. Therefore, integrability implies that the trajectories are constrained to an $N$-dimensional manifold in phase space, whose geometry is equivalent to an invariant $N$-dimensional torus. The geometric constraints and the uniqueness of the solutions of the equations of motion ensure that two near initial conditions will separate at most  linearly  during their time evolution, that is, an ensemble of initial conditions located in a small region of phase space will spread linearly as a function of time throughout their available phase space~\cite{Santhanam2022}. 
 
From the KAM theory, we know that the separation between regularity and chaos is not sharp, and in the same system, regular and chaotic trajectories can coexist. This is the case of the classical Dicke Hamiltonian. It can be expanded up to second order  around the minimal energy configuration, which gives rise to a quadratic Hamiltonian that is integrable. As the energy increases and the small-oscillations expansion is no longer valid, chaotic motion can appear and coexist with regular trajectories.   

The classical anisotropic Dicke Hamiltonian [Eq.~\eqref{eqn:generalized_classical_dicke_hamiltonian}] has two ($N=2$) pairs of canonical position-momentum variables, the bosonic degrees of freedom $(q,p)$ and the atomic degrees of freedom $(Q,P)$. Thus, the Hamilton's equations of motion are given by a set of four first-order coupled nonlinear differential equations,
\begin{gather}
    \label{eqn:hamilton_dicke_q}
    \dot{q} = \frac{\partial h_{\text{GD}}(\mathbf{x})}{\partial p} = \omega p + (\gamma_{-}-\gamma_{+})Pf(Q,P) , \\
    \label{eqn:hamilton_dicke_p}
    \dot{p} = -\frac{\partial h_{\text{GD}}(\mathbf{x})}{\partial q} = -\omega q - (\gamma_{-}+\gamma_{+})Qf(Q,P) , \\
    \label{eqn:hamilton_dicke_Q}
    \dot{Q} = \frac{\partial h_{\text{GD}}(\mathbf{x})}{\partial P} = \omega_{0}P + (\gamma_{-}-\gamma_{+})pf(Q,P) - \frac{P[(\gamma_{-}+\gamma_{+})qQ + (\gamma_{-}-\gamma_{+})pP]}{4f(Q,P)}, \\
    \label{eqn:hamilton_dicke_P}
    \dot{P} = -\frac{\partial h_{\text{GD}}(\mathbf{x})}{\partial Q} = -\omega_{0}Q - (\gamma_{-}+\gamma_{+})qf(Q,P) + \frac{Q[(\gamma_{-}+\gamma_{+})qQ + (\gamma_{-}-\gamma_{+})pP]}{4f(Q,P)} ,
\end{gather}
where $f(Q,P) = \sqrt{1-\left(Q^{2}+P^{2}\right)/4}$. The case $\gamma_{-}=\gamma_{+}=\gamma$ gives the Hamilton's equations of motion for the classical Dicke Hamiltonian introduced in Eq.~\eqref{eqn:classical_dicke_hamiltonian}. Since the Dicke Hamiltonian and its anisotropic version have only one integral of motion, the Hamiltonian itself, the equations above can only be solved numerically.

\subsubsection{Poincar\'e sections and Lyapunov exponents}
\label{subsubsec:poincare}

Two tests are commonly employed to diagnose whether a Hamiltonian system exhibits chaos~\cite{Ott2002Book,Goldstein2002Book}. The Poincar\'e sections offer a qualitative diagnostic~\cite{Gutzwiller1990Book,Reichl2004Book,Ott2002Book,Baker1996Book,Goldstein2002Book,Santhanam2022}, because they allow us to visually identify the loss of integrability in the system’s phase space. The Lyapunov exponent  quantifies the system’s extreme sensitivity to minor variations of initial conditions~\cite{Reichl2004Book,Ott2002Book,Baker1996Book,Goldstein2002Book,Santhanam2022,Oseledets1968}, a positive exponent followed by mixing being a hallmark of chaos.

The Poincar\'e surfaces of section provide a way to deal with the often large dimensions of the phase spaces. For a system with $N$-degrees of freedom, the phase space is $2N$-dimensional. The trajectories are parametric curves moving in a $2N-I$ hypersurface, where $I$ is the number of independent integrals of motion.  Instead of analyzing the whole trajectories, the Poincar\'e surface of section focuses on the points where they intersect at a properly chosen surface. We then obtain  a $(2N-I-1)$-dimensional map, where the motion is caught by points distributed according to some pattern.  If the trajectory is regular, the intersections of the surface of section  with the  $N$ dimensional  invariant torus  produce a  regular distribution of points. On the other hand, if the points are spread over the map without any identifiable  pattern, that indicates loss of integrability  and the motion is chaotic. Many Poincar\'e maps can be plotted simultaneously for a given surface of section, which allows for a global view of the kind of dynamics of the system, whether regular or chaotic~\cite{Santhanam2022}.

Poincar\'e sections are particularly useful for $N=2$ degrees-of-freedom chaotic Hamiltonians, for which the only integral of motion is the Hamiltonian itself, such as the classical Dicke model. In  this case, the  Poincar\'e maps consist of points  in a 2-dimensional space. Poincar\'e sections have been extensively studied for the classical Dicke Hamiltonian~\cite{Altland2012PRL,Bakemeier2013,Chavez2016,Wang2018,Wang2020} and its variants, such as the classical anisotropic Dicke Hamiltonian~\cite{DeAguiar1992,Song2008,Song2012}. The results got better over the years with the improvement of computational resources.

\begin{figure}[t!]
    \centering
    \begin{tabular}{cc}
        \includegraphics[height=0.35\textwidth]{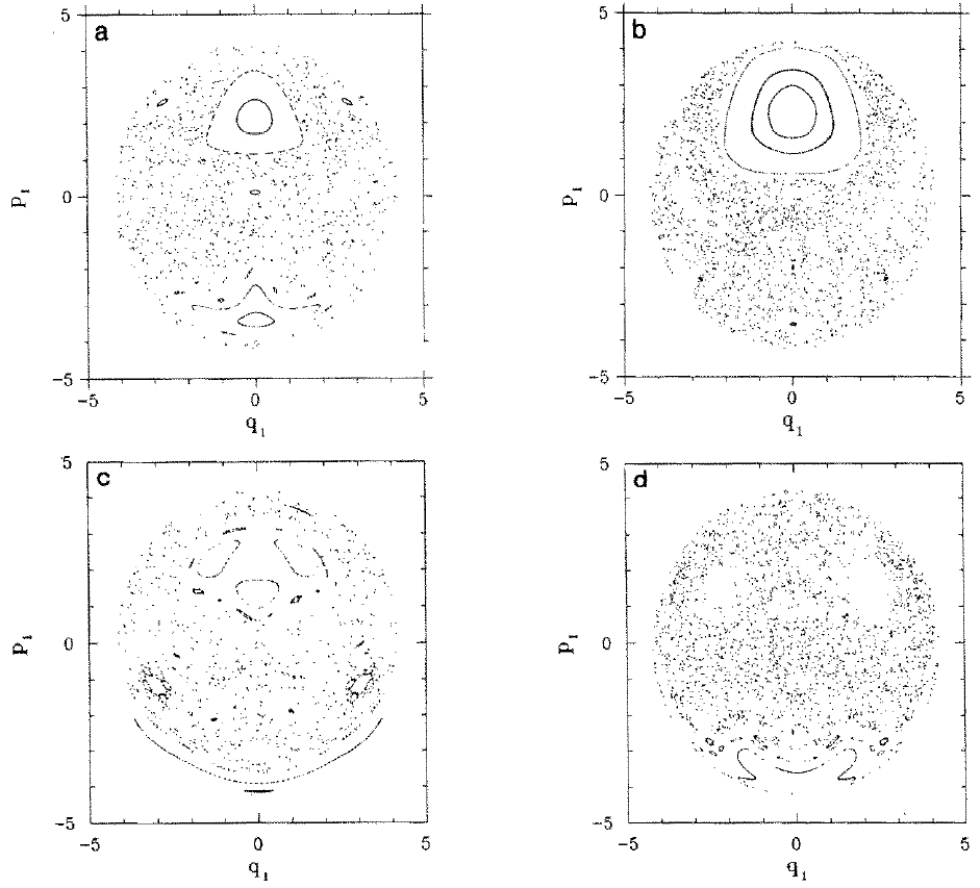} &
        \includegraphics[height=0.35\textwidth]{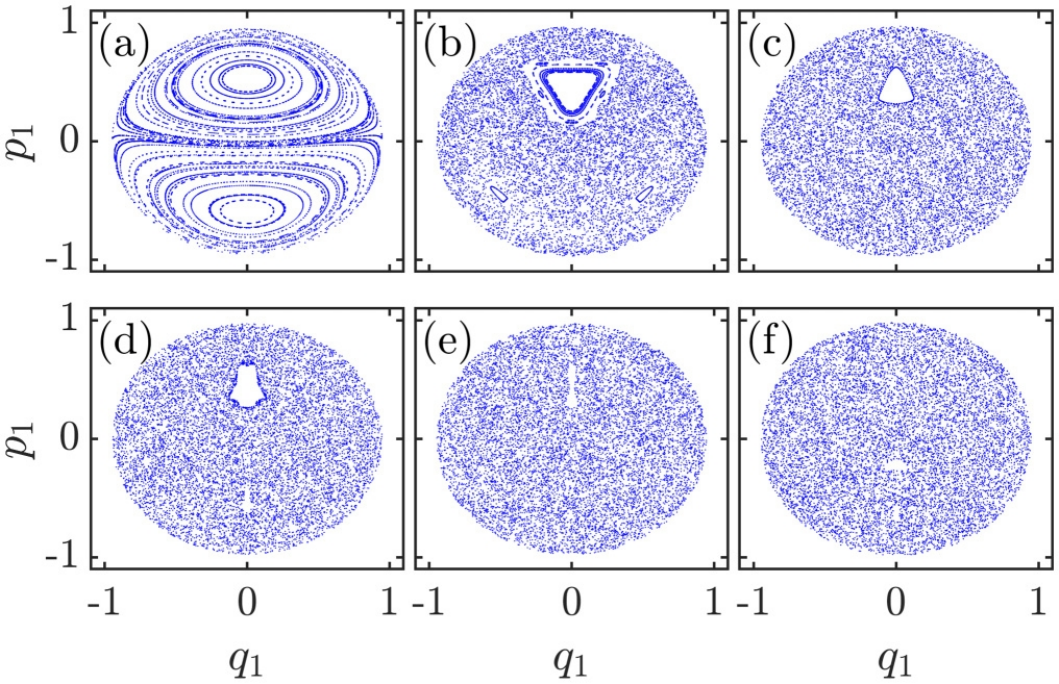}
    \end{tabular}
    \caption{(Left panels) Poincar\'e sections for the classical anisotropic Dicke Hamiltonian [Eq.~\eqref{eqn:generalized_classical_dicke_hamiltonian}]. The projections are shown on the atomic plane $Q$-$P$. Each panel refers to a different classical energy shell: (a) $\epsilon=1.38$, (b) $\epsilon=1.89$, (c) $\epsilon=3.33$, and (d) $\epsilon=5.56$. The correspondence between the labels in panels (a-d) and the main text is: $q_{1}=\sqrt{j}P$ and $p_{1}=\sqrt{j}Q$. Hamiltonian parameters: $\omega=\omega_{0}=1$, $\gamma_{-}=0.5$, $\gamma_{+}=0.2$, and $j=9/2$. Figure taken from Ref.~\cite{DeAguiar1992}.
    (Right panels) Poincar\'e sections for the classical Dicke Hamiltonian [Eq.~\eqref{eqn:classical_dicke_hamiltonian}]. The projections are shown on the atomic plane $Q$-$P$. Each panel refers to a different coupling parameter: (a) $\gamma=0.1$, (b) $\gamma=0.4$, (c) $\gamma=0.46$, (d) $\gamma=0.5$, (e) $\gamma=0.54$, and (f) $\gamma=0.7$. The correspondence between the labels in panels (a)-(f) and the main text is: $q_{1}=P/2$ and $p_{1}=Q/2$. All panels (a)-(f) were computed using the classical energy shell $\epsilon = 0.8$. Hamiltonian parameters: $\omega=\omega_{0}=1$. Figure taken from Ref.~\cite{Wang2020}.
    }
    \label{fig:poincare_anisotropic_dicke}
\end{figure}

In Fig.~\ref{fig:poincare_anisotropic_dicke}, we show, on the left hand side, early computations of the Poincar\'e sections in terms of the atomic phase space $Q$-$P$ variables for the classical anisotropic Dicke Hamiltonian [Eq.~\eqref{eqn:generalized_classical_dicke_hamiltonian}], and, on the right hand side, a more detailed version obtained later for the classical Dicke Hamiltonian [Eq.~\eqref{eqn:classical_dicke_hamiltonian}]. Both sets of Poincar\'e sections detect the characteristic transition from regularity to chaos.

\begin{figure}[t!]
    \centering
    (a) \hspace{0.14\textwidth} (b) \hspace{0.14\textwidth} (c) \\
    \includegraphics[height=0.35\textwidth]{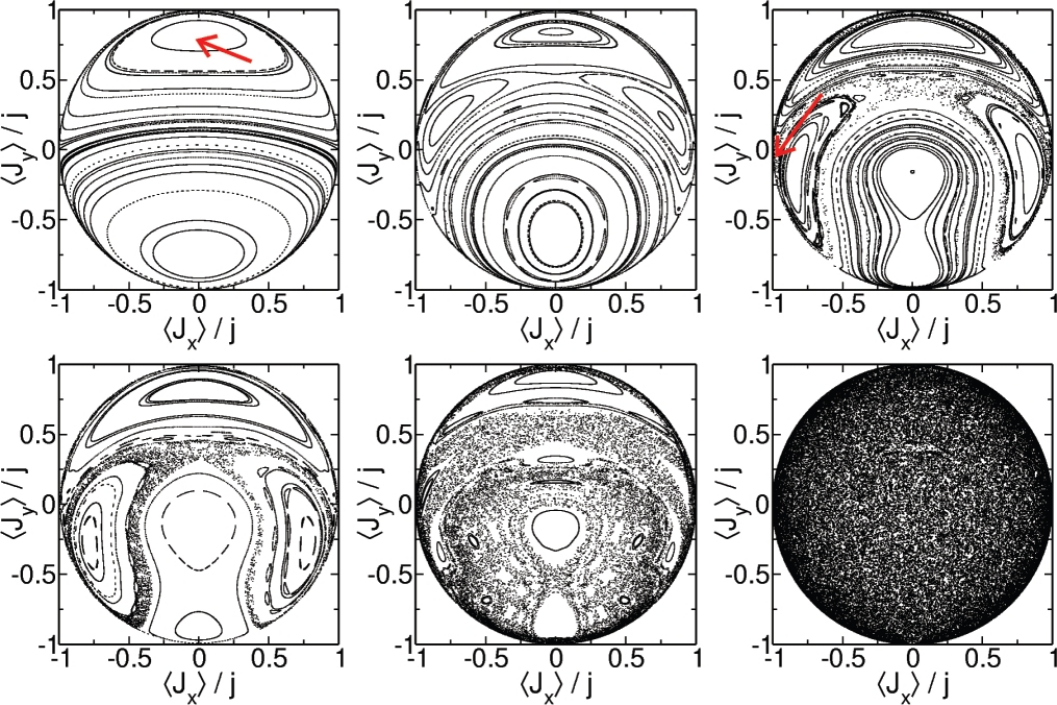}
    \caption{Poincar\'e sections for the classical Dicke Hamiltonian [Eq.~\eqref{eqn:classical_dicke_hamiltonian}]. The projections are shown on the Bloch sphere plane $j_{x}$-$j_{y}$ scaled to the system size $j$. Each panel in the top row refers to a different scaled coupling parameter: (a) $\widetilde{\gamma}=\gamma^{2}/\gamma_{\text{c}}^{2}=0.1$, (b) $\widetilde{\gamma}=0.5$, and (c) $\widetilde{\gamma}=0.6$. All panels (a)-(c) in the top row were computed using the classical energy shell $\epsilon=-0.5$. Each panel in the bottom row refers to a different classical energy shell: (a) $\epsilon=-1$, (b) $\epsilon=1$, and (c) $\epsilon=9.9$. All panels (a)-(c) in the bottom row were computed using the scaled coupling parameter $\widetilde{\gamma}=\gamma^{2}/\gamma_{\text{c}}^{2}=0.6$. Figure taken from Ref.~\cite{Bakemeier2013}.
    }
    \label{fig:poincare_dicke}
\end{figure}

In Fig.~\ref{fig:poincare_dicke}, we show Poincar\'e sections in terms of variables of the Bloch sphere, $j_{x}$-$j_{y}$, for the classical Dicke Hamiltonian [Eq.~\eqref{eqn:classical_dicke_hamiltonian}]. In this figure, we observe the transition from regularity to chaos for a set of Hamiltonian parameters and excitation energies.

The Lyapunov exponent measures the degree of divergence in time of two trajectories in phase space with near initial conditions~\cite{Reichl2004Book,Ott2002Book,Baker1996Book,Goldstein2002Book,Santhanam2022}. The exponent is obtained as follows.
The Hamiltonian evolution of two initial conditions $\mathbf{x}_{1}(0)$ and $\mathbf{x}_{2}(0)$ yields the trajectories $\mathbf{x}_{1}(t)$ and $\mathbf{x}_{2}(t)$ at an arbitrary time $t$, where the separation between them at this time is given by $\Delta\mathbf{x}(t)=\mathbf{x}_{2}(t)-\mathbf{x}_{1}(t)$. For a Hamiltonian system with $N$ degrees of freedom, the norm of the separation between the two trajectories provides a measure of their divergence, 
\begin{equation}
    ||\Delta\mathbf{x}(t)|| = \sqrt{\sum_{i}\delta x_{i}^{2}(t)},
\end{equation}
where $\delta x_{i}$ is the variation to first order of one trajectory with respect to the other~\cite{Tabor1988Book,Santhanam2022}. Sensitivity to initial conditions means that the divergence between the two trajectories grows exponentially,
\begin{equation}
    ||\Delta\mathbf{x}(t)|| \sim e^{\lambda t}||\Delta\mathbf{x}(0)||,
    \label{Eq:Lyapu}
\end{equation}
where  $\lambda$ is the maximal Lyapunov exponent. If $\lambda>0$, the system is chaotic, provided the evolution is followed by mixing. From Eq.~\eqref{Eq:Lyapu}, the maximal Lyapunov exponent can be defined by taking the  double limit
\begin{equation}
    \label{eqn:lyapunov_exponent}
    \lambda = \lim_{t\to+\infty}\lim_{||\Delta\mathbf{x}(0)||\to0}\frac{1}{t}\ln\left(\frac{||\Delta\mathbf{x}(t)||}{||\Delta\mathbf{x}(0)||}\right).
\end{equation}

For a Hamiltonian system with $N$ degrees of freedom, there are $2N$ orthogonal directions in phase space with different separation rates, which gives a set of $2N$ Lyapunov exponents, $\{\lambda_{1},\ldots,\lambda_{2N}\}$. For conservative Hamiltonians that preserve the phase space volume, the spectrum of Lyapunov exponents must satisfy the relation $\lambda_{i}=-\lambda_{2N-i+1}$ for $i=1,\ldots,2N$. The maximal Lyapunov exponent in Eq.~\eqref{eqn:lyapunov_exponent} corresponds to $\lambda=\text{Max}\{\lambda_{i}\}$. For regular motion, since the divergence between trajectories only grows linearly, all elements of the Lyapunov spectrum are zero,  so $\lambda=0$. 
Chaotic behavior requires that at least the maximal value of the Lyapunov spectrum, $\lambda\equiv \text{Max}\{\lambda_i\}$, be positive. We notice, however, that a positive Lyapunov exponent can  be found in regular systems at unstable points, but these cases are not followed by mixing.

The methods to calculate numerically the Lyapunov spectrum of Hamiltonian systems can be quite complicated and challenging. A detailed explanation on how this can be done is given in Refs.~\cite{Benettin1980Ma,Benettin1980Mb}. For conservative Hamiltonian systems, two values of the Lyapunov spectrum are zero, therefore for a two-degrees-of-freedom Hamiltonian, such as  the Dicke model, the Lyapunov spectrum is $\{\lambda, 0,0,-\lambda\}$. Because of this, from here on, every time we write ``Lyapunov exponent'', we mean the maximal Lyapunov exponent.

\begin{figure}[t!]
    \centering
    \begin{tabular}{c}
    (a) \hspace{0.16\textwidth} (b) \hspace{0.16\textwidth} (c) \\
    \includegraphics[width=0.6\textwidth]{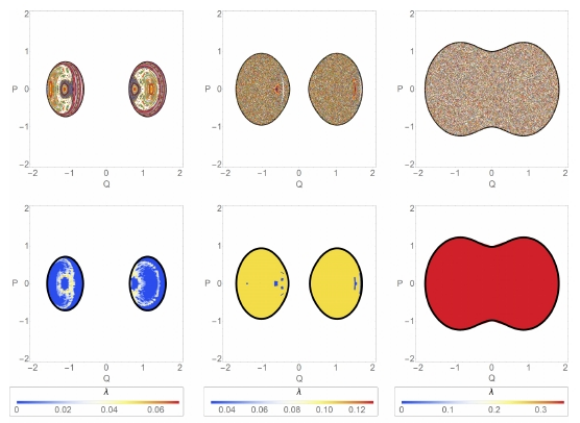}
    \end{tabular}
    \caption{(Top row of panels) Poincar\'e sections for the classical Dicke Hamiltonian [Eq.~\eqref{eqn:classical_dicke_hamiltonian}] and (bottom row of panels) maximal Lyapunov exponents computed for the classical trajectories of the model. The projections are shown on the atomic plane $Q$-$P$. Each column refers to a different classical energy shell: (a) $\epsilon=-1.4$, (b) $\epsilon=-1.1$, and (c) $\epsilon=-0.5$. In each panel, from (a) to (c) in the top and bottom row, the black solid border represents the available phase space at each energy shell. The color scales at the bottom indicate the value of the Lyapunov exponent: dark blue represents regularity, while other colors indicate chaos. Hamiltonian parameters: $\omega=\omega_{0}=1$ and $\gamma=2\gamma_{\text{c}}$. Figure taken from Ref.~\cite{Chavez2016}.
    }
    \label{fig:poincare_lyapunov}
\end{figure}

Reference~\cite{Chavez2016} gives detailed information about the ranges of Hamiltonian parameters and excitation energies where chaos appears in the Dicke model. The correspondence between the Poincar\'e sections and the Lyapunov exponents for this model is shown in Fig.~\ref{fig:poincare_lyapunov}, where the transition from integrability to chaos is detected  by both indicators of classical chaos.

\subsubsection{Maps of chaos}
\label{subsubsec:chaos_map}

\begin{figure}[t!]
    \centering
    \begin{tabular}{c}
    \includegraphics[width=0.9\textwidth]{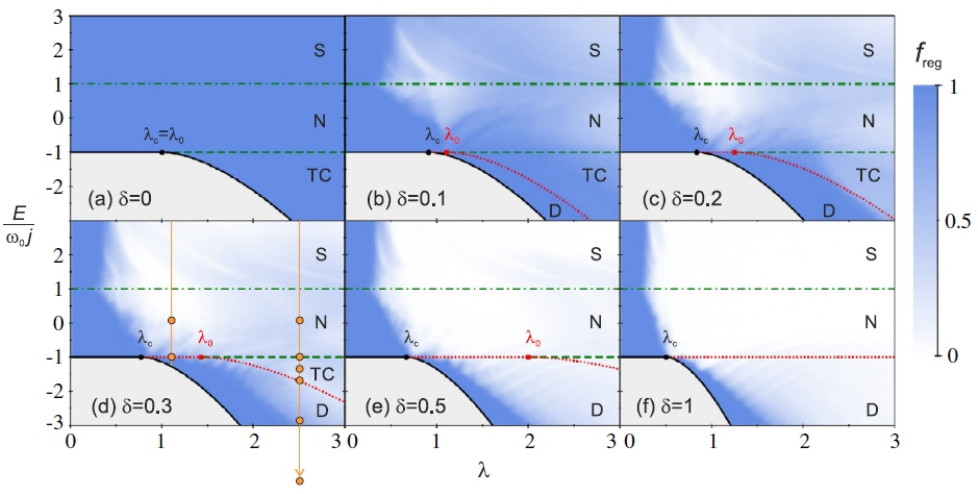} \\
    \includegraphics[width=0.9\textwidth]{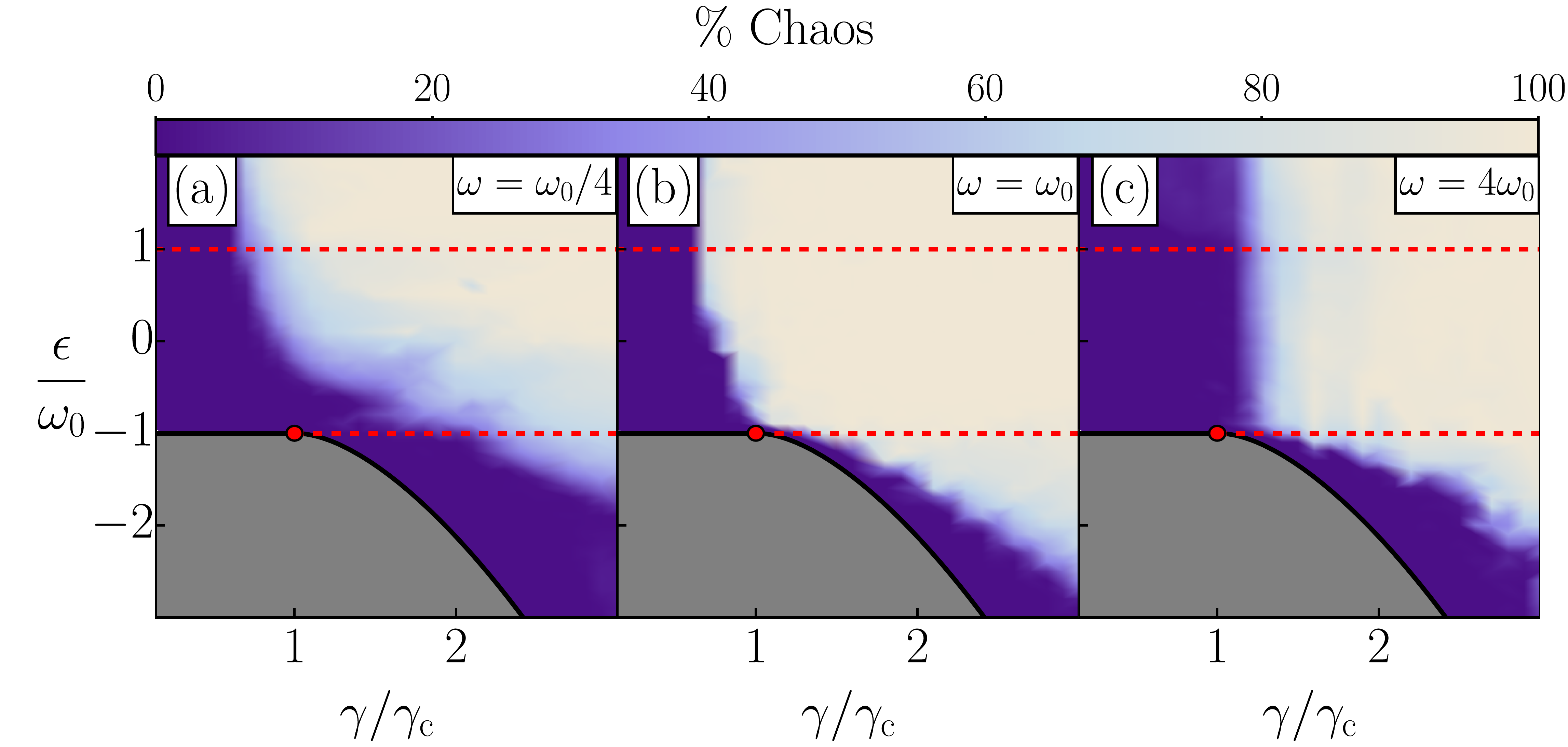}
    \end{tabular}
    \caption{Map of chaos for (top panels) the classical anisotropic Dicke Hamiltonian [Eq.~\eqref{eqn:generalized_classical_dicke_hamiltonian}] and (bottom panels) the classical Dicke Hamiltonian [Eq.~\eqref{eqn:classical_dicke_hamiltonian}]. The top figure was taken from Ref.~\cite{Kloc2017}. The bottom figure was modified from a figure shown in Ref.~\cite{Chavez2016}. 
    Each of the top panels refers to a different counter-rotating coupling parameter: (a) $\gamma_{+}=0$, (b) $\gamma_{+}=0.1\gamma_{-}$, (c) $\gamma_{+}=0.2\gamma_{-}$, (d) $\gamma_{+}=0.3\gamma_{-}$, (e) $\gamma_{+}=0.5\gamma_{-}$, and (f) $\gamma_{+}=\gamma_{-}$. The correspondence between the labels in panels (a)-(f) and the main text is: $\lambda=\gamma_{-}$ and $\delta=\gamma_{+}/\gamma_{-}$. The color scale on the right indicates chaos (lighter tones) and high fraction of regularity (darker tones) in phase space. In each panel, from (a) to (f), the black solid line represents the ground-state energy, while the green dotted-dashed, green dashed, and red dotted lines represent the energies of the ESQPTs. The black dot marks the QPT at the ground-state energy in the thermodynamic limit ($\mathcal{N} \to \infty$) for all panels (a)-(f). Hamiltonian parameters: $\omega=\omega_{0}=1$. 
    Each of the bottom panels refers to a different value of the field frequency: (a) $\omega=\omega_{0}/4$, (b) $\omega=\omega_{0}$, and (c) $\omega=4\omega_{0}$. The color scale at the top indicates regularity (darker tones) and high percentage of chaos (lighter tones) in phase space. In each panel, from (a) to (c), the black solid line represents the ground-state energy, while the horizontal red dashed lines represent the energies of the ESQPTs. The red dot marks the QPT at the ground-state energy in the thermodynamic limit ($\mathcal{N} \to \infty$) for all panels (a)-(c). Hamiltonian parameters: $\omega_{0}=1$ and $\gamma=2\gamma_{\text{c}}$.
    }
    \label{fig:percentage_chaos} 
\end{figure}

It is possible to obtain the  percentage of chaos for a given energy shell by using the Lyapunov exponents for a large set of classical trajectories. This provides a global view of the onset of chaos as a function of energy and  of the parameters of the model. These maps of classical chaos are also useful for the comparison with quantum chaos, as discussed in Sec.~\ref{subsec:quantum_classical}.

To obtain a map of chaos,  we select a set of initial conditions homogeneously distributed in phase space for a given classical energy shell and fixed Hamiltonian parameters. These initial conditions are evolved in time and those with a positive Lyapunov exponent are identified. The fraction of chaotic trajectories over the total number of initial trajectories gives the percentage of chaos for the  energy shell. This procedure is repeated for different  energy shells and Hamiltonian parameters. The result is a map of the percentage of chaos in the parameter-energy  space of the model.

Figure~\ref{fig:percentage_chaos} shows the maps of percentage of chaos for the classical anisotropic Dicke model~\cite{Kloc2017} in Eq.~\eqref{eqn:generalized_classical_dicke_hamiltonian} (top six panels) and for the classical Dicke model~\cite{Chavez2016} in Eq.~\eqref{eqn:classical_dicke_hamiltonian} (bottom three panels). The panels at the top are in resonance, $\omega=\omega_{0}$. They interpolate from the integrable Tavis-Cummings model (marked $\delta=0$  in the figure) to the chaotic Dicke model ($\delta=1$ in the figure).

The maps  at the bottom of Fig.~\ref{fig:percentage_chaos} depict three scenarios: $\omega<\omega_{0}$ (left),  $\omega=\omega_{0}$ (middle), and $\omega>\omega_{0}$ (right).  By examining  these panels, one can conclude   that the onset of chaos in the Dicke model is not directly related to the QPT, as it  was suggested in Ref.~\cite{Brandes2013}. This wrong impression arises if we focus on the resonant case,  $\omega=\omega_{0}$  (as considered in Ref.~\cite{Brandes2013}), in the middle panel, and especially for $\omega>\omega_{0}$, as in the right panel. In  these cases, one might infer   that  in the normal phase ($\gamma < \gamma_{\text{c}}$)  the dynamics is regular, while in the superradiant phase ($\gamma \geq \gamma_{\text{c}}$), the dynamics becomes  chaotic as  energy increases. However, the leftmost panel, corresponding to  $\omega<\omega_{0}$, invalidates this generalization. There, we observe  that chaos  also emerges  in the normal phase ($\gamma < \gamma_{\text{c}}$) when  energies are high. Subsequent studies~\cite{PerezFernandez2011,Bastarrachea2014PRAb} suggested  that the onset of chaos in the Dicke model  was related to the  ESQPT appearing in the superradiant phase at energy $\epsilon/\omega_0=-1$ (indicated by  horizontal lines in  the panels). However,  the  panels at  the bottom  of Fig.~\ref{fig:percentage_chaos} reveal  that this is not true either. In the leftmost panel, chaos appears above the ESQPT for $\gamma/\gamma_{\text{c}} < 2.3$, and   below the ESQPT for  larger $\gamma$. In the other two panels, chaos occurs below the ESQPT for almost  the entire  range $\gamma/\gamma_{\text{c}}>1$. Thus, in the Dicke model,  the relationship between the onset of chaos, QPT and ESQPT  is more subtle than  a simple  direct connection. This assertion can also be extended  to the anisotropic Dicke Hamiltonian, by  observing the similarities between the chaos maps at the top and those at the bottom in Fig.~\ref{fig:percentage_chaos}.

\subsection{Quantum chaos}
\label{subs:quantchaos}

Classically chaotic behavior arises in the presence of nonlinearities~\cite{Casati1980,Tabor1988Book,Baker1996Book,Ott2002Book,Reichl2004Book}. Quantum mechanics, on the other hand, is linear and concepts such as trajectory and phase space are not well defined due to the Heisenberg uncertainty principle. The term ``quantum chaos'' refers to signatures found in the spectrum and eigenstates of a quantum system that indicate whether it is chaotic in the classical limit~\cite{Bohigas1984,Haake1991Book,Casati1995Book,Stockmann2006Book,Wimberger2014Book}. This difference from chaos in classical systems motivated some authors to use the term ``quantum chaology'' instead of quantum chaos~\cite{Berry1987,Berry1989PS}. However, the use of ``quantum chaos'' became widespread and nowadays, it refers also to systems without a well-defined classical limit, such as spin-1/2 models, provided they exhibit correlated eigenvalues~\cite{Haake1991Book,Reichl2004Book,Stockmann2006Book,Wimberger2014Book} similar to what one sees in random matrix theory~\cite{Mehta1991Book,Guhr1998} and eigenstates that ergodically fill their energy shell~\cite{Borgonovi1996}.

\subsubsection{Spectral fluctuations and random matrix theory}

Random matrices were employed in Statistics since the first half of the last century~\cite{Forrester2003,Fyodorov2011}. Their use in Physics was strongly influenced by E. P. Wigner, who in the 1950s used real and symmetric matrices filled with random numbers to describe statistically the spectra of heavy nuclei~\cite{Fyodorov2011,Wigner1951PCPS,Wigner1958}. The success of this approach prompted the use of full random matrices to analyze  spectral fluctuations in various other complex quantum many-body systems~\cite{Dyson1962JMPa,Dyson1962JMPb,Dyson1962JMPc,Dyson1963,Mehta1963,Porter1965Book,Mehta1991Book,Brody1981}, from atoms and molecules~\cite{Zimmermann1988} to quantum dots~\cite{Beenakker1997} and 
other systems.

Ensembles of $d$-dimensional random matrices  with random numbers drawn from a Gaussian distribution, $P_{d,\beta}(\hat{H}) \propto \exp\left[-\beta d\text{Tr}\left(\hat{H}^{2}\right)\right]$, were classified according to their symmetries~\cite{Dyson1962JMPa,Dyson1962JMPb,Dyson1962JMPc,Wigner1959Book,Mehta1991Book,Guhr1998}. In a Gaussian orthogonal ensemble (GOE), the Hamiltonian matrices are invariant under an orthogonal transformation, they are real and symmetric, and $\beta=1$ in $P_{d,\beta}(\hat{H})$. In a Gaussian unitary ensemble (GUE), $\beta=2$ and the Hamiltonian matrices are Hermitian,  being invariant under a unitary transformation. In a Gaussian symplectic ensemble (GSE), $\beta=4$, and the Hamiltonian matrices are  invariant under a symplectic transformation. In this review article, we focus on the isolated Dicke model, so  its level statistics is compared with that of GOE.

The conjecture that in physical systems, level statistics similar to those in random matrix theory (RMT) indicates chaos in the classical limit came much later than Wigner's studies of nuclei. The conjecture was put forward in the 1980s with the studies of billiards in the quantum and classical domain~\cite{Casati1980,Bohigas1984}, and it became known as the {\it quantum chaos conjecture} or the Bohigas-Giannoni-Schmit conjecture. Since then, the traditional analysis of quantum chaos has focused on spectral fluctuations. A rigid spectrum is characteristic of correlated eigenvalues, as those of full random matrices and of chaotic physical quantum systems. Various ways to detect spectral correlations have been explored in different quantum systems~\cite{Bohigas1975,Seligman1984,Seligman1985JPAa,Seligman1985JPAb,Alhassid1989,Friedrich1989,Izrailev1990,Zelevinsky1996PR}, as we now explain in the context of the Dicke model.

{\it Nearest-neighbor spacing distribution:} The distribution $P(s)$ of spacings $s$ of nearest-neighbor levels is a standard test to detect short-range spectral correlations~\cite{Stockmann2006Book,Wimberger2014Book,Haake1991Book,Dyson1962JMPa,Dyson1962JMPb,Dyson1962JMPc,Mehta1991Book,Guhr1998}. The analysis requires separating the levels by symmetry sectors and unfolding the spectrum, so that the density of states of the renormalized eigenvalues is unity~\cite{Gubin2012}. 

For quantum chaotic Hamiltonians that are time-reversal invariant, the level spacing distribution agrees with the GOE predictions from RMT, thus following the Wigner-Dyson surmise~\cite{Hillery1984},
\begin{equation}
    \label{eqn:wigner_dyson_surmise}
    P_{\text{WD}}(s) = \frac{\pi}{2}s\,e^{-\pi s^{2}/4}.
\end{equation}
In 1977, M. V. Berry and M. Tabor found that the energy levels of most regular quantum systems are not correlated and follow a Poisson distribution, 
\begin{equation}
    \label{eqn:poisson_distribution}
    P_{\text{P}}(s) = e^{-s},
\end{equation}
which indicates that the levels are distributed at random. However, for regular systems with nearly equidistant levels, in what became known as ``picket-fence'' spectrum, the distribution can exhibit a peak away from zero~\cite{Berry1977}.

The standard way to unfold an energy spectrum considers the computation of the cumulative spectral function,
\begin{equation}
    \eta(E) = \int_{-\infty}^{E}dE'\rho(E') = \sum_{k}\Theta(E-E_{k}),
\end{equation}
where $\rho(E) = \sum_{k}\delta(E-E_{k})$ is the density of states introduced in Eq.~\eqref{eqn:level_density} and $\Theta$ is the Heaviside step function. 
Similarly to the density of states, $\rho(E) = \nu(E) + \nu_{\text{f}}(E)$, the cumulative spectral function, $\eta(E) = \xi(E)+\xi_{\text{f}}(E)$, can be separated in a smooth term $\xi(E)=\int_{-\infty}^{E}dE'\nu(E')$ and a fluctuating term $\xi_{\text{f}}(E)=\int_{-\infty}^{E}dE'\nu_{\text{f}}(E')$. The unfolding procedure consists of mapping the set of energy levels $E_{k}$ to the new set $\xi_{k}$ scaled by the smooth term $\xi_{k}=\xi(E_{k})$  with $k=1,\ldots,d$. This procedure ensures that the level spacing distribution detects only the local fluctuations. The unfolding procedure for the Dicke model can be done using the semiclassical approximation to the smooth term of the density of states $\nu(E)$ given in Eq.~\eqref{eqn:semiclassical_density_states}, after ordering the $N_{\text{c}}$ converged energy levels of the model.

An alternative test of short-range correlations that avoids the unfolding procedure is the ratio of consecutive energy levels~\cite{Oganesyan2007,Atas2013}
\begin{equation}
    \label{eqn:ratio_energy_levels}
    \widetilde{r}_{k} = \frac{\min(s_{k}, s_{k-1})}{\max(s_{k}, s_{k-1})},
\end{equation}
where $s_{k} = E_{k+1} - E_{k}$. For uncorrelated eigenvalues, whose level spacing distribution follows the Poisson distribution, the mean value of $\widetilde{r}_k$ is $\langle \widetilde{r} \rangle_{\text{P}} \approx 0.386$. However, for correlated levels, whose level spacing distribution is Wigner–Dyson, the mean value is $\langle \widetilde{r} \rangle_{\text{WD}} \approx 0.536$.

\begin{figure}[t!]
    \centering
    \includegraphics[width=0.7\textwidth]{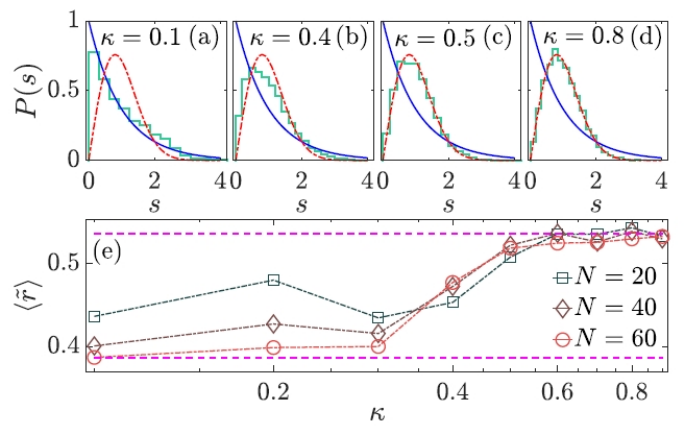}
    \caption{[(a)-(d)] Level spacing distribution for eigenvalues of the Dicke Hamiltonian with positive parity. Each panel refers to a different coupling parameter: (a) $\gamma=0.1$, (b) $\gamma=0.4$, (c) $\gamma=0.5$, and (d) $\gamma=0.8$. The correspondence between the labels in panels (a)-(d) and the main text is: $\kappa=\gamma$. All panels (a)-(d) were computed using the system size $j=30$. In each panel, from (a) to (d), the blue solid line represents the Poisson distribution [Eq.~\eqref{eqn:poisson_distribution}] and the red dashed line indicates the Wigner-Dyson surmise [Eq.~\eqref{eqn:wigner_dyson_surmise}].
    (e) Average ratio of consecutive energy levels [Eq.~\eqref{eqn:ratio_energy_levels}] as a function of the coupling parameter. Each line represents a different system size: $j=10$ (squares), $j=20$ (diamonds), and $j=30$ (circles). The correspondence between the labels in panel (e) and the main text is: $\kappa=\gamma$ and $N=2j$. The horizontal magenta dashed lines represent the limiting mean values of the ratio for Poisson ($\langle \widetilde{r} \rangle_{\text{P}} = 0.386$) and GOE ($\langle \widetilde{r} \rangle_{\text{WD}} = 0.536$) distributions.
    Hamiltonian parameters: $\omega=\omega_{0}=1$. Figure taken from Ref.~\cite{Wang2020}.
    }
    \label{fig:nns_dicke}
\end{figure}

In Figs.~\ref{fig:nns_dicke}(a)-\ref{fig:nns_dicke}(d), we show the level spacing distribution for the Dicke Hamiltonian, as the coupling parameter $\gamma$ grows and the system moves from the regular [Fig.~\ref{fig:nns_dicke}(a)] to a chaotic regime [Fig.~\ref{fig:nns_dicke}(d)]. In Fig.~\ref{fig:nns_dicke}(e), we show the average ratio of consecutive energy levels  as a function of $\gamma$ for different system sizes. The agreement between the numerical results and the limiting values of $\langle \widetilde{r} \rangle$ improves as the system size increases. The analysis of level statistics for the isolated Dicke model has been done in various works~\cite{Emary2003PRL,Emary2003PRE,Bastarrachea2014PRAb,Bastarrachea2016PRE,Bastarrachea2017PS,Wang2020}.

{\it Spectral form factor:} A more complete analysis of the spectrum requires the use of quantities that detect both short- and long-range correlations, such as rigidity, level number variance~\cite{Guhr1998} and the spectral form factor~\cite{Mehta1991Book}. The advantage of the spectral form factor is that it does not require unfolding and detects correlations even in the presence of symmetries~\cite{Cruz2020,Santos2020}.

The two-level spectral form factor is defined as
\begin{equation}
    S_{\text{FF}}(t) =  \frac{1}{d^2}\sum_{k, k'}e^{-i(E_{k}-E_{k'})t}.
    \label{Eq:SFFnoaverage}
\end{equation}
The average over a GOE gives the following expression~\cite{Torres2018,Das2024PRR}
\begin{equation}
	\label{eq_SFF_GOE}
	\langle S_{\text{FF}}(t) \rangle \simeq \frac{\mathcal{J}_1^2(2 \Gamma t)}{\Gamma^{2} t^{2}} + \frac{1}{d}\left[ 1 -  b_2\left(\frac{\Gamma t}{2d}\right) \right],
\end{equation}
where the first term on the right-hand side is obtained from the Fourier transform of the density of states, which is a semicircle for the GOE, resulting in the Bessel function of the first kind, $\mathcal{J}_1$, and $\Gamma$ is the width of the level density. The $b_2$ function on the second term is the two-level form factor~\cite{Mehta1991Book},
\begin{equation}
    \label{eqn:two_level_form_factor_goe}
    b_{2}(t) =  \left\{\begin{array}{ll} 1-2 t + t \ln(2 t+1) & \text{if } 0< t\leq1 \\ -1+ t \ln\left(\dfrac{2 t+1}{2 t-1}\right) & \text{if } t >1\end{array}\right. ,
\end{equation}
which only appears if the eigenvalues are correlated as in RMT. The last term is the infinite-time average. The spectral form factor provides an analysis of the spectrum in the time domain. It is contained in the survival probability, which is a dynamical quantity described for the Dicke model in Sec.~\ref{subsec:survival_probability}.

\subsubsection{Ergodicity and finite-time dynamics}
\label{subsubsec:FTDynamics}

In the study of quantum chaos, the traditional method of using spectral statistics to infer different classical dynamical properties falls short in typical scenarios. This includes situations like slow relaxation processes, dynamical localization, or other forms of behavior that deviate from ergodicity. Extensive research has identified these deviations in fundamental models, such as billiard systems~\cite{Richens1981,Cheon1989,Borgonovi1996,DeAguiar2008,Tuan2012,Araujo2013}. However, these deviations have also been observed in other systems under unusual conditions~\cite{Casati1985,Relano2004}, and even in non-physical systems~\cite{Benet2003}.

A recent novel approach has been proposed to address the challenges associated with spectral analysis. This approach involves linking a quantum energy shell to classical finite-time trajectories~\cite{Chen2025}. Initially, this method was validated in billiard systems and later tested in systems with a few degrees of freedom, such as the kicked top and the Dicke model~\cite{Wang2025}. While this innovative technique offers new insights into quantum chaos and warrants further exploration, spectral analysis remains fundamental to the traditional characterization of quantum chaos~\cite{Haake1991Book,Casati1995Book,Stockmann2006Book,Wimberger2014Book}. Thus, in this work, the characterization of quantum chaos in the Dicke model is thoroughly described using conventional spectral analysis.

\subsubsection{Eigenstate thermalization hypothesis}
\label{subsubsec:ETH}

The eigenstate thermalization hypothesis (ETH) states that if the eigenstate expectation values of a few-body observable do not fluctuate much, then its infinite-time average,
\begin{equation}
    \overline{O} = \lim_{t\to+\infty}\frac{1}{t}\int_{0}^{t}dt'O(t')  = \sum_{k}|c_{k}|^{2}\langle E_{k}|\hat{O}|E_{k}\rangle,
\end{equation}
where $c_k = \langle E_k|\Psi(0)\rangle $ and $|\Psi(0)\rangle $ is the initial state, agrees with its thermodynamic (microcanonical) average~\cite{Dalessio2016}, 
\begin{equation}
     O_{\text{mic}} = \frac{1}{W_{E,\Delta E}}\sum_{k}\langle E_{k}|\hat{O}|E_{k}\rangle,
\end{equation}
where $W_{E,\Delta E}$ is the number of eigenstates  in an energy window $[E-\Delta E,E+\Delta E]$. The agreement should be perfect in the thermodynamic limit. The ETH is valid when the system exhibits chaotic features~\cite{Jensen1985,Deutsch1991,Srednicki1994}, that is, the eigenvalues are correlated and the eigenstates behave as random vectors~\cite{Zelevinsky1996PR,Borgonovi2016}.

\begin{figure}[t!]
    \centering
    \includegraphics[width=0.7\textwidth]{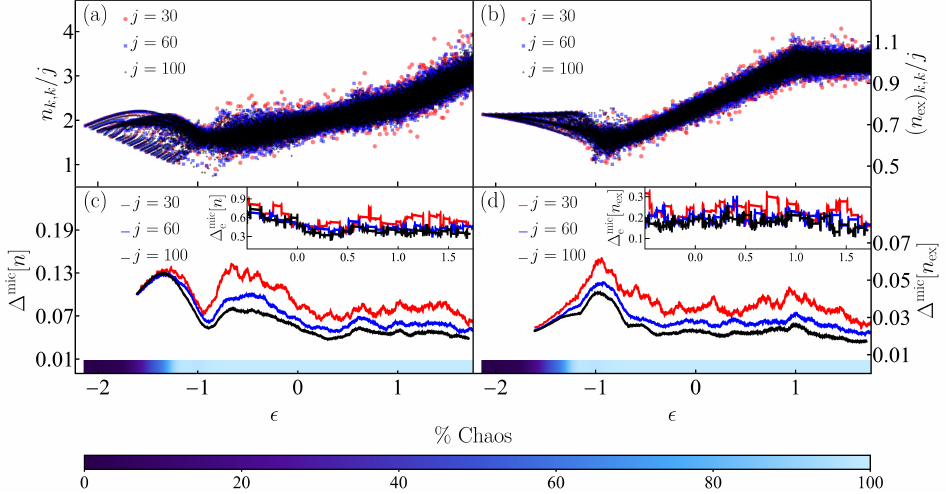}
    \caption{[(a)-(b)] Expectation values of (a) the number of photons $\hat{n}=\hat{a}^{\dag}\hat{a}$ and (b) the number of excited atoms $\hat{n}_{\text{ex}}=\hat{J}_{z}+j\hat{\mathbb{I}}$ for eigenstates of the Dicke Hamiltonian with positive parity. The expectation values are scaled to the system size $j$.
    [(c)-(d)] Deviations of the eigenstate expectation values from their microcanonical value [Eq.~\eqref{eqn:delta_MIC}] for (c) the number of photons and (d) the number of excited atoms. The insets in panels (c) and (d) show the corresponding extremal fluctuations [Eq.~\eqref{eqn:delta_MICe}] for each observable.
    For all panels (a)-(d), three system sizes are shown: $j=30$ (red), $j=60$ (blue), and $j=100$ (black). The color scale at the bottom indicates regularity (darker tones) and high percentage of chaos (lighter tones) in phase space with respect to the value of the energy shell $\epsilon$. Hamiltonian parameters: $\omega = \omega_0 = 1$ and $\gamma = 2\gamma_{\text{c}}$. Figure taken from Ref.~\cite{Villasenor2023}.
    }
    \label{fig:eth_dicke}
\end{figure}

The validity of the ETH for the chaotic regime of the Dicke model was studied in Ref.~\cite{Villasenor2023}. The Dicke model possesses two natural observables, the number of photons, $\hat{n}=\hat{a}^{\dagger}\hat{a}$, and the number of excited atoms, $\hat{n}_{\text{ex}}=\hat{J}_{z}+j\hat{1}$.  Figure~\ref{fig:eth_dicke}(a) [Fig.~\ref{fig:eth_dicke}(b)] shows the eigenstate expectation values for $\hat{n}$ [$\hat{n}_{\text{ex}}$] for a coupling parameter  where the model is chaotic ($\gamma=2\gamma_{\text{c}}$). This sort of figure is known as Peres lattice~\cite{Peres1984PRL}. Larger fluctuations are found for low energies, where the eigenstates are not chaotic. We also see that the fluctuations in the chaotic region ($\epsilon \gtrsim -0.8$) decrease as the system size increases.

To quantify the size of the fluctuations, we show in the main panel of Figs.~\ref{fig:eth_dicke}(c) and~\ref{fig:eth_dicke}(d), the deviation of the eigenstate expectation values with respect to the microcanonical value~\cite{RigolSantos2010,Santos2010PREb},
\begin{equation}
    \label{eqn:delta_MIC}
    \Delta^{\text{mic}}[O] = \frac{\sum_{k}|\langle E_{k}|\hat{O}|E_{k}\rangle-O_{\text{mic}}|}{\sum_{k}\langle E_{k}|\hat{O}|E_{k}\rangle}, 
\end{equation}
as a function of energy
and, in the insets, we also show the normalized extremal fluctuations
\begin{equation}
    \label{eqn:delta_MICe}
    \Delta^{\text{mic}}_{e}[O] = \left|\frac{\max(O) - \min(O)}{O_{\text{mic}}}\right|,
\end{equation}
where $\max(O)$ [$\min(O)$] is the maximum [minimum] value in a selected energy window around $\epsilon$. The figures confirm that in the chaotic energy region, the size of the fluctuations for both observables decreases as the system size increases, indicating that the ETH should hold.

\subsubsection{Structure of the eigenstates}

The validity of the ETH is closely related to the presence of chaotic eigenstates that spread ergodically in their energy shell. To analyze the structure of the eigenstates, we can resort to the entanglement entropy, which is the von Neumann entropy of the reduced density matrix of a subsystem of the system,
\begin{equation}
    \label{eqn:entanglement_entropy}
    S_{\text{E}} = -\text{Tr}[\hat{\rho}_{\text{A}}\ln(\hat{\rho}_{\text{A}})] = -\text{Tr}[\hat{\rho}_{\text{B}}\ln(\hat{\rho}_{\text{B}})],
\end{equation}
where the reduced density matrix $\hat{\rho}_{\text{A}} = \text{Tr}_{\text{B}} (\hat{\rho})$ is obtained by tracing out the degrees of freedom of the complementary subspace B. Large entanglement (large $S_{\text{E}}$) implies a delocalized state. The entanglement entropy of random vectors as in random matrices was obtained in Ref.~\cite{Page1993}. 

The Dicke model is a bipartite system and its Hilbert space is a tensor product of two subspaces, $\mathcal{H}_{\text{D}}=\mathcal{H}_{\text{A}}\otimes\mathcal{H}_{\text{B}}$. The subspace $\mathcal{H}_{\text{A}}$ defines the atomic sector of the model and has a finite dimension of magnitude $2j+1$, while the subspace $\mathcal{H}_{\text{B}}$ defines the bosonic sector of the model and has infinite dimension. Tracing out the bosonic subspace from a state of the Dicke model gives the reduced density matrix over the atomic subspace,
\begin{equation}
    \hat{\rho}_{\text{A}} = \text{Tr}_{\text{B}}(\hat{\rho}) = \sum_{m_{z},m'_{z}}\left(\sum_{n}c_{n,m_{z}}c_{n,m'_{z}}^{\ast}\right)|j,m_{z}\rangle\langle j,m'_{z}|,
\end{equation}
where $c_{n,m_{z}}=\langle n;j,m_{z}|\Psi\rangle$ and $| n;j,m_{z}\rangle $ is the Fock basis in Eq.~\eqref{eqn:fock_basis}. A derivation of the entanglement entropy for the ground state of the Dicke model using a Holstein-Primakoff approximation can be found in Ref.~\cite{Vidal2007}.

\begin{figure}[t!]
    \centering
    \includegraphics[width=0.5\textwidth]{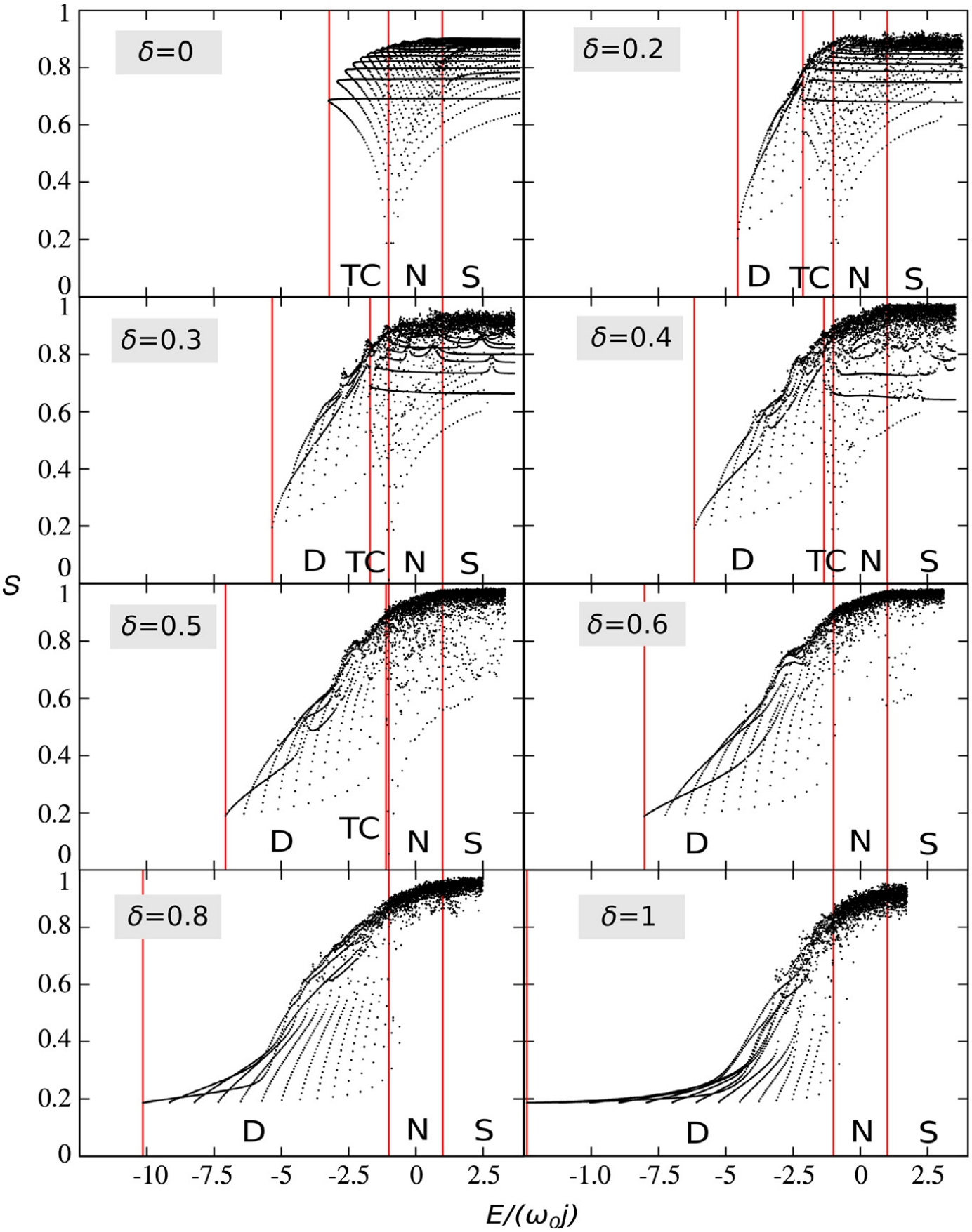} 
    \caption{Von Neumann entanglement entropy [Eq.~\eqref{eqn:entanglement_entropy}] for eigenstates of the anisotropic Dicke Hamiltonian [Eq.~\eqref{eq:HD_generalized}] with positive parity. Each panel refers to a different counter-rotating coupling parameter: $\gamma_{+}=0,0.5,0.75,1,1.25,1.5,2$, and 2.5. The correspondence between the labels in all panels and the main text is: $S=S_{\text{E}}$, $\lambda=\gamma_{-}$, and $\delta=\gamma_{+}/\gamma_{-}$. In each panel, the vertical red solid lines represent the ground-state and ESQPT energies. Hamiltonian parameters: $\omega=\omega_{0}=1$, $\gamma_{-}=2.5$, and $j=20$. Figure taken from Ref.~\cite{Kloc2017}.
    }
    \label{fig:entanglement_anisotropic_dicke}
\end{figure}

Figure~\ref{fig:entanglement_anisotropic_dicke} shows the entanglement entropy for the anisotropic Dicke Hamiltonian for a set of coupling parameters~\cite{Kloc2017}. The top right panel ($\delta=\gamma_{+}/\gamma_{-}=0$) gives $S_{\text{E}}$ for the integrable limit of the Dicke model, the Tavis-Cummings model in Eq.~\eqref{eqn:tavis_cummings_hamiltonian}. There is a clear pattern, with lines of points separated from each other. As the interaction increases and the model moves to the chaotic regime, from $\delta=0$ to $\delta=1$, the pattern disintegrates, persisting only at low energies, where chaos does not set in.

\begin{figure}[t!]
    \centering
    \includegraphics[width=0.7\textwidth]{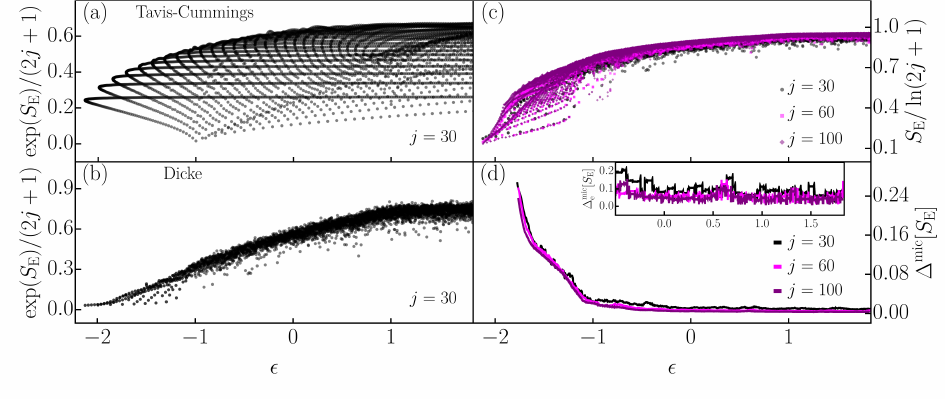}
    \caption{[(a)-(b)] Exponential of the von Neumann entanglement entropy [Eq.~\eqref{eqn:entanglement_entropy}] for eigenstates of (a) the Tavis-Cummings Hamiltonian [Eq.~\eqref{eqn:tavis_cummings_hamiltonian}] and (b) the Dicke Hamiltonian [Eq.~\eqref{eqn:dicke_hamiltonian}]. The exponential of each entropy is scaled to the value $2j+1$. Panels (a) and (b) were computed using the system size $j=30$.
    (c) Von Neumann entanglement entropy for eigenstates of the Dicke Hamiltonian for three system sizes: $j=30$ (black), $j=60$ (magenta), and $j=100$ (purple). Each entropy is scaled to the value $\ln(2j+1)$.
    (d) Deviation of the entanglement entropy from its microcanonical value [Eq.~\eqref{eqn:delta_MIC}]. The inset in panel (d) shows the deviation using the extremal fluctuations [Eq.~\eqref{eqn:delta_MICe}].
    Hamiltonian parameters: $\omega=\omega_{0}=1$ and $\gamma=2\gamma_{\text{c}}$.
    }
    \label{fig:entanglement_dicke}
\end{figure}

Figure~\ref{fig:entanglement_dicke}(a) shows the exponential of the von Neumann entanglement entropy for eigenstates of the Tavis-Cummings model, where an organized pattern of these values can be identified. In contrast, Fig.~\ref{fig:entanglement_dicke}(b) shows the exponential of the entanglement entropy for eigenstates of the Dicke model, where the pattern disappears for $\epsilon \gtrsim -0.8$ and the fluctuations are small. In Fig.~\ref{fig:entanglement_dicke}(c), the entanglement entropy is computed for different system sizes of the Dicke model and Fig.~\ref{fig:entanglement_dicke}(d) shows the deviation $\Delta^{\text{mic}}[S_E]$ [Eq.~\eqref{eqn:delta_MIC}] in the main panel and the extreme deviation $\Delta_e^{\text{mic}}[S_E]$ in Eq.~\eqref{eqn:delta_MICe} in the inset. These figures support the analysis in Fig.~\ref{fig:eth_dicke}, confirming that chaotic eigenstates lead to small fluctuations of the entanglement entropy and, consequently, to eigenstate expectation values of few-body observables, and that these fluctuations decrease as the system size increases.

\subsection{Quantum-classical correspondence}
\label{subsec:quantum_classical}

\begin{figure}[t!]
    \centering
    \includegraphics[width=0.9\textwidth]{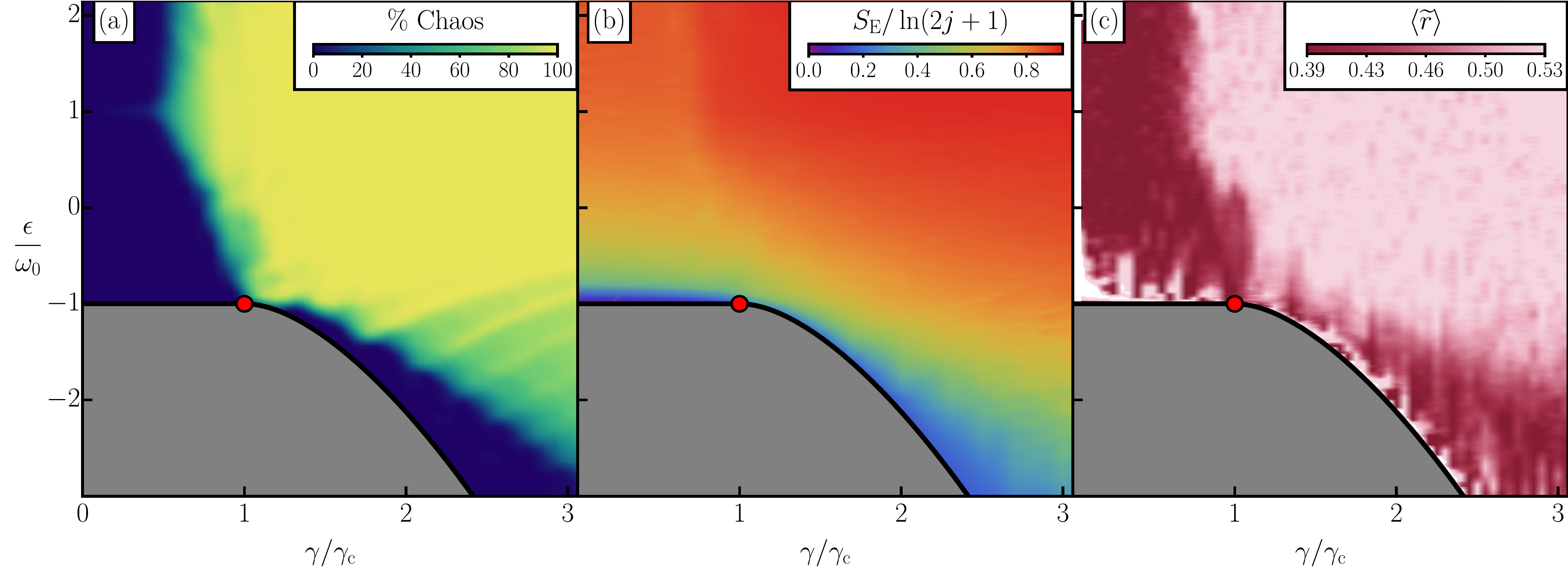}
    \caption{(a) Map of chaos for the classical Dicke Hamiltonian [Eq.~\eqref{eqn:classical_dicke_hamiltonian}]. The color scale in the upper right corner indicates regularity (darker tones) and high percentage of chaos (lighter tones) in phase space.
    (b) Map of the average von Neumann entanglement entropy [Eq.~\eqref{eqn:entanglement_entropy}] for eigenstates of the quantum Dicke Hamiltonian [Eq.~\eqref{eqn:dicke_hamiltonian}] with positive parity. The average was computed taking energy windows centered in $\epsilon$ and using the system size $j=30$. The color scale in the upper right corner contrasts low entanglement (darker blue tones) with high entanglement (darker red tones) values scaled to the value $\ln(2j+1)$.
    (c) Map of the average ratio of consecutive energy levels [Eq.~\eqref{eqn:ratio_energy_levels}] for eigenstates of the quantum Dicke Hamiltonian with positive parity. The average was computed taking energy windows centered in $\epsilon$ and using the system size $j=100$. The color scale in the upper right corner distinguishes uncorrelated levels (darker tones) from correlated levels (lighter tones). In each panel, from (a) to (c), the black solid line represents the ground-state energy and the red dot marks the QPT at the ground-state energy in the thermodynamic limit ($\mathcal{N} \to \infty$).
    Hamiltonian parameters: $\omega=\omega_{0}=1$ and $\gamma=2\gamma_{\text{c}}$.
     }
    \label{fig:lyapunov_entropy_factorR}
\end{figure}

The results in Fig.~\ref{fig:entanglement_dicke} indicate that the correspondence between chaos in the classical limit and correlated eigenvalues in the quantum domain can be extended to the appearance of nearly ergodic eigenstates and thus large values of the entanglement entropy. This correspondence is shown explicitly in Fig.~\ref{fig:lyapunov_entropy_factorR}, where we show the map of chaos and regularity in phase space, obtained with the Lyapunov exponents of the classical Dicke Hamiltonian, in Fig.~\ref{fig:lyapunov_entropy_factorR}(a), 
the map of entanglement entropy as a function of the coupling parameter $\gamma$ and the energies $\epsilon$ in Fig.~\ref{fig:lyapunov_entropy_factorR}(b), and the map of the average ratio of consecutive energy levels in Fig.~\ref{fig:lyapunov_entropy_factorR}(c). Overall, regions of classical chaos in Fig.~\ref{fig:lyapunov_entropy_factorR}(a) coincide with high entanglement in Fig.~\ref{fig:lyapunov_entropy_factorR}(b) and correlated eigenvalues in Fig.~\ref{fig:lyapunov_entropy_factorR}(c), while classical regular regions coincide with regions of low entanglement and uncorrelated eigenvalues. There are details, however, that are not equally captured by all three panels. For example, there is an interesting region in Fig.~\ref{fig:lyapunov_entropy_factorR}(a) for  $\gamma \sim 3\gamma_{\text{c}}$ and low energies, $-1 \lesssim \epsilon \lesssim -2$,  where a rib-like structure emerges, suggesting  mixed regimes in phase space, which is not clearly identified in the quantum domain [Figs.~\ref{fig:lyapunov_entropy_factorR}(b)-\ref{fig:lyapunov_entropy_factorR}(c)].

A parallel between the quantum and classical features of the Dicke model can also be drawn by comparing Peres lattices [plots of eigenstate expectation values versus eigenvalues, as in Figs.~\ref{fig:eth_dicke}(a)-\ref{fig:eth_dicke}(b)] with the Poincar\'e sections for the classical model [as in right panels of Fig.~\ref{fig:poincare_anisotropic_dicke}, Fig.~\ref{fig:poincare_dicke}, and top panels of Fig.~\ref{fig:poincare_lyapunov}]~\cite{Bastarrachea2014JP,Bastarrachea2014PRAb,Bastarrachea2015,Bastarrachea2016PRE,Bastarrachea2017PS}. The Peres lattice can be interpreted as a quantum analog of the Poincar\'e section, both providing visual information about where the system is regular or chaotic depending on the Hamiltonian parameters and energies.

\begin{figure}[t!]
    \centering
    \includegraphics[width=0.95\textwidth]{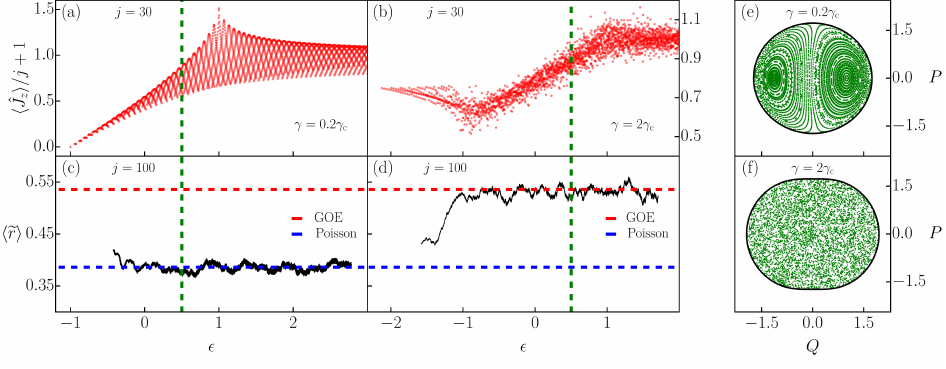}
    \caption{[(a)-(b)] Expectation values (Peres lattice) of the number of excited atoms $\hat{n}_{\text{ex}}=\hat{J}_z+j\hat{\mathbb{I}}$ for eigenstates of the Dicke Hamiltonian with positive parity. The expectation values are scaled to the system size $j$. The expectation values were calculated in (a) the normal phase ($\gamma=0.2\gamma_{\text{c}}$) and (b) the superradiant phase ($\gamma=2\gamma_{\text{c}}$) using the system size $j=30$.
    [(c)-(d)] Average ratio of consecutive energy levels [Eq.~\eqref{eqn:ratio_energy_levels}] for eigenvalues in (c) the normal phase and (d) the superradiant phase using the system size $j=100$. In panels (c) and (d), horizontal dashed lines represent the limit ratios from Poisson (bottom blue line) and GOE (top red line) distributions.
    [(e)-(f)] Poincar\'e sections in (e) the normal phase and (f) the superradiant phase using the classical energy shell $\epsilon=0.5$, which is depicted by the vertical green dashed lines in panels (a)-(d). The projections are shown on the atomic plane $Q$-$P$. In panels (e) and (f), the black solid border represents the available phase space at the energy shell.
    Hamiltonian parameters: $\omega=\omega_{0}=1$.
    }
    \label{fig:peres_poincare}
\end{figure}

Figure~\ref{fig:peres_poincare}(a) shows identifiable patterns in the Peres lattice of excited atoms $\hat{J}_{z}+j\hat{\mathbb{I}}$ for a coupling parameter in the normal phase ($\gamma<\gamma_{\text{c}}$). The whole spectrum shows correspondence with the average ratio of consecutive energy levels [Eq.~\eqref{eqn:ratio_energy_levels}] for the Poisson distribution $\langle \widetilde{r} \rangle_{\text{P}} \approx 0.39$  in Fig.~\ref{fig:peres_poincare}(c). Figure~\ref{fig:peres_poincare}(e) shows regular behavior (structured patterns) in the Poincar\'e section computed for the energy shell $\epsilon=0.5$. A similar analysis performed for a coupling parameter in the superradiant phase ($\gamma>\gamma_{\text{c}}$) is shown in Figs.~\ref{fig:peres_poincare}(b),~\ref{fig:peres_poincare}(d), and~\ref{fig:peres_poincare}(f). Figure~\ref{fig:peres_poincare}(b) shows an organized Peres lattice for low energies that becomes disordered as energy increases. The energy spectrum captures this behavior in Fig.~\ref{fig:peres_poincare}(d), transiting from an average ratio of consecutive energy levels for the Poisson distribution $\langle \widetilde{r} \rangle_{\text{P}} \approx 0.39$ to an average ratio for the GOE distribution $\langle \widetilde{r} \rangle_{\text{WD}} \approx 0.53$. Figure~\ref{fig:peres_poincare}~(f) exhibits disordered points in the Poincar\'e section for the same energy shell ($\epsilon=0.5$) as in Fig.~\ref{fig:peres_poincare}(e).

\section{Dynamical signatures of quantum chaos}
\label{sec:ChaosDynamics}
This section explains that dynamical manifestations of quantum chaos in the Dicke model can be detected at short times through the evolution of out-of-time-ordered correlators and entropies, and at long times, via the evolution of the survival probability. The exponential growth rate of the out-of-time-ordered correlator coincides with the Lyapunov exponent and the survival probability exhibits a ``correlation hole'' (``ramp'') when the eigenvalues are correlated.

\subsection{Out-of-time-ordered correlators}

The out-of-time-ordered correlator (OTOC) measures the degree of non-commutativity in time between two quantum operators $\hat{W}$ and $\hat{V}$, as
\begin{equation}
    C(t) = -\langle[\hat{W}(t),\hat{V}(0)]^{2}\rangle,
\end{equation}
where $\hat{W}(t)=e^{i\hat{H}t}\hat{W}e^{-i\hat{H}t}$ is the time-evolved operator in the Heisenberg picture and $\langle\bullet\rangle$ can indicate a thermal average or the expectation value under a chosen state. The OTOC was introduced in Ref.~\cite{Larkin1969} and has received significant attention recently~\cite{Maldacena2016PRD,Rozenbaum2017PRL,Hashimoto2017,Garcia2018,Jalabert2018,Pappalardi2018,Garcia2018,Hummel2019,Chavez2019,Rozenbaum2019,Niknam2020,Pilatowsky2020,Xu2020,Hashimoto2020,Garcia2023,Chavez2023} due to its relation with chaos. The idea is that $C(t)$ should grow exponentially if the system is chaotic and the exponential growth rate $2\Lambda_{\text{Q}}$ should coincide with twice the Lyapunov exponent $\lambda$.

An understanding of the connection between OTOC and chaos can be achieved by writing the OTOC in terms of the position and momentum operators for an eigenstate $|E_k\rangle$,
\begin{equation}
    C^{qp}_{k}(t) = -\langle E_k|[\hat{q}(t),\hat{p}(0)]^2|E_k\rangle.
    \label{eq:cqp}
\end{equation}
In the semiclassical limit, the commutator becomes the Poisson bracket. If the system is classically chaotic, then $\{q(t),p(0)\}=\partial q(t)/\partial q (0)\sim e^{\lambda t}$, where $\lambda$ is the classical Lyapunov exponent. For times shorter than the Ehrenfest time, the quantum and classical dynamics coincide, thus $\Lambda_{\text{Q}} \sim \lambda$, and one refers to $\Lambda_{\text{Q}}$ as the quantum Lyapunov exponent. Notice, however, that the exponential growth of the OTOC can also emerge in regular systems at a critical point, where a positive Lyapunov exponent emerges due to instability~\cite{Pappalardi2018,Hummel2019,Pilatowsky2020,Xu2020,Hashimoto2020,Chavez2023}. 

The analysis of the OTOC for the Dicke model was done in Ref.~\cite{Chavez2019}. Using the temporal evolution of the operator $\hat{q}(t)=e^{i\hat{H}_{\text{D}}t}\hat{q}e^{-i\hat{H}_{\text{D}}t}$, Eq.~\eqref{eq:cqp} can be expressed as~\cite{Hashimoto2020}
\begin{equation}
    \label{eqn:otoc}
    C^{qp}_{k}(t)= \sum_{k'} b_{k,k'}(t)b^*_{k,k'}(t),
\end{equation}
where for the Dicke Hamiltonian~\cite{Chavez2019}
\begin{equation}
    \label{eq:bnl}
    b_{k,k'}(t) = \frac{1}{\omega}\sum_{l} q_{k,l}q_{l,k'}\left(\Omega_{l,k'}e^{i\Omega_{k,l}t} - \Omega_{k,l}e^{i\Omega_{l,k'}t} \right),
\end{equation}
with $q_{k,l}=\langle E_{k}|\hat{q}|E_{l} \rangle$ and $\Omega_{k,l}=E_{k}-E_{l}$. The OTOC is then obtained by computing numerically the matrix elements of $\hat{q}$ in the energy eigenbasis.

\begin{figure}[t!]
    \centering
    \begin{tabular}{cc}
        (a) & (b) \\
        \includegraphics[height=0.25\textwidth]{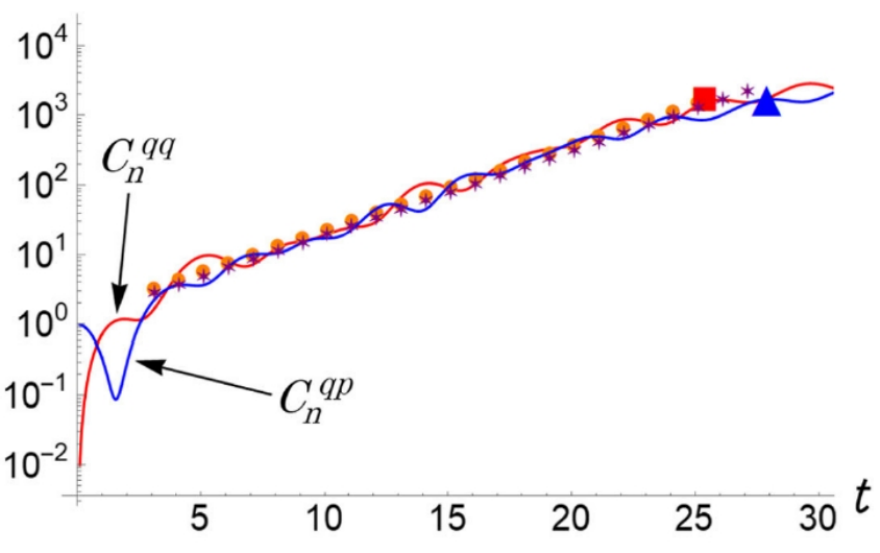}  & 
        \includegraphics[height=0.25\textwidth]{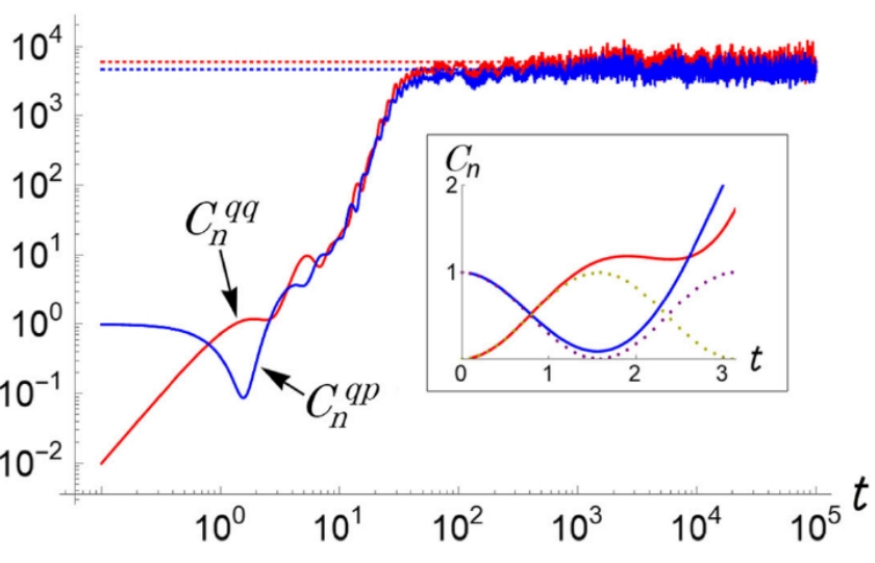} 
    \end{tabular}
    \caption{(a) Exponential growth of the OTOC [Eq.~\eqref{eqn:otoc}] for an eigenstate with energy in the chaotic region of the Dicke model, $\epsilon_{1625}/\omega_{0}\approx -1.1$. Numerical results for $C^{qp}_{k}(t)$ (blue solid line) and $C^{qq}_{k}(t)$ (red solid line). Numerical fit for $C^{qp}_{k}(t)$ (purple stars) and for $C^{qq}_{k}(t)$ (orange circles). The blue triangle and red square indicate the end of the exponential growth.
    (b) Exponential growth of the OTOC in logarithmic scale and the saturation value for $C^{qp}_{k}(t)$ (horizontal blue dotted line) and $C^{qq}_{k}(t)$ (horizontal red dotted line). The inset shows the short time behavior compared with functions $\sin^2(t)$ and $\cos^2(t)$ (dotted lines). The correspondence between the labels in panels (a)-(b) and the main text is: $n=k$.
    Hamiltonian parameters: $\omega=\omega_{0}=1$, $\gamma=2\gamma_{\text{c}}$, and $j=100$. Figures taken from Ref.~\cite{Chavez2019}.
    }
    \label{fig:OTOC}
\end{figure}

In Fig.~\ref{fig:OTOC}(a), we show that the behavior of the OTOC for the chaotic Dicke model for both $C^{qq}_{k}(t)$ and $C^{qp}_{k}(t)$ is indeed exponential. The growth rate $\Lambda_{\text{Q}}$ is obtained by fitting the curve and it coincides with the Lyapunov exponent obtained with the classical model. The log-log plot in Fig.~\ref{fig:OTOC}(b) shows that at very short times, $t < \pi/\omega_0$, the OTOC shows a sinusoidal behavior (see inset), before growing exponentially. This happens because at very short times, the dynamics is controlled by the diagonal elements $b_{k,k}(t)=(2/\omega)\sum_{k'}  q^2_{k',k}\Omega_{k',k}\cos(\Omega_{k',k}t)$.

At long times, the dynamics finally saturates to the infinite-time average
\begin{equation}
    \overline{C^{qp}_{k}}=\frac{1}{\omega^2}\sum_{l,l'}q^2_{k,l}q^2_{l,l'}\left(\Omega_{k,l}^2 + \Omega_{l,l'}^2\right),
\end{equation}
which scales with $j^2$ and with the number of bosons in the system. After saturation, the OTOC fluctuates around its asymptotic value, as seen in Fig.~\ref{fig:OTOC}(b). It has been argued that the ratio between the standard-deviation of the fluctuations and the mean value of the OTOC  can be used as a measure of chaos. The temporal fluctuations of the OTOC after the saturation  have also been used to differentiate chaos and integrability in Ref.~\cite{Fortes2019}. However, while the fluctuations for a fixed system size tend to be smaller in chaotic models than in integrable models, scaling analysis shows that the fluctuations can decrease exponentially  with system size in both integrable and chaotic systems~\cite{Lezama2023}.

A variation of the OTOC, known as fidelity out-of-time-ordered correlator (FOTOC)~\cite{Lewis2019,Pilatowsky2020}, has been used to assess the spread in time of an initial wave packet. The FOTOC is given by
\begin{equation}
    C_{\text{F}}(t) = \langle\hat{W}(t)^{\dagger}\hat{V}^{\dagger}\hat{W}(t)\hat{V}\rangle \propto \sigma_{O}^{2}(t),
\end{equation}
where $\hat{W}=e^{i\delta\phi\hat{O}}$, $\delta\phi\ll1$, $\hat{O}$ is a Hermitian operator with variance $\sigma_{O}^{2}(t)=\langle\hat{O}^{2}(t)\rangle-\langle\hat{O}(t)\rangle^{2}$, and $\hat{V}=|\Psi(0)\rangle\langle\Psi (0)|$ is the projector of an arbitrary initial state $|\Psi (0) \rangle$. When the operator $\hat{O}$ is position or momentum, the FOTOC provides a way to analyze  quantum evolution of the variance of canonical variables in phase space.

\begin{figure}[t!]
    \centering
    \begin{tabular}{ccc}
        (a) & (b) & (c) \\
        \includegraphics[height=0.18\textwidth]{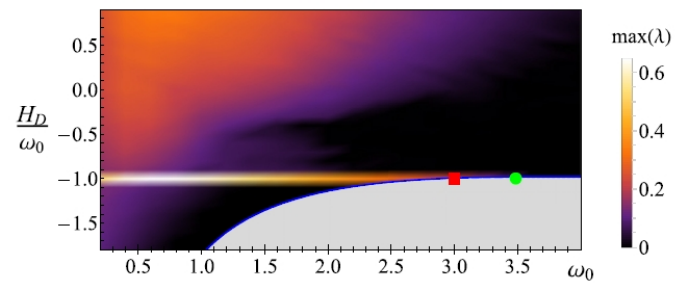}  & 
        \includegraphics[height=0.18\textwidth]{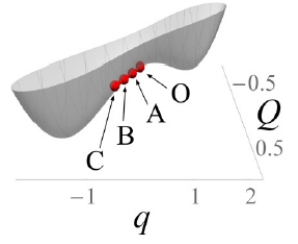} & 
        \includegraphics[height=0.18\textwidth]{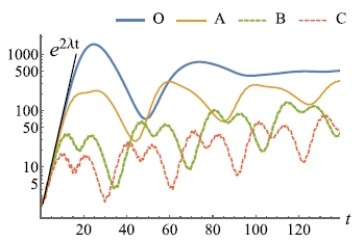} 
    \end{tabular}
    \caption{(a) Map of the Lyapunov exponent for the classical Dicke Hamiltonian [Eq.~\eqref{eqn:classical_dicke_hamiltonian}]. The color scale on the right distinguishes regularity (black) from chaos (lighter tones). The green dot marks the QPT. The red square at $\omega_{0}=3$ is an unstable point on the line of the ESQPT at $\epsilon/\omega_0$. The correspondence between the labels in panel (a) and the main text is: $H_{D}=\epsilon$.
    (b) Energy surface of the classical Dicke Hamiltonian for $P=p=0$. The red circles indicate the center of the coherent states used for the evolution of the FOTOC, $\sigma_{q}^{2}(t) + \sigma_{p}^{2}(t) + \sigma_{Q}^{2}(t) + \sigma_{P}^{2}(t)$, in panel (c) using the atomic frequency $\omega_{0}=3$.
    Hamiltonian parameters: $\omega=0.5$, $\gamma=0.66$, and $j=500$. Figure taken from Ref.~\cite{Pilatowsky2020}.
    }
    \label{fig:FOTOC}
\end{figure}

We use the FOTOC in Fig.~\ref{fig:FOTOC}(c) to show that the exponential growth of OTOCs can also happen  at  critical points of regular systems. The color gradient in Fig.~\ref{fig:FOTOC}(a) indicates the values of the Lyapunov exponent, which is zero in the black region. Within this region, one sees at energy $H_D/\omega_0$ a bright horizontal line that marks the ESQPT. The positive Lyapunov exponent on this line is associated with instability, not with chaos. From this line, we select the point at $\omega_0=3$ (red square) to analyze the FOTOC. This is the point O in Fig.~\ref{fig:FOTOC}(b), which depicts the energy surface of the Dicke model.  Points O, A, B, and C in Fig.~\ref{fig:FOTOC}(b) are the centers of the coherent states used for the evolution of the FOTOC in Fig.~\ref{fig:FOTOC}(c). For all 4 points, the FOTOC grows exponentially with a rate given by $2\Lambda_{\text{Q}} \sim 2\lambda$. This figure shows that the exponential growth due to instability happens not only exactly at the critical point, but also at its vicinity due to the quantum uncertainty of the position and momentum of the initial states. In Refs.~\cite{Rozenbaum2017PRL,Chavez2019}, it was noted that differences arise if instead of evaluating the quantum Lyapunov exponent as the logarithm of the OTOC, we  evaluate it  as the expectation value of the logarithm of the squared commutator. Recently, this observation has been   further explored~\cite{Trunin2023a,Trunin2023b},  and it has been found that the latter option yields a zero quantum Lyapunov exponent in the case of regular systems with unstable fixed points,  thus avoiding  false positives when diagnosing quantum chaos.

\subsection{Evolution of the entanglement entropy}

Entanglement grows fast in chaotic systems taken out of equilibrium. This relationship has been explored for decades~\cite{Zureck1993,Zureck1995,Tanaka1996,Angelo1999} aiming to establish that the entropy growth could be used as a quantum test of classical chaotic motion. However, this connection is not universal~\cite{Tanaka1996,Angelo1999}, since integrable systems can also exhibit fast entropy growth~\cite{Santos2012PRE,Znidaric2020}. 

\begin{figure}[t!]
    \centering
    \begin{tabular}{ccc}
        \includegraphics[height=0.25\textwidth]{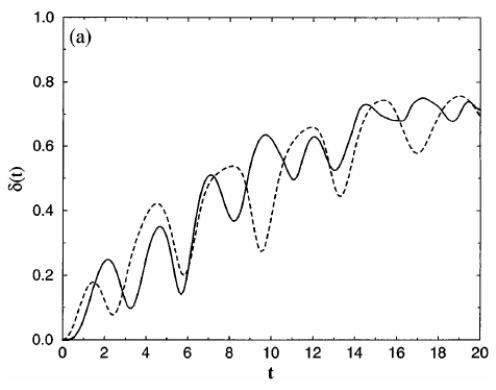} &
        \includegraphics[height=0.25\textwidth]{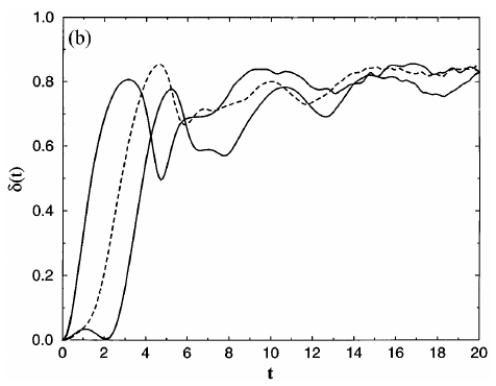} &
        \includegraphics[height=0.25\textwidth]{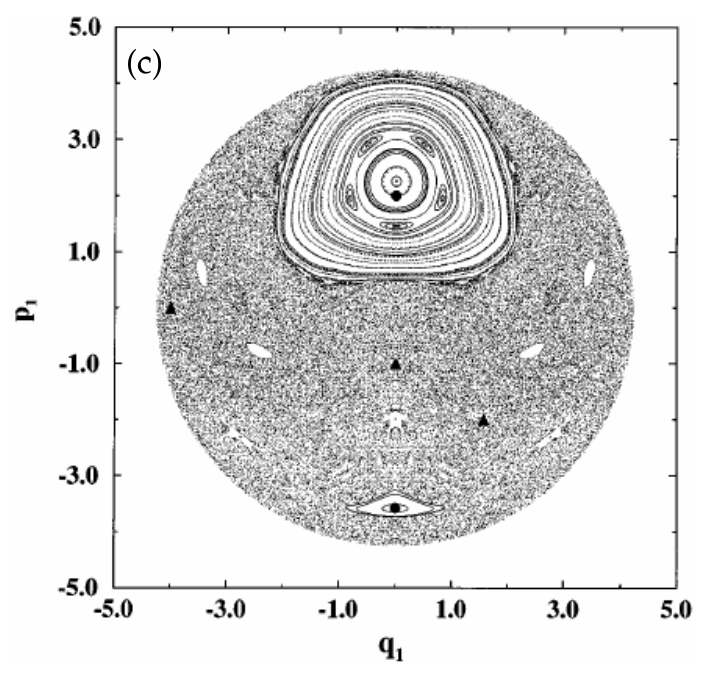}
    \end{tabular}
    \caption{[(a)-(b)] Evolution of the linear entanglement entropy [Eq.~\eqref{eqn:evolved_entanglement_linear_entropy}] for coherent states located in (a) stability islands and (b) the chaotic sea in the phase space of the classical anisotropic Dicke Hamiltonian [Eq.~\eqref{eqn:generalized_classical_dicke_hamiltonian}]. In panels (a) and (b), each line represents a different initial condition with $q_{2}=0$ and $p_{2}>0$. For panel (a): $(q_{1},p_{1})=(0,2)$ (solid line) and $(q_{1},p_{1})=(0,-3.577)$ (dashed line). For panel (b): $(q_{1},p_{1})=(-4,0)$ (upper solid line), $(q_{1},p_{1})=(0,-1)$ (dashed line), and $(q_{1},p_{1})=(1.57,-2)$ (lower solid line). The correspondence between the labels in panels (a)-(b) and the main text is: $\delta=S_{\text{L}}$.
    (c) Poincar\'e section for the classical anisotropic Dicke Hamiltonian using the classical energy shell $\epsilon=1.89$. The projection is shown on the atomic plane $Q$-$P$. The initial conditions from panel (a) are marked as dots and from panel (b) as triangles. The correspondence between the labels in panel (c) and the main text is: $q_{1}=\sqrt{j}P$, and $p_{1}=\sqrt{j}Q$.
    Hamiltonian parameters: $\omega=\omega_{0}=1$, $\gamma_{-}=0.5$, $\gamma_{+}=0.2$, and $j=9/2$. Figures taken from Ref.~\cite{Furuya1998}.
    }
    \label{fig:linear_entropy}
\end{figure}

The growth of entropy for the isotropic~\cite{Hou2004} and anisotropic Dicke models was analyzed in Refs.~\cite{Furuya1998,Song2008,Song2012} using the linear entanglement entropy, 
\begin{equation}
    \label{eqn:evolved_entanglement_linear_entropy}
    S_{\text{L}}(t) = 1-\text{Tr}\left[\hat{\rho}_{\text{A}}^{2}(t)\right],
\end{equation}
where $\hat{\rho}_{\text{A}}(t) = \text{Tr}_{\text{B}} \left[\hat{\rho}(t)\right]$ and $\hat{\rho}(t) = e^{i\hat{H}t}\hat{\rho}e^{-i\hat{H}t}$ is the time-evolved density matrix. Figures~\ref{fig:linear_entropy}(a) and~\ref{fig:linear_entropy}(b) show the evolution of the linear entanglement entropy for initial coherent states centered in regular and chaotic regions of phase space of the anisotropic Dicke model [Eq.~\eqref{eq:HD_generalized}]. In Fig.~\ref{fig:linear_entropy}(c), we see the position of the coherent states in phase space. Entropy grows slower for states located near a stability island  than for states located on a separatrix or in the chaotic sea. States in the regular region develop periodic oscillations that are captured at short times, while the chaotic states quickly saturate to an asymptotic value without traces of periodicity.

\begin{figure}[t!]
    \centering
    \includegraphics[width=0.7\textwidth]{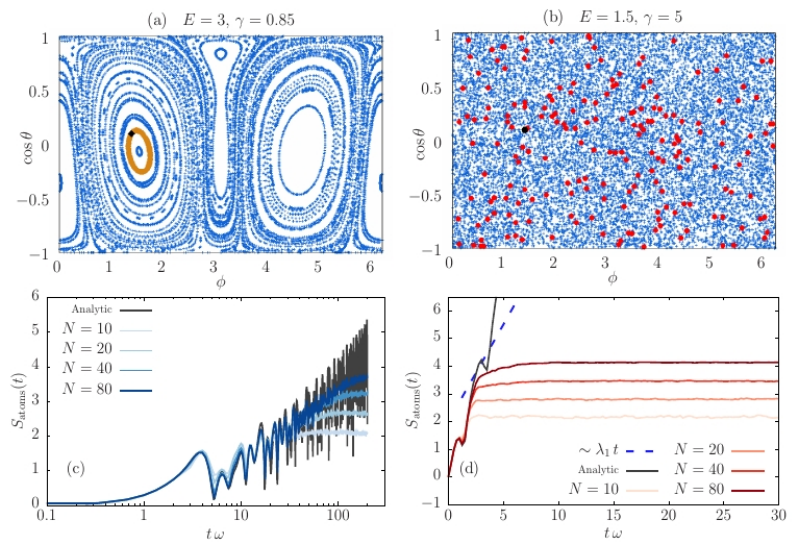}
    \caption{[(a)-(b)] Poincar\'e sections for the classical Dicke Hamiltonian [Eq.~\eqref{eqn:classical_dicke_hamiltonian}] for (a) regular $(E,\gamma)=(3,0.85)$ and (b) chaotic $(E,\gamma)=(1.5,5)$ regimes. The projections are shown on the plane $\phi-\cos\theta$. The relations between angular and position-momentum atomic variables are given by $\phi=\tan(-P/Q)$ and $\cos\theta=1-(Q^{2}+P^{2})/2$. In panels (a) and (b), the black dot represents an initial state with coordinates $(\phi_{0},\theta_{0})=(1.4,0.1)$.
    [(c)-(d)] Evolution of the von Neumann entanglement entropy [Eq.~\eqref{eqn:evolved_entanglement_entropy}] for the initial coherent state in (c) the regular regime and (d) the chaotic regime. In panels (c) and (d), each line represents a different system size ($j=5,10,20$, and 40) and the black solid line is an analytical expression in the thermodynamic limit ($\mathcal{N}\to\infty$). The correspondence between the labels in panels (c)-(d) and the main text is: $S_{\text{atoms}}=S_{\text{E}}$ and $N=2j$. The blue dashed line in panel (d) represents a linear curve with slope given by the maximum Lyapunov exponent $\lambda=\lambda_{1}=0.7$. Figure taken from Ref.~\cite{Lerose2020}.
    }
    \label{fig:poincare_entropy}
\end{figure}

Entropy growth was studied for the Dicke model in Ref.~\cite{Lerose2020} using the von Neumann entanglement entropy for the time-evolved density matrix $\hat{\rho}(t)$,
\begin{equation}
    \label{eqn:evolved_entanglement_entropy}
    S_{\text{E}}(t) = -\text{Tr}\left[\hat{\rho}_{\text{A}}(t)\ln(\hat{\rho}_{\text{A}}(t))\right],
\end{equation}
where $\hat{\rho}_{\text{A}}(t) = \text{Tr}_{\text{B}} \left[\hat{\rho}(t)\right]$. The Poincar\'e sections in  Fig.~\ref{fig:poincare_entropy}(a) exhibit regions of regularity of the Dicke model. The black dot marks the position of the initial coherent state considered for the evolution of the entanglement entropy in Fig.~\ref{fig:poincare_entropy}(c). The results for different system sizes are compared with an analytical expression in the thermodynamic limit. Entanglement grows slowly and the curve shows oscillations. This behavior is contrasted with the fast linear growth of entanglement in Fig.~\ref{fig:poincare_entropy}(d), where the coherent state chosen for the evolution is marked with a black dot in the chaotic sea of Fig.~\ref{fig:poincare_entropy}(b). The results in Fig.~\ref{fig:poincare_entropy}(d) are compared with a linear curve whose slope deviates from the Lyapunov exponent.

In addition to the entanglement entropy, the relationship between the quantum Fisher information and chaos was studied for the anisotropic Dicke model~\cite{Song2012}. The quantum Fisher information~\cite{Wootters1981,Braunstein1994} is defined as
\begin{equation}
    \mathcal{F}(\hat{\rho},\hat{O}) = 2\sum_{k,k'}\frac{(p_{k}-p_{k'})^{2}}{p_{k}+p_{k'}}|\langle E_{k}|\hat{O}|E_{k'}\rangle|^{2} \leq 4\langle\Delta\hat{O}^{2}\rangle,
\end{equation}
where $\langle\Delta\hat{O}^{2}\rangle = \text{Tr}\left(\hat{\rho}\hat{O}^{2}\right) - \text{Tr}\left(\hat{\rho}\hat{O}\right)^{2}$, $\hat{\rho}=\sum_{k}p_{k}|E_{k}\rangle\langle E_{k}|$ is an arbitrary mixed state with weights $p_{k}$, and $\hat{O}$ is an arbitrary operator. The equality in the equation above holds for pure states $\hat{\rho}= |\Psi\rangle\langle\Psi|$. Considering the pseudospin operator $\hat{O} = \hat{J}_{n} = \hat{\mathbf{J}} \cdot \vec{n}$ with $\vec{n}=(n_{x},n_{y},n_{z})$, a maximum Fisher information can be defined as
\begin{equation}
    \mathcal{F}_{\max}(t) = 4\max_{n}\frac{\langle\Delta\hat{J}_{n}^{2}(t)\rangle}{\mathcal{N}},
\end{equation}
which is normalized to the number of particles $\mathcal{N}=2j$ and can be averaged in time,
\begin{equation}
    \label{eqn:mean_quantum_fisher_information}
    \overline{\mathcal{F}}_{\max} = \frac{1}{T}\int_{0}^{T}dt\,\mathcal{F}_{\max}(t).
\end{equation}

Reference~\cite{Song2012} shows the explicit quantum-classical correspondence between the Poincar\'e sections of the classical anisotropic Dicke Hamiltonian [Eq.~\eqref{eqn:generalized_classical_dicke_hamiltonian}], including maps of the time-averaged quantum Fisher information [Eq.~\eqref{eqn:mean_quantum_fisher_information}] for initial coherent states in regular and mixed energy regimes. The reference also shows maps of the time-averaged linear entanglement entropy $\overline{S}_{\text{L}}=T^{-1}\int_{0}^{T}dt\,S_{\text{L}}(t)$. The correspondence between the Poincar\'e sections and the quantum Fisher information is better than the correspondence between the Poincar\'e sections and the entanglement entropy.

\subsection{Survival probability}
\label{subsec:survival_probability}

The survival probability is the probability of finding the system still in its initial state  $|\Psi(0) \rangle$ at a later time. It is a dynamical quantity that contains the spectral form factor [Eq.~\eqref{Eq:SFFnoaverage}], thus providing the means for detecting spectral correlations through quench dynamics. The survival probability  has been studied since the early years of quantum mechanics~\cite{Khalfin1958} and may be experimentally accessible~\cite{Das2024PRR}. It is defined as
\begin{equation}
    \label{eqn:survival_probability}
    S_{\text{P}}(t) = |\langle\Psi(0)|\Psi(t)\rangle|^{2} = |\langle\Psi(0)|e^{-i\hat{H} t}|\Psi(0)\rangle|^{2} =  \left| \sum_{k}|c_{k}|^{2}e^{-iE_{k}t} \right|^2,
\end{equation}
where  $c_{k}=\langle E_{k}|\Psi(0) \rangle$. Equation~\eqref{eqn:survival_probability} can also be written in an integral form,
\begin{equation}
    S_{\text{P}}(t) =  \left| \int_{E_{\min}}^{E_{\max}}dE\,\rho_{0}(E)e^{-iEt} \right|^2,
\end{equation}
where $E_{\min}$ and $E_{\max}$ are the bounds of the spectrum and 
\begin{equation}
    \rho_{0}(E) = \sum_{k}|c_{k}|^{2}\delta(E-E_{k})
\end{equation}
is the  local density of states (LDOS), which corresponds to the energy distribution of the initial state. One sees that the survival probability is the absolute square of the Fourier transform of the LDOS. If the components satisfy $|c_{k}|^2=1/d$, the LDOS coincides with the DOS and $S_{\text{P}}(t)$ becomes the spectral form factor $S_{\text{FF}}(t)$. This is why in chaotic systems taken far from equilibrium, where $|c_{k}|^2$ are nearly random numbers, we expect the survival probability to detect spectral correlations just as the spectral form factor does.

At very short times, $t\ll\sigma_{0}^{-1}$, where $\sigma_{0}$ is the width of the LDOS, the survival probability shows a universal quadratic behavior, obtained by Taylor expanding Eq.~\eqref{eqn:survival_probability}, that is,
\begin{equation}
    \label{eqn:survival_probability_quadratic_behavior}
    S_{\text{P}}(t) \approx 1-\sigma_{0}^{2}t^{2}, 
\end{equation}
where 
\begin{equation}
    \sigma_{0}^{2} = \langle\Psi(0)|\hat{H}^{2}|\Psi(0) \rangle-\langle\Psi(0)|\hat{H}|\Psi(0)  \rangle^{2} = \sum_{k}|c_{k}|^{2}(E_{k}-E_{\text{c}})^{2} 
\end{equation}
is the variance of the LDOS and
\begin{equation}
    E_{\text{c}}  = \langle\Psi(0) |\hat{H}|\Psi(0) \rangle = \sum_{k}|c_{k}|^{2}E_{k}
\end{equation}
is the (conserved) energy expectation value of the initial state $|\Psi(0)\rangle$. 

After the quadratic decay, the behavior of the survival probability up to intermediate times is determined by the shape and borders of the LDOS and the properties of the coefficients $c_{k}$~\cite{Vanicek2003,Flambaum2001a,Flambaum2001b,Izrailev2006,Torres2014PRAa,Torres2014PRAb,Torres2014NJP,Tavora2016,Tavora2017,Lerma2018JPA}, while at long times, when the dynamics resolves the discreteness of the spectrum, features associated with the presence of correlated eigenvalues can emerge~\cite{Alhassid1992,Torres2018,Schiulaz2019,Das2024PRR,Lerma2019,Villasenor2020}.

At the Heisenberg time (which is the inverse of the mean level spacing), if the spectrum does not have many degeneracies, the survival probability,
\begin{equation}
    \label{eqn:survival_probability_participation_ratio}
    S_{\text{P}}(t) = \sum_{k\neq k'}|c_{k}|^{2}|c_{k'}|^{2}e^{-i(E_{k}-E_{k'})t}+\sum_{k}|c_{k}|^{4},
\end{equation}
saturates to its asymptotic value given by the infinite-time average
\begin{equation}
    \label{eqn:pr_asymptotic_value}
    S_{\text{P}}^{\infty}=\sum_{k}|c_{k}|^{4}= \frac{1}{P_{\text{R}}} = I_{\text{PR}},
\end{equation} 
where 
\begin{equation}
    \label{eqn:participation_ratio}
    P_{\text{R}} = \left(\sum_{k}|\langle  E_{k}|\Psi (0) \rangle|^{4}\right)^{-1}
\end{equation}
is the participation ratio of the initial state written in the energy eigenbasis and $I_{\text{PR}}$ defines the inverse participation ratio. The participation ratio is bounded from below at 1, when $|\Psi (0) \rangle$ coincides with a basis vector and there is no evolution, and from above at the size of the Hilbert space.

\subsubsection{Survival probability of random states and correlation hole}
\label{Sec:SPrandom}

In Ref.~\cite{Lerma2019}, the analysis of the survival probability for the Dicke model assumed a random superposition of energy eigenstates as the initial state, 
\begin{equation}
    \label{eqn:random_state}
    |\Psi_{\text{R}} (0) \rangle = \sum_{k}c_{k}|E_{k}\rangle,
\end{equation}
with   coefficients $c_{k}=\langle E_{k}|\Psi_{\text{R}} (0) \rangle$ being random numbers following the definition
\begin{equation}
    \label{eqn:random_state_coefficients}
    c_{k} = \sqrt{\frac{r_{k}\rho(E_{k})}{M\nu(E_{k})}}e^{i\theta_{k}},
\end{equation}
where the weights $r_{k}$ are sampled from a distribution of uncorrelated random numbers, the phases $\theta_{k}$ are sampled from a uniform distribution defined in the interval $[0,2\pi)$,  $\rho(E)$ is the (assumed unimodal but otherwise arbitrary) envelope  of the LDOS, $\nu(E)$ is the semiclassical density of states given in Eq.~\eqref{eqn:semiclassical_density_states}, and $M=\sum_{k}r_{k}\rho(E_{k})/\nu(E_{k})$ is a normalization constant.

In addition to numerical results, in Ref.~\cite{Lerma2019} an analytical expression was also obtained  for the survival probability averaged over an ensemble of random states as in Eq.~\eqref{eqn:random_state} evolved under the Dicke model in the chaotic regime. The expression is given by
\begin{equation}
    \label{eqn:survival_probability_b2_goe}
    \langle S_{\text{P}}(t)\rangle = \frac{1-\langle I_{\text{PR}}\rangle}{\eta-1}\left[\eta S_{\text{P}}^{\text{st}}(t)-b_{2}\left(\frac{Dt}{2\pi}\right)\right] + \langle I_{\text{PR}}\rangle.
\end{equation}
In the equation above, $\langle\bullet\rangle$ defines the ensemble average, $S_{\text{P}}^{\text{st}}(t)$ describes the behavior of the survival probability at short times and depends on the envelope of the LDOS, the term
\begin{equation}
    \label{eqn:effective_dimension}
    \eta = \frac{[\sum_{k}\rho(E_{k})/\nu(E_{k})]^{2}}{\sum_{k}\rho^{2}(E_{k})/\nu^{2}(E_{k})} \approx \frac{\nu_{\text{c}}}{\int_{-\infty}^{+\infty} dE\,\rho^{2}(E)} \approx \frac{\langle r_{k}^{2} \rangle}{\langle r_{k} \rangle^{2}}\frac{1}{\langle I_{\text{PR}}\rangle}
\end{equation}
is an effective dimension that depends on the ensemble of initial states, $\nu_{\text{c}}=\nu(E_{\text{c}})$ is the semiclassical density of states evaluated at the energy center $E_{\text{c}}$ of the LDOS envelope, $b_{2}(t)$ is the GOE two-level form factor in Eq.~\eqref{eqn:two_level_form_factor_goe}, $D = \langle s\rangle$ is the mean level spacing of correlated eigenvalues, and $I_{\text{PR}}$ is the inverse participation ratio of the initial state. 

The expression in Eq.~\eqref{eqn:survival_probability_b2_goe} describes the entire evolution of the survival probability, from $t=0$ to saturation.  We stress that the $b_2(t)$ function only appears, because the model is chaotic. Next, we explain each term of the equation in the context of  particular  choices for the ensembles of initial states.

\begin{figure}[t!]
    \centering
    \begin{tabular}{ccc}
        (a) & (b) & (c) \\
        \includegraphics[width=0.3\textwidth]{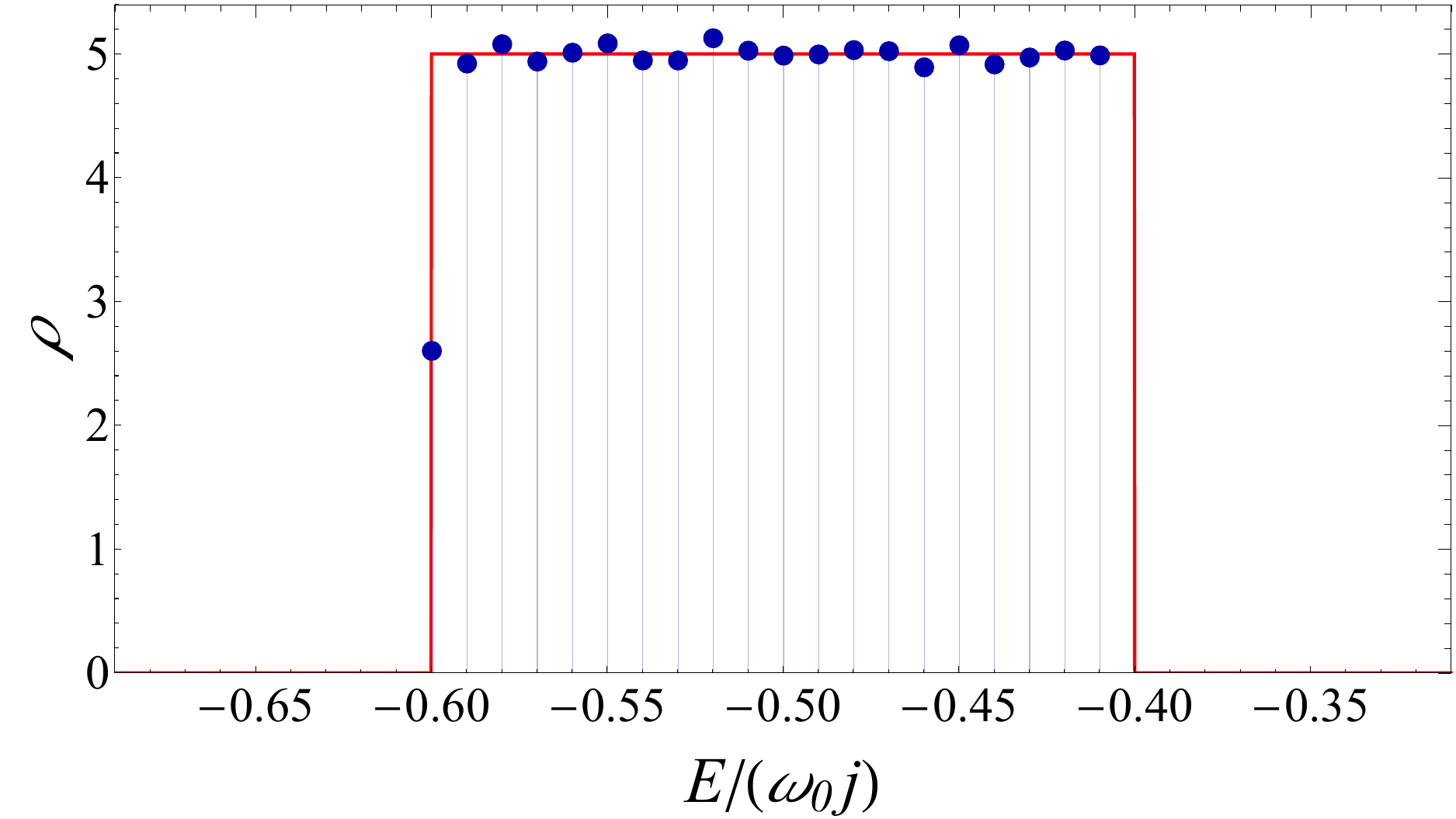} &
        \includegraphics[width=0.3\textwidth]{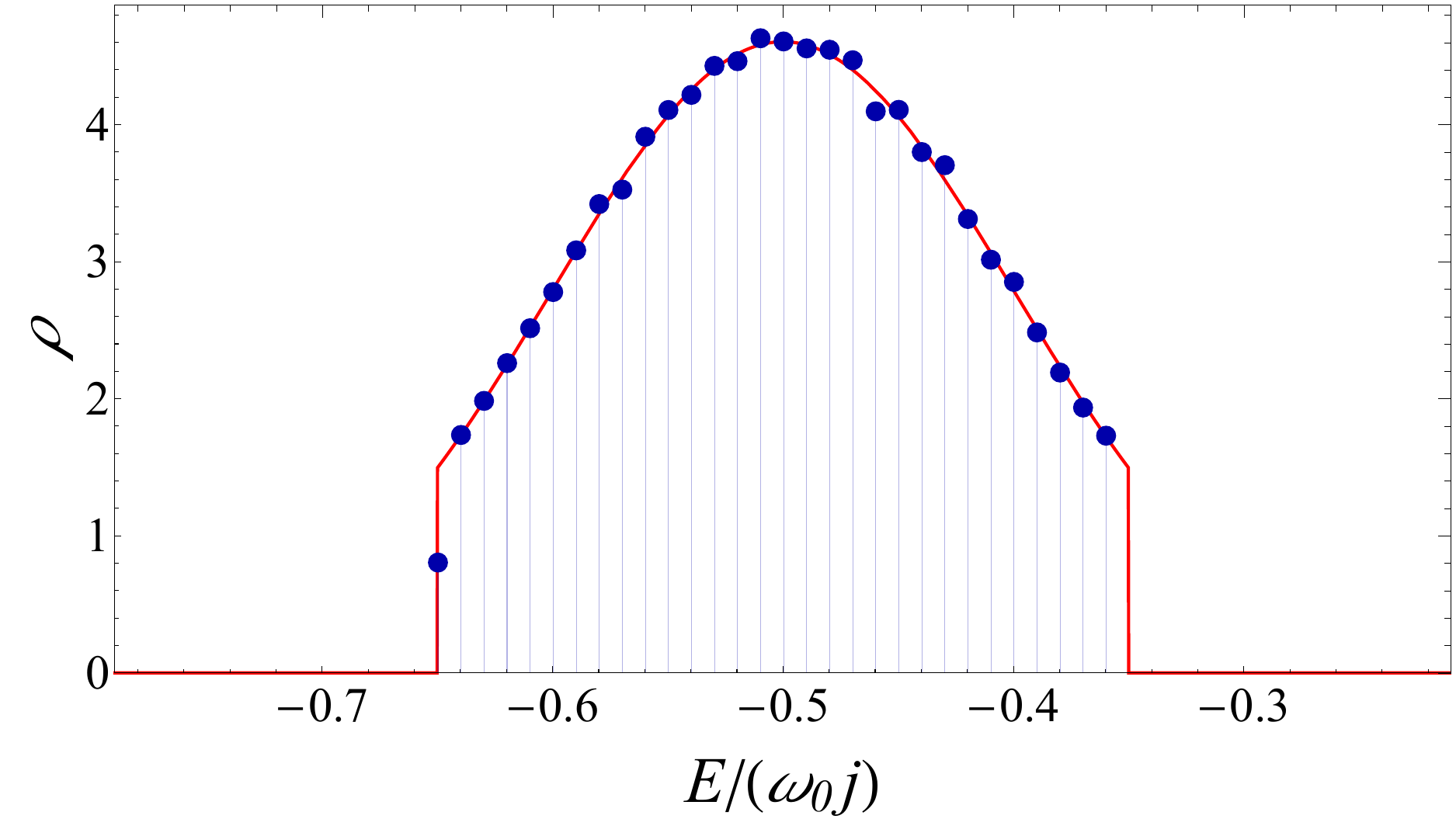} &
        \includegraphics[width=0.3\textwidth]{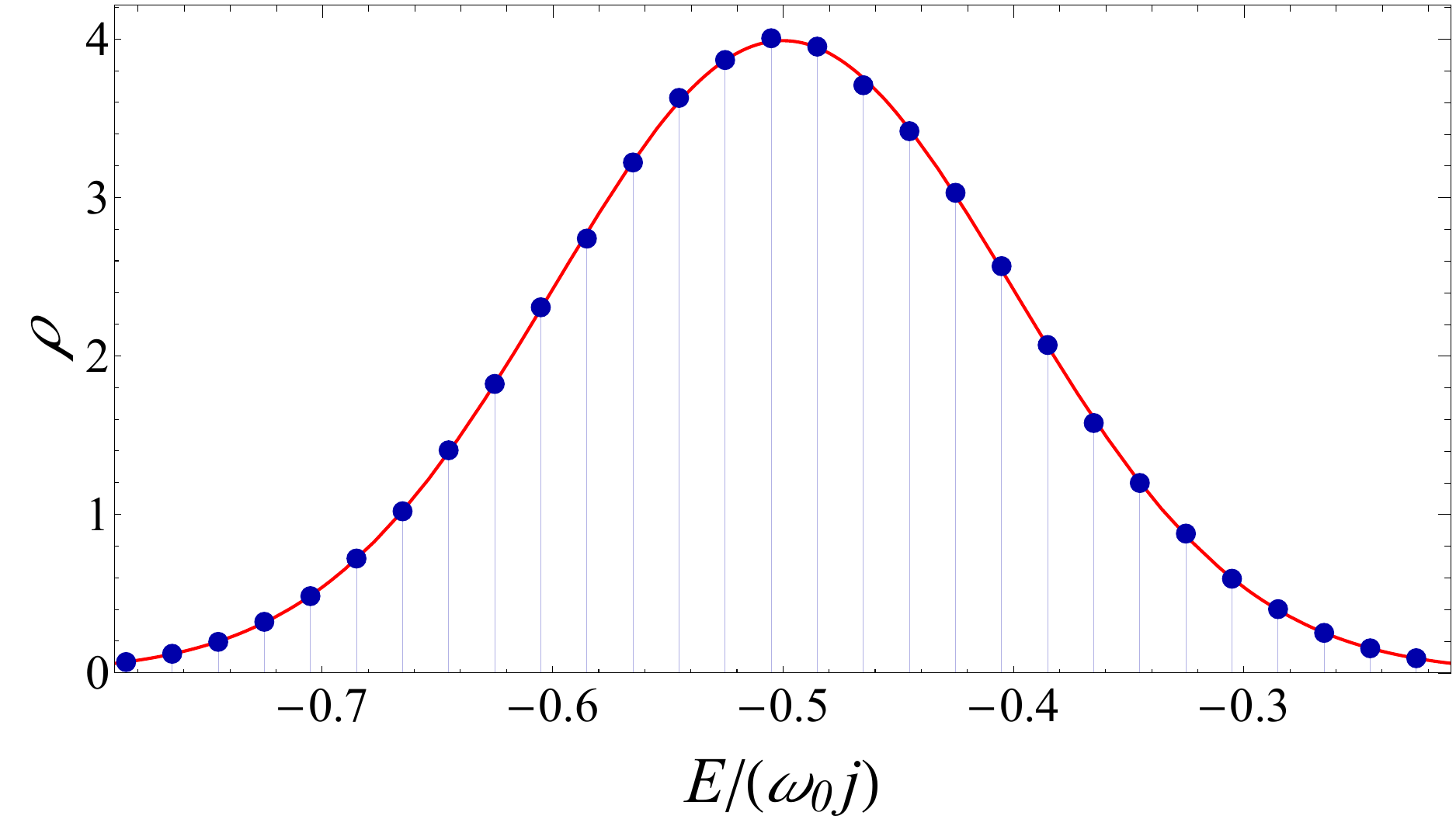}
    \end{tabular}
    \caption{Average LDOS envelope (blue dots) for an ensemble of 500 random states. Each panel refers to a different LDOS envelope shape [Eqs.~\eqref{eqn:ldos_rectangular_function}-~\eqref{eqn:ldos_gaussian_function}] (red solid line): (a) rectangular, (b) bounded Gaussian, and (c) Gaussian. In each panel, from (a) to (c), the random states are centered at the chaotic energy shell $\epsilon=-0.5$. Hamiltonian parameters: $\omega=\omega_{0}=1$, $\gamma=2\gamma_{\text{c}}$, and $j=100$. Figure taken from Ref.~\cite{Lerma2019}.
    }
    \label{fig:ldos_random_states}
\end{figure}

In Fig.~\ref{fig:ldos_random_states}, $r_{k}\in[0,1]$ for the coefficients $c_{k}$ in Eq.~\eqref{eqn:random_state_coefficients} were taken from a real-number uniform distribution with moments $\langle r_{k}^{n}\rangle=(n+1)^{-1}$, and three envelopes for the LDOS centered in a chaotic energy regime were chosen: rectangular [Fig.~\ref{fig:ldos_random_states}(a) and $\rho_{\text{R}}(E)$ below], bounded Gaussian [Fig.~\ref{fig:ldos_random_states}(b) and $\rho_{\text{BG}}(E)$ below], and Gaussian [Fig.~\ref{fig:ldos_random_states}(c) and $\rho_{\text{G}}(E)$ below],
\begin{gather}
    \label{eqn:ldos_rectangular_function}
    \rho_{\text{R}}(E) = \left\{\begin{array}{ll} \frac{1}{2\sigma_{\text{R}}} & \text{if } E_{\text{c}}-\sigma_{\text{R}}\leq E\leq E_{\text{c}}+\sigma_{\text{R}} \\ 0 & \text{otherwise} \end{array}\right., \\
    \label{eqn:ldos_bounded_gaussian_function}
    \rho_{\text{BG}}(E) = \left\{\begin{array}{ll} \frac{1}{\mathcal{C}\sqrt{2\pi}\sigma_{\text{BG}}}e^{-(E-E_{\text{c}})^{2}/(2\sigma_{\text{BG}}^{2})} & \text{if } E_{a}\leq E\leq E_{b} \\ 0 & \text{otherwise} \end{array}\right., \\
    \label{eqn:ldos_gaussian_function}
    \rho_{\text{G}}(E) = \frac{1}{\sqrt{2\pi}\sigma_{\text{G}}}e^{-(E-E_{\text{c}})^{2}/(2\sigma_{\text{G}}^{2})},
\end{gather}
where $E_{\text{c}}$ is the LDOS mean energy, $\sigma_{\text{R}}$, $\sigma_{\text{BG}}$, and $\sigma_{\text{G}}$ identify the width of each distribution, and the constant $\mathcal{C}$ normalizes the bounded Gaussian distribution.

The Fourier transform of the LDOS envelope describes the behavior of the survival probability at short times, characterized by $S_{\text{P}}^{\text{st}}(t)$ in Eq.~\eqref{eqn:survival_probability_b2_goe}. For the three envelopes in Eqs.~\eqref{eqn:ldos_rectangular_function}-~\eqref{eqn:ldos_gaussian_function}, one gets
\begin{gather}
    \label{eqn:short_times_rectangular_function}
    S_{\text{P}}^{\text{st,R}}(t) = \frac{\sin^{2}(\sigma_{\text{R}}t)}{\sigma_{\text{R}}^{2}t^{2}}, \\
    \label{eqn:short_times_bounded_gaussian_function}
    S_{\text{P}}^{\text{st,BG}}(t) = \frac{e^{-\sigma_{\text{BG}}^{2}t^{2}}}{4\mathcal{C}^{2}}\left|\text{erf}\left(\frac{E_{\text{c}}-E_{a}-i\sigma_{\text{BG}}^{2}t}{\sqrt{2}\sigma_{\text{BG}}}\right)-\text{erf}\left(\frac{E_{\text{c}}-E_{b}-i\sigma_{\text{BG}}^{2}t}{\sqrt{2}\sigma_{\text{BG}}}\right)\right|^{2}, \\
    \label{eqn:short_times_gaussian_function}
    S_{\text{P}}^{\text{st,G}}(t) = e^{-\sigma_{\text{G}}^{2}t^{2}},
\end{gather}
where erf is the error function.

\begin{figure}[t!]
    \centering
    \begin{tabular}{cc}
        (a) & (b) \\
        \includegraphics[width=0.45\textwidth]{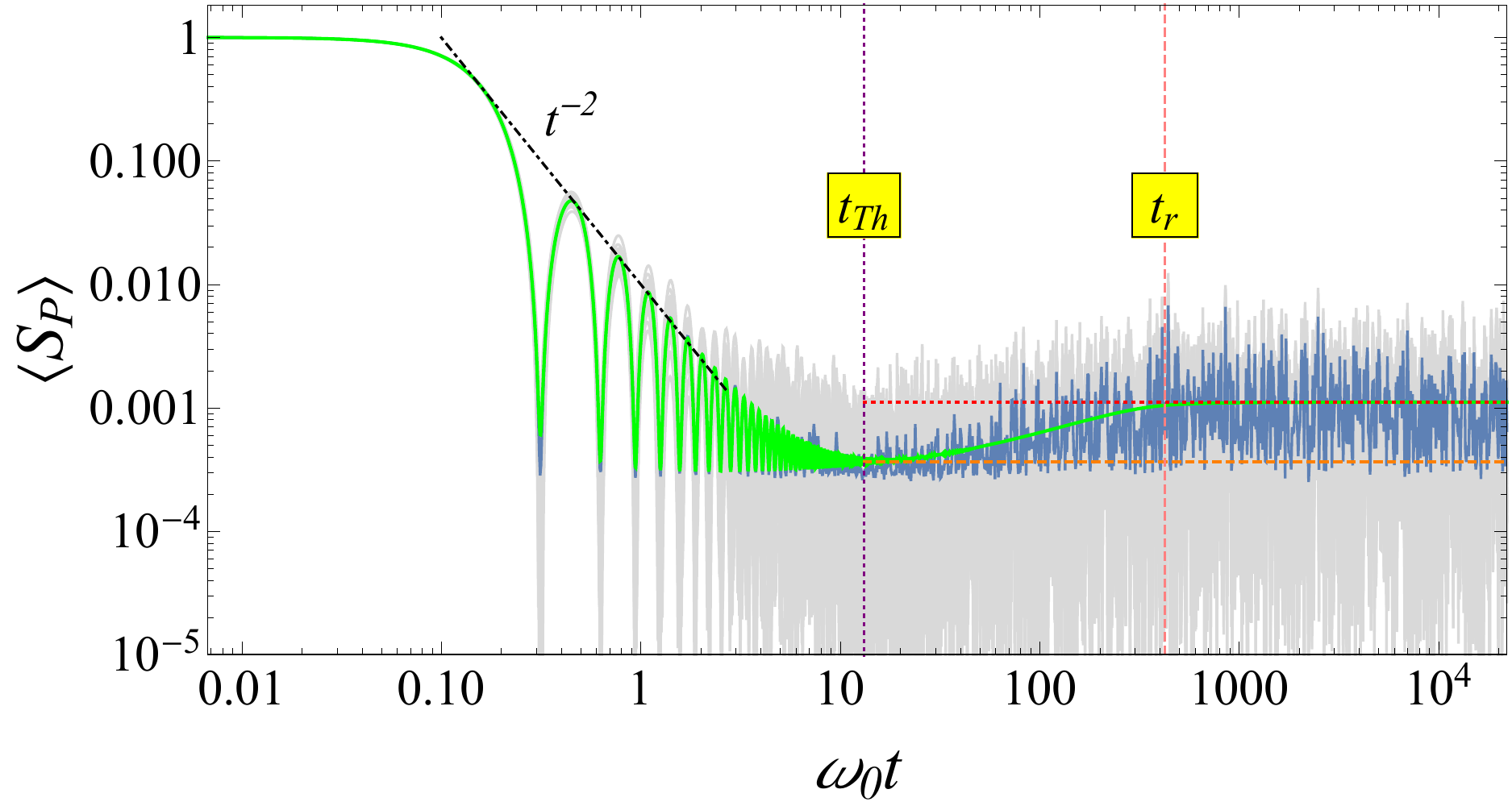} &
        \includegraphics[width=0.45\textwidth]{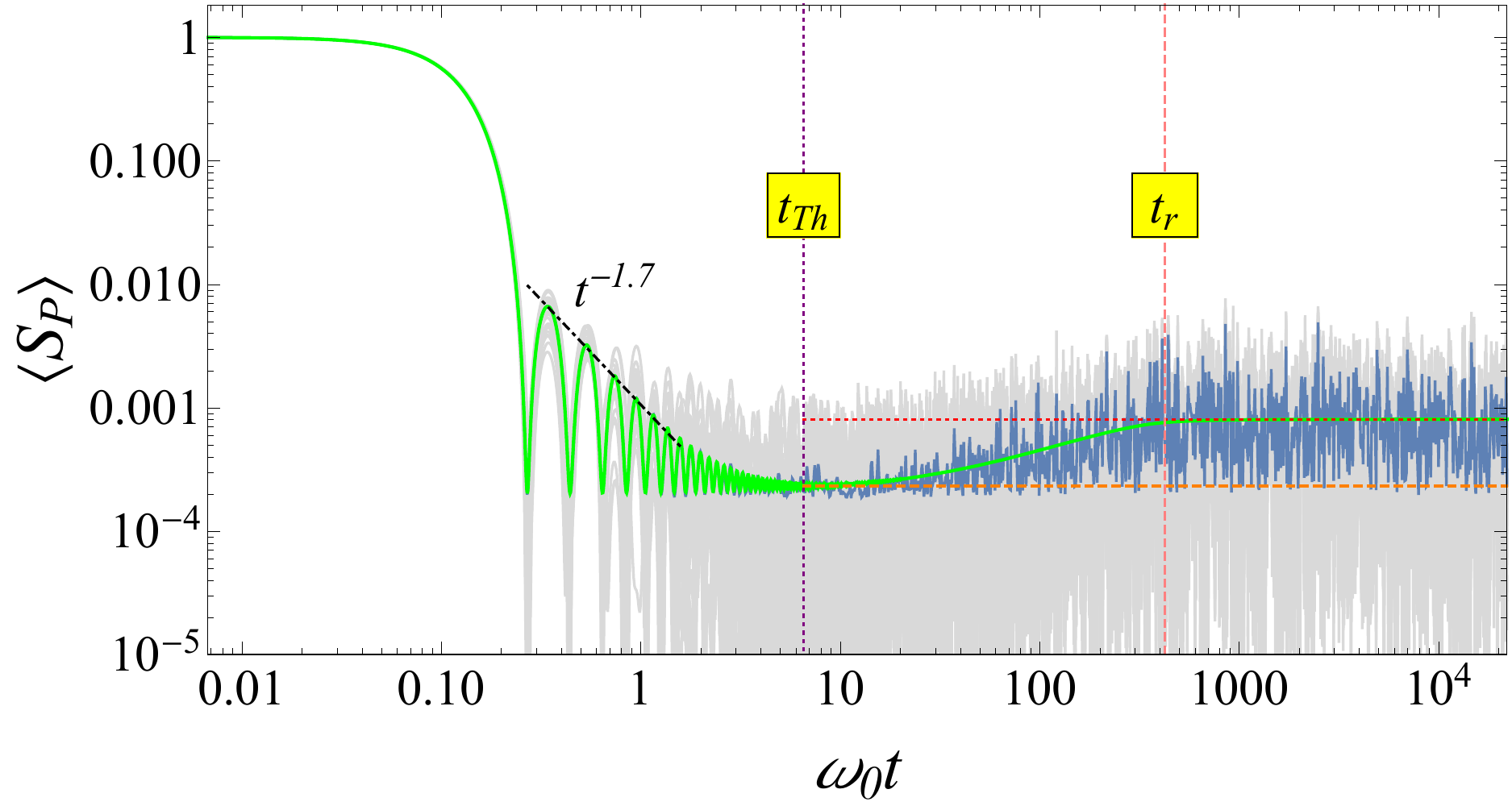} \\
        (c) & (d) \\
        \includegraphics[width=0.45\textwidth]{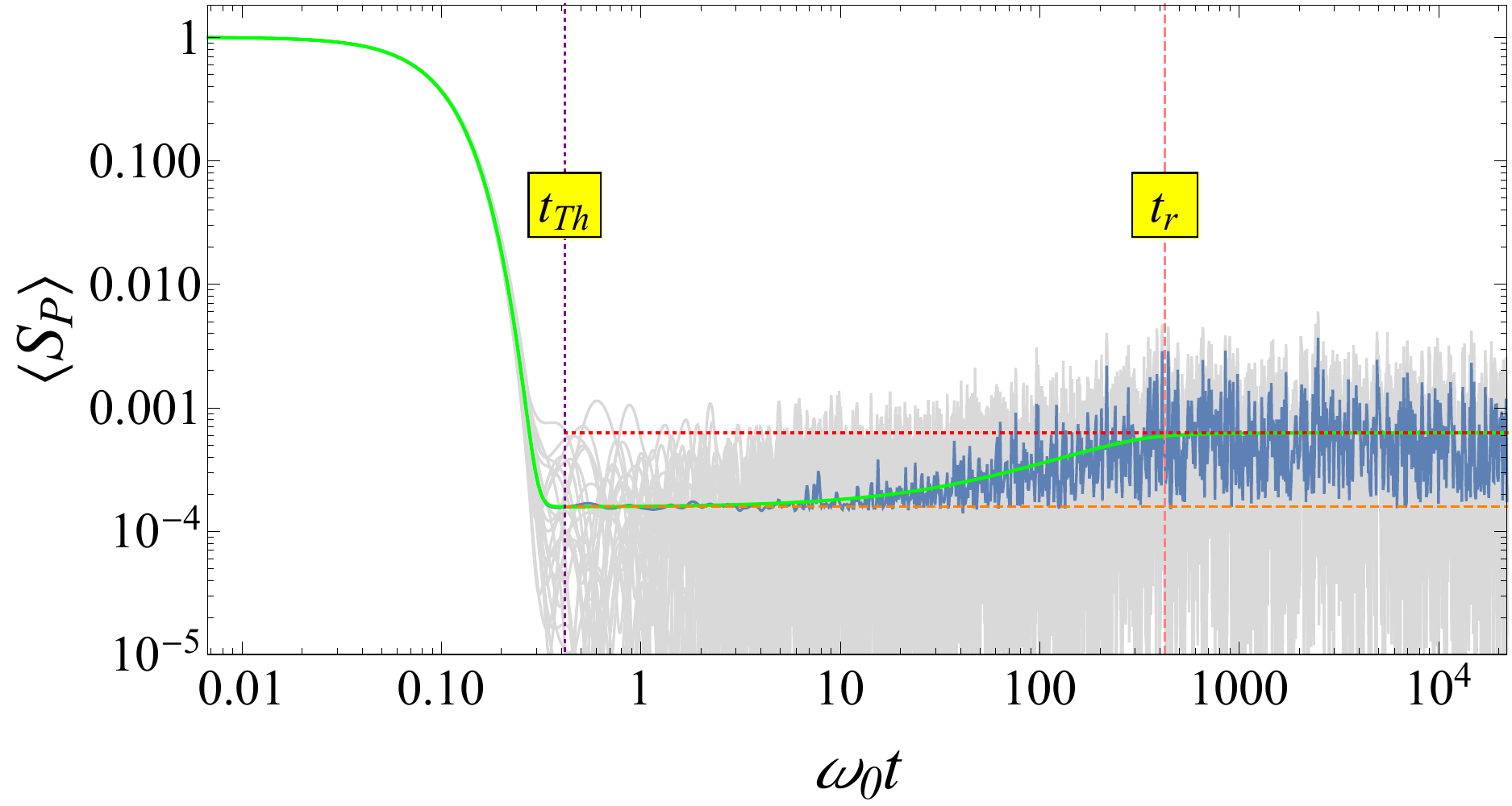} &
        \includegraphics[width=0.45\textwidth]{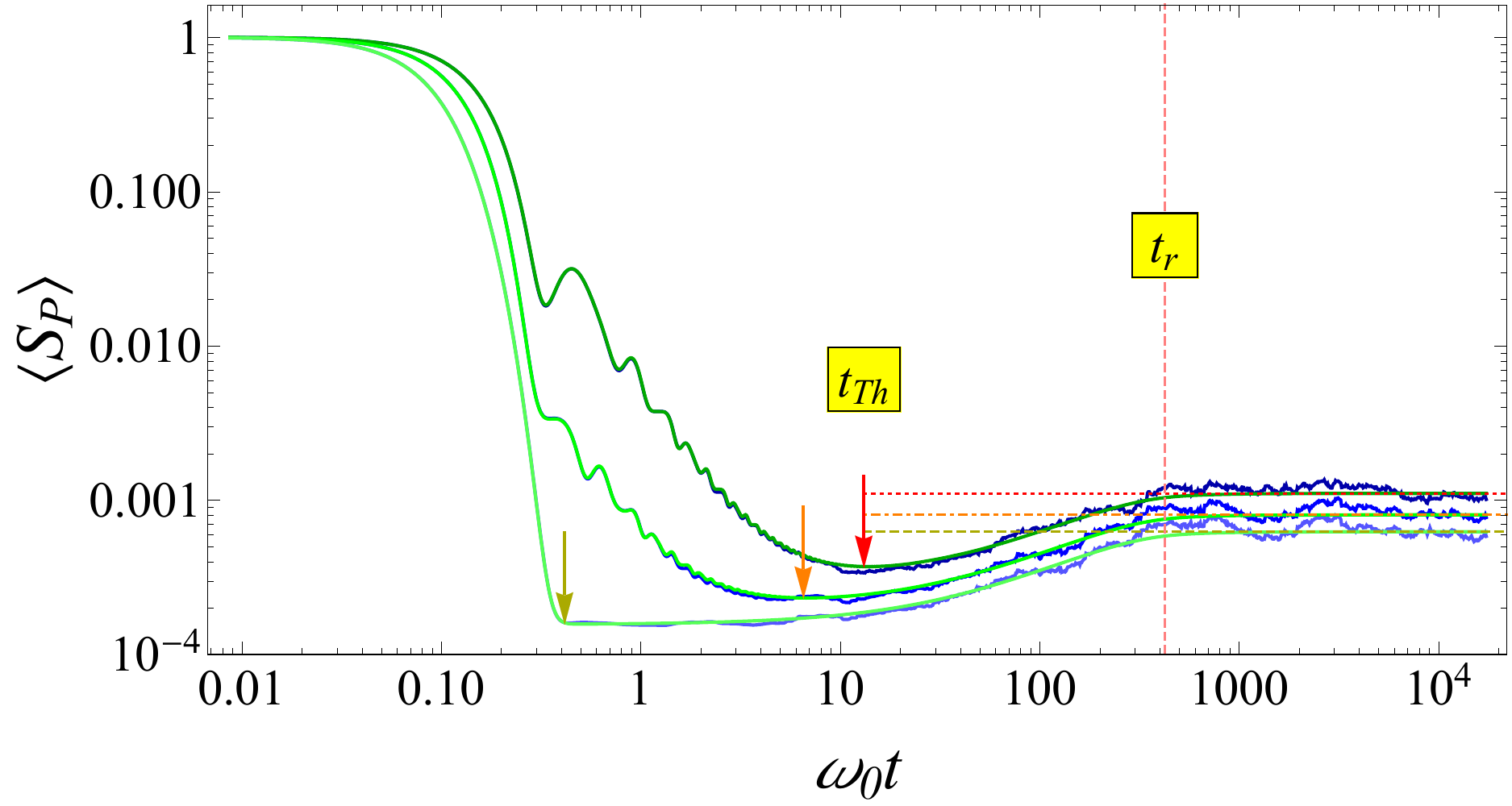}
    \end{tabular}
    \caption{[(a)-(c)] Survival probability [Eq.~\eqref{eqn:survival_probability}] (blue solid line) averaged over an ensemble of 500 random states. Each panel refers to a different ensemble with average LDOS envelope shape [Eqs.~\eqref{eqn:ldos_rectangular_function}-~\eqref{eqn:ldos_gaussian_function}]: (a) rectangular, (b) bounded Gaussian, and (c) Gaussian. In each panel, from (a) to (c), the gray solid lines represent the survival probability for single random states and the green solid line is the analytical expression in Eq.~\eqref{eqn:survival_probability_b2_goe}. Horizontal lines give the asymptotic (red dotted line) and minimum (orange dashed line) value of the survival probability. Vertical lines represent the Thouless (purple dotted line) and relaxation (pink dashed line) time. In panels (a) and (b), the black dotted-dashed line gives the power-law decay for the bounded LDOS.
    (d) Running-time average for the numerical (blue solid lines) and analytical (green solid lines) results of the survival probabilities in panels (a)-(c).
    Hamiltonian parameters: $\omega=\omega_{0}=1$, $\gamma=2\gamma_{\text{c}}$, and $j=100$. Only the positive-parity sector of the model is considered. Figure taken from Ref.~\cite{Lerma2019}.
    }
    \label{fig:survival_probability_random_states}
\end{figure}

Figures~\ref{fig:survival_probability_random_states}(a)-\ref{fig:survival_probability_random_states}(c) compare the numerical results for the survival probability (blue line) obtained for an ensemble of random initial states using~\eqref{eqn:survival_probability} with the analytical expression (green line) in Eq.~\eqref{eqn:survival_probability_b2_goe} for the three LDOS envelopes in Eqs.~\eqref{eqn:ldos_rectangular_function}-~\eqref{eqn:ldos_gaussian_function}. For these cases, $\eta = 4/(3\langle I_{\text{PR}} \rangle)$, $D \approx 2/\nu_{\text{c}}$, and $\nu_{\text{c}} = \nu(E_{\text{c}})$ where $\nu(E)$ is the density of states in Eq.~\eqref{eqn:semiclassical_density_states}. In all three cases the analytical expression accurately describes the numerical results at all time scales.

The short-time decay is faithfully described by $S_{\text{P}}^{\text{st}}(t)$ in Eqs.~\eqref{eqn:short_times_rectangular_function}-\eqref{eqn:short_times_gaussian_function}. In the case of the rectangular and bounded Gaussian envelopes, because of the bounds in the LDOS, a power-law decay arises~\cite{Khalfin1958} (see also Refs.~\cite{Tavora2016,Tavora2017} and references therein).

After the behavior described by Eqs.~\eqref{eqn:short_times_rectangular_function}-\eqref{eqn:short_times_gaussian_function}, the survival probability falls within a hole that has values below the saturation at $\langle I_{\text{PR}} \rangle$ and it rises toward saturation following a ``ramp'' determined by the $b_2(t)$ function. This ``dip-plateau'' feature of the evolution is the correlation hole, which is a manifestation of the  presence of spectral correlations, and thus an indicator of quantum chaos. 

The term ``correlation hole'' was introduced in 1986 by L. Leviandier et al.~\cite{Leviandier1986}. They proposed an alternative to the analysis of level statistics based on the Fourier transform of the spectra. They had in mind the spectra of molecules, for which the levels are often not as well resolved as in nuclei. Even if the levels overlap, the Fourier transform should still be able to capture the presence of correlations through the hole. This indicator of quantum chaos was widely studied in the 1990s using the spectral form factor and survival probability~\cite{Leviandier1986,Pique1987,Guhr1990,Wilkie1991,Hartmann1991,Delon1991,Alhassid1992,Alhassid1993,Lombardi1993,Kudrolli1994,Alt1997,Michaille1999,Gorin2002,Alhassid2006} and has regained attention recently in the context of many-body quantum systems~\cite{Torres2017AP,Torres2017PTRSA,Santos2017PROC,Torres2018,DelCampo2018,Torres2019,Schiulaz2019,Suntajs2020,Xu2021,Das2024PRR}.

The time where the survival probability in Eq.~\eqref{eqn:survival_probability_b2_goe} attains its minimum value can be obtained from 
\begin{equation}
    \label{eqn:thouless_time}
    \left. \frac{d}{dt}\langle S_{\text{P}}(t) \rangle \right|_{t=t_{\text{Th}}} = 0.
\end{equation}
This time, $t_{\text{Th}}$, marks the beginning of the ramp toward saturation and has been called Thouless time in analogy with the Thouless energy obtained in studies of level statistics in disordered systems. The time $t_{\text{Th}}$ is the point where the decaying function $S_{\text{P}}^{\text{st}}(t)$ for $t\gg\sigma_{0}^{-1}$ meets the beginning of the ramp described by $b_{2}(t) \approx 1-2t$ for $t\ll 1$ (see Ref.~\cite{Lerma2019}). The Thouless time is indicated in Figs.~\ref{fig:survival_probability_random_states}(a)-\ref{fig:survival_probability_random_states}(c) with a vertical line. Beyond $t_{\text{Th}}$, the dynamics no longer has information about the initial state. It follows the behavior of random matrices from a GOE, exhibiting thus universal behavior. 

For even longer times, the ramp of the survival probability becomes described by the term $b_{2}(t) \approx 1/(12 t^{2})$ and finally reaches the saturation value at the relaxation (Heisenberg) time. This time is given by~\cite{Lerma2019}
\begin{equation}
t_\text{r}=\frac{\pi \nu_{\text{c}}}{4\sqrt{\delta}}=\frac{\pi \nu_{\text{o}}}{2\omega\sqrt{\delta}}j,
\end{equation}
where Eq.~\eqref{eqn:semiclassical_density_states} was used  to determine the scaling of $\nu_{\text{c}}$ with $j$,  $\nu_{\text{c}}=2\nu_{\text{o}} j/\omega$, and $\delta$ is a small quantity that ensures that the survival probability lies inside the fluctuations of its asymptotic value. The relaxation time is shown with a vertical line in Figs.~\ref{fig:survival_probability_random_states}(a)-\ref{fig:survival_probability_random_states}(c). It scales linearly with the number of atoms in the Dicke model and grows faster than $t_{\text{Th}}$~\cite{Lerma2019}.

Despite the advantages of the survival probability as a detector of dynamical manifestations of quantum chaos, this quantity is non-self-averaging~\cite{Prange1997,Schiulaz2020}. This implies that large averages need to be performed, no matter how large the system size is. This becomes clear by comparing the averaged survival probability (blue line) with the result for a single initial state (gray line) in Figs.~\ref{fig:survival_probability_random_states}(a)-\ref{fig:survival_probability_random_states}(c). The averages can be done over initial states, as in  Figs.~\ref{fig:survival_probability_random_states}(a)-\ref{fig:survival_probability_random_states}(c), over disorder realizations in systems with disorder, and it can also be a running-time average. In Fig.~\ref{fig:survival_probability_random_states}(d), both averages, over initial states and the time average, are performed simultaneously leading to excellent agreement with the analytical expression.

\subsection{Survival probability of coherent states}

We discussed in Sec.~\ref{sec:ClassicalDickeModel} that Glauber-Bloch coherent states link the classical and quantum realms and provide the classical  limit of the Dicke model. They are the most ``classical'' of the quantum states in the sense that they lead to quantum fluctuations at a limit determined by the Heisenberg uncertainty relation. Contrary to the random states considered in Sec.~\ref{Sec:SPrandom}, coherent states are physical and could be experimentally prepared for the analysis of the dynamics of the Dicke model. In this subsection, we present results for the evolution of the survival probability of coherent states and compare them with results obtained from a semiclassical approximation.

A classical representation of the survival probability for coherent states evolving under the Dicke Hamiltonian can be constructed using the TWA~\cite{Ozorio1998,Klimov2002JPA,Case2008,Dittrich2010,Polkovnikov2010} explained in Sec.~\ref{subsubsec:TWA}.  The basic idea is to write the survival probability in terms of Wigner functions,
\begin{equation}
    \label{eqn:glauber_bloch_wigner_function_arbitrary_state}
    \mathcal{W}_{\rho}(\mathbf{x}) = \mathcal{W}_{\rho}(\alpha)\mathcal{W}_{\rho}(z) = \mathcal{W}_{\rho}(q,p)\mathcal{W}_{\rho}(Q,P),
\end{equation}
where  $\hat{\rho}$ is an arbitrary state
and
$\mathbf{x}=(q,p;Q,P)$ are the coordinates of the four-dimensional phase space $\mathcal{M}$ of the Dicke model. For initial Glauber-Bloch coherent states, $\hat{\rho}_{\mathbf{x}_{0}}
=|\mathbf{x}_{0}\rangle\langle\mathbf{x}_{0}|$, the Wigner function is given in Eq.~\eqref{eqn:glauber_bloch_wigner_function}. The quantum survival probability in Eq.~\eqref{eqn:survival_probability} can  be written in terms of Wigner functions as
\begin{equation}
    S_{\text{P}}(t) = \left(2\pi\hbar_{\text{eff}}\right)^{2} \int_{\mathcal{M}} d\mathbf{x}\,\mathcal{W}_{\rho}(\mathbf{x},0)\mathcal{W}_{\rho}(\mathbf{x},t),
    \label{Eq:SPWignerWi}
\end{equation}
where $\hbar_{\text{eff}}=1/j$ and each factor $2\pi\hbar_{\text{eff}}$ is due to one degree of freedom of the system.

Using the TWA, the short-time behavior  of the Wigner function can be expressed  in terms of  the classical  Hamiltonian flow $\varphi^{t}:\mathcal{M} \to \mathcal{M}$, giving the relation $\mathcal{W}_{\rho}(\mathbf{x},t) \approx \mathcal{W}_{\rho}(\mathbf{x}(-t),0)$ [Eq.~\eqref{eqn:time_evolved_wigner_function}]. Using this relation in Eq.~\eqref{Eq:SPWignerWi} and considering the expressions $\mathcal{W}_{\rho}(\mathbf{x}) \equiv \mathcal{W}_{\rho}(\mathbf{x},0)$ and $\mathcal{W}_{\rho}(\mathbf{x}(-t)) \equiv \mathcal{W}_{\rho}(\mathbf{x}(-t),0)$, the classical limit of the survival probability can be defined as~\cite{Villasenor2020}
\begin{equation}
    \label{eqn:classical_survival_probability}
    \mathfrak{S}_{\text{P}}(t) = \left(\frac{2\pi}{j}\right)^{2} \int_{\mathcal{M}} d\mathbf{x}\,\mathcal{W}_{\rho}(\mathbf{x})\mathcal{W}_{\rho}(\mathbf{x}(-t)),
\end{equation}
whose asymptotic value is
\begin{equation}    \label{eqn:classical_survival_probability_infinite_time_average}
    \mathfrak{S}_{\text{P}}^{\infty} = \lim_{t\to+\infty}\frac{1}{t}\int_{0}^{t}dt'\mathfrak{S}_{\text{P}}(t').
\end{equation}

\subsubsection{Coherent states in regular and chaotic energy regimes}

\begin{figure}[t!]
    \centering
    \includegraphics[width=0.6\textwidth]{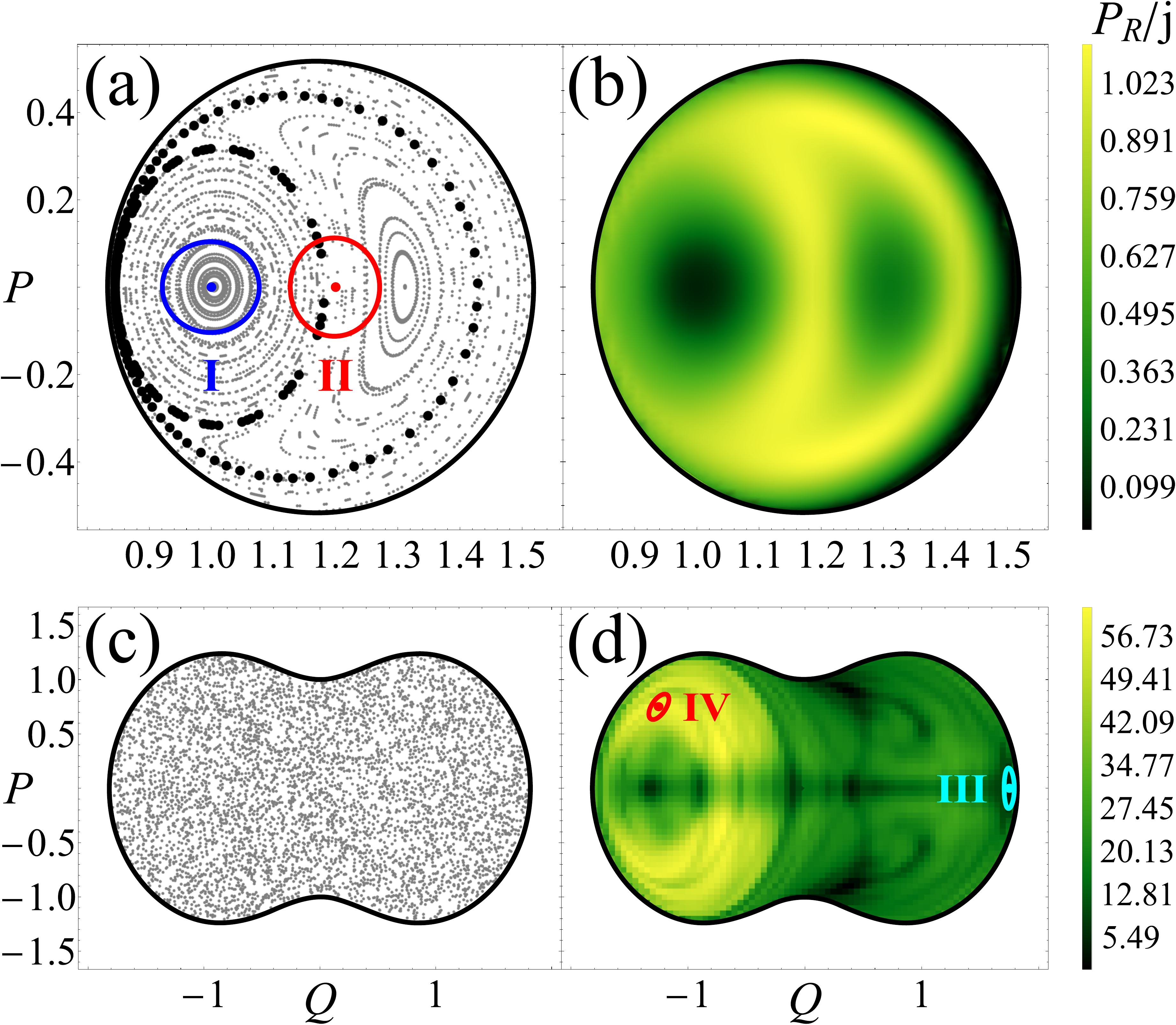}
    \caption{[(a) and (c)] Poincar\'e sections for the (a) regular $\epsilon=-1.8$  and (c) chaotic $\epsilon=-0.5$  energy shell. The projections are shown on the atomic plane $Q$-$P$. The black dots in panel (a) mark a separatrix.
    [(b) and (d)] Participation ratio map for coherent states with centers in (b) the regular region from panel (a) and (d) the chaotic region from panel (c). The projections are shown on the atomic plane $Q$-$P$. The color scales on the right indicate low (darker tones) and high (lighter tones) values of the participation ratio scaled to the system size $j$.
    In each panel, from (a) to (d), the black solid border represents the available phase space at each energy shell. In panel (a), two initial conditions are shown: state I (blue circle) is far from a separatrix and state II (red circle) is close to a separatrix. In panel (d), two initial conditions are also shown: state III (cyan circle) has $P_{\text{R}}=1066$ and state IV (red circle) has $P_{\text{R}}=5743$. Center of the coherent states $\mathbf{x}_{0}=(q_{+}(\epsilon),0;Q_{0},P_{0})$: $(Q_{0},P_{0})=(1,0)$ (state I), $(Q_{0},P_{0})=(1.2,0)$ (state II), $(Q_{0},P_{0})=(1.75,0)$ (state III), and $(Q_{0},P_{0})=(-1.25,0.75)$ (state IV). Hamiltonian parameters: $\omega=\omega_{0}=1$ and $\gamma=2\gamma_{\text{c}}$. The system size in panels (b) and (d) is $j=100$. Figure taken from Ref.~\cite{Villasenor2020}.
    }
    \label{fig:poincare_section_participation_ratio_map}
\end{figure}

To study the survival probability, a set of Glauber-Bloch coherent states is selected according to their center $\mathbf{x}_{0}=(q_{+}(\epsilon),0;Q_{0},P_{0})$ in phase space, where $q_{+}$ is the positive root of the second-degree equation $h_{\text{D}}(\mathbf{x})=\epsilon$ and $\epsilon$ identifies a classical energy shell. The states are selected in the regular and chaotic regime as explained next. 

We identify regular and chaotic regions in phase space using Poincar\'e sections of the classical trajectories. Figure~\ref{fig:poincare_section_participation_ratio_map}(a) shows the Poincar\'e section for a regular energy region of the Dicke model and Fig.~\ref{fig:poincare_section_participation_ratio_map}(c) shows the Poincar\'e section for a chaotic energy region with the same Hamiltonian parameters as in Fig.~\ref{fig:poincare_section_participation_ratio_map}(a).  Figure~\ref{fig:poincare_section_participation_ratio_map}(b) [\ref{fig:poincare_section_participation_ratio_map}(d)] gives the participation ratio for coherent states expanded in the energy eigenbasis of the Dicke model and centered at the initial conditions used in Fig.~\ref{fig:poincare_section_participation_ratio_map}(a) [\ref{fig:poincare_section_participation_ratio_map}(c)]. There is a clear correspondence between the 
Poincar\'e sections (left panels) and the participation ratio maps (right panels). For the regular regions of low energies, states far from a separatrix have small values of the participation ratio, while close to the separatrix, the participation ratio increases.  For the chaotic region of high energies, the classical trajectories fill the available phase space [Fig.~\ref{fig:poincare_section_participation_ratio_map}(c)] and the participation ratio reaches large values [Fig.~\ref{fig:poincare_section_participation_ratio_map}(d)]. Notice, however, that Fig.~\ref{fig:poincare_section_participation_ratio_map}(d)  reveals structures that are not visible in the Poincar\'e section. Thus, when selecting states in the chaotic regime,  the participation ratio map is more useful than the Poincar\'e  sections~\cite{Villasenor2020}.

In Fig.~\ref{fig:poincare_section_participation_ratio_map}(a), two coherent states in the regular energy regime are marked, one centered close to an elliptic point and  far from a separatrix (state I) and one close to a separatrix (state II). In the chaotic regime, we resort to Fig.~\ref{fig:poincare_section_participation_ratio_map}(d)
and select two coherent states with different values of participation ratio, one with small participation ratio (state III) and one with high participation ratio (state IV).

\begin{figure}[t!]
    \centering
    \includegraphics[width=0.9\textwidth]{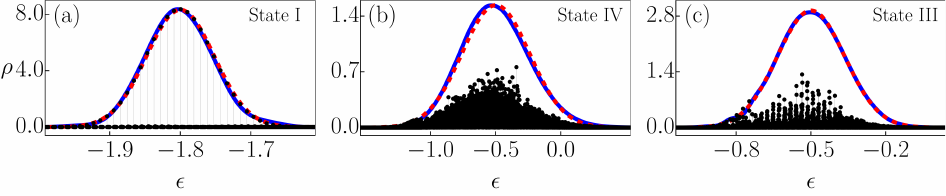}
    \caption{LDOS envelope (blue solid line) for coherent states. Each panel refers to a different state: (a) state I, (b) state IV, and (c) state III. The states are shown in Figs.~\ref{fig:poincare_section_participation_ratio_map}(a) and~\ref{fig:poincare_section_participation_ratio_map}(d). In each panel, from (a) to (c), the red dashed line represents a Gaussian function [Eqs.~\eqref{eqn:ldos_gaussian_function} and~\eqref{eqn:gaussian_ldos_gb}] and the black dots represent the LDOS (components $|c_k|^2$ versus $E_k/j$) with different structure for each state. The coherent states are centered at the (a) regular $\epsilon=-1.8$ and (b)-(c) chaotic energy shell $\epsilon=-0.5$, respectively. Hamiltonian parameters: $\omega=\omega_{0}=1$, $\gamma=2\gamma_{\text{c}}$, and $j=100$.
    }
    \label{fig:ldos_coherent_states}
\end{figure}

The way to calculate the  projection of Glauber-Bloch coherent states $\hat{\rho}_{\mathbf{x}}=|\mathbf{x}\rangle\langle\mathbf{x}|$ in the energy eigenbasis is presented in~\ref{app:coherent_state_efficient_basis}. Using this projection, one can see that the envelope of the LDOS for coherent states  is approximately  Gaussian  for all energy regimes and sets of coordinates in phase space~\cite{Lerma2018AIPCP,Lerma2018JPA,Villasenor2020},
\begin{equation}
    \label{eqn:gaussian_ldos_gb}
    \rho_{\text{GB}}(E) = \frac{1}{\sqrt{2\pi}\sigma_{0}}e^{-(E-E_{\text{c}})^{2}/(2\sigma_{0}^{2})},
\end{equation}
where $E_{\text{c}}$ and $\sigma_{0}$ are the mean energy and width of the LDOS. In Fig.~\ref{fig:ldos_coherent_states}, we present the LDOS envelope, calculated as described in Ref.~\cite{Villasenor2020}, for the coherent states I, IV, and III. Notably,  despite the different LDOS structures, all states exhibit an envelope well described by a Gaussian function. This implies that  the initial decay of the survival probability, governed by the Fourier transform of the LDOS envelope, is Gaussian, $S_{\text{P}}^{\text{st}}(t) = e^{-\sigma_{0}^{2}t^{2}}$,  independently of the energy of the initial coherent state.

Beyond the Gaussian decay, the analytical expression in Eq.~\eqref{eqn:survival_probability_b2_goe} accurately reproduces the behavior of the survival probability for most initial states located in chaotic energy regimes. For the regular energy regimes, an analytical expression of the survival probability was derived in Ref.~\cite{Lerma2018JPA}. For a coherent state centered close to an  elliptic point, such as state I in Fig.~\ref{fig:poincare_section_participation_ratio_map}(a),  this analytical expression is obtained  from  the fact that only a few energy eigenstates have large components and they conform to a Gaussian distribution as can be seen in Fig.~\ref{fig:ldos_coherent_states}(a). Let the indexes of this subset of energy levels with large components be  $\mathcal{K}=\{k_1, k_2,\dots,k_i,\dots\}$. The energy  components of  such a coherent state are  
\begin{equation}
|c_{k}|^{2} \approx\left\{\begin{array}{ll}  \frac{\omega_{1}}{\sqrt{2\pi}\sigma_{0}}e^{-(E_{k}-E_{\text{c}})^{2}/(2\sigma_{0}^{2})} & \text{for } k\in\mathcal{K} \\
0 & \text{otherwise}
\end{array}\right. ,
\end{equation}
where $\omega_{1}\approx E_{k_{\max+1}}-E_{k_{\max}}$ is the fundamental frequency given by  the energies of the two energy states  with largest components, those from the set $\mathcal{E}=\{E_k | k\in\mathcal{K}\}$ closest to the center of the Gaussian function, $E_{\text{c}}$. The analytical expression for the survival probability is  
given by
\begin{equation}
    \label{eqn:survival_probability_regular_regime}
    S_{\text{P}}(t) \approx  \frac{\omega_{1}}{2\sigma_{0}\sqrt{\pi}}\left[1+2\sum_{n}\left(e^{-\omega_{1}^{2}/(4\sigma_{0}^{2})-t^{2}/t_{\text{D}}^{2}}\right)^{n^{2}}\cos(n\omega_1 t)\right],
\end{equation}
where the decay time is $t_{\text{D}}=\omega_{1}/(\sigma_{0}|e_{2}|)$,  and  the anharmonicity $e_{2}=\frac{1}{2}(E_{k_{\max+1}}+E_{k_{\max-1}})-E_{k_{\max}}$ measures the deviation from an energy spectrum whose energies are equidistant.

\begin{figure}[t!]
    \centering
    \includegraphics[width=0.4\textwidth]{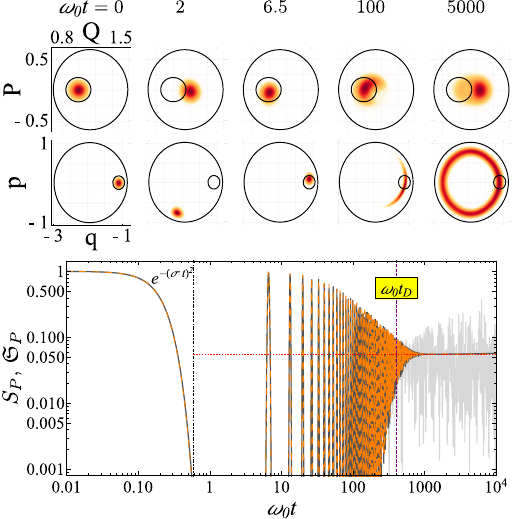}
    \caption{(Top panels) TWA time-evolved Wigner function  [Eq.~\eqref{eqn:glauber_bloch_wigner_function}] (colored shape) for state I [see Fig.~\ref{fig:poincare_section_participation_ratio_map}(a)] at different time steps. The projections are shown on the atomic plane $Q$-$P$ (top row of panels) and the bosonic plane $q$-$p$ (bottom row of panels). In both top and bottom rows of panels, the inner small black border represents the initial volume occupied by the Wigner function and the outer black border represents the available phase space for the regular energy shell $\epsilon=-1.8$. Darker tones inside each atomic and bosonic projection represent higher concentration regions of the Wigner function.
    (Bottom panel) Survival probability (light gray solid line) and its classical limit (dark gray solid line). The orange line is the analytical expression for the regular energy regime [Eq.~\eqref{eqn:survival_probability_regular_regime}]. The horizontal red dashed line gives the asymptotic value for the numerical survival probability and the classical result. The vertical purple dashed line represents the decay time.
    Hamiltonian parameters: $\omega=\omega_{0}=1$, $\gamma=2\gamma_{\text{c}}$, and $j=100$. Figure taken from Ref.~\cite{Villasenor2020}.
    }
    \label{fig:survival_probability_coherent_state_I}
\end{figure}

The survival probability for state I is shown in the bottom panel of Fig.~\ref{fig:survival_probability_coherent_state_I}, where the initial Gaussian decay is indicated. The numerical result (light gray line) agrees with the classical limit obtained with the TWA in Eq.~\eqref{eqn:classical_survival_probability} (dark gray line) up to long times, ceasing to hold when quantum fluctuations associated with the saturation of the dynamics start.
Since state I is in the regular energy regime, close to an elliptic point  and away from a separatrix, the numerical result also agrees with the analytical expression presented in Eq.~\eqref{eqn:survival_probability_regular_regime} (orange line) up to the decay time, $t\sim t_{\text{D}}$, beyond which the  analytical expression is unable to reproduce the quantum fluctuations. As typical of regular dynamics, the survival probability shows periodic revivals before saturation and does not develop a correlation hole, since the energy levels are uncorrelated.

The two rows in the top panel of Fig.~\ref{fig:survival_probability_coherent_state_I} show the classical (TWA) evolution of the  Wigner function (colored shape) projected in the atomic $Q$-$P$ plane (top row) and in the bosonic $q$-$p$ plane (bottom row), where each column is for a given time, as indicated in the figure. These panels help us better understand features of the evolution of the survival probability. They have two concentric black circles, where the inner one represents the initial volume occupied by the Wigner function. At time $\omega_0 t=2$ (second column), the Wigner function is entirely outside its initial volume, which explains why the survival probability goes to zero. At long times, $\omega_{0}t=5000$ (fifth column), the Wigner function reaches its maximum level of delocalization in phase space and the evolution of the survival probability saturates.

\begin{figure}[t!]
    \centering
    \begin{tabular}{cc}
        (a) & (b) \\
        \includegraphics[width=0.4\textwidth]{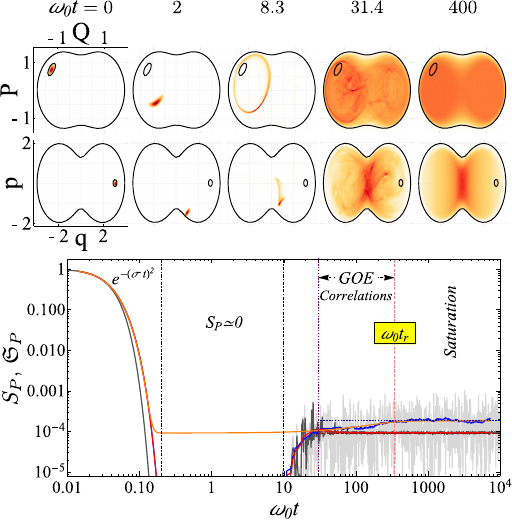} &
        \includegraphics[width=0.4\textwidth]{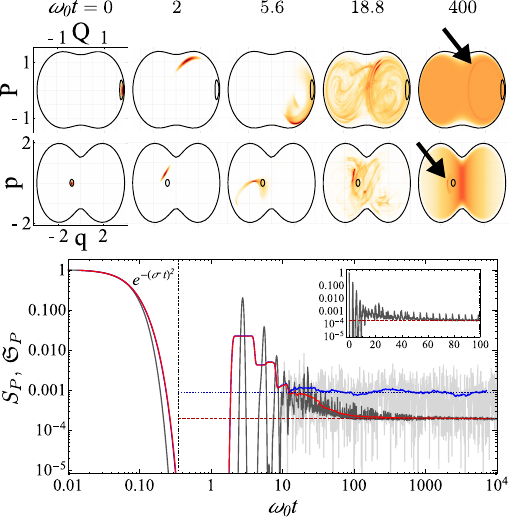} 
    \end{tabular}
    \caption{(Top panels) TWA time-evolved Wigner function  [Eq.~\eqref{eqn:glauber_bloch_wigner_function}] (colored shape) for state (a) IV and (b) III [see Fig.~\ref{fig:poincare_section_participation_ratio_map}(d)] at different time steps. The projections are shown on the atomic plane $Q$-$P$ (top row of panels) and the bosonic plane $q$-$p$ (bottom row of panels). In both top and bottom rows of panels (a) and (b), the inner small black border represents the initial volume occupied by the Wigner function and the outer black border represents the available phase space for the chaotic energy shell $\epsilon=-0.5$. Darker tones inside each atomic and bosonic projection of panels (a) and (b) represent higher concentration regions of the Wigner function. The black arrow in panel (b) at $\omega_{0}t = 400$ indicates the presence of an unstable periodic orbit.
    (Bottom panel) Survival probability (light gray solid line) and its classical limit (dark gray solid line) for states (a) IV and (b) III. In panels (a) and (b), a running time-average is performed for the numerical survival probability (light blue solid line) and its classical limit (light red solid line). The horizontal line gives the asymptotic value for the numerical survival probability (dark blue dotted line) and the asymptotic  value of the classical result (dark red dashed line). In panel (a), the orange line is the analytical expression for the chaotic energy regime [Eq.~\eqref{eqn:survival_probability_b2_goe} with $\eta = 2 P_{\text{R}}$] and the vertical pink dashed line represents the relaxation time. The inset in panel (b) shows the classical limit of the survival probability in a linear time scale where periodicity can be clearly seen.
    Hamiltonian parameters: $\omega=\omega_{0}=1$, $\gamma=2\gamma_{\text{c}}$, and $j=100$. Figures taken from Ref.~\cite{Villasenor2020}.
    }
    \label{fig:survival_probability_coherent_state_III_IV}
\end{figure}

The bottom panel of Fig.~\ref{fig:survival_probability_coherent_state_III_IV}(a) compares the survival probability of the coherent state IV for the following cases: numerical evolution (light gray line) and its time-average (blue line), classical limit under the TWA (dark gray line) and its time average (red line), and the analytical expression for chaotic states in Eq.~\eqref{eqn:survival_probability_b2_goe} with $\eta=2P_{\text{R}}$ instead of the result for $\eta$ from Eq.~\eqref{eqn:effective_dimension}, which was for random states. The classical limit reproduces the numerical results up to the point where the correlation hole emerges. The correlation hole is caused by the correlated levels of the discrete spectrum of the system, being a purely quantum feature. The analytical expression describes the whole evolution of the survival probability, including the correlation hole, but excluding the time interval where the survival probability goes to zero. Analogously to the description of Fig.~\ref{fig:survival_probability_coherent_state_I}, during this time interval the TWA evolving Wigner function lies outside its initial volume, as seen in the top rows of Fig.~\ref{fig:survival_probability_coherent_state_III_IV}(a).
Notice that for this chaotic state IV, the last top panel for the TWA evolution of the Wigner function in the atomic plane $Q$-$P$ ($\omega_{0}t=400$) shows  complete delocalization of the Wigner function, which fills the available phase space homogeneously.

The bottom panel of Fig.~\ref{fig:survival_probability_coherent_state_III_IV}(b) shows the survival probability for state III, which is in a chaotic region, but has low participation ratio. The figure shows the survival probability obtained numerically (light gray line) and its time average (blue line), the classical limit (dark gray line) and its time average (red line). The validity of the TWA is confirmed for short times, before the emergence of quantum fluctuations. Notice that the survival probability for this state exhibits periodic revivals similar to what one finds for regular dynamics. This behavior is associated with quantum scarring, where  unstable periodic orbits in the classical domain affect the quantum dynamics~\cite{Villasenor2020}. See more details about scarring in Sec.~\ref{sec:QuantumScarring}. 
In the two top rows of Fig.~\ref{fig:survival_probability_coherent_state_III_IV}(b), by examining the snapshot of the TWA evolving Wigner function at $\omega_{0}t=400$, we can indeed discern a structure (indicated with  black arrows) associated with a periodic orbit embedded in the phase space.

Figures~\ref{fig:survival_probability_coherent_state_I} and~\ref{fig:survival_probability_coherent_state_III_IV} illustrate three of the general four scenarios for the survival probability of initial coherent states~\cite{Villasenor2020}. They are described below.

\begin{itemize}
    \item For coherent states in the regular energy regime:
    \begin{enumerate}
        \item If the center of the initial coherent state is far from a separatrix in phase space, the survival probability and its classical limit agree up to saturation, where the classical dynamics reaches an asymptotic value and the quantum dynamics fluctuates around it due to the discreteness of the energy spectrum.
        \item If the initial state is located close to a separatrix in phase space, the survival probability and its classical limit agree up to a time where  tunneling occurs between classically disconnected phase-space regions (stable centers) affecting the quantum dynamics only. In this case, the  quantum asymptotic value of the dynamics is lower than the classical one. An example for this case can be found in Ref.~\cite{Villasenor2020}.
    \end{enumerate}
    \item For coherent states  in the chaotic energy regime:
    \begin{enumerate}
        \item If the initial state has a large participation ratio, the survival probability and its classical limit agree up to the point where the correlation hole emerges, which appears only for the quantum dynamics. In this case, the saturation of the quantum dynamics happens at a value that is  higher than  the classical saturation value, and the quantum saturation happens later than the classical one.
        \item If the initial state has small participation ratio, the survival probability and its classical limit exhibit periodic revivals and agree up to the saturation of the quantum dynamics, which happens earlier than the classical saturation. The early quantum saturation is caused by the small number of eigenstates that participate in the quantum evolution.
    \end{enumerate}
\end{itemize}

A useful quantity to distinguish between the two previous dynamical scenarios of coherent states in the chaotic regime is the ratio between the asymptotic value of the quantum survival probability and that of the classical survival probability. This ratio turns out to be equal to the fraction of phase space explored by the coherent states~\cite{Heller1991COLL} and is given by
\begin{equation}
    \label{eqn:factor_r}
    R=2\frac{\mathfrak{S}_{\text{P}}^{\infty}}{S_{\text{P}}^{\infty}}.
\end{equation}
The value of $R$ is bounded in the interval $[0,1]$, where $R\approx 0$ implies a coherent state whose dynamics exhibits periodic revivals and a small number of participating eigenstates, while $R\approx 1$ implies an initial coherent state  highly  delocalized in the energy eigenbasis  that exhibits manifestations of spectral correlations in the evolution of its survival probability. The ratio in Eq.~\eqref{eqn:factor_r} was calculated in Ref.~\cite{Villasenor2020} for the same coherent states in a chaotic energy regime as those used in  Fig.~\ref{fig:poincare_section_participation_ratio_map}(d), confirming that  coherent states with  a low $R$-ratio exhibit  periodic revivals and   do not show a correlation hole. As discussed  in Sec.~\ref{sec:QuantumScarring}, this particular dynamics is related to the fact that coherent states with low $R$-ratio   are centered close to a classical unstable periodic orbit; consequently, they are identified as quantum scarred states.  In contrast,  coherent states with  high $R$-ratio show  spectral correlations in the form of a correlation hole in the evolution of their survival probability. These states are thus  identified as ergodic states and they are more numerous than scarred states.  Intermediate cases may arise, where the competition between periodic revivals and  the emergence of the correlation hole is assessed based on  the value of $R$.

\section{Quantum localization}
\label{sec:QuantumLocalization}
Quantum localization in association with dynamical localization denotes a quantum feature that limits the classical diffusion in chaotic regimes. Dynamical localization was first observed in the kicked rotor~\cite{Chirikov1981,Fishman1982,Izrailev1990}. A review about this system, its relation to chaos and the influence of localization on its spectral statistics is available in Ref.~\cite{Izrailev1990}. Many quantum systems have shown the localization phenomenon, such as the hydrogen atom in a monochromatic field and Rydberg atoms~\cite{Casati1984PRLb,Casati1987,Blumel1987}, band-random-matrices and quantum billiards ~\cite{Casati1993,Borgonovi1996,Batistic2010,Batistic2013JPA,Batistic2013PRE,Batistic2019,Robnik2020}, among others. Moreover, dynamical localization has been related to Anderson localization~\cite{Anderson1958,Fishman1982,Lee1985,Fishman2010}, which is caused by quantum interferences that suppress the classical diffusion of particles in real space. A significant portion of the recent papers on localization have focused on quantum many-body systems, because of the increasing interest in the many-body localized phase~\cite{Sierant2025}.

We review studies of quantum localization performed in the Dicke model. First, we introduce the concept of localization measures using the R\'enyi entropy in the Hilbert space of the Dicke model and then extend the concept to continuous bases. Subsequently, we present the R\'enyi occupations, which are localization measures defined over bounded phase-space subspaces of the Dicke model.

\subsection{Localization measures}

Measures of localization are used in research areas as diverse as ecology, information science, and linguistics~\cite{Jelinek1977,Brown1992,Jost2006}.
The exponential of an entropy is a way to measure spreading (delocalization) in a mathematical sense~\cite{Campbell1966}. In physics, localization is defined with respect to a given space. When a discrete space is chosen, the delocalization degree of a quantum state is based on its spreading over a finite basis, as quantified by measures such as the participation ratio and the R\'enyi entropy~\cite{Edwards1972,Izrailev1990,Atas2012,Atas2014}. The concept can be extended to bounded or unbounded continuous spaces. When unbounded continuous spaces are chosen, these measures delocalize themselves arbitrarily, reaching very large values. For this reason, to define them appropriately, a finite space region must be selected as a reference. 

A useful feature of measuring localization of quantum states in phase space is that connections between the structures of classical and quantum dynamics can be created~\cite{Gorin1997}. Early studies explored the concept of localization in phase space using the generalized R\'enyi, also known as R\'enyi-Wehrl, entropy~\cite{Wehrl1978,Gnutzmann2001}. The R\'enyi entropy is defined with quasiprobability distributions, such as Wigner or Husimi functions~\cite{Hillery1984,Lee1995}, which are common ways to represent quantum states in phase space.

\subsubsection{R\'enyi entropy in discrete Hilbert space}

Generalized localization measures in a discrete Hilbert space may be defined through the R\'enyi entropy of order $\alpha$, as
\begin{equation}
    \label{eqn:discrete_renyi_entropy}
    S_{\alpha} = \frac{1}{1-\alpha}\ln\left(\sum_{k}p_{k}^{\alpha}\right),
\end{equation}
where $\alpha\geq0$ ($\alpha\neq1$). The term $p_{k}=|\langle E_{k}|\Psi\rangle|^{2}$, which satisfies the normalization condition $\sum_{k}p_{k}=1$, defines the probability to find an arbitrary quantum state $|\Psi\rangle$ in a basis state $|E_{k}\rangle$. For the Dicke model, the discrete basis corresponds to the eigenbasis of its Hamiltonian, $\hat{H}_{\text{D}}|E_{k}\rangle=E_{k}|E_{k}\rangle$, with dimension given by the number of converged eigenstates, $N_{\text{c}}$. In the limiting case $\alpha=1$, the above entropy corresponds to the Shannon entropy,
\begin{equation}
    S_{1} = \lim_{\alpha\to1} S_{\alpha} = -\sum_{k}p_{k}\ln(p_{k}).
\end{equation}
The exponential of the entropies $S_{\alpha}$,
\begin{gather}
    \label{eqn:generalized_participation_ratio}
    L_{\alpha} = e^{S_{\alpha}} = \left(\sum_{k}p_{k}^{\alpha}\right)^{1/(1-\alpha)}, \\
    L_{1} = \lim_{\alpha\to1}L_{\alpha} = e^{S_{1}} = e^{-\sum_{k}p_{k}\ln(p_{k})},
\end{gather}
are known as generalized participation ratios of order $\alpha$~\cite{Ott2002Book,Murphy2011}.  For the Dicke model, they are bounded in the interval $[1,N_{\text{c}}]$. The particular case $\alpha=2$ corresponds to the usual participation ratio, $L_{2}=e^{S_{2}}=P_{\text{R}}=\left(\sum_{k}p_{k}^{2}\right)^{-1}$, given in Eq.~\eqref{eqn:participation_ratio}.

The R\'enyi entropies normalized to an effective dimension give the so-called fractal (R\'enyi) dimensions, used in the analysis of multifractality of quantum systems~\cite{Hentschel1983,Halsey1986}. The investigation of multifractality in the Dicke model~\cite{Bastarrachea2024} was shown to complement the classical analysis of phase space, providing a sensitive tool to detect structures in phase space, such as regularity, chaos, and localization features.

\subsubsection{R\'enyi entropy in phase space}
\label{Sec:RenyiEntropyPhaseSpace}

The concept of localization can be extended to a bounded continuous space $\mathcal{S}$ with finite volume $\mathcal{V}_{\mathcal{S}} = \int_{\mathcal{S}}d\mathcal{V}(\upsilon)$. To obtain measures of localization in $\mathcal{S}$, we replace the discrete probabilities in Eq.~\eqref{eqn:discrete_renyi_entropy} by a probability distribution $p_{k} \to \Phi(\upsilon)$, which is normalized $\int_{\mathcal{S}}d\mathcal{V}(\upsilon)\,\Phi(\upsilon)=1$, and replace the sums by integrals $\sum\bullet \to \int d\mathcal{V}(\upsilon)\bullet$. With this procedure, the R\'enyi entropies take the continuous form~\cite{Gnutzmann2001}
\begin{gather}
    \label{eqn:continuous_renyi_entropy}
    \widetilde{S}_{\alpha}(\mathcal{S},\Phi) = \frac{1}{1-\alpha}\ln\left[\int_{\mathcal{S}}d\mathcal{V}(\upsilon)\,\Phi^{\alpha}(\upsilon)\right], \\
    \widetilde{S}_{1}(\mathcal{S},\Phi) = \lim_{\alpha\to1} \widetilde{S}_{\alpha}(\mathcal{S},\Phi) = -\int_{\mathcal{S}}d\mathcal{V}(\upsilon)\,\Phi(\upsilon)\ln[\Phi(\upsilon)],
\end{gather}
and their exponential defines the generalized localization measures
\begin{gather}
    \label{eqn:renyi_volume_alpha}
    \mathcal{V}_{\alpha}(\mathcal{S},\Phi) = e^{\widetilde{S}_{\alpha}(\mathcal{S},\Phi)} = \left[\int_{\mathcal{S}}d\mathcal{V}(\upsilon)\,\Phi^{\alpha}(\upsilon)\right]^{1/(1-\alpha)}, \\
    \label{eqn:renyi_volume_alpha1}
    \mathcal{V}_{1}(\mathcal{S},\Phi) = \lim_{\alpha\to1}\mathcal{V}_{\alpha}(\mathcal{S},\Phi) 
    = e^{\widetilde{S}_{1}(\mathcal{S},\Phi)} = e^{-\int_{\mathcal{S}}d\mathcal{V}(\upsilon)\,\Phi(\upsilon)\ln[\Phi(\upsilon)]},
\end{gather}
bounded in the interval $(0,\mathcal{V}_{\mathcal{S}}]$. We call $\mathcal{V}_{\alpha}(\mathcal{S},\Phi)$ the {\it R\'enyi volume of order $\alpha$}, since it defines the spreading of a quantum state over the finite volume $\mathcal{V}_{\mathcal{S}}$ of the continuous space $\mathcal{S}$. By scaling the R\'enyi volume by the volume $\mathcal{V}_{\mathcal{S}}$, new generalized and normalized localization measures can be defined as~\cite{Villasenor2021}
\begin{gather}
    \label{eqn:renyi_occupation_alpha}
    \mathfrak{L}_{\alpha}(\mathcal{S},\Phi) = \frac{\mathcal{V}_{\alpha}(\mathcal{S},\Phi)}{\mathcal{V}_{\mathcal{S}}}, \\
    \label{eqn:renyi_occupation_alpha1}
    \mathfrak{L}_{1}(\mathcal{S},\Phi) = \lim_{\alpha\to1}\mathfrak{L}_{\alpha}(\mathcal{S},\Phi) = \frac{\mathcal{V}_{1}(\mathcal{S},\Phi)}{\mathcal{V}_{\mathcal{S}}},
\end{gather}
which are bounded in the interval (0,1]. We call $\mathfrak{L}_{\alpha}(\mathcal{S},\Phi)$ the {\it R\'enyi occupation of order $\alpha$}. It defines the fraction of the finite volume $\mathcal{V}_{\mathcal{S}}$ occupied by a given quantum state.
 
The four-dimensional phase space $\mathcal{M}$ of the Dicke model, usually analyzed in terms of the coordinates $\mathbf{x}=(q,p;Q,P)$, is an unbounded space with infinite volume $\mathcal{V}_{\mathcal{M}}=\int_{\mathcal{M}}d\mathcal{V}(\mathbf{x})=\infty$. As a result, the probability distribution $\Phi(\mathbf{x})$ can be arbitrarily large and the R\'enyi occupations $\mathfrak{L}_{\alpha,1}(\mathcal{M},\Phi)$ are not well defined. To circumvent this problem, the R\'enyi occupations are defined within a bounded subspace $\mathcal{S}\subset\mathcal{M}$ with finite volume $\mathcal{V}_{\mathcal{S}}<\mathcal{V}_{\mathcal{M}}$. The selection of the bounded subspace is not universal and different choices can lead to different interpretations about the level of localization of quantum states in phase space.

The complete phase space of the Dicke model is a tensor product between a bounded atomic subspace and an unbounded bosonic subspace. Therefore, the atomic subspace represents a natural space to be studied. It is given by~\cite{Wang2020,Villasenor2021}
\begin{equation}
    \label{eqn:atomic_subspace}
    \mathcal{A} = \{(Q,P)|Q^{2}+P^{2}\leq4\},
\end{equation}
and its phase-space volume is $\mathcal{V}_{\mathcal{A}} = \int_{\mathcal{A}} dQ\,dP = 4\pi$. By taking the probability distribution as $\Phi_{\rho}(Q,P) = C_{\text{A}}^{-1}\widetilde{\mathcal{Q}}_{\rho}(Q,P) = C^{-1}\int\int dq\,dp\,\mathcal{Q}_{\rho}(\mathbf{x})$ [see Eq.~\eqref{eqn:projected_exact_atomic_husimi_function}], the R\'enyi occupations $\mathfrak{L}_{\alpha,1}(\mathcal{A},\hat{\rho})$ can be found using Eqs.~\eqref{eqn:renyi_occupation_alpha} and~\eqref{eqn:renyi_occupation_alpha1},
\begin{gather}
    \label{eqn:atomic_subspace_renyi_occupation_alpha}
    \mathfrak{L}_{\alpha}(\mathcal{A},\hat{\rho}) = \frac{C_{\text{A}}^{\alpha/(\alpha-1)}}{\mathcal{V}_{\mathcal{A}}}\left[\int_{\mathcal{A}}dQ\,dP\,\widetilde{\mathcal{Q}}_{\rho}^{\alpha}(Q,P)\right]^{1/(1-\alpha)}, \\
    \label{eqn:atomic_subspace_renyi_occupation_alpha1}
    \mathfrak{L}_{1}(\mathcal{A},\hat{\rho}) = \frac{C_{\text{A}}}{\mathcal{V}_{\mathcal{A}}}e^{-(1/C_{\text{A}})\int_{\mathcal{A}}dQ\,dP\,\widetilde{\mathcal{Q}}_{\rho}(Q,P)\ln[\widetilde{\mathcal{Q}}_{\rho}(Q,P)]},
\end{gather}
where the constant $C = C_{\text{A}}C_{\text{B}} = \int_{\mathcal{M}}d\mathbf{x}\,\mathcal{Q}_{\rho}(\mathbf{x})$, with $C_{\text{A}}=4\pi/(2j+1)$ and $C_{\text{B}}=2\pi/j$, ensures the normalization of the Husimi function in the whole phase space $\mathcal{M}$.

Another natural subspace for the Dicke model is the space of the classical energy shell $\epsilon=E/j$~\cite{Villasenor2021},
\begin{equation}
    \label{eqn:classical_energy_shell_subspace}
    \mathcal{M}_{\epsilon} = \{\mathbf{x}=(q,p;Q,P)|h_{\text{D}}(\mathbf{x})=\epsilon\} ,
\end{equation}
with phase-space volume
\begin{equation}
    \label{eqn:classical_energy_shell_volume}
    \mathcal{V}_{\epsilon} = \int_{\mathcal{M}_{\epsilon}}d\mathbf{s} = \int_{\mathcal{M}}d\mathbf{x}\,\delta[h_{\text{D}}(\mathbf{x})-\epsilon]  = (2\pi\hbar_{\text{eff}})^{2}\nu(\epsilon),
\end{equation}
where $d\mathbf{s}=\delta[h_{\text{D}}(\mathbf{x})-\epsilon]d\mathbf{x}$ is a surface element, $\hbar_{\text{eff}}=1/j$, and $\nu(\epsilon)$ is the semiclassical density of states in Eq.~\eqref{eqn:semiclassical_density_states}. For this case, the probability distribution is taken as the standard Husimi function $\Phi_{\epsilon,\rho}(\mathbf{x}) = C_{\epsilon}^{-1}\mathcal{Q}_{\rho}(\mathbf{x})$, where the constant $C_{\epsilon} = \int_{\mathcal{M}_{\epsilon}}d\mathbf{s}\,\mathcal{Q}_{\rho}(\mathbf{x})$ ensures its normalization within the phase space of the classical energy shell $\mathcal{M}_{\epsilon}$. For this chosen subspace, the R\'enyi occupations become
\begin{gather}
    \label{eqn:classical_energy_shell_renyi_occupation_alpha}
    \mathfrak{L}_{\alpha}(\epsilon,\hat{\rho}) = \frac{C_{\epsilon}^{\alpha/(\alpha-1)}}{\mathcal{V}_{\epsilon}}\left[\int_{\mathcal{M}_{\epsilon}}d\mathbf{s}\,\mathcal{Q}_{\rho}^{\alpha}(\mathbf{x})\right]^{1/(1-\alpha)}, \\
    \label{eqn:classical_energy_shell_renyi_occupation_alpha1}
    \mathfrak{L}_{1}(\epsilon,\hat{\rho}) = \frac{C_{\epsilon}}{\mathcal{V}_{\epsilon}}e^{-(1/C_{\epsilon})\int_{\mathcal{M}_{\epsilon}}d\mathbf{s}\,\mathcal{Q}_{\rho}(\mathbf{x})\ln[\mathcal{Q}_{\rho}(\mathbf{x})]}.
\end{gather}

The Husimi function profiles for eigenstates of the Dicke model are  well described by Gaussian functions, which implies that the R\'enyi occupations remain almost constant for these eigenstates around the corresponding classical energy shells~\cite{Villasenor2021}. Therefore, the R\'enyi occupation of second order, $\mathfrak{L}_{2}(\epsilon,\hat{\rho})$, can be seen as the phase-space analog of the discrete participation ratio given in Eq.~\eqref{eqn:participation_ratio}.

\subsection{Phase-space localization of eigenstates and coherent states}
\label{subsubsec:renyi_occupation_eigenstate_coherent_state}

\begin{figure}[t!]
    \centering
    \begin{tabular}{c}
        \includegraphics[width=0.7\textwidth]{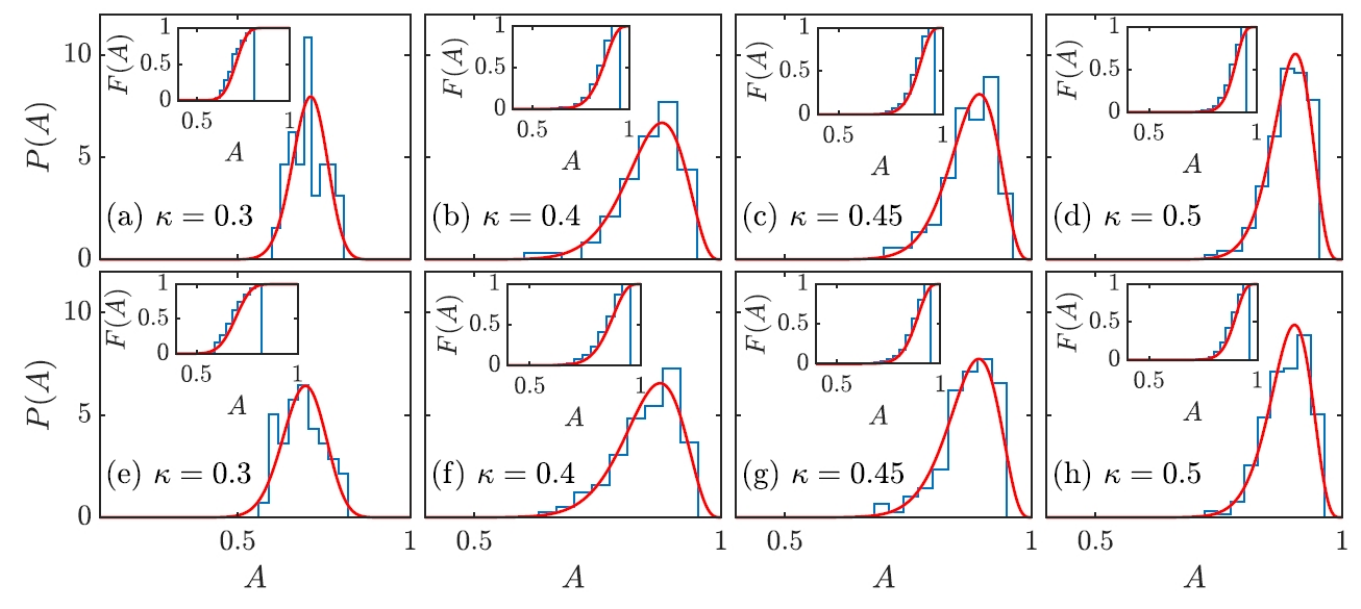} \\
        \includegraphics[width=0.7\textwidth]{figures/PRE2021_fig2.pdf}
    \end{tabular}
    \caption{(Top panels) Probability distribution of R\'enyi occupation $\mathfrak{L}_{1}(\mathcal{A},\hat{\rho})$  [Eq.~\eqref{eqn:atomic_subspace_renyi_occupation_alpha1}] (blue bars) for eigenstates $\hat{\rho}_{k}$ of the Dicke Hamiltonian in the chaotic energy interval $\epsilon_{k}\in(0.44,0.95)$. Each panel refers to a different coupling parameter: (a) $\gamma=0.3$, (b) $\gamma=0.4$, (c) $\gamma=0.45$, and (d) $\gamma=0.5$. All panels (a)-(d) were computed using the system size $j=20$. Panels (e)-(h) represent the same information as panels (a)-(d) for the system size $j=24$. The correspondence between the labels in panels (a)-(h) and the main text is: $A=\mathfrak{L}_{1}$, $\kappa=\gamma$, and $N=2j$. In each panel, from (a) to (h), the red solid line represents a fitting to the beta distribution [Eq.~\eqref{eq:beta_distribution}] and the inset shows the cumulative probability distribution of each beta distribution. The fitting parameters $(a,b)$ for each panel can be found in Ref.~\cite{Wang2020}. The field frequency is $\omega=\omega_{0}=1$. Figure taken from Ref.~\cite{Wang2020}.
    (Bottom panel) R\'enyi occupations $\mathfrak{L}_{2}(\mathcal{A},\hat{\rho})$ [Eq.~\eqref{eqn:atomic_subspace_renyi_occupation_alpha} with $\alpha=2$] (blue bars) and $\mathfrak{L}_{2}(\epsilon,\hat{\rho})$ [Eq.~\eqref{eqn:classical_energy_shell_renyi_occupation_alpha} with $\alpha=2$] (green bars) for eigenstates $\hat{\rho}_{k}$ of the Dicke Hamiltonian in the chaotic energy interval $\epsilon_{k}\in(1,1.274)$. The inset shows the cumulative probability distribution of each probability distribution. Hamiltonian parameters: $\omega=\omega_{0}=1$, $\gamma=2\gamma_{\text{c}}$, and $j=30$. Figure taken from Ref.~\cite{Villasenor2021}.
    }
    \label{fig:probability_distribution_renyi_occupation}
\end{figure}

To elucidate how the choices of subspaces above affect the results of the R\'enyi occupations, we consider first eigenstates of the Dicke model. The top panels (a)-(h) in Fig.~\ref{fig:probability_distribution_renyi_occupation} show the probability distribution of the atomic R\'enyi occupation of order $\alpha=1$, $\mathfrak{L}_{1}(\mathcal{A},\hat{\rho})$ [Eq.~\eqref{eqn:atomic_subspace_renyi_occupation_alpha1}], for a set of eigenstates $\hat{\rho}_{k}$ of the Dicke model located in a chaotic energy regime. The insets in each of the latter panels (a)-(h) show the corresponding cumulative probability distribution. There is good agreement between the numerical data and a fit to the beta distribution,
\begin{equation}
    \label{eq:beta_distribution}
    P(\mathfrak{L}_{1}) = \frac{\mathfrak{L}_{1}^{a-1}(1-\mathfrak{L}_{1})^{b-1}}{B(a,b)},
\end{equation}
where $a$ and $b$ are two positive fitting parameters and $B(a,b)$ is the beta function, which normalizes the distribution. The agreement holds for different values of the coupling parameter $\gamma$ (increasing from Fig.~\ref{fig:probability_distribution_renyi_occupation}(a) [\ref{fig:probability_distribution_renyi_occupation}(e)] to Fig.~\ref{fig:probability_distribution_renyi_occupation}(d) [\ref{fig:probability_distribution_renyi_occupation}(h)]) and  different system sizes [$j=20$ in panels (a)-(d) and $j=24$ in panels (e)-(h)]. For eigenstates in the chaotic region, the mean value of the R\'enyi occupation gets closer to unity as the coupling parameter increases, indicating a high level of delocalization.

The bottom panel of Fig.~\ref{fig:probability_distribution_renyi_occupation} compares the R\'enyi occupation of order $\alpha=2$ for the atomic subspace, $\mathfrak{L}_{2}(\mathcal{A},\hat{\rho})$ [Eq.~\eqref{eqn:atomic_subspace_renyi_occupation_alpha} with $\alpha=2$], with the R\'enyi occupation for the energy-shell subspace, $\mathfrak{L}_{2}(\epsilon,\hat{\rho})$ [Eq.~\eqref{eqn:classical_energy_shell_renyi_occupation_alpha} with $\alpha=2$], for a set of eigenstates $\hat{\rho}_{k}$ of the Dicke model in a chaotic energy regime. The mean value of the atomic R\'enyi occupation is located near unity, $\mathfrak{L}_{2}(\mathcal{A},\hat{\rho}_{k})\sim0.9$, as it happens for the atomic R\'enyi occupation of first order in Figs.~\ref{fig:probability_distribution_renyi_occupation}(d) and~\ref{fig:probability_distribution_renyi_occupation}(h), $\mathfrak{L}_{1}(\mathcal{A},\hat{\rho}_{k})\sim0.9$. This finding confirms that most eigenstates are highly delocalized in the atomic Bloch sphere and that in the regime of strong chaos, the order of the measure does not play an important role. In contrast, the mean value of the energy-shell R\'enyi occupation is $\mathfrak{L}_{2}(\epsilon_{k},\hat{\rho}_{k})\sim0.4$, which means that most eigenstates occupy less than half of the phase-space volume of the classical energy shell.

Each choice of the subspace in which the R\'enyi occupation is defined leads to a quantity that is sensitive to different properties in phase space. Coherent states can help us better understand the dependence of the R\'enyi occupation on the subspace chosen for its calculation. In what follows, we consider a Glauber-Bloch coherent state $\hat{\rho}_{\mathbf{x}_{0}}=|\mathbf{x}_{0}\rangle\langle\mathbf{x}_{0}|$ with initial condition $\mathbf{x}_{0}=(q_{+}(\epsilon),0;Q_{0},P_{0})$
evolving in time as $\hat{\rho}_{\mathbf{x}}(t) = e^{-i\hat{H}_{\text{D}}t}\hat{\rho}_{\mathbf{x}_{0}}e^{i\hat{H}_{\text{D}}t}$. 

\begin{figure}[t!]
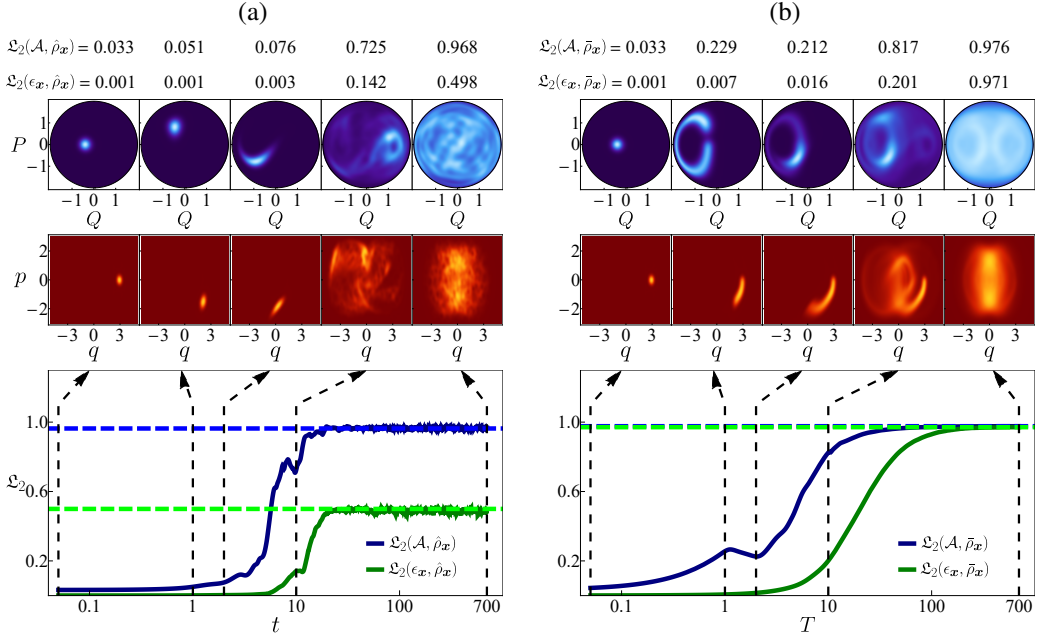

    \centering
    \begin{tabular}{cc}
        (a) & (b) \\
        \includegraphics[width=0.4\textwidth]{figures/PRE2021_fig3a.pdf} &
        \includegraphics[width=0.4\textwidth]{figures/PRE2021_fig3b.pdf} 
    \end{tabular}
    \caption{
    (Top panels) Husimi function for (a) the time-evolved coherent state $\hat{\rho}_{\mathbf{x}}(t)$  and (b) its time average $\overline{\rho}_{\mathbf{x}}$ at different time steps. The projections are shown on the atomic plane $Q$-$P$ [Eq.~\eqref{eqn:projected_exact_atomic_husimi_function}] (top row of panels) and the bosonic plane $q$-$p$ [Eq.~\eqref{eqn:projected_exact_bosonic_husimi_function}] (bottom row of panels). In the top row of panels (a) and (b), the black solid border represents the available phase space at the chaotic energy shell $\epsilon=1$. Lighter tones inside each atomic and bosonic projection represent higher concentration values of the Husimi function, while dark blue (atomic) and dark red (bosonic) correspond to zero.
    (Bottom panel) R\'enyi occupations $\mathfrak{L}_{2}(\mathcal{A},\hat{\rho})$ [Eq.~\eqref{eqn:atomic_subspace_renyi_occupation_alpha} with $\alpha=2$] (dark blue solid line) and $\mathfrak{L}_{2}(\epsilon,\hat{\rho})$ [Eq.~\eqref{eqn:classical_energy_shell_renyi_occupation_alpha} with $\alpha=2$] (dark green solid line) for (a) the evolved coherent state $\hat{\rho}_{\mathbf{x}}(t)$ and (b) its time-average $\overline{\rho}_{\mathbf{x}}$. In panels (a) and (b), the horizontal dashed lines represent the corresponding asymptotic values of each quantity and the vertical black dashed lines represent the time steps where the projections are shown in the top panels. Initial condition: $\mathbf{x}_{0}=(q_{+}(\epsilon),0;-0.4,0)$.
    Hamiltonian parameters: $\omega=\omega_{0}=1$, $\gamma=2\gamma_{\text{c}}$, and $j=30$. Figures taken from Ref.~\cite{Villasenor2021}. 
    }
    \label{fig:renyi_occupation_evolved_coherent_states}
\end{figure}

The bottom panel of Fig.~\ref{fig:renyi_occupation_evolved_coherent_states}(a) shows $\mathfrak{L}_{2}(\mathcal{A},\hat{\rho})$ (dark blue solid line) and $\mathfrak{L}_{2}(\epsilon,\hat{\rho})$ (dark green solid line) for a time-evolved coherent state $\hat{\rho}_{\mathbf{x}}(t)$ located in a chaotic energy region and the bottom panel of Fig.~\ref{fig:renyi_occupation_evolved_coherent_states}(b) shows their values for the time average $\overline{\rho}_{\mathbf{x}} = T^{-1}\int_{0}^{T}dt\,\hat{\rho}_{\mathbf{x}}(t)$ for a finite time $T$. The bottom panel in Fig.~\ref{fig:renyi_occupation_evolved_coherent_states}(a) shows that the atomic R\'enyi occupation $\mathfrak{L}_{2}(\mathcal{A},\hat{\rho}_{\mathbf{x}})$ quickly saturates to unity, indicating that the coherent state becomes completely delocalized over the atomic phase space, while the energy-shell R\'enyi occupation saturates to the value $\mathfrak{L}_{2}(\epsilon,\hat{\rho}_{\mathbf{x}})\sim1/2$, indicating that the coherent state never occupies more than half of the available phase space of the classical energy shell. This limit is fulfilled for any pure state, because quantum interferences of the wave function prevent complete delocalization over the classical energy shell~\cite{Pilatowsky2021NC}.

The behavior of the atomic R\'enyi occupation $\mathfrak{L}_{2}(\mathcal{A},\overline{\rho}_{\mathbf{x}})$ (dark blue line) for the time average $\overline{\rho}_{\mathbf{x}}$ in the bottom panel of Fig.~\ref{fig:renyi_occupation_evolved_coherent_states}(b) is similar to that of $\mathfrak{L}_{2}(\mathcal{A},\hat{\rho}_{\mathbf{x}})$ in Fig.~\ref{fig:renyi_occupation_evolved_coherent_states}(a), reaching unity fast. The energy-shell R\'enyi occupation $\mathfrak{L}_{2}(\epsilon,\overline{\rho}_{\mathbf{x}})$ (dark green line) also reaches unity, which contrasts with the behavior of the instantaneous occupation $\mathfrak{L}_{2}(\epsilon,\hat{\rho}_{\mathbf{x}})$ in Fig.~\ref{fig:renyi_occupation_evolved_coherent_states}(a). The fact that $\mathfrak{L}_{2}(\epsilon,\overline{\rho}_{\mathbf{x}})$ reaches 1 is a manifestation of quantum ergodicity, which only holds as an ensemble property, as in the case of the temporal average performed in Fig.~\ref{fig:renyi_occupation_evolved_coherent_states}(b) ~\cite{Pilatowsky2021NC}. More discussions about quantum ergodicity are found in Sec.~\ref{subsec:dynamical_scarring_quantum_ergodicity}.

The top panels of Figs.~\ref{fig:renyi_occupation_evolved_coherent_states}(a) and~\ref{fig:renyi_occupation_evolved_coherent_states}(b) show the time evolution of the Husimi function of a coherent state $\hat{\rho}_{\mathbf{x}}(t)$ and the time average $\overline{\rho}_{\mathbf{x}}$, respectively. The time-evolved Husimi function is projected in the atomic plane $Q$-$P$ (top row) and in the bosonic plane $q$-$p$ (bottom row) using the expressions presented in Eqs.~\eqref{eqn:projected_exact_atomic_husimi_function} and~\eqref{eqn:projected_exact_bosonic_husimi_function} for reduced density operators. 
Notice that the top panels in Figs.~\ref{fig:renyi_occupation_evolved_coherent_states}(a) and~\ref{fig:renyi_occupation_evolved_coherent_states}(b) present the projections over all energy shells, and the localization measure $\mathfrak{L}_{2}(\epsilon,\hat{\rho})$ is defined over a single shell. The state becomes completely delocalized at long times in the atomic plane $Q$-$P$, consistent with the results for the atomic R\'enyi occupations $\mathfrak{L}_{2}(\mathcal{A},\hat{\rho}_{\mathbf{x}})$ and $\mathfrak{L}_{2}(\mathcal{A},\overline{\rho}_{\mathbf{x}})$, while in the bosonic plane $q$-$p$ the state spreads within a confined region of phase space. This feature can be detected by the energy-shell R\'enyi occupation, $\mathfrak{L}_{2}(\epsilon,\hat{\rho}_{\mathbf{x}})$, because it contains information about both atomic and bosonic spaces. This is why $\mathfrak{L}_{2}(\epsilon,\hat{\rho}_{\mathbf{x}})$ saturates to $\sim 1/2$. The atomic R\'enyi occupation $\mathfrak{L}_{2}(\mathcal{A},\hat{\rho}_{\mathbf{x}})$, on the other hand, is unable to detect information from the bosonic plane, because of the integration over the bosonic variables $(q,p)$, which explains why $\mathfrak{L}_{2}(\mathcal{A},\hat{\rho}_{\mathbf{x}})$ saturates to unity and $\mathfrak{L}_{2}(\mathcal{A},\overline{\rho}_{\mathbf{x}})$ behaves in the same way.

The analysis of the time evolution of Husimi functions for coherent states has also been done in Ref.~\cite{Bakemeier2013} for different system sizes $j$ and values of the coupling parameter $\gamma$. As the system size increases, the Husimi distribution becomes thinner, better coinciding with the classical phase space structures, such as periodic orbits.

Valuable information can be obtained by mixing two coherent states, whose distance $D$ between their centers is changed over the planes $(Q,P)$ or $(q,p)$, keeping the energy constant:
$\hat{\rho}_{\text{M}}(D) = \left(\hat{\rho}_{\mathbf{x}_{0}}+\hat{\rho}_{\mathbf{x}}\right)/2$. $D$ is the phase-space separation given by Eq.~\eqref{eqn:phase_space_separation}. One state $\hat{\rho}_{\mathbf{x}_{0}}$ is fixed in phase space, while the second one $\hat{\rho}_{\mathbf{x}}$ is separated from it gradually in the atomic $Q$-$P$ or bosonic $q$-$p$ plane. Top panels in Figs.~\ref{fig:renyi_occupation_atomic_bosonic_planes}(a) and~\ref{fig:renyi_occupation_atomic_bosonic_planes}(b) show the projection of the Husimi function for each mixed state where the phase-space separation can be gradually observed in each of the atomic and bosonic planes.

The bottom panel of Fig.~\ref{fig:renyi_occupation_atomic_bosonic_planes}(a) shows $\mathfrak{L}_{2}(\mathcal{A},\hat{\rho})$ and $\mathfrak{L}_{2}(\epsilon,\hat{\rho})$ for a mixed coherent state $\hat{\rho}_{\text{M}}$ gradually separated in the atomic plane $Q$-$P$ and located in a chaotic energy regime. The bottom panel of Fig.~\ref{fig:renyi_occupation_atomic_bosonic_planes}(b) shows the same comparison for a mixed state gradually separated in the bosonic $q$-$p$ plane. Both R\'enyi occupations $\mathfrak{L}_{2}(\mathcal{A},\hat{\rho}_{\text{M}})$ and $\mathfrak{L}_{2}(\epsilon,\hat{\rho}_{\text{M}})$ are sensitive to changes in the atomic plane $Q$-$P$ and increase their initial value and saturate in Fig.~\ref{fig:renyi_occupation_atomic_bosonic_planes}(a). Nevertheless, the situation is different in Fig.~\ref{fig:renyi_occupation_atomic_bosonic_planes}(b), where only the energy-shell R\'enyi occupation $\mathfrak{L}_{2}(\epsilon,\hat{\rho}_{\text{M}})$ detects changes in the bosonic plane $q$-$p$ and also increases its initial value and saturates, while the atomic R\'enyi occupation $\mathfrak{L}_{2}(\mathcal{A},\hat{\rho}_{\text{M}})$ remains unaltered. The atomic R\'enyi occupation $\mathfrak{L}_{2}(\mathcal{A},\hat{\rho}_{\text{M}})$ is unable to detect changes in the bosonic plane $q$-$p$ because the integration of the bosonic variables $(q,p)$ erases information in this plane.

\begin{figure}[t!]
    \centering
    \begin{tabular}{cc}
        (a) & (b) \\
        \includegraphics[width=0.4\textwidth]{figures/PRE2021_fig4a.pdf} &
        \includegraphics[width=0.4\textwidth]{figures/PRE2021_fig4b.pdf}
    \end{tabular}
    \caption{(Top panels) Husimi function for a mixed coherent state $\hat{\rho}_{\text{M}}$ at different phase-space separations $D$ in (a) the atomic plane $Q$-$P$ and (b) the bosonic plane $q$-$p$. The projections are shown on the atomic plane $Q$-$P$ [Eq.~\eqref{eqn:projected_exact_atomic_husimi_function}] (top row of panels) and the bosonic plane $q$-$p$ [Eq.~\eqref{eqn:projected_exact_bosonic_husimi_function}] (bottom row of panels). In the top row of panels (a) and (b), the black solid border represents the available phase space at the chaotic energy shell $\epsilon=1$. Lighter tones inside each atomic and bosonic projection represent higher concentration values of the Husimi function, while dark blue (atomic) and dark red (bosonic) correspond to zero.
    (Bottom panel) R\'enyi occupations $\mathfrak{L}_{2}(\mathcal{A},\hat{\rho})$ [Eq.~\eqref{eqn:atomic_subspace_renyi_occupation_alpha} with $\alpha=2$] (dark blue solid line) and $\mathfrak{L}_{2}(\epsilon,\hat{\rho})$ [Eq.~\eqref{eqn:classical_energy_shell_renyi_occupation_alpha} with $\alpha=2$] (dark green solid line) scaled to their initial value at $D=0$ for the state $\hat{\rho}_{\text{M}}$. In panels (a) and (b), the vertical black dashed lines represent the phase-space separations where the projections are shown in the top panels.
    Hamiltonian parameters: $\omega=\omega_{0}=1$, $\gamma=2\gamma_{\text{c}}$, and $j=30$. Figures taken from Ref.~\cite{Villasenor2021}.
    }
    \label{fig:renyi_occupation_atomic_bosonic_planes}
\end{figure}

A similar study is shown in Ref.~\cite{Villasenor2021}, where the state is defined as a mixture of initial coherent states gradually saturating the atomic plane.
Since the R\'enyi occupation $\mathfrak{L}_{2}(\mathcal{A},\hat{\rho}_{\text{M}})$ is only sensitive to changes in the atomic subspace, it grows until it reaches unity as the number of added coherent states increases, indicating that the mixed state is completely delocalized over the available atomic phase space. On the other hand, the energy-shell R\'enyi occupation $\mathfrak{L}_{2}(\epsilon,\hat{\rho}_{\text{M}})$ only covers a fraction of the available phase space of the energy shell as discussed above.

\section{Quantum scarring}
\label{sec:QuantumScarring}
Berry’s conjecture~\cite{Berry1977JPA} asserts that the eigenstates of a quantum system that is classically chaotic are 
superpositions of random plane waves. Eigenstates of real physical systems do not satisfy this conjecture, but can get close to it~\cite{Zelevinsky1996PR}, which guarantees the validity of the ETH~\cite{Santos2010PRE}. Nevertheless, quantum chaotic systems may also present scarred states~\cite{Heller1984,Heller1986PROC,Heller1991COLL,Shapiro1984,McDonald1988}. Their origin lies in the existence of quantum interferences and unstable periodic orbits in the classical phase space. The phase-space structure of chaotic energy eigenstates is not only determined by energy, but also by individual unstable periodic orbits that densely populate the phase space. The unstable periodic orbits can leave imprints in some eigenstates, corresponding to regions of high probability. These concentrations of large quantum probability density along the paths of unstable classical periodic orbits were first noticed for the Bunimovich stadium billiard in Ref.~\cite{McDonald1983Thesis} and were later denominated
quantum scars~\cite{Heller1984,Heller1986PROC,Heller1991COLL}. 

Studies of quantum scarring were initially focused on one-body systems~\cite{Heller1987,Bogomolny1988,Wintgen1989,Kus1991,DAriano1992,Bohigas1993,Agam1993,Agam1994,Muller1994,Kaplan1998,Kaplan1999,Wisniacki2006,Porter2017}, although early studies for the Dicke model were also performed~\cite{DeAguiar1991,DeAguiar1992,Furuya1992}. Recent studies on scarring have considered two-dimensional harmonic oscillators~\cite{Keski2019PRL,Keski2019JPCM}, as well as time-dependent systems~\cite{Pai2019}, where the phenomenon is dubbed dynamical scarring. Furthermore, long-lived oscillations in many-body quantum systems taken out of equilibrium have recently been referred to as many-body quantum scars~\cite{Turner2018NP}.

Below, we review the studies of quantum scars in the Dicke model. We start by describing, in the classical limit, the families of stable and unstable periodic orbits that are known, and then explain how to quantify scars by means of the Husimi function. Next, we discuss the connection between quantum scarring and phase-space localization,  present the phenomenon of dynamical scarring, and provide a possible definition for quantum ergodicity.

\subsection{Families of periodic orbits}
\label{subsec:families_orbits}

A periodic orbit of period $T$ in the classical Dicke model is a set of points $\{\mathbf{x}(t)|t\in[0,T]\}$ in the four-dimensional phase space of the  model, $\mathbf{x}=(q,p;Q,P)$, that satisfies the condition $\mathbf{x}(0)=\mathbf{x}(T)$. In chaotic systems, periodic orbits are isolated within a single energy shell, with neighboring trajectories being non-periodic. However, when directions perpendicular to the energy shell are considered, the periodic orbits are no longer isolated. The set of all neighboring periodic orbits at all energies forms a family of periodic orbits, which can be pictured as a two-dimensional manifold spanning multiple energies in the phase space. The intersection of this manifold with an energy shell gives a single periodic orbit in the family. A numerical method to find these families consists in slightly perturbing the initial condition of a known periodic orbit or of a stationary point to another energy and then using the monodromy method~\cite{Weinstein1973,DeAguiar1988,Baranger1988,Gaspard1998,DeAguiar1992,Simonovic1999} to converge the initial condition to a new orbit. This iterative procedure was first applied to the Dicke model in Ref.~\cite{DeAguiar1992} and later refined in Ref.~\cite{Pilatowsky2021NJP}. 

Starting from the stationary point $q=p=Q=P=0$,  two families of periodic orbits, labeled $\textit{DIAGM}$ and $\textit{DIAGP}$, were found in a generalized anisotropic Dicke model~\cite{DeAguiar1992}. Later, two families of periodic orbits  emanating from the stationary point $\mathbf{x}_{\text{gs}}$ associated with the ground-state energy in the superradiant phase ($\gamma>\gamma_{\text{c}}$) were found in Ref.~\cite{Pilatowsky2021NJP}. These families were labeled $\mathcal{A}$ and $\mathcal{B}$ and called ``fundamental'' families of periodic orbits, because each originates from one of the two fundamental modes of the ground state. Due to the parity symmetry of the Dicke Model, one can obtain a new pair of families, $\widetilde{\mathcal{A}}$ and $\widetilde{\mathcal{B}}$, by mirroring the positions $(q,Q) \mapsto (-q,-Q)$ in $\mathcal{A}$ and $\mathcal{B}$. The same procedure does not work for the families $\textit{DIAGM}$ and $\textit{DIAGP}$, because they are invariant under the parity transformation. 
Other periodic orbits of the Dicke model were found in Ref.~\cite{Pilatowsky2022Q} (see Sec.~\ref{subsubsec:scarring_localization} for more details).

Whether a periodic orbit is stable or unstable is determined by the maximal Lyapunov exponent of any initial point in the orbit.  For a periodic orbit, the expression for the Lyapunov exponent in Eq.~\eqref{eqn:lyapunov_exponent} can be simplified to
\begin{equation}
    \lambda=\frac{1}{T} \max_i \log \left |m_i\right|,
\end{equation}
where $m_i$ are the eigenvalues of the monodromy matrix obtained by integrating the fundamental matrix over one period $T$~\cite{Pilatowsky2021NJP}. In the Dicke model, families of stable periodic orbits become unstable as the energy increases and the system becomes chaotic.

\begin{figure}[t!]
    \centering
    \begin{tabular}{cc}
        \includegraphics[height=0.7\textwidth]{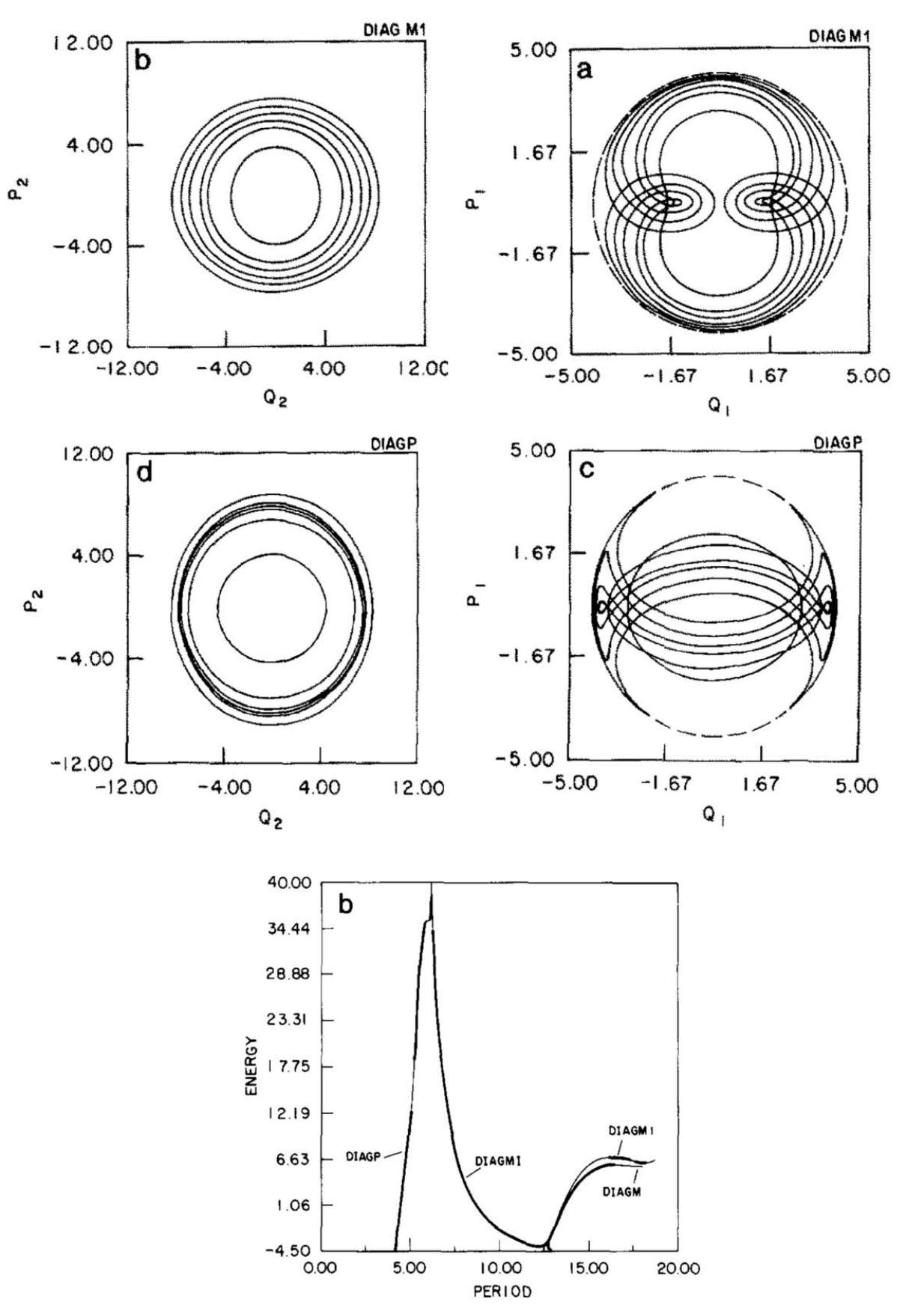} &
        \includegraphics[height=0.7\textwidth]{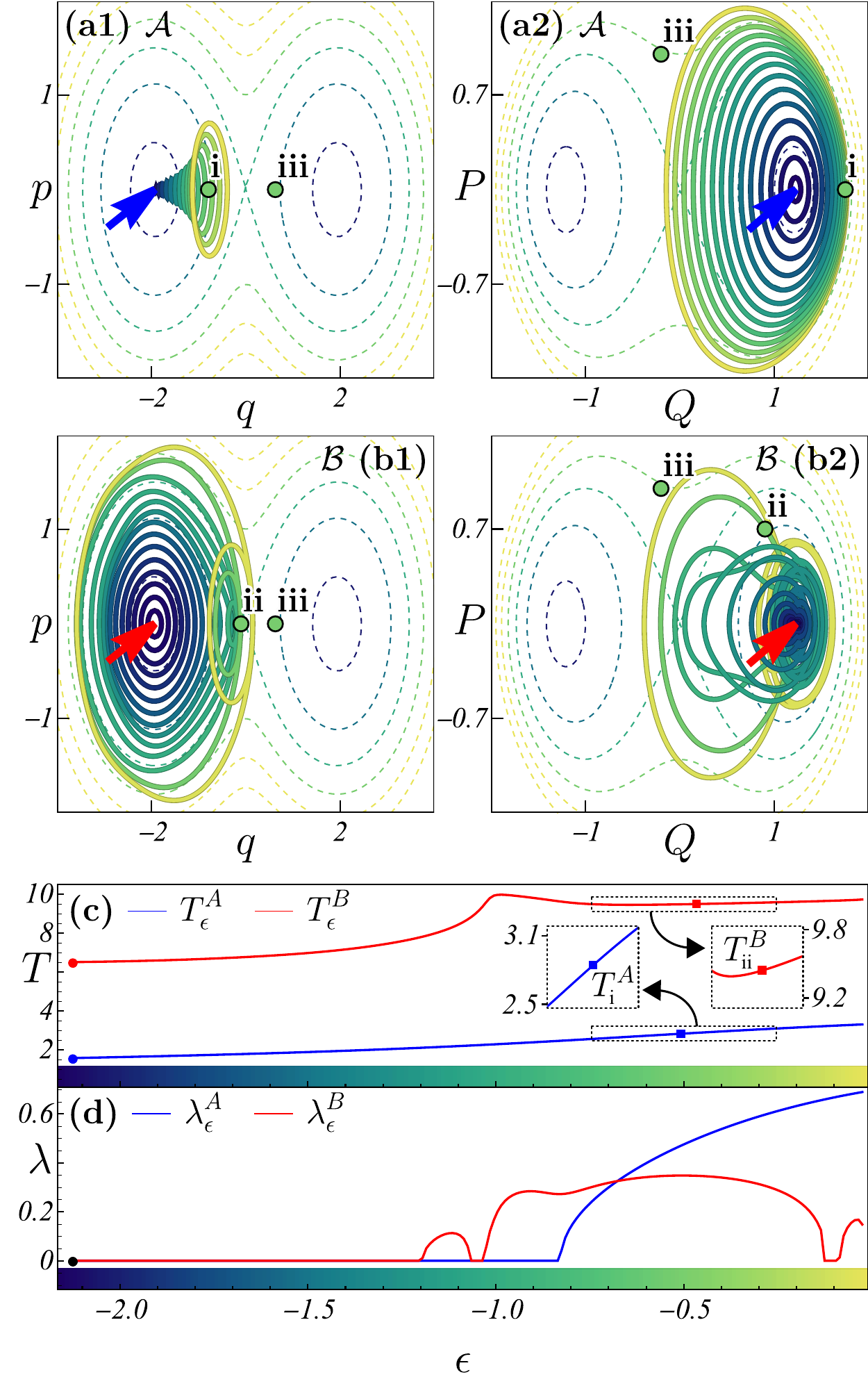}
    \end{tabular}
        \caption{(Left panels) (a-d) Periodic orbits (solid lines) from families $DIAGM$ and $DIAGP$ of the classical anisotropic Dicke Hamiltonian [Eq.~\eqref{eqn:generalized_classical_dicke_hamiltonian}]. The projections are shown on (b and d) the bosonic plane $q$-$p$ and (a and c) the atomic plane $Q$-$P$. The correspondence between the labels in panels (a-d) and the main text is: $Q_{1}=\sqrt{j}P$, $P_{1}=\sqrt{j}Q$, $Q_{2}=\sqrt{j}p$, and $P_{2}=\sqrt{j}q$.
        (b at bottom) Energy versus period for the periodic orbits from families $DIAGM$ and $DIAGP$. Thick (thin) lines represent stable (unstable) periodic orbits. Hamiltonian parameters: $\omega=\omega_{0}=1$, $\gamma_{-}=0.5$, $\gamma_{+}=0.2$, and $j=9/2$. Figures taken from Ref.~\cite{DeAguiar1992}.
        (Right panels) [(a1)-(b2)] Periodic orbits (solid lines) from families $\mathcal{A}$ and $\mathcal{B}$ of the classical Dicke Hamiltonian [Eq.~\eqref{eqn:classical_dicke_hamiltonian}]. The projections are shown on [(a1) and (b1)] the bosonic plane $q$-$p$ and [(a2) and (b2)] the atomic $Q$-$P$ plane. In each panel, from (a1) to (b2), the dashed lines represent the available phase space for different classical energy shells. Darker colors define low energy values. The blue [(a1) and (a2)] and red [(b1) and (b2)] arrows define the stationary point $\mathbf{x}_{\text{gs}}$ at the ground-state energy.
        (c) Period and (d) maximum Lyapunov exponent of periodic orbits from families $\mathcal{A}$ (blue solid line) and $\mathcal{B}$ (red solid line) as a function of energy. In panels (a1)-(b2), three initial conditions for the chaotic energy shell $\epsilon=-0.5$ are shown (see Ref.~\cite{Pilatowsky2021NJP} for further details): state i (near family $\mathcal{A}$), state ii (near family $\mathcal{B}$), and state iii (far from both families). Initial conditions $\mathbf{x}_{0}=(q_{+}(\epsilon),0;Q_{0},P_{0})$: $(Q_{0},P_{0})=(1.75,0)$ (state i), $(Q_{0},P_{0})=(0.9,0.7)$ (state ii), and $(Q_{0},P_{0})=(-0.2,1)$ (state iii). Hamiltonian parameters: $\omega=\omega_{0}=1$, $\gamma=2\gamma_{\text{c}}$, and $j=30$. Figure taken from Ref.~\cite{Pilatowsky2021NJP}.
        }
    \label{fig:stationary_points_families_A_B}
\end{figure}

The left top panels (a-d) of Fig.~\ref{fig:stationary_points_families_A_B} display the families $\textit{DIAGM}$ and $\textit{DIAGP}$ and the right top panels (a1)-(b2) show the families $\mathcal{A}$ and $\mathcal{B}$  projected in both the bosonic $q$-$p$ and atomic $Q$-$P$ planes. The bottom panel on the left displays the energy as a function of the period, while the bottom panels on the right show the (c) period and (d) the maximal Lyapunov exponent as a function of the energy. We see that the period of family $\mathcal{A}$ in Fig.~\ref{fig:stationary_points_families_A_B}(c) grows monotonically, while the period of family $\mathcal{B}$ displays a maximum value near the energy value $\epsilon\sim-1$, where the model exhibits an ESQPT. The Lyapunov exponent of family $\mathcal{A}$ transitions from zero to a positive value at energy $\epsilon\sim-0.8$, while $\lambda$ for family $\mathcal{B}$ moves away from zero at different points, indicating a complex transition to chaos for the periodic orbits of this family.

\subsection{Scarring measures}

Quantum scars are stationary states localized  around unstable periodic orbits in the phase space, so they can be identified using quasiprobability distributions, such as Wigner and Husimi functions. In Sec.~\ref{subsubsec:husimi_functions}, we reviewed numerical procedures to construct the Husimi function of eigenstates of the Dicke model. For each eigenstate, the procedure produces a two-dimensional plot in phase space, which can be visually compared with the periodic orbits. All the structures visible in Figs.~\ref{fig:projected_husimi_function_scarred_eigenstates} and~\ref{fig:projected_husimi_function_scarred_eigenstates2}, for example, correspond to quantum scars. 

\begin{figure}[t!]
    \centering
    \includegraphics[width=0.9\textwidth]{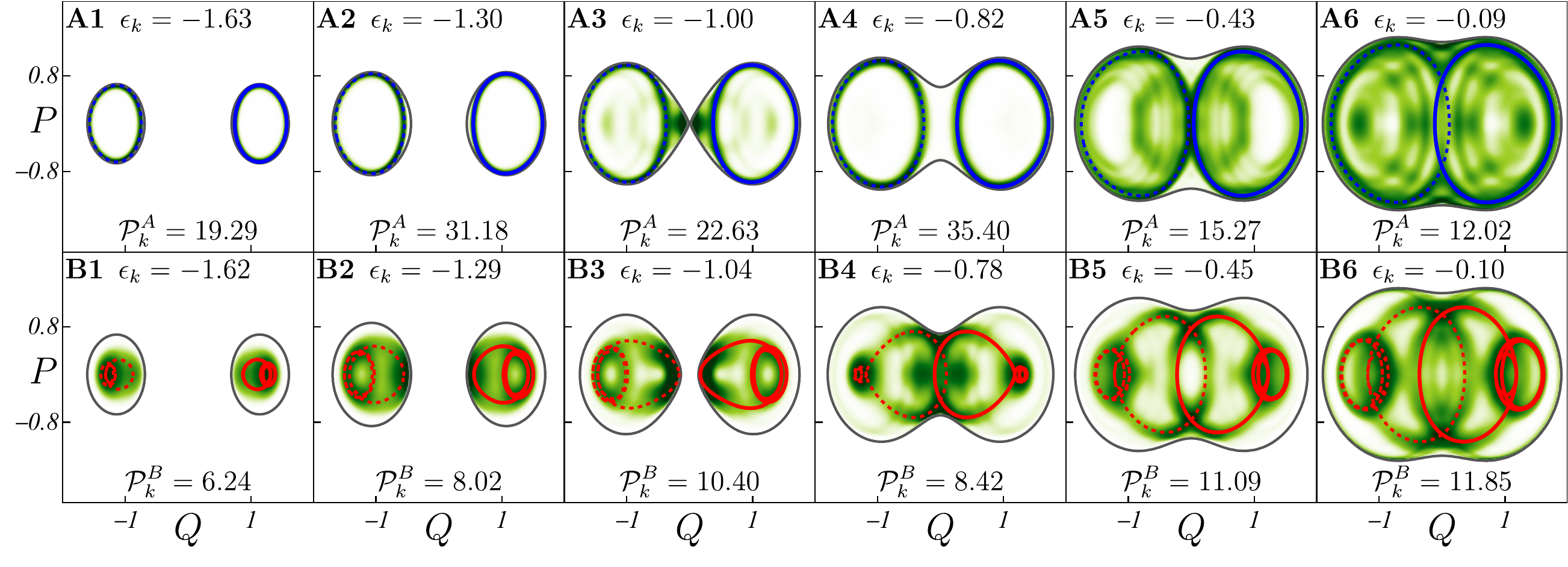} 
    \caption{Atomic projection of the Husimi function [Eq.~\eqref{eqn:projected_husimi_function}] for selected eigenstates of the Dicke Hamiltonian. Each panel, from A1 to B6, refers to a different eigenenergy $\epsilon_{k}$ covering the energy spectrum from regular to chaotic ($\epsilon>-0.8$) energy regimes. The values of the eigenenergy $\epsilon_{k}$ and the scarring measure [Eq.~\eqref{eqn:scarring_measure_family_A_B}] of families $\mathcal{A}$ and $\mathcal{B}$ are indicated in each panel. Darker tones inside the projections represent higher concentration values of the Husimi function, while white corresponds to zero. In each panel, from A1 to A6, the blue solid (dashed) line represents a periodic orbit (mirrored periodic orbit) from family $\mathcal{A}$ ($\widetilde{\mathcal{A}}$). In each panel, from B1 to B6, the red solid (dashed) line represents a periodic orbit (mirrored periodic orbit) from family $\mathcal{B}$ ($\widetilde{\mathcal{B}}$). Hamiltonian parameters: $\omega=\omega_{0}=1$, $\gamma=2\gamma_{\text{c}}$, and $j=30$. Figure taken from Ref.~\cite{Pilatowsky2021NJP}.
    }
    \label{fig:families_andHusimies_A_B}
\end{figure}

In Fig.~\ref{fig:families_andHusimies_A_B}, we show the atomic projections of Husimi functions for eigenstates of the Dicke model scarred by the orbits of families $\mathcal{A}$ and $\mathcal{B}$. 
The eigenstates sample the energy spectrum from regular (where the periodic orbits are stable) to chaotic (where the periodic orbits are unstable) energy regimes. Quantum scarring is clearly visible in all projections, although for some eigenstates the region of high probability density is not confined exclusively to the periodic orbits, which means that these states are also scarred by some other unidentified periodic orbits. Below, we discuss a systematic approach to match the scars in an eigenstate with the corresponding known unstable periodic orbits and how to measure
the state's degree of scarring~\cite{Pilatowsky2021NC}.

Identifying which eigenstate is scarred by which periodic orbit and to what degree is not straightforward. One possible approach consists of defining scarring measures. The basic idea is to construct a quantum state localized around a known periodic orbit and then compare its overlap with an energy eigenstate. 

To construct a state $\hat{\rho}_{\mathcal{O}}$ localized around the periodic orbit $\mathcal{O}$ consisting of points $\mathbf{x}(t)$ in the phase space, one can use an incoherent mixture of all coherent states centered at points in $\mathcal{O}$, 
\begin{equation}
    \hat{\rho}_\mathcal{O}\propto \int_{\mathcal{O}} d{\mathbf x} |\mathbf{x}\rangle\langle\mathbf{x}| .
    \label{Eq:RhoXX}
\end{equation}
Another approach, not used here, is to construct a pure state localized around the orbit, which  is called a scar function~\cite{Vergini2001}.
The integral over $\mathcal{O}$ in Eq.~\eqref{Eq:RhoXX} requires a parametrization of the orbit. This can be a geometrical parametrization~\cite{Kaplan1998}  or the natural temporal parametrization, which gives~\cite{Pilatowsky2021NJP,Evrard2024}
\begin{equation}
    \label{eqn:tubular_state}
    \hat{\rho}_{\mathcal{O}}=\frac{1}{T}\int_{0}^{T}dt\,|\mathbf{x}(t)\rangle\langle\mathbf{x}(t)|,
\end{equation}
where $\mathbf{x}(t)\in \mathcal{O}$. 

The overlap between an arbitrary state $\hat{\rho}$ and  $\hat{\rho}_\mathcal{O}$ is large if $\hat{\rho}$ is localized around $\mathcal{O}$ and is  equal to the average of the Husimi function $\mathcal{Q}_{\rho}(\mathbf{x}(t))$ evaluated at the points in the periodic orbit,
\begin{equation}
    \label{eqn:husimi_average_periodic_orbit}
    \text{Tr}\left(\hat{\rho}\hat{\rho}_{\mathcal{O}}\right)  = \frac{1}{T}\int_{0}^{T}dt\,\mathcal{Q}_{\rho}(\mathbf{x}(t)).
\end{equation}
To compare this average with a benchmark value, Ref.~\cite{Pilatowsky2021NJP} constructs a  delocalized state, 
\begin{equation}
    \hat{\rho}_{\epsilon} =  \frac{1}{\mathcal{V}_{\epsilon}}\int_{\mathcal{M}}d\mathbf{x}\,\delta[h_{\text{D}}(\mathbf{x})-\epsilon]|\mathbf{x}\rangle\langle\mathbf{x}|,
\end{equation}
which is composed of all coherent states within the classical energy shell at $\epsilon=h_{\text{D}}(\mathcal{O})$, which has volume ${\mathcal{V}_{\epsilon}}$ given by Eq.~\eqref{eqn:classical_energy_shell_volume}. The scarring measure is then defined by the ratio
\begin{equation}
    \label{eqn:scarring_measure}
    \mathcal{P}(\mathcal{O},\hat{\rho})=\frac{\text{Tr}(\hat{\rho}\hat{\rho}_{\mathcal{O}})}{\text{Tr}(\hat{\rho}_{\epsilon}\hat{\rho}_{\mathcal{O}})}.
\end{equation} 
The value of $\mathcal{P}(\mathcal{O},\hat{\rho})$
equals $n$ if $\hat{\rho}$ is $n$ times more localized around $\mathcal{O}$ than the delocalized state $\rho_{\mathcal{\epsilon}}$. In Ref.~\cite{Evrard2024}, a similar measure is defined for a spinor-condensate system, but instead of the delocalized state $\hat{\rho}_\epsilon$, a pure superposition of coherent states centered along a chaotic orbit is taken.

In Ref.~\cite{Pilatowsky2021NJP}, the measure $\mathcal{P}$ was calculated for the fundamental families $\mathcal{A}$ and $\mathcal{B}$ for the energy eigenstates of the Dicke model, $\hat{\rho}=\hat{\rho}_k=|E_k\rangle\langle E_k|$. Since these states are invariant under parity inversion, the symmetrized measures 
\begin{equation}
    \label{eqn:scarring_measure_family_A_B}
    \mathcal{P}^{\text{A},\text{B}}_k = \mathcal{P}(\mathcal{O}_{\epsilon_k}^{\text{A},\text{B}},\hat{\rho}_k)+\mathcal{P}(\widetilde{\mathcal{O}}_{\epsilon_k}^{\text{A},\text{B}},\hat{\rho}_k),
\end{equation}
were introduced, where $\mathcal{O}_{\epsilon_k}^{\text{A},\text{B}} (\widetilde{\mathcal{O}}_{\epsilon_k}^{\text{A},\text{B}})$ is the orbit in families $\mathcal{A}$ and $\mathcal{B}$ ($\widetilde{\mathcal{A}}$ and $\widetilde{\mathcal{B}}$) at energy $\epsilon_k=E_k/j$. The value of this quantity is indicated at the bottom of each panel of Fig.~\ref{fig:families_andHusimies_A_B}.

\subsection{Scarring and phase-space localization}

The notions of quantum scarring and phase-space localization are intrinsically related. A state that is strongly scarred on a periodic orbit is phase-space localized. This connection can be made quantitative by using the R\'enyi occupations described in Sec.~\ref{Sec:RenyiEntropyPhaseSpace} for bounded subspaces of the Dicke model. Since eigenstates belong to single energy shells, it is natural to use the R\'enyi occupation within an energy shell, defined in Eq.~\eqref{eqn:classical_energy_shell_renyi_occupation_alpha}.

\begin{figure}[t!]
    \centering
    \includegraphics[width=0.9\textwidth]{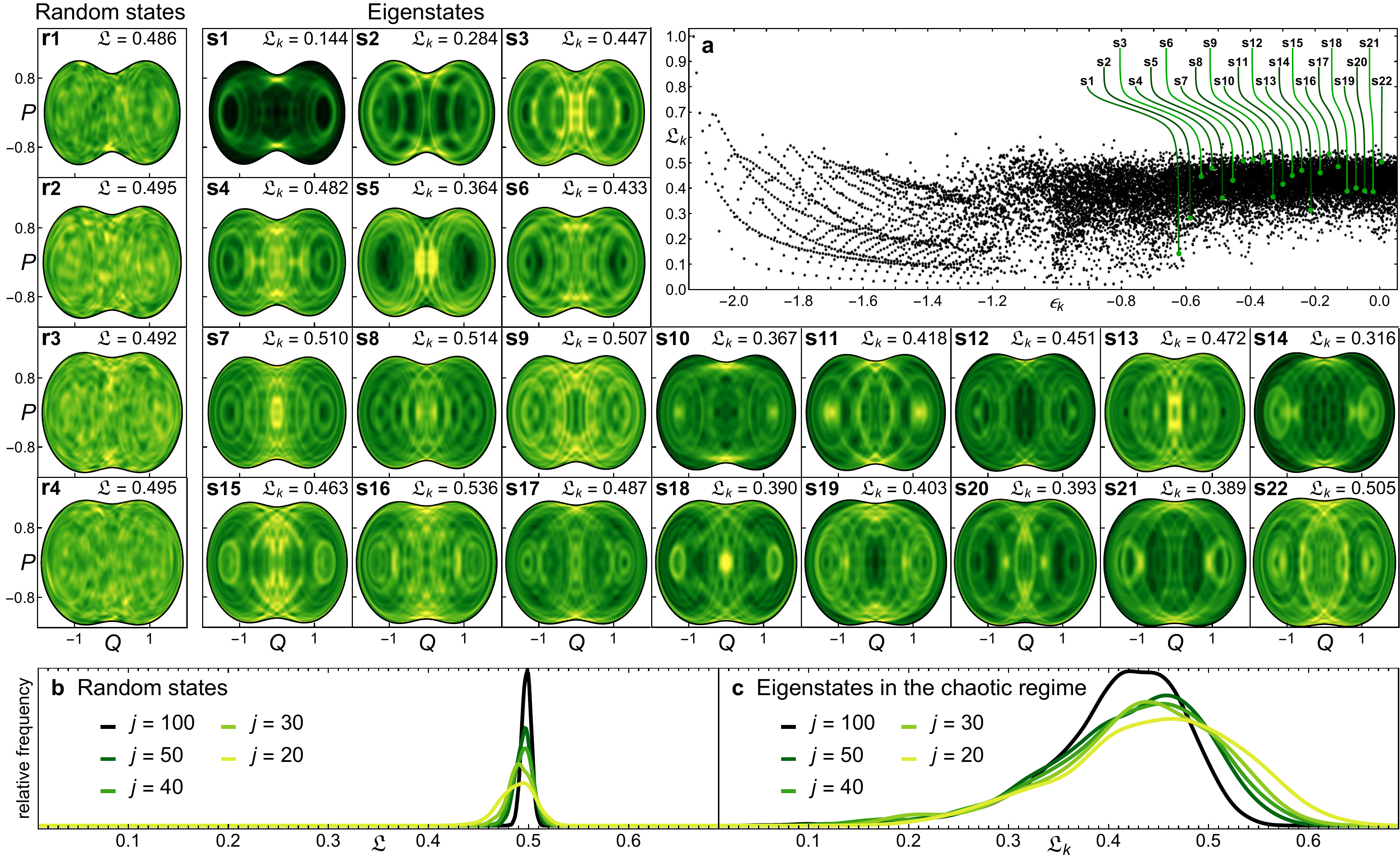}
    \caption{(a) R\'enyi occupation $\mathfrak{L}_{2}(\epsilon,\hat{\rho})$ [Eq.~\eqref{eqn:classical_energy_shell_renyi_occupation_alpha} with $\alpha=2$] (black dots) for eigenstates $\hat{\rho}_{k}$ of the Dicke Hamiltonian, contained in the energy interval $\epsilon_{k}\in[\epsilon_{\text{gs}}=-2.125,0.1]$.
    (s1-s22) Atomic projection of the Husimi function [Eq.~\eqref{eqn:projected_husimi_function}] for selected eigenstates of the Dicke Hamiltonian in the chaotic regime ($\epsilon>-0.8$). Lighter tones inside the projections represent higher concentration values of the Husimi function, while black corresponds to zero. The R\'enyi occupation [Eq.~\eqref{eqn:renyi_occupation_alpha} with $\alpha=2$] for each eigenstate s1-s22 is indicated in each panel.
    (r1-r4) Atomic projection of the Husimi function for random states centered at different classical energy shells: (r1) $\epsilon=-0.6$, (r2) $\epsilon=-0.4$, (r3) $\epsilon=-0.2$, and (r4) $\epsilon=-0.1$.
    (b) Probability distribution of R\'enyi occupations $\mathfrak{L}_{2}(\epsilon,\hat{\rho})$ for 20000 random states $\hat{\rho}_{\text{R}}$ centered at the chaotic energy shell $\epsilon=-0.5$. (c) Probability distribution of R\'enyi occupations $\mathfrak{L}_{2}(\epsilon,\hat{\rho})$ for eigenstates $\hat{\rho}_{k}$ contained in the chaotic energy interval $\epsilon_{k}\in[-0.8,0]$. In panels (b and c), each line represents a different system size ($j=20,30,40,50$, and 100).
    Hamiltonian parameters: $\omega=\omega_{0}=1$, $\gamma=2\gamma_{\text{c}}$, and $j=100$. Figure taken from Ref.~\cite{Pilatowsky2021NC}.
    }
    \label{fig:ubiquitous_quantum_scarring}
\end{figure}

Figure~\ref{fig:ubiquitous_quantum_scarring}(a) shows the energy-shell R\'enyi occupation of order $\alpha=2$, $\mathfrak{L}_{2}(\epsilon,\hat{\rho})$ [Eq.~\eqref{eqn:classical_energy_shell_renyi_occupation_alpha}], for eigenstates $\hat{\rho}_{k}$ of the Dicke model ranging from the ground-state energy to a chaotic energy regime. The eigenstates exhibit varying degrees of localization in phase space. In the low-energy regime, the points are arranged along lines that are related to quasi-integrals of motion associated with classical periodic orbits. In the high-energy regime, the points show no pattern and are densely concentrated  around a mean value of approximately $\mathfrak{L}_{2}(\epsilon_{k},\hat{\rho}_{k}) \sim 1/2$. This value, discussed in Sec.~\ref{subsubsec:renyi_occupation_eigenstate_coherent_state} for time-evolved coherent states [see Fig.~\ref{fig:renyi_occupation_evolved_coherent_states}(a)] is an upper limit on the spreading of any pure state at chaotic energy regimes~\cite{Pilatowsky2021NC}.

Figure~\ref{fig:ubiquitous_quantum_scarring} also displays atomic projections of the Husimi function [Eq.~\eqref{eqn:projected_husimi_function}] for selected eigenstates in the chaotic energy regime [panels (s1-s22)], and  for comparison, atomic projections of the Husimi function for pure random states centered at a chaotic energy shell [panels (r1-r4)]. In the chaotic energy regime of the Dicke model, all eigenstates are affected by unstable periodic orbits, which can be noticed through closed patterns seen in all Figs.~\ref{fig:ubiquitous_quantum_scarring}(s1)-\ref{fig:ubiquitous_quantum_scarring}(s22). This is not the case for random states, for which patterns are non-existent.

Figure~\ref{fig:ubiquitous_quantum_scarring}(b) shows the probability distribution of R\'enyi occupations for random states $\hat{\rho}_{\text{R}}$ and Fig.~\ref{fig:ubiquitous_quantum_scarring}(c) shows the probability distribution of R\'enyi occupations for eigenstates $\hat{\rho}_{k}$ located in the chaotic energy regime. Different  system sizes $j$ are considered. For random states, the probability distribution is centered at the limit value $\mathfrak{L}_{2}(\epsilon,\hat{\rho}_{\text{R}}) \sim 1/2$ and it becomes narrower as the system size increases. In contrast, for the eigenstates, the probability distribution is skewed and broader. In this case, there is always a set of highly localized eigenstates, no matter how large the system is. Moreover, the fraction of eigenstates most delocalized decreases and the probability distribution approaches the limit value $\mathfrak{L}_{2}(\epsilon_{k},\hat{\rho}_{k}) \sim 1/2$ from the right when the system size increases.

Scarring is ubiquitous for the eigenstates of the 
Dicke model, which are located in the chaotic regime, as seen in Fig.~\ref{fig:ubiquitous_quantum_scarring}. This observation may be counterintuitive, since one might have expected scarred states to be isolated cases. The fact is that eigenstates can be scarred by more than one unstable periodic orbit, presenting different degrees of scarring, which contrasts with random pure states, which are non-scarred~\cite{Pilatowsky2021NC}.

Random pure states provide the limit value for the R\'enyi occupation of order $\alpha$, which was analytically derived in Ref.~\cite{Pilatowsky2022Q},
\begin{gather}
    \label{eqn:limit_value_alpha}
    \mathfrak{L}_{\alpha}^{\max} \equiv \mathfrak{L}_{\alpha}(\epsilon,\hat{\rho}_{\text{R}}) = \Gamma^{1/(1-\alpha)}(\alpha+1), \\
    \label{eqn:limit_value_alpha1}
    \mathfrak{L}_{1}^{\max} = \lim_{\alpha\to1} \mathfrak{L}_{\alpha}^{\max} = e^{\gamma-1},
\end{gather}
where $\gamma \approx 0.5772$ is the Euler-Mascheroni constant. The equations above impose a limit value for the R\'enyi occupation that depends only on the order $\alpha$ and which cannot be crossed by any pure quantum state. Notice that Eq.~\eqref{eqn:limit_value_alpha} gives the exact value for $\alpha=2$, $\mathfrak{L}_{2}^{\max}=1/\Gamma(3)=1/2$. The only way to cross the limit imposed by Eqs.~\eqref{eqn:limit_value_alpha} and~\eqref{eqn:limit_value_alpha1} is 
by considering mixed states. Reference~\cite{Pilatowsky2022Q}
shows alternative localization measures incorporating Eqs.~\eqref{eqn:limit_value_alpha} and~\eqref{eqn:limit_value_alpha1} as a benchmark to characterize the degree of delocalization of generic states.

\subsubsection{Identification of unstable periodic orbits}
\label{subsubsec:scarring_localization}

The study of the $\alpha$-moments  of the Husimi function, $\mathcal{Q}_{\rho}^{\alpha}(\mathbf{x}) = \langle\mathbf{x}|\hat{\rho}|\mathbf{x}\rangle^{\alpha}$, is a useful tool to identify highly concentrated regions in phase space. By using them as initial conditions, one can then single out periodic orbits. Basically, low powers $\alpha<1$ tend to homogenize small and large contributions to the Husimi function, while high powers $\alpha>1$ tend to erase small contributions in its projection, enhancing high concentration regions, where the periodic orbits lie. The atomic projection of the $\alpha$-moments of the Husimi function can be defined as
\begin{equation}
    \label{eqn:projected_husimi_function_alpha}
    \widetilde{\mathcal{Q}}_{\epsilon,\rho}^{\alpha}(Q,P) = \int\int dq\,dp\,\delta[h_{\text{D}}(\mathbf{x})-\epsilon]\mathcal{Q}_{\rho}^{\alpha}(\mathbf{x}),
\end{equation}
where $\delta$ is the Dirac delta function and $h_{\text{D}}(\mathbf{x})$ is the classical Dicke Hamiltonian [Eq.~\eqref{eqn:classical_dicke_hamiltonian}].

\begin{figure}[t!]
    \centering
    \includegraphics[width=0.8\textwidth]{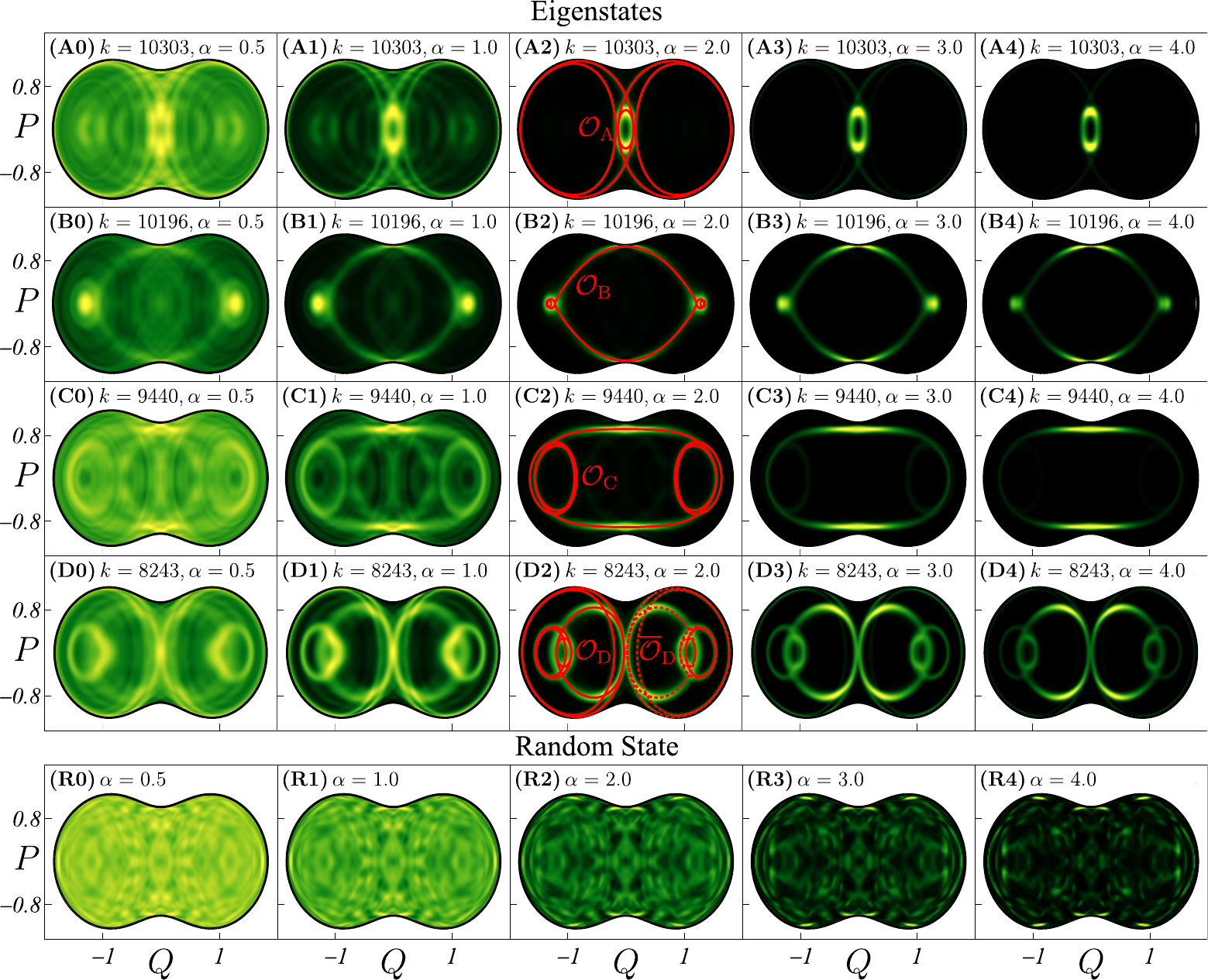}
    \caption{[(A0)-(D4)] Atomic projection of the Husimi function for selected eigenstates of the Dicke Hamiltonian in the chaotic energy regime ($\epsilon>-0.8$). Each column refers to a different $\alpha$-moment of the Husimi function [Eq.~\eqref{eqn:projected_husimi_function_alpha}]: (A0),(B0),(C0),(D0) $\alpha=0.5$, (A1),(B1),(C1),(D1) $\alpha=1$, (A2),(B2),(C2),(D2) $\alpha=2$, (A3),(B3),(C3),(D3) $\alpha=3$, and (A4),(B4),(C4),(D4) $\alpha=4$. Lighter tones inside the projections represent higher concentration values of the Husimi function, while black corresponds to zero. The center column shows different classical unstable periodic orbits (red solid lines): (A2) $\mathcal{O}_{\text{A}}$, (B2) $\mathcal{O}_{\text{B}}$, (C2) $\mathcal{O}_{\text{C}}$, and (D2) $\mathcal{O}_{\text{D}}$, respectively. In panel (D2), the mirrored image (red dashed line) of the periodic orbit $\mathcal{O}_{\text{D}}$ is also shown. The spectrum label $k$ for each eigenstate is indicated in each panel.
    [(R0)-(R4)] Atomic projection of the Husimi function for random states centered at the chaotic energy shell $\epsilon=-0.5$. Each panel refers to a different $\alpha$-moment of the Husimi function [Eq.~\eqref{eqn:projected_husimi_function_alpha}]: (R0) $\alpha=0.5$, (R1) $\alpha=1$, (R2) $\alpha=2$, (R3) $\alpha=3$, and (R4) $\alpha=4$.
    Hamiltonian parameters: $\omega=\omega_{0}=1$, $\gamma=2\gamma_{\text{c}}$, and $j=100$. Figure taken from Ref.~\cite{Pilatowsky2022Q}.
    }
    \label{fig:localization_measure_alpha_scarred_eigestates}
\end{figure}

In Figs.~\ref{fig:localization_measure_alpha_scarred_eigestates}(A0)-\ref{fig:localization_measure_alpha_scarred_eigestates}(D4),  we present the atomic projection of the Husimi function for eigenstates in the chaotic energy region and in Figs.~\ref{fig:localization_measure_alpha_scarred_eigestates}(R0)-\ref{fig:localization_measure_alpha_scarred_eigestates}(R4) for random states centered at a chaotic energy shell. The projections are computed for different $\alpha$-moments, increasing from left to right. All projections for the eigenstates show structures resembling periodic orbits, while for the random states, granular structures are seen. As expected, the projections for the eigenstates are more homogeneous for low moments ($\alpha<1$), while high moments ($\alpha>1$) enhance the largest concentration regions, where the unstable periodic orbits lie.

In Figs.~\ref{fig:localization_measure_alpha_scarred_eigestates}(A2),~\ref{fig:localization_measure_alpha_scarred_eigestates}(B2),~\ref{fig:localization_measure_alpha_scarred_eigestates}(C2), and~\ref{fig:localization_measure_alpha_scarred_eigestates}(D2), the projections of the Husimi function [Eq.~\eqref{eqn:projected_husimi_function_alpha}] are overlaid with unstable periodic orbits, which were identified by taking as initial conditions the high concentration regions of the highest moment ($\alpha=4$) of the Husimi function. These periodic orbits do not correspond to the fundamental families of periodic orbits $\mathcal{A}$ and $\mathcal{B}$ presented in Sec.~\ref{subsec:families_orbits}. They belong to other unidentified families~\cite{Pilatowsky2022Q}.

\subsubsection{Dynamics of unstable periodic orbits}

In Eq.~\eqref{eqn:tubular_state}, we constructed a quantum state $\hat{\rho}_{\mathcal{O}}$ by doing a time-average over coherent states  $|\mathbf{x}(t)\rangle\langle\mathbf{x}(t)|$,
centered at a point $\mathbf{x}(t)$ of a periodic orbit $\mathcal{O}$ and parametrized by the time. These states are known as tubular states. By comparing them with the classical dynamics of unstable periodic orbits  $\mathcal{O}_{\epsilon}$ with energy $\epsilon$ and period $T$ we gain further insight into the structure of the tubular states.

\begin{figure}[t!]
    \centering
    \begin{tabular}{c}
    \includegraphics[width=0.9\textwidth]{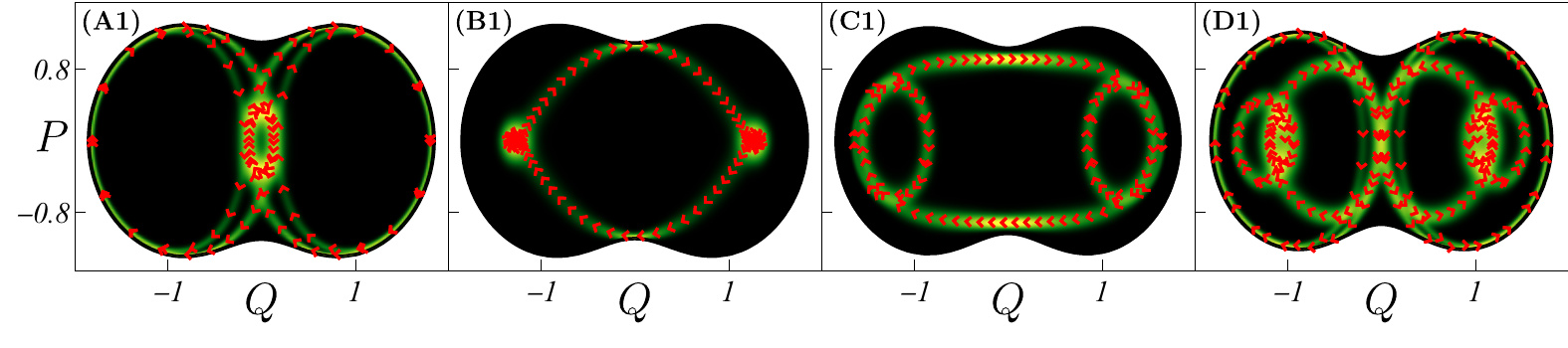} \\
    \includegraphics[width=0.9\textwidth]{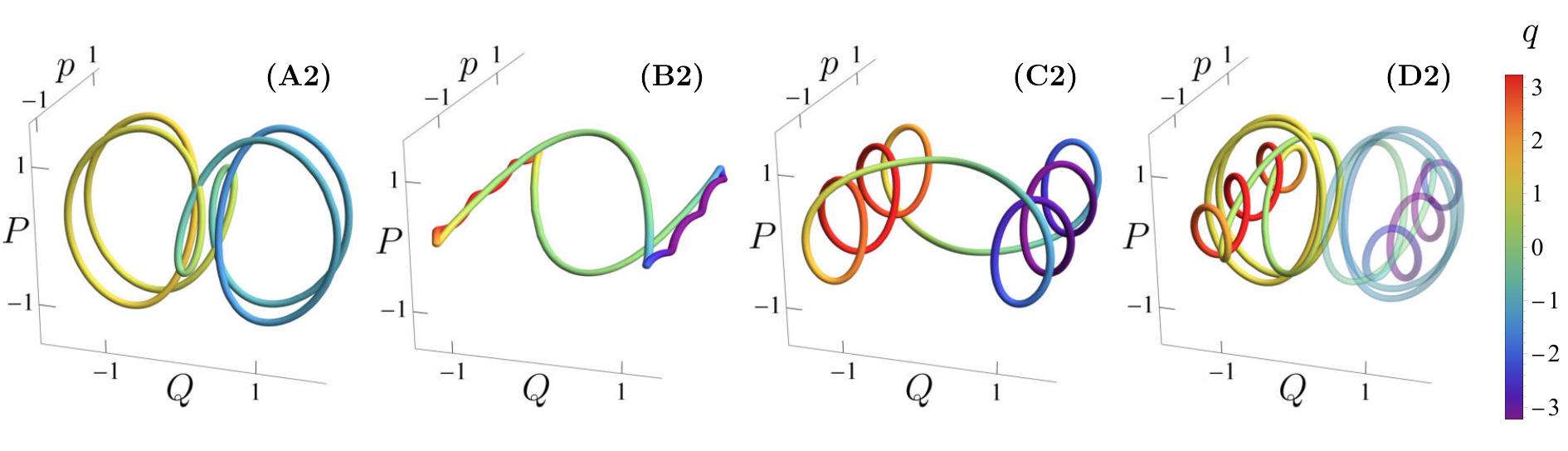}
    \end{tabular}
    \caption{[(A1)-(D1)] Atomic projection of the Husimi function [Eq.~\eqref{eqn:projected_husimi_function}] for tubular states [Eq.~\eqref{eqn:tubular_state}] centered at a chaotic energy shell $\epsilon=-0.5$. Each panel refers to a different tubular state that resembles the classical unstable periodic orbits shown in the central column of Fig.~\ref{fig:localization_measure_alpha_scarred_eigestates}: (A1) $\mathcal{O}_{\text{A}}$, (B1) $\mathcal{O}_{\text{B}}$, (C1) $\mathcal{O}_{\text{C}}$, and (D1) $\mathcal{O}_{\text{D}}$. Lighter tones inside the projections represent higher concentration values of the Husimi function, while black corresponds to zero. In each panel, from (A1) to (D1), the red arrows represent the dynamics of the unstable periodic orbit and are placed at constant time intervals.
    [(A2)-(D2)] Three-dimensional unstable periodic orbits shown in the central column of Fig.~\ref{fig:localization_measure_alpha_scarred_eigestates}: (A2) $\mathcal{O}_{\text{A}}$, (B2) $\mathcal{O}_{\text{B}}$, (C2) $\mathcal{O}_{\text{C}}$, and (D2) $\mathcal{O}_{\text{D}}$. The color scale on the right indicates the value of the $q$-coordinate.
    Hamiltonian parameters: $\omega=\omega_{0}=1$, $\gamma=2\gamma_{\text{c}}$, and $j=100$. Figure taken from Ref.~\cite{Pilatowsky2022Q}.
    }
    \label{fig:localization_measure_alpha_tubular_states}
\end{figure}

Figures~\ref{fig:localization_measure_alpha_tubular_states}(A1)-\ref{fig:localization_measure_alpha_tubular_states}(D1) show the atomic projection of the Husimi function [Eq.~\eqref{eqn:projected_husimi_function}] of tubular states built as follows. For the eigenstates of the Dicke model shown in the previous section and scarred by some unstable periodic orbits, we select the coherent states centered along these orbits and use them to build  the tubular states. These projections for the tubular states agree with the projections of the Husimi function for the scarred eigenstates in Figs.~\ref{fig:localization_measure_alpha_scarred_eigestates}(A2),~\ref{fig:localization_measure_alpha_scarred_eigestates}(B2),~\ref{fig:localization_measure_alpha_scarred_eigestates}(C2), and~\ref{fig:localization_measure_alpha_scarred_eigestates}(D2). The  comparison of the projections for the tubular states with the dynamics of the periodic orbits [red arrows in Figs.~\ref{fig:localization_measure_alpha_tubular_states}(A1)-\ref{fig:localization_measure_alpha_tubular_states}(D1)], shows that regions of slow dynamics (concentrated arrows) coincide with regions of high concentration in the Husimi function (lighter tones of yellow), while for regions of fast dynamics (spaced arrows), the Husimi function shows low concentration (darker tones of green).

The three-dimensional shapes of the classical unstable periodic orbits shown in Figs.~\ref{fig:localization_measure_alpha_scarred_eigestates}(A2)-\ref{fig:localization_measure_alpha_scarred_eigestates}(D2) are presented in Figs.~\ref{fig:localization_measure_alpha_tubular_states}(A2)-\ref{fig:localization_measure_alpha_tubular_states}(D2). These orbits have a complex structure in phase space which is helpful to understand the behavior of the classical orbits in the regions of slow (low) and fast (high) dynamics (concentration).

\subsection{Dynamical scarring}
\label{subsec:dynamical_scarring_quantum_ergodicity}

\begin{figure}[t!]
    \centering
    \includegraphics[width=0.9\textwidth]{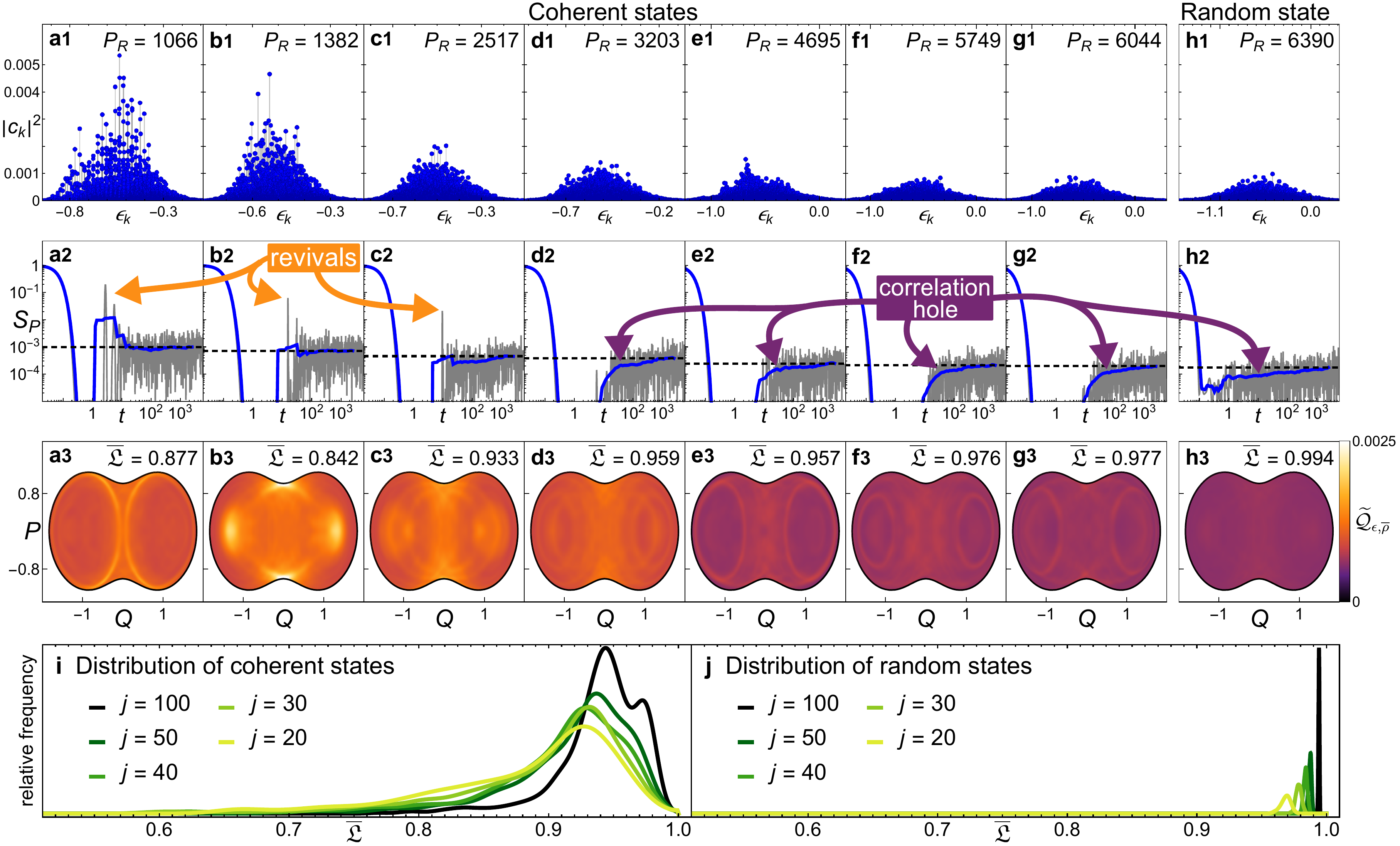}
    \caption{(a1-g1) LDOS of coherent states projected in the energy eigenbasis of the Dicke model and centered at the chaotic energy shell $\epsilon=-0.5$. 
    Each panel refers to a different initial condition $\mathbf{x}_{0}=(q_{+}(\epsilon),0;Q_{0},P_{0})$: (a1) $(Q_{0},P_{0})=(1.75,0)$, (b1) $(Q_{0},P_{0})=(0.5,-0.75)$, (c1) $(Q_{0},P_{0})=(0.75,0.5)$, (d1) $(Q_{0},P_{0})=(1.25,-0.25)$, (e1) $(Q_{0},P_{0})=(-1.25,-1)$, (f1) $(Q_{0},P_{0})=(-1.25,0.75)$, and (g1) $(Q_{0},P_{0})=(-0.75,0.5)$.
    (h1) LDOS for a random state. 
    The participation ratio [Eq.~\eqref{eqn:participation_ratio}] for each state a1-h1 is indicated in each panel. 
    (a2-h2) Survival probability [Eq.~\eqref{eqn:survival_probability}] (gray solid line), its time average (blue solid line), and its asymptotic value (horizontal black dashed line) for states a1-h1.
    (a3-h3) Atomic projection of the Husimi function [Eq.~\eqref{eqn:projected_husimi_function}] for time-averaged ensembles [Eq.~\eqref{eqn:time_averaged_ensemble}] of states a1-h1. Lighter tones inside the projections represent higher concentration values of the Husimi function, while dark purple corresponds to zero. The R\'enyi occupation $\mathfrak{L}_{2}(\epsilon,\hat{\rho})$ [Eq.~\eqref{eqn:classical_energy_shell_renyi_occupation_alpha} with $\alpha=2$] for each ensemble is indicated in each panel. (i) Probability distribution of R\'enyi occupations $\mathfrak{L}_{2}(\epsilon,\hat{\rho})$ for a time-averaged ensemble of 1551 coherent states $\hat{\rho}_{\mathbf{x}}$ distributed equally in the atomic plane $Q$-$P$. (j) Probability distribution of R\'enyi occupations $\mathfrak{L}_{2}(\epsilon,\hat{\rho})$ for a time-averaged ensemble of 500 random states $\hat{\rho}_{\text{R}}$ centered at the chaotic energy shell $\epsilon=-0.5$. In panels (i and j), each line represents a different system size ($j=20,30,40,50$, and 100).
    Hamiltonian parameters: $\omega=\omega_{0}=1$, $\gamma=2\gamma_{\text{c}}$, and $j=100$. Figure taken from Ref.~\cite{Pilatowsky2021NC}.
    }
    \label{fig:dynamical_quantum_scarring}
\end{figure}

In analogy with the scars of a single eigenstate, ``dynamical scars'' correspond to structures that resemble periodic orbits and are identified in the projections of the Husimi functions of states obtained from time-averaged ensembles~\cite{Tomiya2019,Pilatowsky2021NC}. Below, we analyze such projections for the infinite-time average of coherent states $\overline{\rho}_{\mathbf{x}} = \lim_{t\to\infty}t^{-1}\int_{0}^{t}dt'\,\hat{\rho}_{\mathbf{x}}(t')$ [Figs.~\ref{fig:dynamical_quantum_scarring}(a3)-\ref{fig:dynamical_quantum_scarring}(g3)] and for a random state $\overline{\rho}_{\text{R}}$ [Fig.~\ref{fig:dynamical_quantum_scarring}(h3)], all computed according to Eq.~\eqref{eqn:time_averaged_ensemble}  and projected onto the eigenstates of the  Dicke model in the chaotic region.  The results are compared with  the survival probability of these states [Figs.~\ref{fig:dynamical_quantum_scarring}(a2)-\ref{fig:dynamical_quantum_scarring}(h2)] and their LDOS [Figs.~\ref{fig:dynamical_quantum_scarring}(a1)-\ref{fig:dynamical_quantum_scarring}(h1)].

The LDOS of coherent states with a small participation ratio [e.g. Fig.~\ref{fig:dynamical_quantum_scarring}(a1)] shows  energy components bunched around specific energy levels (comb structure), while the LDOS for states with large participation ratios are more similar to the LDOS of the random state in Fig.~\ref{fig:dynamical_quantum_scarring}(h1), where the majority of the energy components participate. The survival probability and its time average exhibit revivals when the coherent state has low participation ratio [e.g. Fig.~\ref{fig:dynamical_quantum_scarring}(a2)]. These revivals are caused by unstable periodic orbits that strongly scar some of the eigenstates that participate in the evolution of the initial state. In contrast, for initial coherent states with a high participation ratio, just as for the random state [Fig.~\ref{fig:dynamical_quantum_scarring}(h2)], the survival probability shows a correlation hole before saturating [Fig.~\ref{fig:dynamical_quantum_scarring}(d2)-\ref{fig:dynamical_quantum_scarring}(g2)] and no revivals.

Similarly to the results for the LDOS and the survival probability, the projections of the Husimi functions for the infinite-time average of the coherent states in Figs.~\ref{fig:dynamical_quantum_scarring}(a3)-\ref{fig:dynamical_quantum_scarring}(g3) show traces of dynamical scarring in the form of structures that resemble periodic orbits. These structures are nonexistent for the random state in Fig.~\ref{fig:dynamical_quantum_scarring}(h3).

\subsection{Quantum ergodicity}

The concept of quantum ergodicity can be extended from classical ergodicity, where the trajectories of an ergodic system cover its phase space homogeneously. In this way, quantum ergodicity can be defined through the infinite-time average~\cite{schnack1996,sunada1997,Pilatowsky2021NC}
\begin{equation}
    \label{eqn:time_averaged_ensemble}
    \overline{\rho} = \lim_{t\to+\infty}\frac{1}{t}\int_{0}^{t}dt'\hat{\rho}(t'),
\end{equation}
where $\hat{\rho}(t) = e^{-i\hat{H}t}\hat{\rho}e^{i\hat{H}t}$ is the time-evolved density matrix. Thus, we call an arbitrary quantum state $\hat{\rho}$ {\it ergodic} if the R\'enyi occupation [Eq.~\eqref{eqn:classical_energy_shell_renyi_occupation_alpha}] of its time-averaged ensemble is unity,
\begin{gather}
    \overline{\mathfrak{L}}_{\alpha}(\epsilon,\hat{\rho}) \equiv \mathfrak{L}_{\alpha}(\epsilon,\overline{\rho}) = 1, \\
    \overline{\mathfrak{L}}_{1}(\epsilon,\hat{\rho}) = \lim_{\alpha \to 1} \overline{\mathfrak{L}}_{\alpha}(\epsilon,\hat{\rho}) = 1,
\end{gather}
which implies that the whole classical energy shell is homogeneously visited by the ensemble.

In Fig.~\ref{fig:dynamical_quantum_scarring}(i), we present the probability distribution of the R\'enyi occupations for the time-averaged ensembles of coherent states $\overline{\rho}_{\mathbf{x}}$ located in a chaotic energy regime. Figure~\ref{fig:dynamical_quantum_scarring}(j) shows the same for a time-averaged ensemble of random states $\overline{\rho}_{\text{R}}$. Different system sizes $j$ are considered. For random states, the distribution becomes narrower and better centered at the limit value $\mathfrak{L}_{2}(\epsilon,\overline{\rho}_{\text{R}})\sim1$ as the system size increases, which implies that the infinite-time average of random states is ergodic. On the one hand,  the distribution for coherent states becomes narrower as the system size increases, but the fraction of states with R\'enyi occupation near unity remains small.

We conclude by summarizing some of the main findings described in this section:

\begin{itemize}
    \item Any pure state (scarred or unscarred) in the chaotic regime of the Dicke model, including random states, is a non-ergodic state presenting some level of localization in phase space. 
    \item This persisting minimum degree of localization is caused by quantum interferences of the wave function and not by quantum scarring. Therefore, quantum ergodicity is an ensemble property only achievable for time averaged ensembles.    
    \item The three concepts, scarring, localization, and lack of ergodicity are related to each other, but they do not mean exactly the same thing. Scarring is a peculiar phenomenon caused by classical structures that restrict the spreading of the wave function, while localization and lack of ergodicity are related to quantum interferences of the wave function, which restrict its degree of spreading. By eliminating these interferences using averages, quantum ergodicity can be attained.
\end{itemize}

\section{Conclusions and perspectives}
\label{sec:Conclusions}
The results reviewed in this article demonstrate the importance and flexibility of the Dicke model. This light-matter interaction model provides the means for investigating a variety of interesting phenomena, from ground-state and excited-state quantum phase transitions to complex quantum behavior associated with entanglement growth, chaos, and ergodicity, as well as constrained dynamics arising from localization and quantum scarring. Due to the collective nature of its interactions, the model has only two degrees of freedom, thereby bridging between the simplicity of one-body systems and the complexity of many-body quantum physics. At the same time, its few degrees of freedom  enable a detailed analysis of the quantum-classical correspondence, offering valuable insights into how classical phase-space structures manifest themselves in spectral properties, eigenstate structures, and quantum dynamics.

The Dicke model has become a paradigmatic platform for exploring collective light-matter interactions across a wide range of experimental systems, from ultracold atoms and trapped ions to superconducting circuits. Its versatility has motivated numerous extensions, including anisotropic, driven, multimode, and dissipative variants, which continue to reveal new forms of critical behavior and dynamical phenomena. In particular, the open Dicke model offers a route for investigating the interplay between dissipation, collective effects, and quantum chaos, a topic that remains far less understood than its counterpart in isolated systems.

\section*{Acknowledgments}
\label{sec:Acknowledgments}
D.V. acknowledges the Instituto de Ciencias Nucleares (ICN) - UNAM, where this work was conceived, and the Instituto de Investigaciones en Matem\'aticas Aplicadas y en Sistemas - UNAM, where this work was developed. D.V. also acknowledges CAMTP - Center for Applied Mathematics and Theoretical Physics, University of Maribor, where this work was completed in its current form. We acknowledge the support of the Computation Center - ICN - UNAM to develop the majority of the results presented in this work, in particular to Enrique Palacios, Luciano D\'iaz, and Eduardo Murrieta.

\textbf{Author contributions }
D.V. provided the first version of the manuscript. All authors have contributed equally to the conceptualization, organization, development, and writing of the final version of the manuscript. All authors have read and agreed to the final version of the manuscript.

\textbf{Funding information }
D.V. acknowledges partial financial support from the PhD fellowship program Becas Nacionales para Estudios de Posgrado - CONACYT and from the postdoctoral fellowship program DGAPA - UNAM. During the last edition of this work at CAMTP, D.V. was supported by the Slovenian Research and Innovation Agency (ARIS) under Grants No. J1-4387 and No. P1-0306.
J.C.-C. was supported by the United States National Science Foundation (NSF, CCI Grant, Award No. 2124511).
L.F.S. was supported by the United States National Science Foundation (NSF, Grant No. DMR-1936006).
J.G.H. acknowledges partial financial support from DGAPA - UNAM projects No. IN109523 and No. IN101526.

\appendix

\section{Coherent states}
\label{app:CoherentStates}
Coherent states were first introduced by E. Schr\"{o}dinger in 1926~\cite{Perelomov1986Book,Schrodinger1926,Zhang1990}, without a defined name, when he was searching for solutions that satisfy the Heisenberg uncertainty principle. In 1963, these states were studied by R. J. Glauber and E. C. G. Sudarshan~\cite{Zhang1990,Glauber1963PRL,Sudarshan1963,Glauber1963PRa,Glauber1963PRb,Klauder1985Book}, and later referred to by R. J. Glauber as coherent states. As a result, they are commonly known as Glauber coherent states or field coherent states, since they are states of the quantum harmonic oscillator whose dynamics resembles the behavior of the classical harmonic oscillator. Coherent states are then seen as a special kind of quantum state that establish connections between classical and quantum mechanics.

\subsection{Glauber coherent states}

The Glauber coherent states can be defined as normalized eigenstates of the annihilation operator
\begin{equation}
    \hat{a}|\alpha\rangle = \alpha|\alpha\rangle,
\end{equation}
where $\alpha\in\mathbb{C}$ is the coherent state parameter. The expectation values of the creation-annihilation operators under these coherent states can be easily obtained as $\langle\alpha|\hat{a}^{\dag}|\alpha\rangle = \alpha^{\ast}$ and $\langle\alpha|\hat{a}|\alpha\rangle = \alpha$.

Another way to define the Glauber coherent states is through a displacement operator $\hat{D}(\alpha) = e^{\alpha\hat{a}^{\dag}-\alpha^{\ast}\hat{a}}$, as follows
\begin{equation}
    \label{eqn:glauber_coherent_state}
    |\alpha\rangle = \hat{D}(\alpha)|0\rangle = e^{-|\alpha|^{2}/2}e^{\alpha\hat{a}^{\dag}}|0\rangle = e^{-|\alpha|^{2}/2}\sum_{n=0}^{\infty}\frac{\alpha^{n}}{\sqrt{n!}}|n\rangle,
\end{equation}
where $|0\rangle$ is the vacuum of the field in the Fock basis $|n\rangle$ and the displacement operator obeys $\hat{D}^{-1}(\alpha)=\hat{D}^{\dag}(\alpha)=\hat{D}(-\alpha)$.

The Glauber coherent states form a basis in the space of Fock states given by the closure relation
\begin{equation}
    \frac{1}{\pi} \iint d^{2}\alpha \, |\alpha\rangle\langle\alpha| = \sum_{n=0}^{\infty}|n\rangle\langle n| = \hat{\mathbb{I}},
\end{equation}
where $d^{2}\alpha=d\Re(\alpha) \, d\Im(\alpha)$ and the integration is computed over the complex plane, which is the analog of the phase space of the field. In addition, the basis spanned by these coherent states is an overcomplete basis,
\begin{equation}
    |\alpha\rangle = \frac{1}{\pi} \iint d^{2}\alpha' \, |\alpha'\rangle \, \langle\alpha'|\alpha\rangle = \frac{1}{\pi} \iint d^{2}\alpha' \, |\alpha'\rangle \, e^{-|\alpha'|^{2}/2-|\alpha|^{2}/2+(\alpha')^{\ast}\alpha}.
\end{equation}
They are normalized states, but
they are non-orthogonal, such that their overlap is a Gaussian function $\langle\alpha|\alpha'\rangle = e^{-|\alpha|^{2}/2-|\alpha'|^{2}/2+\alpha^{\ast}\alpha'}$, with probability
\begin{equation}
    \label{eqn:glauber_coherent_state_square_norm}
    |\langle\alpha|\alpha'\rangle|^{2} = e^{-|\alpha-\alpha'|^{2}}.
\end{equation}

Another feature of the Glauber coherent states is that they minimize the Heisenberg uncertainty principle and can be considered as the quantum states most classically accessible~\cite{Klimov2009Book}. Setting $\hbar=1$, the creation-annihilation operators $\hat{a}^{\dagger}$ and $\hat{a}$ can be expressed in terms of the position-momentum operators of a quantum harmonic oscillator with unitary mass as
\begin{gather}
    \hat{q} = \frac{1}{\sqrt{2}}\left(\hat{a}^{\dag}+\hat{a}\right), \\
    \hat{p} = \frac{i}{\sqrt{2}}\left(\hat{a}^{\dag}-\hat{a}\right),
\end{gather}
which satisfy the commutation relation $[\hat{q},\hat{p}]=i\hat{\mathbb{I}}$. In the last expressions we defined the scaled operators as $\sqrt{\omega}\hat{q} \to \hat{q}$ and $\hat{p}/\sqrt{\omega} \to \hat{p}$. Under the position-momentum variables, the generalized uncertainty principle, known as the Schr\"{o}dinger-Robertson uncertainty principle~\cite{Klimov2009Book}, takes the form
\begin{equation}
    \sigma_{q}^{2}\sigma_{p}^{2}-\sigma_{pq}^{4} \geq \frac{1}{4}.
\end{equation}
The last expression becomes the Heisenberg uncertainty principle and at the same time an equality $\sigma_{q}\sigma_{p} = 1/2$ for Glauber coherent states $|\alpha\rangle$, where $\sigma_{pq}=0$, showing that these states are states of minimum uncertainty.

\subsection{Bloch coherent states}

The atomic or spin coherent states are an extension of the Glauber coherent states~\cite{Zhang1990,Klauder1985Book,Radcliffe1971,Arecchi1972,Arecchi1972PROC}. They are commonly known as the Bloch coherent states and can be defined through a rotation operator $\hat{\Omega}(\tau) = e^{-i\theta\left(\sin\phi\hat{J}_{x}-\cos\phi\hat{J}_{y}\right)} = e^{\tau\hat{J}_{+}-\tau^{\ast}\hat{J}_{-}}$, as follows
\begin{equation}
    |\tau\rangle = \hat{\Omega}(\tau)|j,-j\rangle,
\end{equation}
where $\tau\in\mathbb{C}$ is the coherent state parameter, $|j,-j\rangle$ is the extreme eigenstate of the collective pseudospin operator $\hat{J}_{z}$ in the angular-momentum basis $|j,m_{z}\rangle$ (also known as Dicke basis), and the rotation operator obeys $\hat{\Omega}^{-1}(\tau)=\hat{\Omega}^{\dag}(\tau)=\hat{\Omega}(-\tau)$. The explicit parameter $\tau=(\theta/2)e^{-i\phi}$ is defined over the Bloch sphere, where the angular variables $(\phi,\theta)$ are the azimuthal and zenith angles of the spherical coordinates, measuring the $\theta$ angle from the negative $z$ axis.

The rotation operator takes the form of a displacement operator and using the Baker-Campbell-Hausdorff formula can be written as $\hat{\Omega}(z) = e^{z\hat{J}_{+}}e^{\ln(1+|z|^{2})\hat{J}_{z}}e^{-z^{\ast}\hat{J}_{-}}$ where the new parameter $z=\tan\left(\theta/2\right)e^{-i\phi}$ is related to $\tau$ through a stereographic projection of the Bloch sphere~\cite{Arecchi1972}. In this way, the Bloch coherent state can be written using the new parameter as
\begin{equation}
    \label{eqn:bloch_coherent_state}
    |z\rangle = \hat{\Omega}(z)|j,-j\rangle = \frac{1}{(1+|z|^{2})^{j}}e^{z\hat{J}_{+}}|j,-j\rangle = \frac{1}{(1+|z|^{2})^{j}}\sum_{m_{z}=-j}^{j}\sqrt{\left(\begin{array}{c}2j\\j+m_{z}\end{array}\right)}z^{j+m_{z}}|j,m_{z}\rangle,
\end{equation}
which form an overcomplete basis in the space of Dicke states using a specific normalization condition~\cite{Radcliffe1971}, given by the closure relation
\begin{equation}
    \frac{2j+1}{\pi}\iint d^{2}z\frac{|z\rangle\langle z|}{(1+|z|^{2})^{2}} = \sum_{m_{z}=-j}^{j}|j,m_{z}\rangle\langle j,m_{z}| = \hat{\mathbb{I}},
\end{equation}
where $d^{2}z = d\Re(z) \, d\Im(z)$. The Bloch coherent states are normalized states and non-orthogonal, such that their overlap $\langle z|z'\rangle = (1+z^{\ast}z')^{2j}(1+|z|^{2})^{-j}(1+|z'|^{2})^{-j}$ has probability
\begin{equation}
    \label{eqn:bloch_coherent_state_square_norm}
    |\langle z|z'\rangle|^{2} = \left[1-\frac{|z-z'|^{2}}{(1+|z|^{2})(1+|z'|^{2})}\right]^{2j}.
\end{equation}

The expectation values of the collective pseudospin operators $\hat{J}_{x,y,z}$ under Bloch coherent states $|z\rangle$ define the components of the classical angular momentum vector $\vec{j} = j(j_{x} , j_{y} , j_{z})$ whose magnitude is $|\vec{j}|=j$. The explicit expressions are given by
\begin{equation}
     \left(\begin{array}{c}\langle z|\hat{J}_{x}|z\rangle \\ \langle z|\hat{J}_{y}|z\rangle \\ \langle z|\hat{J}_{z}|z\rangle\end{array}\right) = \frac{j}{1+|z|^{2}}\left(\begin{array}{c}2\Re(z) \\ -2\Im(z) \\ |z|^{2}-1\end{array}\right) = j\left(\begin{array}{c}\cos\phi\sin\theta \\ \sin\phi\sin\theta \\ -\cos\theta\end{array}\right) = j\left(\begin{array}{c}j_{x} \\ j_{y} \\ j_{z}\end{array}\right).
\end{equation}
In the same way, the expectation values for the corresponding raising-lowering operators $\hat{J}_{\pm}=\hat{J}_{x}\pm i\hat{J}_{y}$ are given by
\begin{equation}
    \left(\begin{array}{c}\langle z|\hat{J}_{+}|z\rangle \\ \langle z|\hat{J}_{-}|z\rangle \end{array}\right) = \frac{2j}{1+|z|^{2}}\left(\begin{array}{c}z^{\ast} \\ z\end{array}\right) = j\sin\theta\left(\begin{array}{c}e^{i\phi} \\ e^{-i\phi}\end{array}\right) = j\left(\begin{array}{c}j_{+} \\ j_{-}\end{array}\right).
\end{equation} 

From the classical angular momentum vector can be extracted the canonical action-angle variables $(\phi,j_{z})$ with $j_{z}=-\cos\theta$ and $\{\phi,j_{z}\}=1$. Through a canonical transformation we can define position-momentum atomic variables,
\begin{equation}
    \label{eqn:QP_transform}
    (Q,P) = \sqrt{2(1+j_{z})}(\cos\phi,-\sin\phi),
\end{equation}
whose Poisson bracket satisfies $\{Q,P\}=1$.

\subsection{Coherent-state representation in the efficient basis}
\label{app:coherent_state_efficient_basis}

For numerical purposes in this work using the Dicke Hamiltonian, it turns out useful to express the tensor product of Glauber-Bloch coherent states $|\mathbf{x}\rangle=|\alpha\rangle\otimes|z\rangle$ in the efficient basis. In the eigenbasis of the Dicke Hamiltonian $|E_{k}\rangle$, the tensor product of Glauber-Bloch coherent states is represented as
\begin{equation}
    |\mathbf{x}\rangle = \sum_{k}C_{k}(\mathbf{x})|E_{k}\rangle,
\end{equation}
where $C_{k}(\mathbf{x})=\langle E_{k}|\mathbf{x}\rangle$, $\mathbf{x}=(q,p;Q,P)$ are the coordinates of the four-dimensional phase space $\mathcal{M}$ of the Dicke model, and the eigenstates $|E_{k}\rangle$ are represented in the efficient basis $|N;j,m_{x}\rangle$ as
\begin{equation}
    |E_{k}\rangle = \sum_{N,m_{x}}C_{N,m_{x}}^{k}|N;j,m_{x}\rangle,
\end{equation}
where the coefficients $C_{N,m_{x}}^{k}=\langle N;j,m_{x}|E_{k}\rangle$ are found numerically~\cite{Bastarrachea2016PRE}. Using the last representation, the coherent state coefficients in the efficient basis are given as
\begin{equation}
    \label{eqn:coherent_state_coefficients}
    C_{k}(\mathbf{x}) = \sum_{N,m_{x}}(C_{N,m_{x}}^{k})^{\ast}C_{N,m_{x}}(\mathbf{x}),
\end{equation}
where
\begin{equation}
    C_{N,m_{x}}(\mathbf{x}) = \langle N;j,m_{x}|\mathbf{x}\rangle = \frac{(\alpha-\alpha_{m_{x}})^{N}}{\sqrt{N!}}\langle\alpha_{m_{x}}|\alpha\rangle\langle j,m_{x}|z\rangle,
\end{equation}
with
\begin{gather}
    \langle\alpha_{m_{x}}|\alpha\rangle = e^{-|\alpha_{m_{x}}|^{2}/2-|\alpha|^{2}/2+\alpha_{m_{x}}^{\ast}\alpha}, \\
    \langle j,m_{x}|w(z)\rangle = \sqrt{\left(\begin{array}{c}2j\\j+m_{x}\end{array}\right)}\frac{w^{j+m_{x}}(z)}{\left(1+|w(z)|^{2}\right)^{j}},
\end{gather}
where $w(z)=(1+z)/(1-z)$ is the rotated atomic parameter with $z$ given by Eq.~\eqref{eqn:atomic_parameter}. Using the latter results, the Glauber-Bloch coherent states have an explicit representation in the energy eigenbasis of the Dicke Hamiltonian using the efficient basis through the coefficients given in Eq.~\eqref{eqn:coherent_state_coefficients}.

An analogous procedure can be derived using the efficient basis with well-defined parity $|N;j,m_{x};p\rangle$, which gives the following coefficients
\begin{equation}
    C_{k,p}(\mathbf{x}) = \sum_{N,m_{x}}(C_{N,m_{x}}^{k,p})^{\ast}C_{N,m_{x}}^{p}(\mathbf{x}),
\end{equation}
where
\begin{equation}
    C_{N,m_{x}}^{p}(\mathbf{x}) = \frac{C_{N,m_{x}}(\mathbf{x})+p(-1)^{N}C_{N,-m_{x}}(\mathbf{x})}{\sqrt{2(1+\delta_{m_{x},0})}}.
\end{equation}

\bibliography{main}

\end{document}